\documentclass[11pt,a4paper]{article}
\usepackage{jheppub}
\usepackage{amsmath,amsfonts,amssymb}
\usepackage[T1]{fontenc}
\usepackage[utf8]{inputenc}
\usepackage[british]{babel}
\usepackage{pdfpages}
\usepackage{lmodern}
\usepackage{tikz}
\usepackage{pgfplots}
\usepackage{float}
\usepackage{placeins}
\usepackage{slashed}
\usepackage{mathrsfs}
\usepackage{dsfont}
\usetikzlibrary{intersections}
\usetikzlibrary{decorations.markings}

\newcommand{\tr}{\mathrm{tr}}  
\renewcommand{\S}{\mathcal S}

\newcommand{\nn}{\nonumber}
\newcommand{\be}{\begin{equation}}
\newcommand{\ee}{\end{equation}}
\newcommand{\ta}{\tilde a}
\newcommand{\tb}{\tilde b}
\newcommand{\tc}{\tilde c}
\newcommand{\td}{\tilde d}
\newcommand{\te}{\tilde e} 
\newcommand{\tf}{\tilde f}
\newcommand{\tg}{\tilde g}
\newcommand{\thh}{\tilde h}
\newcommand{\ti}{\tilde \imath}
\newcommand{\tj}{\tilde \jmath}
\newcommand{\tk}{\tilde k}
\newcommand{\tl}{\tilde l}
\newcommand{\tm}{\tilde m}
\newcommand{\tn}{\tilde n}
\newcommand{\too}{\tilde o}
\newcommand{\tp}{\tilde p}
\newcommand{\tq}{\tilde q}
\newcommand{\trr}{\tilde r}
\newcommand{\ts}{\tilde s}
\newcommand{\ttt}{\tilde t}
\newcommand{\tu}{\tilde u}
\newcommand{\tv}{\tilde v}
\newcommand{\tw}{\tilde w}
\newcommand{\tx}{\tilde x}
\newcommand{\ty}{\tilde y}
\newcommand{\tz}{\tilde z}
\newcommand{\vphi}{\varphi}
\newcommand{\vep}{\varepsilon}
\newcommand{\pr}{\partial}
\newcommand{\bphi}{{\bar \varphi}}
\newcommand{\by}{{\bar y}}
\newcommand{\bt}{{\bar t}}
\newcommand{\N}{{\mathcal N}}
\newcommand{\A}{{\mathcal A}}
\renewcommand{\S}{{\mathcal S}}
\newcommand{\rmd}{{\rm d}}

\newcommand{\cD}{{\cal D}}
\newcommand{\cG}{{\cal G}}
\newcommand{\cL}{{\cal L}}
\newcommand{\R}{{\cal R}}
\newcommand{\I}{\mathds{1}}
\newcommand{\vv}{{\rm v}}
\newcommand{\ev}{{\rm e}}
\newcommand{\gv}{{\rm g}}
\newcommand{\tsig}{\tilde \sigma}
\newcommand{\tA}{\tilde A}
\newcommand{\tB}{\tilde B}
\pgfplotsset{compat=1.13}
\title{Explorations in Scalar Fermion Theories: $\beta$-functions, Supersymmetry and Fixed Points}
\author[a]{Ian Jack,}
\author[b]{Hugh Osborn}
\author[c]{and Tom Steudtner}
\affiliation[a]{Department of Mathematical Sciences, University of Liverpool, {Liverpool L69 3BX}}
\affiliation[b]{Department of Applied Mathematics and Theoretical Physics, Wilberforce Road, Cambridge CB3 0WA}
\affiliation[c]{Fakult\"{a}t f\"{u}r Physik, TU Dortmund, Otto-Hahn-Str. 4, D-44221 Dortmund, Germany}

\abstract{
Results for $\beta$-functions and anomalous dimensions in general scalar fermion theories are presented
to three loops. Various constraints on the individual coefficients for each diagram
following from supersymmetry are analysed. 
The results are used to discuss potential fixed points in the $\vep$-expansion
for scalar fermion theories, with arbitrary numbers of scalar fields, and where there are just two scalar couplings and one
Yukawa coupling. For different examples the fixed points follow a similar pattern as the numbers of fermions 
is varied. For diagrams with subdivergences there are extensive consistency constraints arising from the existence
of a perturbative $a$-function and these are analysed in detail.
Further arbitrary scheme variations which preserve the form of $\beta$ functions
and anomalous dimensions in terms of 1PI diagrams are also discussed.
The existence of linear and quadratic scheme  invariants is demonstrated and the consistency condition are
shown to be expressible in terms of these invariants.
}

\preprint{DO-TH 22/06}
\arxivnumber{2301.10903}

\begin{document}

\maketitle
\section{Introduction}

Whatever the role of supersymmetry in the phenomenological description of the world
at accessible energies there is no doubt that supersymmetric quantum field theories in
various dimensions have enhanced our understanding of quantum field theories
more generally. This is especially true non perturbatively where the duality between different
theories was first developed and the existence of conformal fixed points in three and higher
dimensions is much better understood. However there are also constraints at the perturbative level
where supersymmetric non-renormalisation theorems have implications for more general $\beta$-functions
and related quantities when they are reduced to the supersymmetric case.

In this paper we explore these, and other, constraints for general scalar fermion theories  in four space-time dimensions 
at up to three loops.
For pure scalar theories, this is hardly state of the art as general three loop results have been known for more than 30 years~\cite{Jack:1990eb}, and higher orders are available \cite{Jack:2018oec,Kazakov,Steudtner:2020tzo,Bednyakov:2021ojn}.
Nevertheless, the corresponding expressions for general scalar fermion theories, allowing for arbitrary Yukawa couplings,
have only been obtained quite recently \cite{Steudtner:2021fzs,Bednyakov:2021qxa,Davies:2021mnc}. While field anomalous dimensions and Yukawa $\beta$-functions were obtained, these depended on results already found
for a variety of special cases.  The  general $\beta$- and $\gamma$-functions are expressed in terms of contractions of generalised coupling tensors with each term corresponding to a specific allowed Feynman  diagram at each loop order.
The associated results for the quartic
scalar $\beta$-function at three loops have not previously been fully determined~\cite{Steudtner:2021fzs}.
Closing this gap would also represent a stepping stone towards complete three-loop renormalisation group equations of any renormalisable QFT, which is now feasible after recent advances in general four-loop gauge and three-loop Yukawa results \cite{Poole:2019kcm,Bednyakov:2021qxa,Davies:2021mnc}. Without gauge interactions, each term corresponds to a one particle irreducible (1PI) diagram, whose numbers increase rapidly with each loop order. With the quartic scalar coupling in standard regularisation schemes all
one vertex reducible diagrams (or \textit{snail diagrams}) can be omitted so that the necessary diagrams are one
vertex irreducible (1VI). 
The unknown coefficients may be partially fixed with direct calculations, e.g. \cite{Chetyrkin:2012rz,Bednyakov:2013eba,Chetyrkin:2013wya,Bednyakov:2013cpa,Bednyakov:2014pia,Zerf:2017zqi,Mihaila:2017ble} in our case.
However, their number can be greatly reduced and literature results cross-checked by applying constraints
arising from special cases such as supersymmetry, which is the exercise undertaken here.  
We are then able to fully determine the three loop beta function
for the quartic scalar couplings in  general scalar fermion theories.

To carry out our analysis for general four-dimensional renormalisable scalar fermion theories,
it is natural to consider a basis with $n_s$ real scalars $\phi^a$ and essentially $n_f$ pseudo-real 
Majorana fermions $\psi$  where the couplings are just 
a symmetric 4 index real tensor $\lambda^{abcd}$ and a Yukawa coupling $y^a$ which is  a symmetric 
$n_f \times n_f$ real matrix
in the non spinorial fermion indices (which are here suppressed)~\cite{Steudtner:2021fzs,Poole:2019kcm}. 
The discussion in subsequent sections then
concerns the beta functions $\beta_\lambda{\!}^{abcd}, \ \beta_y{\!}^a$ as well as associated anomalous 
dimensions $\gamma_\phi{\!}^{ab},$ and  $\gamma_\psi$. These quantities completely determine $\beta$-functions  when  superrenormalisable  couplings, corresponding to operators with dimension three or less,
are introduced, if background field methods are used and $\beta_\lambda{\!}^{abcd}$ is extended 
to $\beta_V (\phi)$  for an arbitrary quartic scalar potential $V$. Equivalently by applying  the so called dummy field technique~\cite{Martin:1993zk,Luo:2002ti,Schienbein:2018fsw}. The results here encompass those for Dirac
fermions, the corresponding reduction is described later.

For each fermion loop graph which leads  to a trace over products
of the Yukawa coupling matrices $y^a$ then  with our conventions the numerical coefficient for each such trace
for a four-dimensional four-component Majorana spinor $\Psi$ 
should  have an additional  factor 2 times the results quoted here. Such four-dimensional 
Majorana spinors reduce in three dimensions to 
two two-component real spinors which belong to inequivalent representations of the three-dimensional
Dirac algebra. 

General theories of course can be restricted by imposing symmetries. With complex fields, $n_s$ even,
 then we may take
$y^a \to (y^i, \by_i) $ with $y^i, \by_i$ not necessarily square matrices but related by hermitian conjugation.
 Imposing a $U(1)$ symmetry where both scalar and fermion fields carry a charge, so that 
all lines in any diagram are directed, the number of diagrams is significantly reduced (for the three loop 1PI
Yukawa vertex diagrams from 52 to 12)~\cite{Jack:2013sha}.

As a special case the $U(1)$ symmetric theory encompasses the Wess-Zumino theory with $\N=1$ 
supersymmetry and four supercharges \cite{Wess:1973kz}. In a superspace
formalism, for the renormalisable theory, there are complex chiral superfields $\Phi_i$,  ${\bar \Phi}{}^i$, 
 with an overall $U(1)$ symmetry{\hskip 0.5pt} and
the general couplings are given by a  symmetric 3 index tensor $Y^{ijk}$ and its conjugate ${\bar Y}_{ijk}$,
which determine the scalar quartic couplings. There are then very strong non-renormalisation theorems \cite{Salam:1974jj,Grisaru:1979wc}
which ensure that the $\beta$-functions $\beta_Y{}^{ijk}$, $\beta_{{\bar Y} \hskip 0.5pt ijk}$ 
are determined just in terms of the anomalous dimensions $\gamma_{\Phi}, \  
\gamma$. 
Moreover, dedicated literature for such supersymmetric QFT's is available to high orders~\cite{Jack:1996qq,Parkes:1985hh,Abbott:1980jk,Sen:1981hk,Avdeev:1982jx,Gracey:2021yvb}.
This yields conditions on the $\beta$-functions and anomalous dimensions
for an arbitrary scalar fermion theory{\hskip 0.5pt} but these do not significantly reduce the number of independent
terms~\cite{Steudtner:2021fzs}.

In three dimensions there are scalar-fermion theories with just two supercharges~\cite{Gates}. In a superfield formalism
the theory is described in terms of a real superfield $\Phi^a$ and for current interest there are just real cubic couplings
given by the symmetric three index tensor $Y^{abc}$. Such theories can emerge at fixed points under 
RG flow~\cite{Thomas,Grover:2013rc,Fei:2016sgs,Gies}
and may be relevant for fixed point exponents in 
some condensed matter systems. For a single scalar field and a ${\mathbb Z}_2$ symmetry
this is a supersymmetric version of the 3d Ising model. 
For several scalar fields then  extending the theory away from three dimensions it is possible
to set up an epsilon expansion determining potential fixed points and their associated 
critical exponents~\cite{Liendo:2021wpo}. 
The 3d supersymmetric Ising model has been  explored
using the bootstrap~\cite{Bashkirov:2013vya,Rong:2018okz,Atanasov:2018kqw,Rong:2019qer,Atanasov:2022bpi},
with extensions to several fields in \cite{Rong:2019qer}.
Of course extending supersymmetric theories away from their natural integer dimension, even
just in a perturbative expansion, is potentially fraught with problems. Various Ward identities necessary for
supersymmetry are no longer valid. These relate contributions with different numbers of fermion loops
and depend on Fierz identities. However these problems do not arise at low loop order, up to three loops in our case.
A discussion of the four dimensional  $\N=1$ supersymmetry algebra 
extended away from four dimensions using a form of dimensional reduction was given in \cite{Bobev}.
The minimal three dimensional supersymmetric scalar-fermion theory
would define an apparent four dimensional theory with $\N= \tfrac12$ supersymmetry\footnote{This is different 
from the $\N= \tfrac12$ supersymmetry discussed in \cite{Seiberg:2003yz,Araki:2003se} which involve 
non anti-commuting $\theta^\prime$s or ${\bar \theta}^\prime$s. The renormalisation 
of  these theories was considered in 
\cite{Grisaru:2003fd,Britto:2003kg,Romagnoni:2003xt,Lunin:2003bm,Jack:2004pq,Jack:2007fb,Jack:2008ae}}. 
These theories of course do not exist as well defined Lorentz invariant  unitary theories though there
exists the possibility of considering such a theory away from three dimensions where the full $d$-dimensional
Lorentz symmetry is broken.  Dimensional
regularisation breaks supersymmetry but for $\N=1$ theories such anomalous contributions breaking supersymmetric
Ward identities should be removable by an appropriate redefinition of the couplings, or essentially a change
of scheme. At one, two or three loops it is sufficient just to ensure that fermion traces are appropriately normalised.
We defer further discussion to the conclusion.

In section 8 we make use of the three loop results to discuss possible fixed points in the $\vep$-expansion 
for fermion scalar theories. We consider generalisations of the Gross-Neveu, Nambu Jona-Lasinio and Heisenberg
theories which have $n_s=1,2,3$ scalar fields and have $O(n_s)$ symmetry. For $n_s\ge 4$ there are theories
with reduced $H\subset O(n_s)$ symmetry which have two scalar couplings and one Yukawa coupling.
For a consistent RG flow with the reduced set of couplings it is necessary to impose completeness relations 
on the matrices defining the Yukawa couplings. For square matrices we identify six different examples where 
these are satisfied. In each case the numbers of fermions $n_f$ can be arbitrarily large.   For vanishing Yukawa 
coupling these theories
generally have just the $O(n_s)$ and Gaussian fixed points though one example is equivalent to the scalar
theory with hypertetrahedral symmetry where there are two further fixed points with ${\cal S}_{n_s+1} \times
{\mathbb Z}_2$ symmetry. Assuming just lowest order $\beta$-functions  the Yukawa $\beta$-function does not
contain the scalar couplings  and is easily solved. For the scalar couplings there  
are then relations between the fixed points for small and large $n_f$. A similar pattern emerges in each example. 
Even if there are four fixed points when $n_f=0$ these reduce to two except  for very tiny or very large $n_f$.
Generally there are two fixed points for low and large $n_f$ and for intermediate $n_f$ either $0$ or $4$.
These  do not necessarily lead to scalar potentials which are bounded below, the Gaussian fixed point becomes
unstable when $n_f>0$, but there is a stable potential for large $n_f$ related to the Gaussian fixed point as $n_f \to 0$. 
We also consider
an example where the Yukawa matrices are not square, corresponding to chiral fermions, and where there is a
$U(r)\times U(s)$ symmetry and $n_s = r s$. The purely scalar theory may have four fixed points for suitable $r,s$
but with a non zero Yukawa coupling there is a similar pattern.

Further constraints relating the coefficients 
 for the contributions of various diagrams to $\beta$-functions and anomalous dimensions
 can be obtained from applying a perturbative
version of the $a$-theorem~\cite{Osborn:1989td,Jack:1990eb,Osborn:1991gm,Jack:2013sha,Poole:2019kcm}. In general this relates certain combinations of $\beta$ and $\gamma$-function
coefficients at a particular loop order to lower order contributions. In the present context this 
provides relations for the coefficients of the three loop Yukawa $\beta$-function and also $\gamma_\phi$
and $\gamma_\psi$. Such conditions were analysed at length by Poole and Thomsen~\cite{Poole:2019kcm}, including also gauge 
couplings. We present their results here without any explicit evaluation of lower order one and two
loop contributions so that the structure of the conditions is more apparent. We also consider the restriction to
$U(1)$ symmetry where results are more tractable. 

The outline of the paper is as follows: In the next four sections we list the diagrams for the scalar and fermion anomalous dimensions and the Yukawa
and quartic scalar $\beta$-functions for the general scalar fermion theory at up to three loops. We also give 
the values for the corresponding coefficients, 143 at three loops,
 which are all consistent with the various relations obtained later. Of course at one and two loops
 results have been known for a long time, we list the coefficients diagram by diagram.
 Our conventions match those in  \cite{Poole:2019kcm} and our numerical results at one and two loops
agree precisely  once they are multiplied by the required factor to ensure overall symmetry.
Similarly the three loop results for the Yukawa $\beta$-function and also the anomalous dimensions 
agree exactly with \cite{Davies:2021mnc}.
In the case of the quartic scalar $\beta$-function the relations obtained here are used to provide  complete 
results for all terms appearing in the general expansion. For simplicity the Yukawa couplings are rescaled  by
$4\pi$ and the scalar quartic couplings by $16\pi^2$. 
These coefficients all correspond to what would be obtained  in a $\overline {M\!S}$ scheme although no explicit
calculation is undertaken here.

The results are simplified in section 6 where a $U(1)$ symmetry is imposed which significantly reduces the
number of terms present in the expansions of the general $\beta$-functions and anomalous dimensions. 
The $U(1)$ restriction contains as a special case $\N=1$ supersymmetry and the various necessary linear
constraints are derived in section 7. We there also consider also the example of what is here termed
$\N=\frac12$ supersymmetry where there are a significant number of linear constraints which are all
satisfied by the explicit results listed earlier in sections 2,3,4,5.

Besides supersymmetry conditions there are also relations for the various coefficients derived from
the existence of a perturbative $a$-function. We list the conditions for the general scalar-fermion theory
which are all derived from \cite{Poole:2019kcm}. For the two loop anomalous dimensions and the Yukawa
$\beta$-function there are 4 relations whereas at three loops there are 42. At three loops it is necessary
to also allow for 5 possible antisymmetric contributions to the anomalous dimensions and 4 relations
for these are obtained. 

Section 8 contains our discussion of scalar fermion fixed points. For multiple scalars we show there
are theories which can be restricted to a single Yukawa coupling and two quartic scalar couplings.
As special cases these include the well known renormalisable Gross-Neveu and Nambu Jona-Lasinio
theories.

In general results for individual coefficients corresponding to particular diagrams are scheme dependent.
In section 10 scheme variations which preserve the structure in terms of contributions from 1PI diagrams
are considered. Coefficients corresponding to primitive diagrams, which have no subdivergences, are
individually invariant but this of course not true in general. We demonstrate how scheme invariants can be
formed and applied in detail to the three loop Yukawa $\beta$-function.
These can be linear or higher order in the coefficients.  The scheme invariance of the $a$-function relations
is also verified.

Some further details are considered in various appendices. Appendix A describes a basis for Majorana fermions
relevant for reduction to three dimensions and their possible extensions away from $d=3$ with broken Lorentz 
invariance. In appendix B we outline some tensorial calculations relevant for the fixed point discussion.
Some figures elucidating how the fixed points in scalar fermion theories vary with differing numbers 
of fermions are given in appendix C.
In  appendix D we describe the derivation of the $a$-function relations at two and three loops 
after restricting to $U(1)$ symmetry, there is then one relation at two loops and 12 at three. 
Finally in appendix E we discuss some general features of scheme changes
which preserve the perturbative structure in terms of contributions corresponding to 1PI diagrams.

\section{Scalar Anomalous Dimension}\label{sec:gamma-s}

The one and two loop 1PI and 1VI diagrams relevant for $\gamma_\phi{\!}^{ab}= \gamma_\phi{\!}^{ba}$ are just

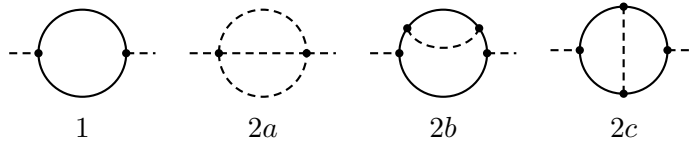
\begin{figure}[hbt!]
  \centering
  \begin{tabular}{ccccc}
  \begin{tikzpicture}
  \draw[black, thick, densely dashed] (1.em, 0.em) -- (2em, 0em);
    \draw[black, thick, densely dashed] (5em, 0.em) -- (6em, 0em);
    \draw[black, thick] (2em, 0em) arc (180:360: 1.5em) ;
    \draw[black, thick] (2em, 0.em) arc (180:0: 1.5em);
    \node at (2em, 0.em)[circle,fill,inner sep=.1em]{};
    \node at (5em, 0.em)[circle,fill,inner sep=.1em]{};
        \end{tikzpicture}
        &
  \begin{tikzpicture}
  \draw[black, thick, densely dashed] (1.em, 0.em) -- (2em, 0em);
    \draw[black, thick, densely dashed] (5em, 0.em) -- (6em, 0em);
    \draw[black, thick, densely dashed] (2em, 0em) arc (180:360: 1.5em) ;
    \draw[black, thick, densely dashed] (2em, 0.em) arc (180:0: 1.5em);
    \draw[black, thick, densely dashed] (2em, 0.em) -- (5em,0);
    \node at (2em, 0.em)[circle,fill,inner sep=.1em]{};
    \node at (5em, 0.em)[circle,fill,inner sep=.1em]{};
        \end{tikzpicture}
        &
   \begin{tikzpicture}
    \draw[black, thick, densely dashed] (1.em, 0.em) -- (2em, 0.em) ;
    \draw[black, thick, densely dashed] (5em, 0.em) -- (6em, 0em);
    \draw[black, thick] (2em, 0em) arc (180:0: 1.5em);
     \draw[black, thick] (2em, 0.em) arc (180:360: 1.5em);
    \draw[white ] (3.5em, 0.em) --+ (35:1.5em) coordinate(n1);
    \draw[white ] (3.5em, 0.em) --+  (145:1.5em) coordinate (n2);
    \draw[black, thick, densely dashed]  (n2) to [out=-70,in =250] (n1);
    \node at (2em, 0.em)[circle,fill,inner sep=.1em]{};
    \node at (5em, 0.em)[circle,fill,inner sep=.1em]{};
    \node at (n1)[circle,fill,inner sep=.1em]{};
    \node at (n2)[circle,fill,inner sep=.1em]{};
        \end{tikzpicture}
        &
         \begin{tikzpicture}
    \draw[black, thick, densely dashed] (1.em, 0.em) -- (2em, 0.em) ;
    \draw[black, thick, densely dashed] (5em, 0.em) -- (6em, 0em);
    \draw[black, thick] (2em, 0em) arc (180:0: 1.5em);
     \draw[black, thick] (2em, 0.em) arc (180:360: 1.5em);
    \draw[white ] (3.5em, 0.em) --+ (90:1.5em) coordinate(n1);
    \draw[white ] (3.5em, 0.em) --+  (270:1.5em) coordinate (n2);
    \draw[black, thick, densely dashed]  (n2) -- (n1);
    \node at (2em, 0.em)[circle,fill,inner sep=.1em]{};
    \node at (5em, 0.em)[circle,fill,inner sep=.1em]{};
    \node at (n1)[circle,fill,inner sep=.1em]{};
    \node at (n2)[circle,fill,inner sep=.1em]{};
        \end{tikzpicture}
         \\ 1 & $2a$ & $2b$ & $2c$ \\[1em]
            \end{tabular}
            \vskip - 18pt
 \caption{One and two loop diagrams giving contributions to the scalar field
anomalous dimensions, containing Yukawa and quartic scalar couplings.
 Fermion lines are solid, scalar lines are dashed.}
  \label{fig:gamma_s1}
\end{figure}

\noindent
while at three loops

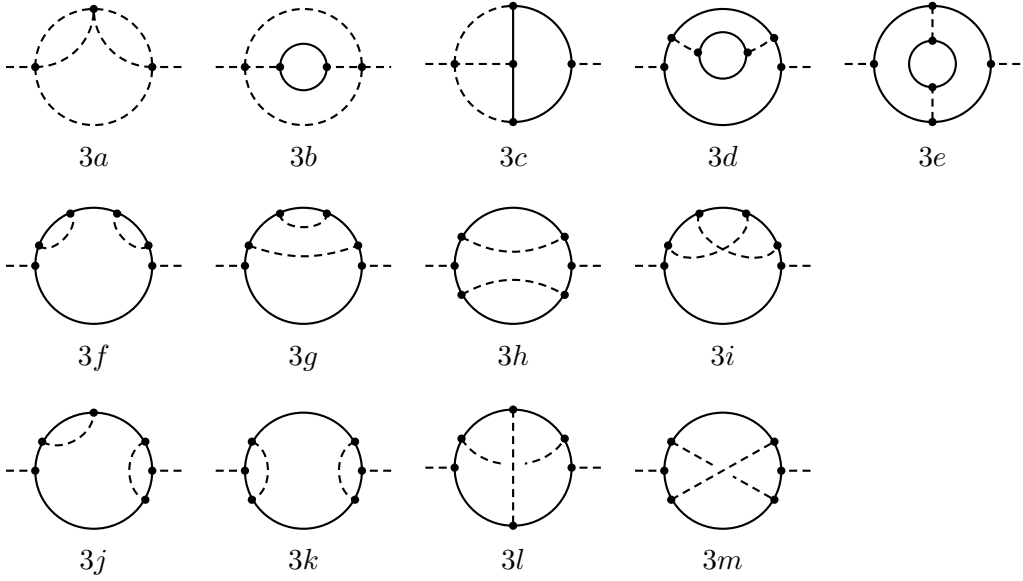
\begin{figure}[hbt!]
  \centering
  \begin{tabular}{ccccc}
  \begin{tikzpicture}
    \draw[black, thick, densely dashed] (1.em, 0.em) -- (2em, 0em);
    \draw[black, thick, densely dashed] (7em, 0.em) -- (6em, 0em);
    \draw[black, thick, densely dashed] (2em, 0.em) arc (180:0:2.em);
    \draw[black, thick, densely dashed] (2em, 0.em) arc (180:360:2.em);
    \draw[black, thick, densely dashed] (2em, 0.em)  to [out= 0, in = 270] (4em, 2em); 
    \draw[black, thick, densely dashed] (6em, 0.em) to [out= 180, in = 270]  (4em, 2em); 
    \node at (2em, 0)[circle,fill,inner sep=.1em]{};
    \node at (6em, 0)[circle,fill,inner sep=.1em]{};
    \node at (4em, 2em)[circle,fill,inner sep=.1em]{};
    \end{tikzpicture}
&
    \begin{tikzpicture}
    \draw[black, thick, densely dashed] (1.em, 0.em) -- (7em, 0em);
    \draw[black, thick, densely dashed] (2em, 0.em) arc (180:0:2.em);
    \draw[black, thick, densely dashed] (2em, 0.em) arc (180:360:2.em);
    \draw[black, thick,fill=white] (4em, 0.em) circle (0.8em);
    \node at (2em, 0)[circle,fill,inner sep=.1em]{};
    \node at (6em, 0)[circle,fill,inner sep=.1em]{};
    \node at (3.2em, 0)[circle,fill,inner sep=.1em]{};
    \node at (4.8em, 0)[circle,fill,inner sep=.1em]{};
    \end{tikzpicture}
        &
    \begin{tikzpicture}
    \draw[black, thick, densely dashed] (1.em, 0.em) -- (4em, 0em);
    \draw[black, thick, densely dashed] (7em, 0.em) -- (6em, 0em);
    \draw[black, thick, densely dashed] (2em, 0.em) arc (180:90:2.em);
    \draw[black, thick, densely dashed] (2em, 0.em) arc (180:270:2.em);
    \draw[black, thick] (6em, 0.em) arc (0:90:2.em);
    \draw[black, thick] (6em, 0.em) arc (0:-90:2.em) -- (4em, 2.em);
    \node at (2em, 0)[circle,fill,inner sep=.1em]{};
    \node at (6em, 0)[circle,fill,inner sep=.1em]{};
    \node at (4em, 0)[circle,fill,inner sep=.1em]{};
    \node at (4em, 2em)[circle,fill,inner sep=.1em]{};
    \node at (4em, -2em)[circle,fill,inner sep=.1em]{};
    \end{tikzpicture}
    &
     \begin{tikzpicture}
    \draw[black, thick, densely dashed] (1.em, 0.em) -- (2em, 0.em);
    \draw[black, thick, densely dashed] (7.em, 0.em) -- (6em, 0.em);
    \draw[black, thick] (4em, 0em) circle (2em);
    \draw[black, thick, densely dashed] (2.3em, 1em) to [bend right=40] (5.7em, 1em);
    \draw[black, thick,fill=white] (4em, 0.4em) circle (0.8em);
    \node at (2em, 0.em)[circle,fill,inner sep=.1em]{};
    \node at (6em, 0.em)[circle,fill,inner sep=.1em]{};
    \node at (2.245em, 1.em)[circle,fill,inner sep=.1em]{};
    \node at (5.755em, 1em)[circle,fill,inner sep=.1em]{};
    \node at (3.15em, .5em)[circle,fill,inner sep=.1em]{};
    \node at (4.85em, .5em)[circle,fill,inner sep=.1em]{};
    \end{tikzpicture}
      &
    \begin{tikzpicture}
    \draw[black, thick, densely dashed] (1.em, 0.em) -- (2em, 0.em);
    \draw[black, thick, densely dashed] (7.em, 0.em) -- (6em, 0.em);
    \draw[black, thick, densely dashed] (4em, 2.em) -- (4em, -2.em);
    \draw[black, thick] (4em, 0em) circle (2em);
    \draw[black, thick,fill=white] (4em,0em) circle (0.8em);
       \node at (2em, 0.em)[circle,fill,inner sep=.1em]{};
    \node at (6em, 0.em)[circle,fill,inner sep=.1em]{};
    \node at (4em, 2.em)[circle,fill,inner sep=.1em]{};
    \node at (4em, -2.em)[circle,fill,inner sep=.1em]{};
    \node at (4em, 0.8em)[circle,fill,inner sep=.1em]{};
    \node at (4em, -0.8em)[circle,fill,inner sep=.1em]{};
    \end{tikzpicture}
            \\ $3a$ & $3b$ & $3c$ & $3d$   &  $3e$ \\[1em]
     \begin{tikzpicture}
    \draw[black, thick, densely dashed] (1.em, 0.em) -- (2em, 0.em);
    \draw[black, thick, densely dashed] (7.em, 0.em) -- (6em, 0.em);
    \draw[black, thick] (4em, 0em) circle (2em);
    \draw[black, thick, densely dashed] (2.13em, .7em) to [bend right=90](3.2em, 1.8em);
    \draw[black, thick, densely dashed] (5.87em, .7em) to [bend left=90](4.8em, 1.8em);
    \node at (2em, 0.em)[circle,fill,inner sep=.1em]{};
    \node at (6em, 0.em)[circle,fill,inner sep=.1em]{};
    \node at (3.2em, 1.8em)[circle,fill,inner sep=.1em]{};
    \node at (2.13em, .7em)[circle,fill,inner sep=.1em]{};
    \node at (4.8em, 1.8em)[circle,fill,inner sep=.1em]{};
    \node at (5.87em, .7em)[circle,fill,inner sep=.1em]{};
    \end{tikzpicture}
&
     \begin{tikzpicture}
    \draw[black, thick, densely dashed] (1.em, 0.em) -- (2em, 0.em);
    \draw[black, thick, densely dashed] (7.em, 0.em) -- (6em, 0.em);
    \draw[black, thick] (4em, 0em) circle (2em);
    \draw[black, thick, densely dashed] (2.13em, .7em) to [bend right=20](5.87em, .7em) ;
    \draw[black, thick, densely dashed] (3.2em, 1.8em) to [bend right=90](4.8em, 1.8em);
    \node at (2em, 0.em)[circle,fill,inner sep=.1em]{};
    \node at (6em, 0.em)[circle,fill,inner sep=.1em]{};
    \node at (3.2em, 1.8em)[circle,fill,inner sep=.1em]{};
    \node at (2.13em, .7em)[circle,fill,inner sep=.1em]{};
    \node at (4.8em, 1.8em)[circle,fill,inner sep=.1em]{};
    \node at (5.87em, .7em)[circle,fill,inner sep=.1em]{};
    \end{tikzpicture}
&
    \begin{tikzpicture}
    \draw[black, thick, densely dashed] (1.em, 0.em) -- (2em, 0.em);
    \draw[black, thick, densely dashed] (7.em, 0.em) -- (6em, 0.em);
    \draw[black, thick] (4em, 0em) circle (2em);
    \draw[black, thick, densely dashed] (2.236em, 1em) to [bend right](5.764em, 1em);
    \draw[black, thick, densely dashed] (2.236em, -1em) to [bend left](5.764em, -1em);
    \node at (2em, 0.em)[circle,fill,inner sep=.1em]{};
    \node at (6em, 0.em)[circle,fill,inner sep=.1em]{};
    \node at (2.236em, 1em)[circle,fill,inner sep=.1em]{};
    \node at (2.236em, -1em)[circle,fill,inner sep=.1em]{};
    \node at (5.764em, 1em)[circle,fill,inner sep=.1em]{};
    \node at (5.764em, -1em)[circle,fill,inner sep=.1em]{};
    \end{tikzpicture}
    &
    \begin{tikzpicture}
    \draw[black, thick, densely dashed] (1.em, 0.em) -- (2em, 0.em);
    \draw[black, thick, densely dashed] (7.em, 0.em) -- (6em, 0.em);
    \draw[black, thick] (4em, 0em) circle (2em);
    \draw[black, thick, densely dashed] (4.8em, 1.8em) to [bend left=90](2.13em, .7em);
    \draw[black, thick, densely dashed] (3.2em, 1.8em) to [bend right=90](5.87em, .7em) ;
    \node at (2em, 0.em)[circle,fill,inner sep=.1em]{};
    \node at (6em, 0.em)[circle,fill,inner sep=.1em]{};
    \node at (3.2em, 1.8em)[circle,fill,inner sep=.1em]{};
    \node at (2.13em, .7em)[circle,fill,inner sep=.1em]{};
    \node at (4.8em, 1.8em)[circle,fill,inner sep=.1em]{};
    \node at (5.87em, .7em)[circle,fill,inner sep=.1em]{};
    \end{tikzpicture}
    \\ $3f$ & $3g$ & $3h$ & $3i$ \\[1em]
    \begin{tikzpicture}
    \draw[black, thick, densely dashed] (1.em, 0.em) -- (2em, 0.em);
    \draw[black, thick, densely dashed] (7.em, 0.em) -- (6em, 0.em);
    \draw[black, thick] (4em, 0em) circle (2em);
    \draw[black, thick, densely dashed] (2.236em, 1em) to [bend right=60] (4em,2em);
    \draw[black, thick, densely dashed] (5.764em, 1em) to [bend right=70] (5.764em,-1em);
    \node at (2em, 0.em)[circle,fill,inner sep=.1em]{};
    \node at (6em, 0.em)[circle,fill,inner sep=.1em]{};
    \node at (2.236em, 1em)[circle,fill,inner sep=.1em]{};
    \node at (5.764em, 1em)[circle,fill,inner sep=.1em]{};
    \node at (4em, 2em)[circle,fill,inner sep=.1em]{};
    \node at (5.764em, -1em)[circle,fill,inner sep=.1em]{};
    \end{tikzpicture}
    &
    \begin{tikzpicture}
    \draw[black, thick, densely dashed] (1.em, 0.em) -- (2em, 0.em);
    \draw[black, thick, densely dashed] (7.em, 0.em) -- (6em, 0.em);
    \draw[black, thick] (4em, 0em) circle (2em);
    \draw[black, thick, densely dashed] (2.236em, 1em) to [bend left=70] (2.236em, -1em);
    \draw[black, thick, densely dashed] (5.764em, 1em) to [bend right=70](5.764em, -1em);
    \node at (2em, 0.em)[circle,fill,inner sep=.1em]{};
    \node at (6em, 0.em)[circle,fill,inner sep=.1em]{};
    \node at (2.236em, 1em)[circle,fill,inner sep=.1em]{};
    \node at (2.236em, -1em)[circle,fill,inner sep=.1em]{};
    \node at (5.764em, 1em)[circle,fill,inner sep=.1em]{};
    \node at (5.764em, -1em)[circle,fill,inner sep=.1em]{};
    \end{tikzpicture}
    &
    \begin{tikzpicture}
    \draw[black, thick, densely dashed] (1.em, 0.em) -- (2em, 0.em);
    \draw[black, thick, densely dashed] (7.em, 0.em) -- (6em, 0.em);
    \draw[black, thick] (4em, 0em) circle (2em);
    \draw[black, thick, densely dashed] (2.236em, 1em) to [bend right=60] (5.764em, 1em);
    \fill [white] (4em,0.15em) circle (0.4em);
    \draw[black, thick, densely dashed] (4em, 2em) -- (4em, -2em);
    \node at (2em, 0.em)[circle,fill,inner sep=.1em]{};
    \node at (6em, 0.em)[circle,fill,inner sep=.1em]{};
    \node at (2.236em, 1em)[circle,fill,inner sep=.1em]{};
    \node at (5.764em, 1em)[circle,fill,inner sep=.1em]{};
    \node at (4em, 2em)[circle,fill,inner sep=.1em]{};
    \node at (4em, -2em)[circle,fill,inner sep=.1em]{};
    \end{tikzpicture}
    &
    \begin{tikzpicture}
    \draw[black, thick, densely dashed] (1.em, 0.em) -- (2em, 0.em);
    \draw[black, thick, densely dashed] (7.em, 0.em) -- (6em, 0.em);
    \draw[black, thick] (4em, 0em) circle (2em);
    \draw[black, thick, densely dashed] (2.236em, 1em) to (5.764em, -1em);
    \fill [white] (4em,0em) circle (0.4em);
    \draw[black, thick, densely dashed] (2.236em, -1em) to (5.764em, 1em);
    \node at (2em, 0.em)[circle,fill,inner sep=.1em]{};
    \node at (6em, 0.em)[circle,fill,inner sep=.1em]{};
    \node at (2.236em, 1em)[circle,fill,inner sep=.1em]{};
    \node at (2.236em, -1em)[circle,fill,inner sep=.1em]{};
    \node at (5.764em, 1em)[circle,fill,inner sep=.1em]{};
    \node at (5.764em, -1em)[circle,fill,inner sep=.1em]{};
    \end{tikzpicture}
        \\ $3j$ & $3k$ & $3l$ & $3m$  \\[1em]
  \end{tabular}
  \caption{Three loop diagrams giving contributions to the scalar field
anomalous dimensions, containing Yukawa and quartic scalar couplings.}
  \label{fig:gamma_s3}
\end{figure}

The corresponding expansions are then
  \begin{align}
   \gamma_{\phi}^{(1)ab} = {}& \gamma_{\phi{\hskip 0.5pt}1} \;  \tr (y^{ab}) \, , \nn \\
    \gamma_{\phi}^{(2)ab} = {}& \gamma_{\phi{\hskip 0.5pt}2a} \; \lambda^{acde} \lambda^{bcde} 
    + \gamma_{\phi{\hskip 0.5pt}2b} \; \tr (y^{abcc}) +  \gamma_{\phi{\hskip 0.5pt}2c} \; \tr (y^{acbc}) \, , \nn\\
    \gamma_{\phi}^{(3)ab} = {}&  \gamma_{\phi{\hskip 0.5pt}3a}\; \lambda^{acde} \lambda^{defg} \lambda^{bcfg} 
+ \gamma_{\phi{\hskip 0.5pt}3b} \; \lambda^{acde} \lambda^{bcdf} \, \tr (y^{ef}) 
   {} + \gamma_{\phi{\hskip 0.5pt}3c} \; \S_2\, \lambda^{acde}\,\tr(y^{bcde})  \nn \\
&{} + \gamma_{\phi{\hskip 0.5pt}3d} \; \tr(y^{abcd})\,  \tr(y^{cd})  + \gamma_{\phi{\hskip 0.5pt}3e} \;
\tr(y^{acbd} ) \, \tr(y^{cd}) 
  + \gamma_{\phi{\hskip 0.5pt}3f} \;  \tr(y^{abccdd})\nn \\
  &{} + \gamma_{\phi{\hskip 0.5pt}3g} \; \tr(y^{abcddc}) 
+  \gamma_{\phi{\hskip 0.5pt}3h} \;\tr(y^{accbdd})  +   \gamma_{\phi{\hskip 0.5pt}3i} \;\tr(y^{abcdcd} ) +  \gamma_{\phi{\hskip 0.5pt}3j} \;
\S_2\, \tr(y^{acbcdd}) \nn \\
    &{}+  \gamma_{\phi{\hskip 0.5pt}3k}\; \tr(y^{acdbdc})  + \gamma_{\phi{\hskip 0.5pt}3l}\; \tr(y^{acbdcd}) 
 + \gamma_{\phi{\hskip 0.5pt}3m }\; \tr(y^{acdbcd}) \, ,
 \label{eq:gamma_s}
  \end{align}
employing the abbreviation $y^{abcd...} = y^a y^b y^c y^d ...$  and where $\S_2$ denotes the sum over
two terms necessary to ensure symmetry for $a\leftrightarrow b$ so that 
$\gamma_{\phi}^{ab} =\gamma_{\phi}^{ba} $ for the three loop expressions.  The normalisations 
of the traces correspond to the fermions having two components, as would be appropriate in three dimensions. 
For four dimensional Majorana fermions 
$y_a \to \left ( \begin{smallmatrix}y_a & 0  \\ 0 & y_a \end{smallmatrix}\right )$ so that
\be
\tr( y^{a_1} y^{a_2} \dots  y^{a_p} ) \big |_{\rm Majorana} = \big ( 1+  (-1)^p\big ) \, \tr  ( y^{a_1 \dots a_p} )\, .
\ee
The coefficients for the trace corresponding to a fermion loop then has an additional factor two and there
are only an even number of Yukawa couplings on any loop.

With this notation the results of calculation for the individual coefficients in the general fermion scalar theory are~\cite{Steudtner:2021fzs,Herren:2017uxn,Bednyakov:2021qxa,Davies:2021mnc}
\begin{align}
& \gamma_{\phi{\hskip 0.5pt}1} = \tfrac12 \, ,  \hskip -3cm
&& \gamma_{\phi{\hskip 0.5pt}2a} = \tfrac{1}{12} \, , &&  \gamma_{\phi{\hskip 0.5pt}2b} =  - \tfrac34 \, , &&
\gamma_{\phi{\hskip 0.5pt}2c} = - \tfrac12 \,  , \nn \\
& \gamma_{\phi{\hskip 0.5pt}3a}  =  - \tfrac{1}{16}\, , &&  \gamma_{\phi{\hskip 0.5pt}3b}  =  - \tfrac{5}{32} \, , &&
 \gamma_{\phi{\hskip 0.5pt}3c}  =   \tfrac58\, , &&   \gamma_{\phi{\hskip 0.5pt}3d}  =   1\, , \qquad   \gamma_{\phi{\hskip 0.5pt}3e}  =   \tfrac{9}{16}\, , \nn \\ 
 & \gamma_{\phi{\hskip 0.5pt}3f}  =   - \tfrac{3}{16} \, , &&   \gamma_{\phi{\hskip 0.5pt}3g}  =   \tfrac{5}{16} \, , &&  
 \gamma_{\phi{\hskip 0.5pt}3h}  =   \tfrac{1}{32}\, , 
 &&   \gamma_{\phi{\hskip 0.5pt}3i}  =   -\tfrac38\, , \nn \\
 &   \gamma_{\phi{\hskip 0.5pt}3j}  =   \tfrac{7}{16}\, , &&  \gamma_{\phi{\hskip 0.5pt}3k}  =   \tfrac{7}{4}\, ,  &&
 \gamma_{\phi{\hskip 0.5pt}3l}  =  - \tfrac{3}{4}\, , &&
 \gamma_{\phi{\hskip 0.5pt}3m}  =   \tfrac32 \zeta_3 -1 \, . 
\end{align}

At three loop order there is the further possibility of 1PI antisymmetric contributions to the anomalous dimension~\cite{Bednyakov:2014pia,Herren:2017uxn,Jack:2016tpp,Herren:2021yur}
which take the form~\cite{Poole:2019kcm}
\begin{align}
\upsilon_{\phi}^{(3)ab} = {}&   \upsilon_{\phi{\hskip 0.5pt}3c} \; \A_2\, \lambda^{acde}\,\tr(y^{bcde})  
+  \upsilon_{\phi{\hskip 0.5pt}3j} \; \A_2\, \tr(y^{acbcdd}) \, ,
\label{upphi3}
\end{align}
where now $\A_2\, \tr(y^{acbcdd})=  \tr(y^{acbcdd}) -  \tr(y^{bcacdd})$. Such terms can usually be 
neglected but they play a role in finding fixed points with vanishing energy momentum tensor
trace. In this context the results~\cite{Davies:2021mnc} are then
\be
\upsilon_{\phi{\hskip 0.5pt}3c} = - \tfrac58 \, , \qquad \upsilon_{\phi{\hskip 0.5pt}3j} = - \tfrac34 \, .
\label{Phiup}
\ee

\vskip 2cm

\section{Fermion Anomalous Dimension}\label{sec:gamma-f}

For $\gamma_\psi=\gamma_\psi{\!}^T $, at one and two loops the 1PI, 1VI diagrams are just

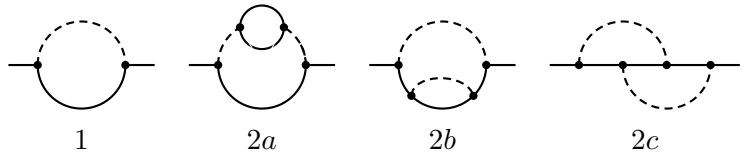
\begin{figure}[ht]
  \centering
  \begin{tabular}{ccccc}
   \begin{tikzpicture}
    \draw[black, thick] (1.em, 0.em) -- (2em, 0.em) arc (180:360: 1.5em) --(6.em, 0.em);
    \draw[black, thick, densely dashed] (2em, 0.em) arc (180:0: 1.5em);
    \node at (2em, 0.em)[circle,fill,inner sep=.1em]{};
    \node at (5em, 0.em)[circle,fill,inner sep=.1em]{};
        \end{tikzpicture}
           &
   \begin{tikzpicture}
    \draw[black, thick] (1.em, 0.em) -- (2em, 0.em) arc (180:360: 1.5em) --(6.em, 0.em);
    \draw[black, thick, densely dashed] (2em, 0.em) arc (180:0: 1.5em)  ;
     \draw[black, thick, densely dashed] (5em, 0.em) arc (0: 35: 1.5em)  ;
    \draw[white ] (3.5em, 0.em) --+ (90:1.3em) coordinate(n1);
      \draw[black, thick, fill=white] (n1) circle (.75em);
      \draw  [white] (3.5em, 0.em) --+ (60:1.5em) coordinate(n2);
      \draw [white] (3.5em, 0.em) --+ (120:1.5em) coordinate(n3);
      \node at (2em, 0.em)[circle,fill,inner sep=.1em]{};
    \node at (5em, 0.em)[circle,fill,inner sep=.1em]{};
    \node at (n2)[circle,fill,inner sep=.1em]{};
    \node at (n3)[circle,fill,inner sep=.1em]{};
        \end{tikzpicture}
        &
         \begin{tikzpicture}
    \draw[black, thick] (1.em, 0.em) -- (2em, 0.em) arc (180:360: 1.5em) --(6.em, 0.em);
    \draw[black, thick, densely dashed] (2em, 0.em) arc (180:0: 1.5em);
    \draw[white ] (3.5em, 0.em) --+ (225:1.5em) coordinate(n1);
    \draw[white ] (3.5em, 0.em) --+  (315:1.5em) coordinate (n2);
    \draw[black, thick, densely dashed]  (n1) to [out=85,in = 105 ] (n2);
    \node at (2em, 0.em)[circle,fill,inner sep=.1em]{};
    \node at (5em, 0.em)[circle,fill,inner sep=.1em]{};
    \node at (n1)[circle,fill,inner sep=.1em]{};
    \node at (n2)[circle,fill,inner sep=.1em]{};
        \end{tikzpicture}
        &
    \begin{tikzpicture}
    \draw[black, thick] (1.em, 0.em) -- (7.5em, 0.em);
    \draw[black, thick, densely dashed] (2em, 0.em) arc (180:0: 1.5em);
    \draw[black, thick, densely dashed] (3.5em, 0.em) arc (180:360: 1.5em);
    \node at (2em, 0.em)[circle,fill,inner sep=.1em]{};
    \node at (6.5em, 0.em)[circle,fill,inner sep=.1em]{};
    \node at (5em, 0.em)[circle,fill,inner sep=.1em]{};
    \node at (3.5em, 0.em)[circle,fill,inner sep=.1em]{};
    \end{tikzpicture}
         \\ 1 & $2a$ & $2b$  & $2c$ \\[1em]
  \end{tabular}
  \caption{One and two loop diagrams giving contributions to the fermion field
anomalous dimensions, containing Yukawa and quartic scalar couplings. }
  \label{fig:gamma_f1}
  \end{figure}
  \noindent
  and at three loops there are 16 1PI diagrams
  
  \begin{figure}[ht]
  \centering
  \begin{tabular}{ccccc}
    \begin{tikzpicture}
      \draw[black, thick] (1.em, 0.em) -- (2em, 0.em) arc (180:360: 2em) --(7.em, 0.em);
      \draw[black, thick]  (6em, 0.em);
      \draw[black, thick, densely dashed] (2.236em, 1em) to [bend right=70](5.764em, 1em);
       \draw[black, thick, densely dashed] (2.236em, 1em) -- (5.764em, 1em);
       \draw[black, thick, densely dashed] (2em, 0.em) arc (180:0: 2em);
      \node at (2em, 0.em)[circle,fill,inner sep=.1em]{};
      \node at (6em, 0.em)[circle,fill,inner sep=.1em]{};
      \node at (2.236em, 1em)[circle,fill,inner sep=.1em]{};
      \node at (5.764em, 1em)[circle,fill,inner sep=.1em]{};
    \end{tikzpicture}
    &
    \begin{tikzpicture}
    \draw[black, thick] (1.em, 0.em) -- (2em, 0.em) arc (180:360: 2em) --(7.em, 0.em);
    \draw[black, thick, densely dashed] (2em, 0.em) arc (180:0: 2em);
    \draw[black, thick, densely dashed] (2.236em, -1em) to [bend right] (4em,2em) 
    to [bend right] (5.764em, -1em);
    \node at (2em, 0.em)[circle,fill,inner sep=.1em]{};
    \node at (6em, 0.em)[circle,fill,inner sep=.1em]{};
    \node at (2.236em, -1em)[circle,fill,inner sep=.1em]{};
    \node at (5.764em, -1em)[circle,fill,inner sep=.1em]{};
    \node at (4em, 2em)[circle,fill,inner sep=.1em]{};
    \end{tikzpicture}
    &
    \begin{tikzpicture}
    \draw[black, thick] (1.em, 0.em) -- (2em, 0.em) arc (180:360: 2em) --(7.em, 0.em);
    \draw[black, thick, densely dashed] (2em, 0.em) arc (180:0: 2em);
    \draw  [white] (4em,0) --+ (55:1.9em)  coordinate (n1) ;
    \draw [white] (4em,0) --+ (125:1.9em)  coordinate (n2) ;
    \draw[black, thick, fill=white] (n1) circle (.7em);
    \draw[black, thick, fill=white] (n2) circle (.7em);
        \node at (2em, 0.em)[circle,fill,inner sep=.1em]{};
    \node at (6em, 0.em)[circle,fill,inner sep=.1em]{};
    \node at (3.55em, 1.9em)[circle,fill,inner sep=.1em]{};
    \node at (2.4em, 1.1em)[circle,fill,inner sep=.1em]{};
    \node at (4.45em, 1.9em)[circle,fill,inner sep=.1em]{};
    \node at (5.6em, 1.1em)[circle,fill,inner sep=.1em]{};
    \end{tikzpicture}
        &
    \begin{tikzpicture}
    \draw[black, thick] (1.em, 0.em) -- (2em, 0.em) arc (180:360: 2em) --(7.em, 0.em);
    \draw[black, thick] (2.236em, 1em) to [bend right=45] (5.764em, 1em);
    \draw[black, thick] (2.236em, 1em) to [bend left=60] (5.764em, 1em);
    \draw[black, thick, densely dashed] (2em, 0em) to [bend left=15] (2.236em,1em);
    \draw[black, thick, densely dashed] (6em, 0em) to [bend right=15] (5.764em,1em);
    \draw[black, thick, densely dashed] (2.236em, -1em) to [bend left=45](5.764em, -1em);
    \node at (2em, 0.em)[circle,fill,inner sep=.1em]{};
    \node at (6em, 0.em)[circle,fill,inner sep=.1em]{};
    \node at (2.236em, 1em)[circle,fill,inner sep=.1em]{};
    \node at (5.764em, 1em)[circle,fill,inner sep=.1em]{};
     \node at (2.236em, -1em)[circle,fill,inner sep=.1em]{};
    \node at (5.764em, -1em)[circle,fill,inner sep=.1em]{};
    \end{tikzpicture}
    &
    \begin{tikzpicture}
    \draw[black, thick] (1.em, 0.em) -- (2em, 0.em) arc (180:360: 2em) --(7.em, 0.em);
    \draw[black, thick, densely dashed] (1.em, 0.em) -- (2em, 0.em) arc (180:0:2em) -- (7.em, 0.em);
    \draw[black, thick, densely dashed] (2.236em, -1em) to [bend left=15](3.15em, -.4em);
    \draw[black, thick, densely dashed] (5.764em, -1em) to [bend right=15](4.85em, -.4em);
    \draw[black, thick] (4.em, -0.35em) circle (0.7em);
    \node at (2em, 0.em)[circle,fill,inner sep=.1em]{};
    \node at (6em, 0.em)[circle,fill,inner sep=.1em]{};
     \node at (2.236em, -1em)[circle,fill,inner sep=.1em]{};
    \node at (5.764em, -1em)[circle,fill,inner sep=.1em]{};
    \node at (3.3em, -.42em)[circle,fill,inner sep=.1em]{};
    \node at (4.7em, -.42em)[circle,fill,inner sep=.1em]{};
    \end{tikzpicture}
     \\ $3a$ & $3b$ & $3c$ & $3d$ & $3e$ \\[0.5em]
    \begin{tikzpicture}
    \draw[black, thick] (1.em, 0.em) -- (7.em, 0.em);
    \draw[black, thick, densely dashed] (2em, 0.em) arc (180:0: 1.5em);
    \draw[black, thick, densely dashed] (3em, 0.em) arc (180:360: 1.5em);
    \draw[black, thick, fill=white] (4.5em, -1.25em) circle (0.7em);
    \node at (2em, 0.em)[circle,fill,inner sep=.1em]{};
    \node at (6em, 0.em)[circle,fill,inner sep=.1em]{};
    \node at (5em, 0.em)[circle,fill,inner sep=.1em]{};
    \node at (3em, 0.em)[circle,fill,inner sep=.1em]{};
    \node at (5.2em, -1.31em)[circle,fill,inner sep=.1em]{};
    \node at (3.8em, -1.31em)[circle,fill,inner sep=.1em]{};
    \end{tikzpicture}
       &
    \begin{tikzpicture}
    \draw[black, thick] (1.em, 0.em) -- (2em, 0.em) arc (180:360: 2em) --(7.em, 0.em);
    \draw[black, thick] (2.236em, 1em) to [bend right=75] (5.764em, 1em);
    \draw[black, thick] (2.236em, 1em) to [bend left=75] (5.764em, 1em);
    \draw[black, thick, densely dashed] (2em, 0em) to [bend left=15] (2.236em,1em);
    \draw[black, thick, densely dashed] (6em, 0em) to [bend right=15] (5.764em,1em);
    \draw[black, thick, densely dashed] (3em, .2em) to [bend left=75] (5em,.2em);
    \node at (2em, 0.em)[circle,fill,inner sep=.1em]{};
    \node at (6em, 0.em)[circle,fill,inner sep=.1em]{};
    \node at (2.236em, 1em)[circle,fill,inner sep=.1em]{};
    \node at (5.764em, 1em)[circle,fill,inner sep=.1em]{};
    \node at (3em, .2em)[circle,fill,inner sep=.1em]{};
    \node at (5em, .2em)[circle,fill,inner sep=.1em]{};
    \end{tikzpicture}
     &
    \begin{tikzpicture}
    \draw[black, thick] (1.em, 0.em) -- (2em, 0.em) arc (180:360: 2em) --(7.em, 0.em);
    \draw[black, thick] (2.236em, 1em) to [bend right=75] (5.764em, 1em);
    \draw[black, thick] (2.236em, 1em) to [bend left=75] (5.764em, 1em);
    \draw[black, thick, densely dashed] (4em, 2em) -- (4em, 0em);
    \draw[black, thick, densely dashed] (2em, 0em) to [bend left=15] (2.236em,1em);
    \draw[black, thick, densely dashed] (6em, 0em) to [bend right=15] (5.764em,1em);
    \node at (2em, 0.em)[circle,fill,inner sep=.1em]{};
    \node at (6em, 0.em)[circle,fill,inner sep=.1em]{};
    \node at (2.236em, 1em)[circle,fill,inner sep=.1em]{};
    \node at (5.764em, 1em)[circle,fill,inner sep=.1em]{};
    \node at (4em, 2em)[circle,fill,inner sep=.1em]{};
    \node at (4em, 0)[circle,fill,inner sep=.1em]{};
    \end{tikzpicture}
    &
    \begin{tikzpicture}
    \draw[black, thick] (1.em, 0.em) -- (7.em, 0.em);
    \draw[black, thick, densely dashed] (2em, 0.em) arc (180:0: 2em);
    \draw[black, thick, densely dashed] (4.25em, 0.em) arc (180:0: .625em);
    \draw[black, thick, densely dashed] (2.5em, 0.em) arc (180:0: .625em);
    \draw[transparent, thick, densely dashed] (2em, 0.em) arc (180:360: 2em);
    \node at (2em, 0.em)[circle,fill,inner sep=.1em]{};
    \node at (2.5em, 0.em)[circle,fill,inner sep=.1em]{};
    \node at (3.75em, 0.em)[circle,fill,inner sep=.1em]{};
    \node at (4.25em, 0.em)[circle,fill,inner sep=.1em]{};
    \node at (5.5em, 0.em)[circle,fill,inner sep=.1em]{};
    \node at (6em, 0.em)[circle,fill,inner sep=.1em]{};
    \end{tikzpicture}
    &
    \begin{tikzpicture}
    \draw[black, thick] (1.em, 0.em) -- (7.em, 0.em);
    \draw[black, thick, densely dashed] (2em, 0.em) arc (180:0: 2em);
    \draw[black, thick, densely dashed] (3em, 0.em) arc (180:0: 1em);
    \draw[black, thick, densely dashed] (3.5em, 0.em) arc (180:0: .5em);
    \draw[transparent, thick, densely dashed] (2em, 0.em) arc (180:360: 2em);
    \node at (2em, 0.em)[circle,fill,inner sep=.1em]{};
    \node at (3.5em, 0.em)[circle,fill,inner sep=.1em]{};
    \node at (3em, 0.em)[circle,fill,inner sep=.1em]{};
    \node at (4.5em, 0.em)[circle,fill,inner sep=.1em]{};
    \node at (5em, 0.em)[circle,fill,inner sep=.1em]{};
    \node at (6em, 0.em)[circle,fill,inner sep=.1em]{};
    \end{tikzpicture}
     \\ $3f$ & $3g$ & $3h$ & $3i$ & $3j$ \\[0.5em]
     \begin{tikzpicture}
    \draw[black, thick] (0.8em, 0.em) -- (7.2em, 0.em);
    \draw[transparent, thick, densely dashed] (2em, 0.em) arc (180:0: 2em);
    \draw[black, thick, densely dashed] (1.5em, 0.em) arc (180:0: 1em);
    \draw[black, thick, densely dashed] (2.9em, 0.em) arc (180:360: 1.2em);
    \draw[black, thick, densely dashed] (4.5em, 0.em) arc (180:0: 1em);
    \draw[transparent, thick, densely dashed] (2em, 0.em) arc (180:360: 2em);
    \node at (1.5em, 0.em)[circle,fill,inner sep=.1em]{};
    \node at (3.5em, 0.em)[circle,fill,inner sep=.1em]{};
    \node at (2.9em, 0.em)[circle,fill,inner sep=.1em]{};
    \node at (4.5em, 0.em)[circle,fill,inner sep=.1em]{};
    \node at (5.3em, 0.em)[circle,fill,inner sep=.1em]{};
    \node at (6.5em, 0.em)[circle,fill,inner sep=.1em]{};
    \end{tikzpicture}
    &
    \begin{tikzpicture}
    \draw[black, thick] (1.em, 0.em) -- (7.em, 0.em);
    \draw[black, thick, densely dashed] (2em, 0.em) arc (180:0: 2em);
    \draw[black, thick, densely dashed] (3em, 0.em) arc (180:360: .75em);
    \draw[black, thick, densely dashed] (3.5em, 0.em) arc (180:0: .75em);
    \draw[transparent, thick, densely dashed] (2em, 0.em) arc (180:360: 2em);
    \node at (2em, 0.em)[circle,fill,inner sep=.1em]{};
    \node at (3.5em, 0.em)[circle,fill,inner sep=.1em]{};
    \node at (3em, 0.em)[circle,fill,inner sep=.1em]{};
    \node at (4.5em, 0.em)[circle,fill,inner sep=.1em]{};
    \node at (5em, 0.em)[circle,fill,inner sep=.1em]{};
    \node at (6em, 0.em)[circle,fill,inner sep=.1em]{};
    \end{tikzpicture}  
    &
    \raisebox{0.75em}{
    \begin{tikzpicture}
    \draw[black, thick] (1.em, 0.em) -- (7.em, 0.em);
    \draw[black, thick, densely dashed] (2em, 0.em) arc (180:0: 1.5em);
    \draw[black, thick, densely dashed] (2.5em, 0.em) arc (180:0: 1.em);
    \draw[black, thick, densely dashed] (3.5em, 0.em) arc (180:360: 1.25em);
    \node at (2em, 0.em)[circle,fill,inner sep=.1em]{};
    \node at (3.5em, 0.em)[circle,fill,inner sep=.1em]{};
    \node at (2.5em, 0.em)[circle,fill,inner sep=.1em]{};
    \node at (4.5em, 0.em)[circle,fill,inner sep=.1em]{};
    \node at (5em, 0.em)[circle,fill,inner sep=.1em]{};
    \node at (6em, 0.em)[circle,fill,inner sep=.1em]{};
     \end{tikzpicture}}
     &
    \begin{tikzpicture}
    \draw[black, thick] (1.em, 0.em) -- (7.em, 0.em);
    \draw[black, thick, densely dashed] (2em, 0.em) arc (180:0: 1.5em);
    \draw[black, thick, densely dashed] (2.5em, 0.em) arc (180:0: .5em);
    \draw[black, thick, densely dashed] (4.em, 0.em) arc (180:360: 1.em);
    \draw[transparent, thick, densely dashed] (2em, 0.em) arc (180:360: 2em);
    \node at (2em, 0.em)[circle,fill,inner sep=.1em]{};
    \node at (3.5em, 0.em)[circle,fill,inner sep=.1em]{};
    \node at (2.5em, 0.em)[circle,fill,inner sep=.1em]{};
    \node at (4.em, 0.em)[circle,fill,inner sep=.1em]{};
    \node at (5em, 0.em)[circle,fill,inner sep=.1em]{};
    \node at (6em, 0.em)[circle,fill,inner sep=.1em]{};
    \end{tikzpicture}
         &
    \begin{tikzpicture}
    \draw[black, thick] (1.em, 0.em) -- (7.em, 0.em);
    \draw[black, thick, densely dashed] (2em, 0.em) arc (180:0: 1.5em);
    \draw[black, thick, densely dashed] (3em, 0.em) arc (180:360: 1.5em);
    \draw[black, thick, densely dashed] (3.5em, 0.em) arc (180:0: .5em);
    \draw[transparent, thick, densely dashed] (2em, 0.em) arc (180:360: 2em);
    \node at (2em, 0.em)[circle,fill,inner sep=.1em]{};
    \node at (3.5em, 0.em)[circle,fill,inner sep=.1em]{};
    \node at (3em, 0.em)[circle,fill,inner sep=.1em]{};
    \node at (4.5em, 0.em)[circle,fill,inner sep=.1em]{};
    \node at (5em, 0.em)[circle,fill,inner sep=.1em]{};
    \node at (6em, 0.em)[circle,fill,inner sep=.1em]{};
    \end{tikzpicture}
         \\ $3k$ & $3l$ & $3m$ & $3n$ & $3o$ \\[0.5em]
    \begin{tikzpicture}
    \draw[black, thick] (1.em, 0.em) -- (7.em, 0.em);
    \draw[black, thick, densely dashed] (2em, 0.em) arc (180:0: 1.25em);
    \draw[black, thick, densely dashed] (2.8em, 0.em) arc (180:360: 1.25em);
    \draw[black, thick, densely dashed] (3.5em, 0.em) arc (180:0: 1.25em);
    \draw[transparent, thick, densely dashed] (2em, 0.em) arc (180:360: 2em);
    \node at (2em, 0.em)[circle,fill,inner sep=.1em]{};
    \node at (3.5em, 0.em)[circle,fill,inner sep=.1em]{};
    \node at (2.8em, 0.em)[circle,fill,inner sep=.1em]{};
    \node at (4.5em, 0.em)[circle,fill,inner sep=.1em]{};
    \node at (5.3em, 0.em)[circle,fill,inner sep=.1em]{};
    \node at (6em, 0.em)[circle,fill,inner sep=.1em]{};
    \end{tikzpicture}
    \\ $3p$
  \end{tabular}
  \caption{Three-loop diagrams giving contributions to the fermion field
anomalous dimensions, containing Yukawa and quartic scalar couplings. }
  \label{fig:gamma_f3}
  \end{figure}
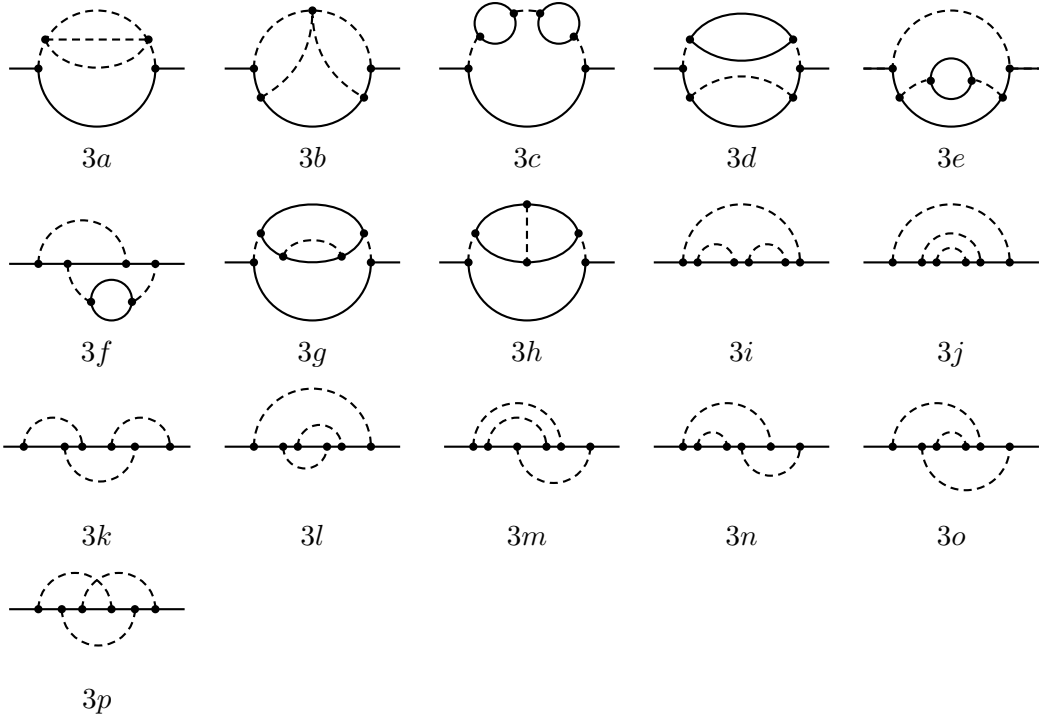
  
  Corresponding to figures 3 and 4 the contributions have the general form
    \begin{align}
  \hskip -0.5cm
  \gamma_{\psi}^{(1)} = {}& \gamma_{\psi{\hskip 0.5pt}1}\; y^{aa} \, , \nn  \\
  \gamma_{\psi}^{(2)} = {}& \gamma_{\psi{\hskip 0.5pt}2a}\; y^{ab}\, \tr(y^{ab}) + \gamma_{\psi{\hskip 0.5pt}2b}\; y^{abba}
  + \gamma_{\psi{\hskip 0.5pt}2c}\; y^{abab}  \, , \nn \\
    \gamma_{\psi}^{(3)} = {}& \gamma_{\psi{\hskip 0.5pt}3a}\; \lambda^{acde} \lambda^{bcde}\,y^{ab} 
+  \gamma_{\psi{\hskip 0.5pt}3b}\; \lambda^{abcd}\, y^{abcd} 
   + \gamma_{\psi{\hskip 0.5pt}3c}\;  y^{ab}\, \tr(y^{ac}) \, \tr(y^{bc})\nn  \\
&   {} 
 +   \big ( \gamma_{\psi{\hskip 0.5pt}3d}\;  y^{accb}  + \gamma_{\psi{\hskip 0.5pt}3e}\; y^{cabc} + 
   \gamma_{\psi{\hskip 0.5pt}3f}\;  \S_2 \, y^{acbc} \big ) \,  \tr(y^{ab})  +\gamma_{\psi{\hskip 0.5pt}3g}\; \,  y^{ab}\, \tr(y^{abcc} )\\
   &{}+ \gamma_{\psi{\hskip 0.5pt}3h}\;   y^{ab}\, \tr(y^{acbc})
   +  \gamma_{\psi{\hskip 0.5pt}3i}\;   y^{abbcca} +  \gamma_{\psi{\hskip 0.5pt}3j}\; y^{abccba}   +
 \gamma_{\psi{\hskip 0.5pt}3k}\; y^{abacbc}  +   \gamma_{\psi{\hskip 0.5pt}3l}\; y^{abcbca} \nn  \\
   &{} +  \gamma_{\psi{\hskip 0.5pt}3m}\; \S_2\, y^{abcbac}  +
   \gamma_{\psi{\hskip 0.5pt}3n}\;  \S_2\, y^{abbcac}   +  \gamma_{\psi{\hskip 0.5pt}3o}\;  y^{abccab}
   +  \gamma_{\psi{\hskip 0.5pt}3p}\;  y^{abcabc} \, .
   \label{eq:gamma_f}
  \end{align}  
where here $\S_2 \, y^{acbc} = y^{acbc} + y^{cbca}, \ \S_2\, y^{abcbac}= y^{abcbac} + y^{cabcba}$ and
similarly{\hskip 0.5pt} as necessary for the symmetry $ \gamma_{\psi}^{(3)} =  \gamma_{\psi}^{(3)T} $.
In this case the coefficients in a $\overline{M\! S}$ scheme are then~\cite{Steudtner:2021fzs,Herren:2017uxn,Bednyakov:2021qxa,Davies:2021mnc}
\begin{align}
& \gamma_{\psi{\hskip 0.5pt}1} = \tfrac12 \, , &&  \gamma_{\psi{\hskip 0.5pt}2a} = -\tfrac38\, ,  &&  \gamma_{\psi{\hskip 0.5pt}2b} =  - \tfrac18 \, , 
&& \gamma_{\psi{\hskip 0.5pt}2c} = 0 \, ,\nn \\
& \gamma_{\psi{\hskip 0.5pt}3a}  =  - \tfrac{11}{96}\, , &&  \gamma_{\psi{\hskip 0.5pt}3b}  =  1 \, , &&
 \gamma_{\psi{\hskip 0.5pt}3c}  =   - \tfrac{3}{32}\, , &&  \gamma_{\psi{\hskip 0.5pt}3d}  =   - \tfrac18\, , \nn \\  
 & \gamma_{\psi{\hskip 0.5pt}3e}  =   \tfrac{9}{32} \, , 
 && \gamma_{\psi{\hskip 0.5pt}3f}  =  - \tfrac{3}{32} \, , &&   \gamma_{\psi{\hskip 0.5pt}3g}  =  1 \, , &&   
 \gamma_{\psi{\hskip 0.5pt}3h}  =  \tfrac12 \, ,  \nn \\
 &   \gamma_{\psi{\hskip 0.5pt}3i}  =   -\tfrac{5}{32}\, , &&
    \gamma_{\psi{\hskip 0.5pt}3j}  =   \tfrac{1}{16}\, , && \gamma_{\psi{\hskip 0.5pt}3k}  =   \tfrac12\, , && 
 \gamma_{\psi{\hskip 0.5pt}3l}  =  - \tfrac{5}{16}\, , \nn \\ 
 & \gamma_{\psi{\hskip 0.5pt}3m}  =   0  \, , 
 &&   \gamma_{\psi{\hskip 0.5pt}3n}  =   \tfrac{3}{32}\, , && \gamma_{\psi{\hskip 0.5pt}3o}  =   \tfrac14\, , && 
 \gamma_{\psi{\hskip 0.5pt}3p}  =   \tfrac{3}{2}\zeta_3 -1   \, .
\end{align}

For potential antisymmetric contributions
\begin{align}
\upsilon_{\psi}^{(3)} =
   \upsilon_{\psi{\hskip 0.5pt}3f}\;  \A_2 \, y^{acbc}  \,  \tr(y^{ab})   +  \upsilon_{\psi{\hskip 0.5pt}3m}\; \A_2\, y^{abcbac}  +
   \gamma_{\psi{\hskip 0.5pt}3n}\;  \A_2\, y^{abbcac}  \, ,
   \label{uppsi3}
  \end{align} 
  with $  \A_2 \, y^{acbc}  \,  \tr(y^{ab}) = (y^{acbc} - y^{cacb} )  \, \tr(y^{ab}), \ \A_2\, y^{abcbac} =  
  y^{abcbac} -  y^{acbabc}$. In this case~\cite{Davies:2021mnc}
  \be
  \upsilon_{\psi{\hskip 0.5pt}3f} = \tfrac{7}{16}\, , \qquad  \upsilon_{\psi{\hskip 0.5pt}3m} = - \tfrac38 \, , \qquad
  \upsilon_{\psi{\hskip 0.5pt}3n} = - \tfrac{5}{16} \, .
  \label{Psiup}
  \ee

  \vskip 2cm

\section{Yukawa Couplings}\label{sec:beta-yuk}

The Yukawa coupling $\beta$-function can be decomposed as
\be
\beta_{y}^{a} = \beta_{y}^{a\hskip 0.5pt T}
= {\tilde \beta}_{y}^{a} + \gamma^{ab}_{\phi}\, y^b +  \gamma_{\psi}\, y^a + y^a \, \gamma_{\psi} 
\, ,
\label{betay}
\ee
where ${\tilde \beta}_{y}^{a}$ is determined solely by the contributions of 1PI diagrams.

At one and two loops the relevant 1PI diagrams are

\begin{figure}[ht]
\hskip - 12pt
  \centering
  \begin{tabular}{ccccccc}
\begin{tikzpicture}
    \draw[black, thick] (1.75em, -0.25em) -- (4em, 2.em) -- (6.25em, -0.25em);
    \draw[black, thick, densely dashed] (4em, 3.5em) -- (4em, 2.em);
      \draw[black, thick, densely dashed] (2.5em, .5em) -- (5.5em, .5em);
    \node at (4em, 2.em)[circle,fill,inner sep=.1em]{};
     \node at (2.5em, .5em)[circle,fill,inner sep=.1em]{};
    \node at (5.5em, .5em)[circle,fill,inner sep=.1em]{};
    \end{tikzpicture}
    &
    \begin{tikzpicture}
    \draw[black, thick] (1.75em, -0.25em) -- (2.5em, .5em) -- (5.5em, .5em) --
(6.25em, -0.25em);
    \draw[black, thick, densely dashed] (4em, .5em) -- (4em, 3.5em);
    \draw[black, thick, densely dashed] (2.5em, .5em) -- (4em, 2.2em) --(5.5em, .5em);
      \node at (4em, .5em)[circle,fill,inner sep=.1em]{};
    \node at (2.5em, .5em)[circle,fill,inner sep=.1em]{};
    \node at (5.5em, .5em)[circle,fill,inner sep=.1em]{};
    \node at (4.em, 2.2em)[circle,fill,inner sep=.1em]{};
     \end{tikzpicture}
    &
    \begin{tikzpicture}
    \draw[black, thick] (1.75em, -0.25em) -- (4em, 2.em) -- (6.25em,-0.25em);
    \draw[black, thick, densely dashed] (4em, 3.5em) -- (4em, 2.em);
    \draw[black, thick] (4em, .5em) circle (.6em);
    \draw[black, thick, densely dashed] (2.5em, .5em) -- (3.4em, .5em);
    \draw[black, thick, densely dashed] (5.5em, .5em) -- (4.9em, .5em);
    \node at (4em, 2.em)[circle,fill,inner sep=.1em]{};
    \node at (3.4em, .5em)[circle,fill,inner sep=.1em]{};
    \node at (4.6em, .5em)[circle,fill,inner sep=.1em]{};
    \node at (3.4em, .5em)[circle,fill,inner sep=.1em]{};
    \node at (2.5em, .5em)[circle,fill,inner sep=.1em]{};
    \node at (5.5em, .5em)[circle,fill,inner sep=.1em]{};
      \end{tikzpicture}
      &
      \begin{tikzpicture}
    \draw[black, thick] (1.75em, -0.25em) -- (4em, 2.em) -- (6.25em, -0.25em);
    \draw[black, thick, densely dashed] (4em, 3.5em) -- (4em, 2.em);
      \draw[black, thick, densely dashed] (2.5em, .5em) -- (5.5em, .5em);
    \node at (4em, 2.em)[circle,fill,inner sep=.1em]{};
     \node at (2.5em, .5em)[circle,fill,inner sep=.1em]{};
    \node at (5.5em, .5em)[circle,fill,inner sep=.1em]{};
    \node at (4.375em, 1.625em)[circle,fill,inner sep=.1em]{};
     \node at (5.125em, 0.875em)[circle,fill,inner sep=.1em]{};
      \draw[thick, densely dashed] (5.125em, 0.875em)  arc (-45:135: 0.53 em); 
    \end{tikzpicture}
     &
      \begin{tikzpicture}
    \draw[black, thick] (1.75em, -0.25em) -- (4em, 2.em) -- (6.25em, -0.25em);
    \draw[black, thick, densely dashed] (4em, 3.5em) -- (4em, 2.em);
      \draw[black, thick, densely dashed] (2.5em, .5em) -- (5.5em, .5em);
    \node at (4em, 2.em)[circle,fill,inner sep=.1em]{};
     \node at (2.5em, .5em)[circle,fill,inner sep=.1em]{};
    \node at (5.5em, .5em)[circle,fill,inner sep=.1em]{};
    \node at (5.15em, 0.85em)[circle,fill,inner sep=.1em]{};
     \node at (5.85em, 0.15em)[circle,fill,inner sep=.1em]{};
      \draw[thick, densely dashed] (5.85em, 0.15em)  arc (-45:135: 0.53 em); 
    \end{tikzpicture}
    &
    \begin{tikzpicture}
    \draw[black, thick] (1.75em, -0.25em) -- (4em, 2.em) -- (6.25em, -0.25em);
    \draw[black, thick, densely dashed] (4em, 3.5em) -- (4em, 2.em);
      \draw[black, thick, densely dashed] (2.4em, .4em) -- (5.6em, .4em);
        \draw[black, thick, densely dashed] (3.2em, 1.2em) -- (4.8em, 1.2em);
    \node at (4em, 2.em)[circle,fill,inner sep=.1em]{};
     \node at (3.2em, 1.2em)[circle,fill,inner sep=.1em]{};
    \node at (4.8em, 1.2em)[circle,fill,inner sep=.1em]{};
     \node at (2.4em, .4em)[circle,fill,inner sep=.1em]{};
    \node at (5.6em, .4em)[circle,fill,inner sep=.1em]{};
    \end{tikzpicture}
     &
    \begin{tikzpicture}
    \draw[black, thick] (1.75em, -0.25em) -- (4em, 2.em) -- (6.25em, -0.25em);
    \draw[black, thick, densely dashed] (4em, 3.5em) -- (4em, 2.em);
     \draw[black, thick, densely dashed] (3.2em, 1.2em) -- (5.6em, .4em);
      \fill[white] (4em,0.9em) circle (0.3em);
       \draw[black, thick, densely dashed] (2.4em, .4em) -- (4.8em, 1.2em);
    \node at (4em, 2.em)[circle,fill,inner sep=.1em]{};
     \node at (3.2em, 1.2em)[circle,fill,inner sep=.1em]{};
    \node at (4.8em, 1.2em)[circle,fill,inner sep=.1em]{};
     \node at (2.4em, .4em)[circle,fill,inner sep=.1em]{};
    \node at (5.6em, .4em)[circle,fill,inner sep=.1em]{};
    \end{tikzpicture}
     \\ 1 & $2a$ & $2b$ & $2c$ & $2d$ & $2e$  & $ 2f$\\ [1em]
 \end{tabular}
  \caption{One and two loop Yukawa  vertex diagrams.}
  \label{fig:yuk1}
\end{figure}
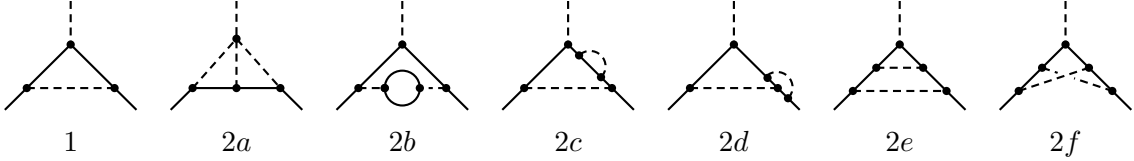
\noindent
where
\begin{align}
 {\tilde  \beta}_{y}^{(1)a}  = {}&\beta_{y{\hskip 0.5pt}1} \; y^{bab} \, , \nn \\
  {\tilde  \beta}_{y}^{(2)a}  = {}& \beta_{y{\hskip 0.5pt}2a} \, \lambda^{abcd}y^{bcd} +  \beta_{y{\hskip 0.5pt}2b} \, y^{bac} \, \tr(y^{bc})
  +  \beta_{y{\hskip 0.5pt}2c} \, \S_2\,  y^{baccb} +  \beta_{y{\hskip 0.5pt}2d} \, \S_2\,  y^{bacbc} \nn \\
  \noalign{\vskip -2pt}
 &{} +  \beta_{y{\hskip 0.5pt}2e} \, y^{bcacb} +  \beta_{y{\hskip 0.5pt}2f} \, y^{bcabc} \, ,
\end{align}
with $\S_2 \, y^{baccb} = y^{baccb} + y^{bccab} $, as necessary for symmetry.
Old results, with our conventions, give
\be
\beta_{y{\hskip 0.5pt}1} = 2\, , \quad  \beta_{y{\hskip 0.5pt}2a} = - 2\, , \ \  \beta_{y{\hskip 0.5pt}2b} =  \beta_{y{\hskip 0.5pt}2c} = -1 \, , 
\ \  \beta_{y{\hskip 0.5pt}2d} = 0 \, ,\ \ \beta_{y{\hskip 0.5pt}2e} = -2\, , \ \ \beta_{y{\hskip 0.5pt}2f} = 2\, .
\ee

At three loops there are 52 distinct diagrams so we use the alphabet twice over as labels
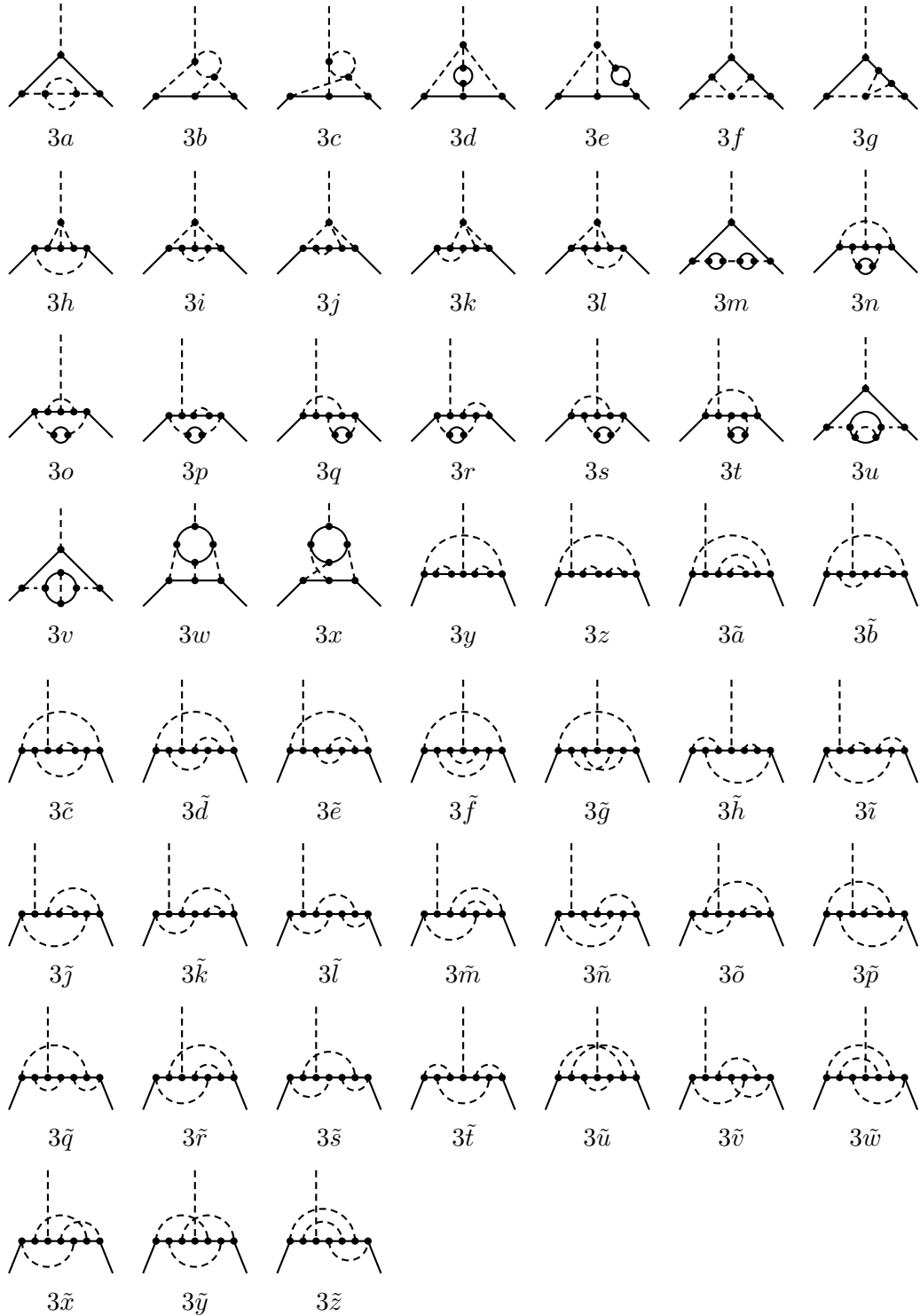
\begin{figure}[H]
  \centering \hskip -0.5cm
\begin{tabular}{ccccccc}
    \begin{tikzpicture}
    \draw[black, thick] (2.em, 0.em) -- (4em, 2.em) -- (6.em, 0.em);
    \draw[black, thick, densely dashed] (4em, 4.em) -- (4em, 2.em);
    \draw[black, thick, densely dashed] (4em, .5em) circle (.6em);
    \draw[black, thick, densely dashed] (2.5em, .5em) -- (5.5em, .5em);
    \node at (4em, 2.em)[circle,fill,inner sep=.1em]{};
    \node at (3.4em, .5em)[circle,fill,inner sep=.1em]{};
    \node at (4.6em, .5em)[circle,fill,inner sep=.1em]{};
    \node at (3.4em, .5em)[circle,fill,inner sep=.1em]{};
    \node at (2.5em, .5em)[circle,fill,inner sep=.1em]{};
    \node at (5.5em, .5em)[circle,fill,inner sep=.1em]{};
    \end{tikzpicture}
    &
    \begin{tikzpicture}
    \draw[black, thick] (2.em, 0.em) -- (2.5em, .5em) -- (5.5em, .5em) --(6.em, 0.em);
    \draw[black, thick, densely dashed] (4em, 4.em) -- (4em, 1.85em) -- (2.5em,.5em);
    \draw[black, thick, densely dashed] (4.5em, 1.75em) circle (.5em);
    \draw[black, thick, densely dashed] (4em, .5em) -- (4.75em, 1.25em) --(5.5em, .5em);
    \node at (4em, .5em)[circle,fill,inner sep=.1em]{};
    \node at (2.5em, .5em)[circle,fill,inner sep=.1em]{};
    \node at (5.5em, .5em)[circle,fill,inner sep=.1em]{};
    \node at (4.75em, 1.25em)[circle,fill,inner sep=.1em]{};
    \node at (4.em, 1.85em)[circle,fill,inner sep=.1em]{};
    \end{tikzpicture}
    &
    \begin{tikzpicture}
    \draw[black, thick] (2.em, 0.em) -- (2.5em, .5em) -- (5.5em, .5em) --(6.em, 0.em);
    \draw[black, thick, densely dashed] (4em, 4.em) -- (4em, .5em);
    \draw[black, thick, densely dashed] (4.5em, 1.75em) circle (.5em);
    \draw[black, thick, densely dashed] (2.5em, .5em) -- (4.75em, 1.25em) --(5.5em, .5em);
    \node at (4em, .5em)[circle,fill,inner sep=.1em]{};
    \node at (2.5em, .5em)[circle,fill,inner sep=.1em]{};
    \node at (5.5em, .5em)[circle,fill,inner sep=.1em]{};
    \node at (4.75em, 1.25em)[circle,fill,inner sep=.1em]{};
    \node at (4.em, 1.85em)[circle,fill,inner sep=.1em]{};
    \end{tikzpicture}
    &
    \begin{tikzpicture}
    \draw[black, thick] (2.em, 0.em) -- (2.5em, .5em) -- (5.5em, .5em) --(6.em, 0.em);
    \draw[black, thick, densely dashed] (4em, 4.em) -- (4em, 2.5em) -- (2.5em,.5em);
    \draw[black, thick, densely dashed] (4em, 1.65em) -- (4em, 2.5em) --(5.5em, .5em);
    \draw[black, thick, densely dashed] (4em, .95em) -- (4em, .5em);
    \draw[black, thick] (4em, 1.3em) circle (.35em);
    \node at (4em, .5em)[circle,fill,inner sep=.1em]{};
    \node at (2.5em, .5em)[circle,fill,inner sep=.1em]{};
    \node at (5.5em, .5em)[circle,fill,inner sep=.1em]{};
    \node at (4.em, 2.5em)[circle,fill,inner sep=.1em]{};
    \node at (4.em, 1em)[circle,fill,inner sep=.1em]{};
    \node at (4.em, 1.6em)[circle,fill,inner sep=.1em]{};
    \end{tikzpicture}
    &
    \begin{tikzpicture}
    \draw[black, thick] (2.em, 0.em) -- (2.5em, .5em) -- (5.5em, .5em) --(6.em, 0.em);
    \draw[black, thick, densely dashed] (4em, 4.em) -- (4em, 2.5em) -- (2.5em,.5em);
    \draw[black, thick, densely dashed] (4em, .5em) -- (4em, 2.5em) -- (5.5em,.5em);
    \draw[black, thick, fill=white] (4.9em, 1.3em) circle (.35em);
    \node at (4em, .5em)[circle,fill,inner sep=.1em]{};
    \node at (2.5em, .5em)[circle,fill,inner sep=.1em]{};
    \node at (5.5em, .5em)[circle,fill,inner sep=.1em]{};
    \node at (4.em, 2.5em)[circle,fill,inner sep=.1em]{};
    \node at (4.7em, 1.6em)[circle,fill,inner sep=.1em]{};
    \node at (5.1em, 1.em)[circle,fill,inner sep=.1em]{};
    \end{tikzpicture}
    &
    \begin{tikzpicture}
    \draw[black, thick] (2.em, 0.em) -- (2.5em, .5em) -- (4em, 2em) -- (5.5em,.5em) -- (6.em, 0.em);
    \draw[black, thick, densely dashed] (4em, 4.em) -- (4em, 2.em);
    \draw[black, thick, densely dashed] (2.5em, .5em)  -- (5.5em, .5em);
    \draw[black, thick, densely dashed] (3.25em, 1.25em)  -- (4em, .5em) --(4.75em, 1.25em);
    \node at (4em, 2em)[circle,fill,inner sep=.1em]{};
    \node at (2.5em, .5em)[circle,fill,inner sep=.1em]{};
    \node at (5.5em, .5em)[circle,fill,inner sep=.1em]{};
    \node at (3.25em, 1.25em)[circle,fill,inner sep=.1em]{};
    \node at (4.75em, 1.25em)[circle,fill,inner sep=.1em]{};
    \node at (4em, .5em)[circle,fill,inner sep=.1em]{};
    \end{tikzpicture}
    &
    \begin{tikzpicture}
    \draw[black, thick] (2.em, 0.em) -- (2.5em, .5em) -- (4em, 2em) -- (5.5em,.5em) -- (6.em, 0.em);
    \draw[black, thick, densely dashed] (4em, 4.em) -- (4em, 2.em);
    \draw[black, thick, densely dashed] (2.5em, .5em)  -- (5.5em, .5em);
    \draw[black, thick, densely dashed] (5em, 1.em)  -- (4em, .5em) -- (4.5em,1.5em);
    \node at (4em, 2em)[circle,fill,inner sep=.1em]{};
    \node at (2.5em, .5em)[circle,fill,inner sep=.1em]{};
    \node at (5.5em, .5em)[circle,fill,inner sep=.1em]{};
    \node at (4.5em, 1.5em)[circle,fill,inner sep=.1em]{};
    \node at (5em, 1.em)[circle,fill,inner sep=.1em]{};
    \node at (4em, .5em)[circle,fill,inner sep=.1em]{};
    \end{tikzpicture}
    \\
\end{tabular}
  \caption{Three loop Yukawa  vertex diagrams.}
  \label{fig:yuk}
\end{figure}
\noindent
Joining the external lines to a single vertex the resulting vacuum diagrams can be either planar or non planar. 
In the above list $3c, 3s, 3x, 3\tg, 3\too, 3\tp, 3\trr, 3\ts, 3\tu, 3 \tv, 3\tw, 3\tx, 3\ty$ are non planar

With this diagrammatic decomposition the three loop Yukawa $\beta$-function is expanded as
\begin{align} \hskip -0.8cm
 {\tilde \beta}_{y}^{(3)a}  = {}& \beta_{y{\hskip 0.5pt}3a} \; \lambda^{bdef}\lambda^{cdef}\, y^{bac} 
  + \lambda^{abef} \lambda^{efcd} \big (\beta_{y{\hskip 0.5pt}3b} \; \S_2 \, y^{bcd} + \beta_{y{\hskip 0.5pt}3c} \; y^{cbd} \big ) \nn \\
& {}  + \lambda^{acde} \,  \tr(y^{eb}) \, \big ( \beta_{y{\hskip 0.5pt}3d} \; y^{cbd} + \beta_{y{\hskip 0.5pt}3e} \; \S_2 \, y^{bcd} \big ) 
  + \lambda^{bcde}\, \big ( \beta_{y{\hskip 0.5pt}3f} \,y^{bcade} + \beta_{y{\hskip 0.5pt}3g} \; \S_2\, y^{bacde} \big ) \nn  \\
  &{} + \lambda^{abcd} \big (  \beta_{y{\hskip 0.5pt}3h} \, y^{ebcde} + \beta_{y{\hskip 0.5pt}3i} \, \, y^{beced}   +
\beta_{y{\hskip 0.5pt}3j} \, \S_2\, y^{beecd} +  \beta_{y{\hskip 0.5pt}3k} \, \S_2 \,  y^{ebecd}  
+ \beta_{y{\hskip 0.5pt}3l} \, \S_2 \, y^{ebced}  \big ) \nn  \\
& {} + \beta_{y{\hskip 0.5pt}3m} \,  y^{bac} \, \tr(y^{bd}) \, \tr(y^{cd})    \nn  \\
  &{}  + \big (  \beta_{y{\hskip 0.5pt}3n} \, y^{dbacd} +  \beta_{y{\hskip 0.5pt}3o} \,y^{bdadc} +  
  \beta_{y{\hskip 0.5pt}3p} \, \S_2 \, y^{baddc}  +  \beta_{y{\hskip 0.5pt}3q} \, \S_2 \, y^{dabdc} \big ) \, \tr(y^{bc})  \nn  \\
    &{} + \big (    \beta_{y{\hskip 0.5pt}3r} \, \S_2\, y^{badcd} + \beta_{y{\hskip 0.5pt}3s} \, \S_2 \,y^{dbadc}  
    + \beta_{y{\hskip 0.5pt}3t} \, \S_2 \, y^{dabcd} \big ) \,\tr(y^{bc})  \nn \\
&{} +   y^{bac} \big ( \beta_{y{\hskip 0.5pt}3u} \, \tr(y^{bcdd}) +  \beta_{y{\hskip 0.5pt}3v} \,\tr(y^{bdcd})  \big )
+ \big ( \beta_{y{\hskip 0.5pt}3w} \, y^{bcd} + \beta_{y{\hskip 0.5pt}3x} \,\S_2 \,  y^{bdc} \big )\,   \tr(y^{abcd}) \nn  \\
 &+ \beta_{y{\hskip 0.5pt}3y} \, y^{bccaddb}  
 +  \beta_{y{\hskip 0.5pt}3z} \; \S_2 \, y^{baccddb} + \beta_{y{\hskip 0.5pt}3\ta} \; \S_2 \, y^{bacddcb}  
 + \beta_{y{\hskip 0.5pt}3\tb} \; \S_2 \, y^{bcacddb} \nn \\
& {}+ \beta_{y{\hskip 0.5pt}3\tc} \; \S_2 \, y^{bcaddcb} 
 + \beta \raisebox{-1.5 pt}{$\scriptstyle y{\hskip 0.5pt}3\td $}\; \S_2 \, y^{bcadcdb} + \beta_{y{\hskip 0.5pt}3\te} \; \S_2 \, y^{bacdcdb}
+   \beta  \raisebox{-1.5 pt}{$\scriptstyle y{\hskip 0.5pt}3\tf $} \; y^{bcdadcb} \nn \\
&{} +  \beta_{y{\hskip 0.5pt}3\tg} \; y^{bcdacdb} 
 +  \beta  \raisebox{-1.5 pt}{$\scriptstyle y{\hskip 0.5pt}3\thh$} \; \S_2\,   y^{bcbaddc} 
 + \beta_{y{\hskip 0.5pt}3\ti} \; \S_2\,   y^{baccdbd} +  \beta_{y{\hskip 0.5pt}3\tj} \;  \S_2\,  y^{bacddbc} \nn \\
&{} + \beta  \raisebox{-1.5 pt}{$\scriptstyle y{\hskip 0.5pt}3\tk$} \, \S_2 \, y^{bacbddc}   + 
\beta  \raisebox{-1.5 pt}{$\scriptstyle y{\hskip 0.5pt}3\tl$} \;  \S_2\,  y^{bacbdcd} + \beta_{y{\hskip 0.5pt}3\tm}\; \S_2\,  y^{bacdbdc} 
 +  \beta_{y{\hskip 0.5pt}3\tn}\; \S_2\,  y^{badcdbc} \nn \\
 &{} + \beta_{y{\hskip 0.5pt}3\too}\; \S_2\,  y^{bcabddc}   + \beta_{y{\hskip 0.5pt}3\tp}\; \S_2\,  y^{bcaddbc} + 
 \beta_{y{\hskip 0.5pt}3\tq}\; \S_2\,  y^{bcacdbd} + \beta_{y{\hskip 0.5pt}3\trr}\; \S_2\,  y^{bcadbdc} \nn  \\
     &{}  + \beta_{y{\hskip 0.5pt}3\ts}\; \S_2\,  y^{bcabdcd}  + \beta  \raisebox{-1.5 pt}{$\scriptstyle y{\hskip 0.5pt}3\ttt$}\;  y^{bcbadcd} + \beta_{y{\hskip 0.5pt}3\tu}\;  y^{bcdadbc} 
    + \beta_{y{\hskip 0.5pt}3\tv}\;  \S_2 \, y^{bacdbcd}  \nn \\
  &   + \beta_{y{\hskip 0.5pt}3\tw}\; \S_2\,  y^{bcdacbd}
     + \beta_{y{\hskip 0.5pt}3\tx}\;  \S_2 \, y^{bcadbcd} 
   + \beta_{y{\hskip 0.5pt}3\ty}\;  y^{bcdabcd}   + \beta_{y{\hskip 0.5pt}3\tz}\;  \S_2 \, y^{bcadcbd}  \, ,
\end{align}
with~\cite{Steudtner:2021fzs,Herren:2017uxn,Bednyakov:2021qxa,Davies:2021mnc}
\begin{align}
& \beta_{y{\hskip 0.5pt}3a}  =  - \tfrac38\, , ~~ &&  \beta_{y{\hskip 0.5pt}3b}  =  \tfrac12 \, , ~~~&&
 \beta_{y{\hskip 0.5pt}3c}  =   \tfrac32 \, , &&  \beta_{y{\hskip 0.5pt}3d}  =   \tfrac32\, , && \beta_{y{\hskip 0.5pt}3e}  =   \tfrac12 \, , \nn \\
 & \beta_{y{\hskip 0.5pt}3f}  =  2  \, , &&   \beta_{y{\hskip 0.5pt}3g}  =  3  \, , &&   
 \beta_{y{\hskip 0.5pt}3h}  =  5  \, ,  
 &&   \beta_{y{\hskip 0.5pt}3i}  =   3\, , && \beta_{y{\hskip 0.5pt}3j}  =   \tfrac12\, ,  \nn \\
 & \beta_{y{\hskip 0.5pt}3k}  =   -1 \, , && \beta_{y{\hskip 0.5pt}3l}  =  2 \, , && 
  \beta_{y{\hskip 0.5pt}3m}  =   -\tfrac12   \, , &&   \beta_{y{\hskip 0.5pt}3n}  =   2 \, , && \beta_{y{\hskip 0.5pt}3o}  =  -1 \, , \nn \\
 &  \beta_{y{\hskip 0.5pt}3p}  =  -\tfrac12   \, , && \beta_{y{\hskip 0.5pt}3q}  =  \tfrac32\, , &&
 \beta_{y{\hskip 0.5pt}3r}  =   - \tfrac32   \, , &&   \beta_{y{\hskip 0.5pt}3s}  =   -\tfrac12 \, , && \beta_{y{\hskip 0.5pt}3t}  =   \tfrac{25}{16} \, , \nn \\
&  \beta_{y{\hskip 0.5pt}3u}  =   \tfrac{25}{8}   \, , &&   \beta_{y{\hskip 0.5pt}3v}  =   \tfrac54 \, , &&   \beta_{y{\hskip 0.5pt}3w}  =   0 \, , 
&&   \beta_{y{\hskip 0.5pt}3x}  =  2 ( 3\zeta_3 - 2 )  \, ,  && \beta_{y{\hskip 0.5pt}3y}  =   - \tfrac12 \, ,  \nn \\
& \beta_{y{\hskip 0.5pt}3z}  =   - \tfrac12 \, , && \beta_{y{\hskip 0.5pt}3\ta}  =   \tfrac7{16} \, , 
&& \beta_{y{\hskip 0.5pt}3\tb}  =  -1 \, , && \beta_{y{\hskip 0.5pt}3\tc}  =  2 \, , && \beta \raisebox{-1.5 pt}{$\scriptstyle y{\hskip 0.5pt}3\td $}  =  - 1 \, , \nn \\
&  \beta_{y{\hskip 0.5pt}3\te}  =  -1 \, , && \beta  \raisebox{-1.5 pt}{$\scriptstyle y{\hskip 0.5pt}3\tf $}  =  4 \, , 
&& \beta_{y{\hskip 0.5pt}3\tg}=  6 \zeta_3 -5  \, ,  && \beta  \raisebox{-1.5 pt}{$\scriptstyle y{\hskip 0.5pt}3\thh$} =  -\tfrac 32  \, , && \beta_{y{\hskip 0.5pt}3\ti} =  - \tfrac32  \, , \nn  \\
& \beta_{y{\hskip 0.5pt}3\tj} =  \tfrac12 \, , && \beta  \raisebox{-1.5 pt}{$\scriptstyle y{\hskip 0.5pt}3\tk$} = \tfrac32 \, , && \beta  \raisebox{-1.5 pt}{$\scriptstyle y{\hskip 0.5pt}3\tl$} =  2 \, ,  && \beta_{y{\hskip 0.5pt}3\tm} =  1\, , &&
 \beta_{y{\hskip 0.5pt}3\tn}  =  -2  \,  \nn \\
& \beta_{y{\hskip 0.5pt}3\too}  =  -\tfrac12 \, ,  && \beta_{y{\hskip 0.5pt}3\tp}  =  - \tfrac32 \, , && \beta_{y{\hskip 0.5pt}3\tq}  = -3   \, ,  
&& \beta_{y{\hskip 0.5pt}3\trr}  =  3(2\zeta_3-1)    \, ,  &&  \beta_{y{\hskip 0.5pt}3\ts}  =  1  \, ,& \nn \\
&   \beta  \raisebox{-1.5 pt}{$\scriptstyle y{\hskip 0.5pt}3\ttt$}  =  -2  \, , & &  \beta_{y{\hskip 0.5pt}3\tu}  =  -3 \, ,  &&  \beta_{y{\hskip 0.5pt}3\tv}  =  3(2\zeta_3-1)  \, , 
&&  \beta_{y{\hskip 0.5pt}3\tw}  =  2( 3\zeta_3 - 1)   \, ,  && \beta_{y{\hskip 0.5pt}3\tx}  =  2( 3\zeta_3 - 1)   \, , \nn \\
&  \beta_{y{\hskip 0.5pt}3\ty}  =  2\, ,  &&  \beta_{y{\hskip 0.5pt}3\tz}  =  -4  \, .
\end{align}

Of the above two and three loop  diagrams $2a$, $2f$, $3f$, $3l$,  $3\tw$, $3\tx$, $3\ty$, $3\tz$ do not have 
subdivergences and are primitive.

\vskip 2cm

\section{Scalar Quartic Couplings}\label{sec:beta-quart}

The scalar quartic coupling is a symmetric 4 index tensor $\lambda^{abcd}$ and the $\beta$-function
has a similar decomposition as for the Yukawa coupling in \eqref{betay}
\be
\beta_{\lambda}^{\, abcd}= {\tilde \beta}_{\lambda}^{\, abcd} + \gamma_\phi^{ae} \lambda^{ebcd} +
\gamma_\phi^{be}\lambda^{aecd} + \gamma_\phi^{ce}
\lambda^{abed} + \gamma_\phi^{de} \lambda^{abce} = {\tilde \beta}_{\lambda}^{\, abcd} + 
\S_4\, \gamma_\phi^{ae} \lambda^{ebcd} 
\label{betalambda}
\ee
with ${\tilde \beta}_{\lambda}^{\, abcd} $ given in terms of 1PI diagrams and $\S_4$ here denoting the
sum over the four terms, each term with unit weight,  necessary to obtain a fully symmetric result.

At one and two loops the relevant diagrams are
\begin{figure}[ht!]
  \centering
  \begin{tabular}{ccccc}
   \begin{tikzpicture}
    \draw[black, thick, densely dashed]  (0.8em, 1.2em) -- (2em, 0em) -- (0.8em,-1.2em);
    \draw[black, thick, densely dashed]  (6.2em, 1.2em) -- (5em, 0em) -- (6.2em,-1.2em);
    \draw[black, thick, densely dashed]  (3.5em,0em) circle (1.5em);
       \node at (2em, 0em)[circle,fill,inner sep=.1em]{};
    \node at (5em, 0em)[circle,fill,inner sep=.1em]{};
    \end{tikzpicture}
    &
    \begin{tikzpicture}
    \draw[black, thick, densely dashed]  (1em, 2.2em) --  (-0.5em, 2.2em);
    \draw[black, thick, densely dashed]  (1em, -0.3em) --  (-0.5em, -0.3em);
    \draw[black, thick]  (3.5em, 2.2em) --  (3.5em, -0.3em) --  (1em, -0.3em) --
(1em, 2.2em) -- cycle;
    \draw[black, thick, densely dashed] (3.5em, 2.2em) -- (5em, 2.2em);
    \draw[black, thick, densely dashed] (3.5em, -0.3em) -- (5em, -0.3em);
        \node at (3.5em, 2.2em)[circle,fill,inner sep=.1em]{};
    \node at (3.5em, -0.3em)[circle,fill,inner sep=.1em]{};
    \node at (1em, 2.2em)[circle,fill,inner sep=.1em]{};
    \node at (1em, -0.3em)[circle,fill,inner sep=.1em]{};
    \end{tikzpicture}
        &
     \begin{tikzpicture}
    \draw[black, thick, densely dashed]  (0.8em, 1.2em) -- (2em, 0em) -- (0.8em,-1.2em);
    \draw[black, thick, densely dashed]  (3.5em,0em) circle (1.5em);
    \draw[black, thick, densely dashed]  (4.6em, 1.15em) to [bend right=60](4.6em, -1.15em);
    \draw[black, thick, densely dashed]  (4.6em, 1.15em) -- (6em, 1.7em);
    \draw[black, thick, densely dashed]  (4.6em, -1.15em) -- (6em, -1.7em);
       \node at (2em, 0em)[circle,fill,inner sep=.1em]{};
    \node at (4.6em, 1.15em)[circle,fill,inner sep=.1em]{};
    \node at (4.6em, -1.15em)[circle,fill,inner sep=.1em]{};
       \end{tikzpicture}
       &
     \begin{tikzpicture}
    \draw[black, thick, densely dashed]  (0.8em, 1.2em) -- (2em, 0em) -- (0.8em,-1.2em);
    \draw[black, thick, densely dashed]  (6.2em, 1.2em) -- (5em, 0em) -- (6.2em,-1.2em);
    \draw[black, thick, densely dashed]  (3.5em,0em) circle (1.5em);
     \draw[black, thick, fill=white]  (3.5em,1.3em) circle (0.7em);
       \node at (2em, 0em)[circle,fill,inner sep=.1em]{};
    \node at (5em, 0em)[circle,fill,inner sep=.1em]{};
     \node at (4.18em, 1.28em)[circle,fill,inner sep=.1em]{};
      \node at (2.82em, 1.28em)[circle,fill,inner sep=.1em]{};
    \end{tikzpicture}
        \\ $1a$ & $1b$ &  $2a$  & $ 2b$  \\ [0.5em]
    \begin{tikzpicture}
    \draw[black, thick, densely dashed]  (1em, 2.2em) --  (-0.4em, 0.95em);
    \draw[black, thick, densely dashed]  (1em, -0.3em) --  (-0.4em, 0.95em);
    \draw[black, thick, densely dashed]  (-0.4em, 0.95em) --+(135: 1.5em);
     \draw[black, thick, densely dashed]  (-0.4em, 0.95em) --+(225: 1.5em);
    \draw[black, thick]  (3.5em, 2.2em) --  (3.5em, -0.3em) --  (1em, -0.3em) --(1em, 2.2em) -- cycle;
    \draw[black, thick, densely dashed] (3.5em, 2.2em) -- (5em, 2.2em);
    \draw[black, thick, densely dashed] (3.5em, -0.3em) -- (5em, -0.3em);
        \node at (3.5em, 2.2em)[circle,fill,inner sep=.1em]{};
    \node at (3.5em, -0.3em)[circle,fill,inner sep=.1em]{};
    \node at (1em, 2.2em)[circle,fill,inner sep=.1em]{};
    \node at (1em, -0.3em)[circle,fill,inner sep=.1em]{};
     \node at (-0.4em, 0.95em)[circle,fill,inner sep=.1em]{};
    \end{tikzpicture}
    &
    \begin{tikzpicture}
    \draw[black, thick, densely dashed]  (1em, 2.2em) --  (-0.4em, 0.95em);
    \draw[black, thick, densely dashed]  (1em, -0.3em) --  (-0.4em, 0.95em);
    \draw[black, thick, densely dashed]  (-0.4em, 0.95em) --+(135: 1.5em);
     \draw[black, thick, densely dashed]  (-0.4em, 0.95em) --+(225: 1.5em);
   \draw[black, thick]  (1em, 2.2em) --  (3.5em, 2.2em)  (1em, 2.2em) --  (3.5em, -0.3em) 
    (1em, -0.3em) --  (3.5em, -0.3em) ;
     \fill [white] (2.25em, 0.95em) circle (0.3em) ;
     \draw[black, thick]  (1em, -0.3em) -- (3.5em, 2.2em);
    \draw[black, thick, densely dashed] (3.5em, 2.2em) -- (5em, 2.2em);
    \draw[black, thick, densely dashed] (3.5em, -0.3em) -- (5em, -0.3em);
        \node at (3.5em, 2.2em)[circle,fill,inner sep=.1em]{};
    \node at (3.5em, -0.3em)[circle,fill,inner sep=.1em]{};
    \node at (1em, 2.2em)[circle,fill,inner sep=.1em]{};
    \node at (1em, -0.3em)[circle,fill,inner sep=.1em]{};
     \node at (-0.4em, 0.95em)[circle,fill,inner sep=.1em]{};
    \end{tikzpicture}
&
     \begin{tikzpicture}
    \draw[black, thick, densely dashed]  (1em, 2.2em) --  (-0.5em, 2.2em);
    \draw[black, thick, densely dashed]  (1em, -0.3em) --  (-0.5em, -0.3em);
    \draw[black, thick]  (3.5em, 2.2em) --  (3.5em, -0.3em) --  (1em, -0.3em) --(1em, 2.2em) -- cycle;
    \draw[black, thick, densely dashed] (3.5em, 2.2em) -- (5em, 2.2em);
    \draw[black, thick, densely dashed] (3.5em, -0.3em) -- (5em, -0.3em);
     \draw[black, thick, densely dashed] (1em, 1.7em) arc (90:-90:.75em);
        \node at (3.5em, 2.2em)[circle,fill,inner sep=.1em]{};
    \node at (3.5em, -0.3em)[circle,fill,inner sep=.1em]{};
    \node at (1em,.2em)[circle,fill,inner sep=.1em]{};
    \node at (1em, 2.2em)[circle,fill,inner sep=.1em]{};
    \node at (1em, -0.3em)[circle,fill,inner sep=.1em]{};
    \node at (1em, 1.7em)[circle,fill,inner sep=.1em]{};
    \node at (1em, 0.2em)[circle,fill,inner sep=.1em]{};
    \end{tikzpicture}
    &
    \begin{tikzpicture}
    \draw[black, thick, densely dashed]  (1em, 2.2em) --  (-0.5em, 2.2em);
    \draw[black, thick, densely dashed]  (1em, -0.3em) --  (-0.5em, -0.3em);
    \draw[black, thick]  (3.5em, 2.2em) --  (3.5em, -0.3em) --  (1em, -0.3em) --(1em, 2.2em) -- cycle;
    \draw[black, thick, densely dashed] (3.5em, 2.2em) -- (5em, 2.2em);
    \draw[black, thick, densely dashed] (3.5em, -0.3em) -- (5em, -0.3em);
     \draw[black, thick, densely dashed] (1em, 0.95em) arc (270:360:1.25em);
        \node at (3.5em, 2.2em)[circle,fill,inner sep=.1em]{};
    \node at (3.5em, -0.3em)[circle,fill,inner sep=.1em]{};
     \node at (1em, 2.2em)[circle,fill,inner sep=.1em]{};
    \node at (1em, -0.3em)[circle,fill,inner sep=.1em]{};
     \node at (1em, 0.95em)[circle,fill,inner sep=.1em]{};
      \node at (2.2em, 2.25em)[circle,fill,inner sep=.1em]{};
    \end{tikzpicture}   
    &
    \begin{tikzpicture}
    \draw[black, thick, densely dashed]  (1em, 2.2em) --  (-0.5em, 2.2em);
    \draw[black, thick, densely dashed]  (1em, -0.3em) --  (-0.5em, -0.3em);
    \draw[black, thick]  (3.5em, 2.2em) --  (3.5em, -0.3em) --  (1em, -0.3em) --(1em, 2.2em) -- cycle;
    \draw[black, thick, densely dashed] (3.5em, 2.2em) -- (5em, 2.2em);
    \draw[black, thick, densely dashed] (3.5em, -0.3em) -- (5em, -0.3em);
    \draw[black, thick, densely dashed]  (1em, 0.95em) --  (3.5em, 0.95em);
        \node at (3.5em, 2.2em)[circle,fill,inner sep=.1em]{};
    \node at (3.5em, -0.3em)[circle,fill,inner sep=.1em]{};
    \node at (1em, 2.2em)[circle,fill,inner sep=.1em]{};
    \node at (1em, -0.3em)[circle,fill,inner sep=.1em]{};
    \node at (1em, 0.95em)[circle,fill,inner sep=.1em]{};
    \node at (3.5em, 0.95em)[circle,fill,inner sep=.1em]{};
    \end{tikzpicture}   
    \\ $2c$ & $2d$ &  $2e$  & $2f$ & $2g$\
   \end{tabular}
  \caption{One and two loop diagrams relevant for the scalar quartic $\beta$-function. }
  \label{fig:quart0}
\end{figure}
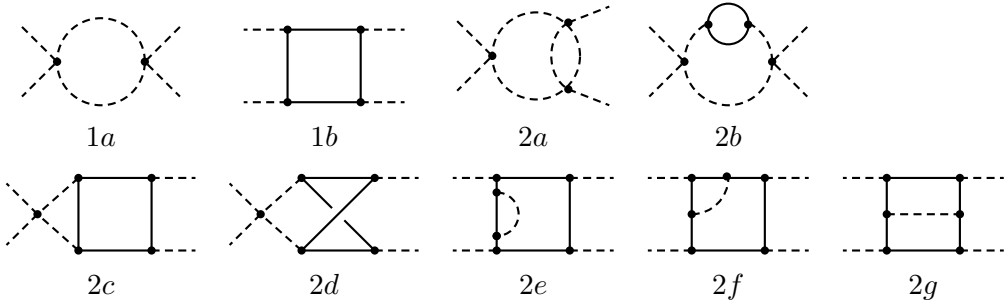
so that 
\begin{align}
 {\tilde  \beta}_{\lambda}^{(1)abcd}  = {}&\beta_{\lambda{\hskip 0.5pt}1a} \; \S_3\, \lambda^{abef} \lambda^{cdef}
  +  \beta_{\lambda{\hskip 0.5pt}1b} \; \S_3\,  \tr( y^{abcd} )\, , \nn \\
  {\tilde  \beta}_{\lambda}^{(2)abcd}  = {}& \beta_{\lambda{\hskip 0.5pt}2a} \;  \S_6\, \lambda^{abef} \lambda^{cfgh} \lambda^{degh}
 +  \beta_{\lambda{\hskip 0.5pt}2b} \;  \S_3\, \lambda^{abef} \lambda^{cdeg} \, \tr(y^{fg})  \nn \\
&{}+  \beta_{\lambda{\hskip 0.5pt}2c} \; \S_{6}  \, \lambda^{abef} \tr(y^{efcd})  
+  \beta_{\lambda{\hskip 0.5pt}2d} \; \S_6  \, \lambda^{abef} \tr(y^{ecfd})   \nn \\
&{}   +  \beta_{\lambda{\hskip 0.5pt}2e} \; \S_{12} \, \tr(y^{aeebcd}) + \beta_{\lambda{\hskip 0.5pt}2f} \; \S_{12} \, \tr(y^{aebcde}) 
 +  \beta_{\lambda{\hskip 0.5pt}2g} \; \S_6\,  \tr(y^{eabecd}) \, ,
\end{align}
with $\S_n$ denoting the sum over $n$ terms necessary to achieve symmetrisation over all 
permutations of  $a,b,c,d$.
Historic results, with our conventions, give
\begin{align}
&\beta_{\lambda{\hskip 0.5pt}1a} = 1\, , && \beta_{\lambda{\hskip 0.5pt}1b} =  - 4  \, , && &&
\beta_{\lambda{\hskip 0.5pt}2a} = - 1\, , && \beta_{\lambda{\hskip 0.5pt}2b} = - 1\, ,    \nn \\
& \beta_{\lambda{\hskip 0.5pt}2c} = 0 \, , && \beta_{\lambda{\hskip 0.5pt}2d} = 2 \, , &&
\beta_{\lambda{\hskip 0.5pt}2e} = 2\, , && \beta_{\lambda{\hskip 0.5pt}2f} = 4  \, , && \beta_{\lambda{\hskip 0.5pt}2g} = 4\, .
\end{align}

At  three loops there are   ${\rm O}(\lambda^4, \lambda^3y^2,
\lambda^2,y^4,\lambda y^6, y^8)$ contributions to the scalar quartic $\beta$-function.
For our discussion it is convenient to isolate related sets of diagrams out of a total of 62.
The ${\rm O}(\lambda^4)$ purely scalar contribution corresponds to the  diagrams
\FloatBarrier
\begin{figure}[ht!]
  \centering
  \begin{tabular}{cccccc}
  \raisebox{0.3em}{
   \begin{tikzpicture}
    \draw[black, thick, densely dashed]  (0.8em, 1.2em) -- (2em, 0em) -- (0.8em,
-1.2em);
    \draw[black, thick, densely dashed]  (6.2em, 1.2em) -- (5em, 0em) -- (6.2em,
-1.2em);
\draw [white] (3.5em,0em) --+  (35:1.5em) coordinate (n1);
\draw  [white] (3.5em,0em) --+  (145:1.5em)  coordinate (n2);
    \draw[black, thick, densely dashed]  (3.5em,0em) circle (1.5em);
    \draw[black, thick, densely dashed]  (n1) -- (n2) ;
 \draw[black, thick, densely dashed]  (n1)  arc (325:215:1.5em)  ;
       \node at (2em, 0em)[circle,fill,inner sep=.1em]{};
    \node at (5em, 0em)[circle,fill,inner sep=.1em]{};
     \node at (n1)[circle,fill,inner sep=.1em]{};
      \node at (n2)[circle,fill,inner sep=.1em]{};
    \end{tikzpicture}}
    &
     \begin{tikzpicture}
    \draw[black, thick, densely dashed]  (0.8em, 1.2em) -- (2em, 0em) -- (0.8em,
-1.2em);
\draw [white] (3.5em,0em) --+  (43:1.5em) coordinate (n1);
\draw  [white] (3.5em,0em) --+  (137:1.5em)  coordinate (n2);
\draw  [white] (3.5em,0em) --+  (-43:1.5em)  coordinate (n3);
    \draw[black, thick, densely dashed]  (3.5em,0em) circle (1.5em);
 \draw[black, thick, densely dashed]  (n1)  arc (317:213:1.5em)  ;
  \draw[black, thick, densely dashed]  (n1) --+ (35:1.5em);
    \draw[black, thick, densely dashed]  (n3) --+ (-35:1.5em);
     \draw[black, thick, densely dashed]  (n2) -- (n3);
       \node at (2em, 0em)[circle,fill,inner sep=.1em]{};
    \node at (n3)[circle,fill,inner sep=.1em]{};
     \node at (n1)[circle,fill,inner sep=.1em]{};
      \node at (n2)[circle,fill,inner sep=.1em]{};
    \end{tikzpicture}
    & 
    \raisebox{0.4em}{
   \begin{tikzpicture}
    \draw[black, thick, densely dashed]  (0.8em, 1.2em) -- (2em, 0em) -- (0.8em,-1.2em);
    \draw[black, thick, densely dashed]  (6.2em, 1.2em) -- (5em, 0em) -- (6.2em,-1.2em);
    \draw[black, thick, densely dashed]  (3.5em,0em) circle (1.5em);
     \draw[black, thick, densely dashed]  (3.5em,-1.5em) to [out=45, in=315]  (3.5em,1.5em);
      \draw[black, thick, densely dashed]  (3.5em,-1.5em) to [out=135, in=225]  (3.5em,1.5em);
       \node at (2em, 0em)[circle,fill,inner sep=.1em]{};
   \node at (5em, 0em)[circle,fill,inner sep=.1em]{};
   \node at (3.5em, -1.5em)[circle,fill,inner sep=.1em]{};
   \node at (3.5em, 1.5em)[circle,fill,inner sep=.1em]{};
    \end{tikzpicture}}
   &
   \begin{tikzpicture}
    \draw[black, thick, densely dashed]  (0.8em, 1.2em) -- (2em, 0em) -- (0.8em,-1.2em);
     \draw[white]  (2em, 0em) --+ (-40:2em) coordinate (n1);
  \draw[white]  (2em, 0em) --+ (40:2em) coordinate (n2);
   \draw[black, thick, densely dashed]  (2em, 0em) --+ (40:3.2em);
   \draw[black, thick, densely dashed]  (2em, 0em) --+ (-40:3.2em);
  \draw [black, thick, densely dashed] (n1) to  [out=40, in=320]  (3.532em,0) ;
   \draw [black, thick, densely dashed] (n1) to  [out=135, in=225]  (3.532em,0) ;
    \draw [black, thick, densely dashed] (3.532em,0)  to  [out=40, in=320]  (n2);
   \draw [black, thick, densely dashed] (3.532em,0)  to  [out=140, in=220]  (n2);
       \node at (2em, 0em)[circle,fill,inner sep=.1em]{};
    \node at (3.532em, 0em)[circle,fill,inner sep=.1em]{};
    \node at (n1)[circle,fill,inner sep=.1em]{};
    \node at (n2)[circle,fill,inner sep=.1em]{};
    \end{tikzpicture}
     &
    \begin{tikzpicture}
    \draw[black, thick, densely dashed]  (1em, 2.2em) --  (-0.5em, 2.8em);
    \draw[black, thick, densely dashed]  (1em, -0.3em) --  (-0.5em, -1.1em);
    \draw[black, thick, densely dashed]   (3.5em, -0.3em) --  (1em, -0.3em);
    \draw[black, thick, densely dashed]   (3.5em, 2.2em) --  (1em, 2.2em);
    \draw[black, thick, densely dashed] (3.5em, 2.2em) -- (5em, 2.8em);
    \draw[black, thick, densely dashed] (3.5em, -0.3em) -- (5em, -1.1em);
     \draw[black, thick, densely dashed] (1em, -0.3em)  to  [out=40, in=320] (1em,  2.2em);
      \draw[black, thick, densely dashed] (1em, -0.3em)  to  [out=140, in=220]  (1em, 2.2em);
       \draw[black, thick, densely dashed] (3.5em, -0.3em)  to  [out=40, in=320] (3.5em,  2.2em);
      \draw[black, thick, densely dashed] (3.5em, -0.3em)  to  [out=140, in=220]  (3.5em, 2.2em);
        \node at (3.5em, 2.2em)[circle,fill,inner sep=.1em]{};
    \node at (3.5em, -0.3em)[circle,fill,inner sep=.1em]{};
    \node at (1em, 2.2em)[circle,fill,inner sep=.1em]{};
    \node at (1em, -0.3em)[circle,fill,inner sep=.1em]{};
    \end{tikzpicture}
     &
    \begin{tikzpicture}
    \draw[black, thick, densely dashed]  (1em, 2.2em) --  (-0.5em, 2.8em);
    \draw[black, thick, densely dashed]  (1em, -0.3em) --  (-0.5em, -1.1em);
    \draw[black, thick, densely dashed]  (3.5em, 2.2em) --  (3.5em, -0.3em) --  (1em, -0.3em) --
(1em, 2.2em) -- cycle;
    \draw[black, thick, densely dashed] (3.5em, 2.2em) -- (5em, 2.8em);
    \draw[black, thick, densely dashed] (3.5em, -0.3em) -- (5em, -1.1em);
     \draw[black, thick, densely dashed] (1em, 2.2em) -- (3.5em, -0.3em);
\fill [white]  (2.25em, 0.95em) circle (0.3em);
      \draw[black, thick, densely dashed] (1em, -0.3em) -- (3.5em, 2.2em);
        \node at (3.5em, 2.2em)[circle,fill,inner sep=.1em]{};
    \node at (3.5em, -0.3em)[circle,fill,inner sep=.1em]{};
    \node at (1em, 2.2em)[circle,fill,inner sep=.1em]{};
    \node at (1em, -0.3em)[circle,fill,inner sep=.1em]{};
    \end{tikzpicture}
   \\ $3a$ & $3b$ &  $3c$ & $3d$ & $3e$ & $3f$ 
   \end{tabular}
  \caption{Three  loop diagrams involving the quartic  scalar coupling contributions to the scalar $\beta$-function. }
  \label{fig:quartA}
\end{figure}
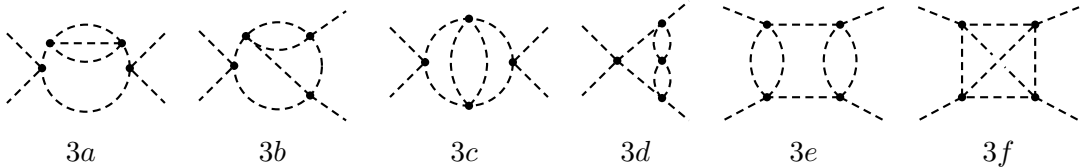 
\FloatBarrier
\noindent
so that 
  \begin{align}
      {\tilde  \beta}_{\lambda{\hskip 0.5pt}A}^{(3)abcd}  = {}&
    \beta_{\lambda{\hskip 0.5pt}3a}\;  \S_3\, \lambda^{abef} \lambda^{ehij} \lambda^{ghij}  \lambda^{cdfg} + 
    \beta_{\lambda{\hskip 0.5pt}3b}\;  \S_{12} \,  \lambda^ {abef}\lambda^{cegh} \lambda^{fgij} \lambda^{dhij} \nn \\
 &{}   +  \beta_{\lambda{\hskip 0.5pt}3c}\;  \S_3\,\lambda^{abef} \lambda^{egij}\lambda^{fhij} \lambda^{cdgh} 
 +  \beta_{\lambda{\hskip 0.5pt}3d}\;  \S_6\, \lambda^{abef} \lambda^{cegh} \lambda^{ghij} \lambda^{dfij} \nn \\
 &    {} +  \beta_{\lambda{\hskip 0.5pt}3e}\;  \S_6\,   \lambda^{afge} \lambda^{bfgh} \lambda^{ceij}\lambda^{dhij} 
 {} +  \beta_{\lambda{\hskip 0.5pt}3f}\;   \lambda^{aefg}\lambda^{behi} \lambda^{cfhj} \lambda^ {dgij} \, .
  \end{align}
  For the purely scalar case the general three loop coefficients have long been known:
  \be
 \beta_{\lambda{\hskip 0.5pt}3a} = - \tfrac38\, , \quad  \beta_{\lambda{\hskip 0.5pt}3b} = 2\, , \quad 
 \beta_{\lambda{\hskip 0.5pt}3c} = \tfrac12\, , \quad \beta_{\lambda{\hskip 0.5pt}3d} = - \tfrac12\, , \quad 
 \beta_{\lambda{\hskip 0.5pt}3e} = - \tfrac12\, , \quad \beta_{\lambda{\hskip 0.5pt}3f} = 12 \zeta_3 \, .
 \ee
  
Diagrams involving two or one insertions of fermion bubbles into internal scalar propagator lines in
 one or two loop diagrams are just
 \FloatBarrier
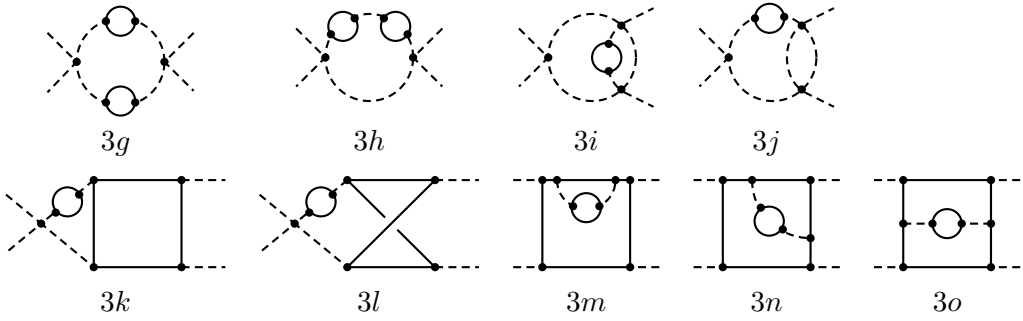
\begin{figure}[ht!]
  \centering
  \begin{tabular}{ccccc}
   \raisebox{0.5em}{
    \begin{tikzpicture}
    \draw[black, thick, densely dashed]  (1.em, 1.5em) -- (2em, 0.5em) -- (1em,-0.55em);
    \draw[black, thick, densely dashed]  (6.em, 1.5em) -- (5em, 0.5em) -- (6em,-0.5em);
    \draw[black, thick, densely dashed]  (3.5em,0.5em) circle (1.5em);
    \draw[black, thick, fill=white] (3.5em, 1.9em) circle (.5em);
    \draw[black, thick, fill=white] (3.5em, -0.85em) circle (.5em);
    \node at (2em, 0.5em)[circle,fill,inner sep=.1em]{};
    \node at (5em, 0.5em)[circle,fill,inner sep=.1em]{};
    \node at (3em, 1.9em)[circle,fill,inner sep=.1em]{};
    \node at (4em, 1.9em)[circle,fill,inner sep=.1em]{};
    \node at (3em, -0.9em)[circle,fill,inner sep=.1em]{};
    \node at (4em, -0.9em)[circle,fill,inner sep=.1em]{};
    \end{tikzpicture}}
     &
    \begin{tikzpicture}
    \draw[black, thick, densely dashed]  (1.em, 1em) -- (2em, 0em) -- (1em,-1em);
    \draw[black, thick, densely dashed]  (6.em, 1em) -- (5em, 0em) -- (6em,-1em);
    \draw[black, thick, densely dashed]  (3.5em,0em) circle (1.5em);
    \draw [white] (3.5em,0) --+(50:1.4em) coordinate (n1);
    \draw  [white] (3.5em,0) --+(130:1.4em) coordinate (n2);
    \draw[black, thick, fill=white] (n1) circle (.5em);
    \draw[black, thick, fill=white] (n2) circle (.5em);
    \draw[transparent, thick] (4em, -2em) circle (.5em);
    \node at (2em, 0em)[circle,fill,inner sep=.1em]{};
    \node at (5em, 0em)[circle,fill,inner sep=.1em]{};
    \node at (3em, 1.35em)[circle,fill,inner sep=.1em]{};
    \node at (4.em, 1.35em)[circle,fill,inner sep=.1em]{};
    \node at (2.19em, 0.8em)[circle,fill,inner sep=.1em]{};
   \node at (4.81em, 0.8em)[circle,fill,inner sep=.1em]{};   
    \end{tikzpicture}
&
    \begin{tikzpicture}
    \draw[black, thick, densely dashed]  (1.em, 1em) -- (2em, 0em) -- (1em,-1em);
    \draw[black, thick, densely dashed]  (3.5em,0em) circle (1.5em);
    \draw[black, thick, densely dashed]  (4.6em, 1.2em) to [bend right=60](4.6em, -1.2em);
    \draw[black, thick, densely dashed]  (4.6em, 1.2em) -- (5.6em, 1.7em);
    \draw[black, thick, densely dashed]  (4.6em, -1.2em) -- (5.6em, -1.7em);
    \draw[black, thick, fill=white] (4em, 0em) circle (.5em);
    \draw[transparent, thick] (4em, -2em) circle (.5em);
    \node at (2em, 0em)[circle,fill,inner sep=.1em]{};
    \node at (4.5em, 1.1em)[circle,fill,inner sep=.1em]{};
    \node at (4.5em, -1.1em)[circle,fill,inner sep=.1em]{};
    \node at (4.07em, .47em)[circle,fill,inner sep=.1em]{};
    \node at (4.07em, -.47em)[circle,fill,inner sep=.1em]{};
    \end{tikzpicture}
    &
    \begin{tikzpicture}
    \draw[black, thick, densely dashed]  (1.em, 1em) -- (2em, 0em) -- (1em,-1em);
    \draw[black, thick, densely dashed]  (3.5em,0em) circle (1.5em);
    \draw[black, thick, densely dashed]  (4.6em, 1.2em) to [bend right=60](4.6em, -1.2em);
    \draw[black, thick, densely dashed]  (4.6em, 1.2em) -- (5.6em, 1.7em);
    \draw[black, thick, densely dashed]  (4.6em, -1.2em) -- (5.6em, -1.7em);
    \draw[black, thick, fill=white] (3.4em, 1.35em) circle (.5em);
    \draw[transparent, thick] (4em, -2em) circle (.5em);
    \node at (2em, 0em)[circle,fill,inner sep=.1em]{};
    \node at (4.5em, 1.1em)[circle,fill,inner sep=.1em]{};
    \node at (4.5em, -1.1em)[circle,fill,inner sep=.1em]{};
    \node at (2.9em, 1.35em)[circle,fill,inner sep=.1em]{};
    \node at (3.9em, 1.4em)[circle,fill,inner sep=.1em]{};
    \end{tikzpicture}
      \\ \noalign{\vskip -8pt}  $3g$ & $3h$ &  $3i$ & $3j$   \\ [0.5em]
    \begin{tikzpicture}
    \draw[black, thick, densely dashed]  (0.em, 1em) -- (1.2em, 0em) -- (0em,-1em);
    \draw[black, thick, densely dashed]  (6em, 1.5em) -- (7.5em, 1.5em);
    \draw[black, thick, densely dashed]  (6em, -1.5em) -- (7.5em, -1.5em);
    \draw[black, thick, densely dashed] (3.em, 1.5em) -- (1.2em, 0em) -- (3.em,-1.5em);
    \draw[black, thick] (3.em, 1.5em) -- (6.em, 1.5em) -- (6.em, -1.5em) --(3.em, -1.5em) -- cycle;
    \draw[black, thick, fill=white] (2.1em, .75em) circle (.5em);
    \node at (1.2em, 0em)[circle,fill,inner sep=.1em]{};
    \node at (6em, -1.5em)[circle,fill,inner sep=.1em]{};
    \node at (6em, 1.5em)[circle,fill,inner sep=.1em]{};
    \node at (3em, -1.5em)[circle,fill,inner sep=.1em]{};
    \node at (3em, 1.5em)[circle,fill,inner sep=.1em]{};
    \node at (1.7em, .38em) [circle,fill,inner sep=.1em]{};
    \node at (2.5em, 1em) [circle,fill,inner sep=.1em]{};
    \end{tikzpicture}
    &
     \begin{tikzpicture}
    \draw[black, thick, densely dashed]  (0.em, 1em) -- (1.2em, 0em) -- (0em,-1em);
    \draw[black, thick, densely dashed]  (6em, 1.5em) -- (7.5em, 1.5em);
    \draw[black, thick, densely dashed]  (6em, -1.5em) -- (7.5em, -1.5em);
    \draw[black, thick, densely dashed] (3.em, 1.5em) -- (1.2em, 0em) -- (3.em,-1.5em);
      \draw[black, thick] (3.em, 1.5em) -- (6.em, -1.5em) ;
       \draw[black, thick] (3.em, 1.5em) -- (6.em, 1.5em);
        \draw[black, thick] (3.em,- 1.5em) -- (6.em, -1.5em);
        \fill [white]  (4.5em, 0em) circle (0.3em);
        \draw[black, thick] (3.em,- 1.5em) -- (6.em, 1.5em);
    \draw[black, thick, fill=white] (2.1em, .75em) circle (.5em);
    \node at (1.2em, 0em)[circle,fill,inner sep=.1em]{};
    \node at (6em, -1.5em)[circle,fill,inner sep=.1em]{};
    \node at (6em, 1.5em)[circle,fill,inner sep=.1em]{};
    \node at (3em, -1.5em)[circle,fill,inner sep=.1em]{};
    \node at (3em, 1.5em)[circle,fill,inner sep=.1em]{};
    \node at (1.7em, .38em) [circle,fill,inner sep=.1em]{};
    \node at (2.5em, 1em) [circle,fill,inner sep=.1em]{};
    \end{tikzpicture}
    &
    \begin{tikzpicture}
    \draw[black, thick, densely dashed]  (2.em, 1.5em) -- (3em, 1.5em);
    \draw[black, thick, densely dashed]  (7.em, 1.5em) -- (6em, 1.5em);
    \draw[black, thick, densely dashed]  (2.em, -1.5em) -- (3em, -1.5em);
    \draw[black, thick, densely dashed]  (7.em, -1.5em) -- (6em, -1.5em);
    \draw[black, thick] (3.em, 1.5em) -- (6.em, 1.5em) -- (6.em, -1.5em) --(3.em, -1.5em) -- cycle;
    \draw[black, thick, densely dashed]  (3.5em, 1.5em) arc (180:360:1.em);
    \draw[black, thick, fill=white] (4.5em, .55em) circle (.5em);
    \node at (6em, -1.5em)[circle,fill,inner sep=.1em]{};
    \node at (6em, 1.5em)[circle,fill,inner sep=.1em]{};
    \node at (3em, -1.5em)[circle,fill,inner sep=.1em]{};
    \node at (3em, 1.5em)[circle,fill,inner sep=.1em]{};
    \node at (3.5em, 1.5em)[circle,fill,inner sep=.1em]{};
    \node at (5.5em, 1.5em)[circle,fill,inner sep=.1em]{};
    \node at (4.05em, .6em)[circle,fill,inner sep=.1em]{};
    \node at (4.95em, .6em)[circle,fill,inner sep=.1em]{};
    \end{tikzpicture}
    &
    \begin{tikzpicture}
    \draw[black, thick, densely dashed]  (2.em, 1.5em) -- (3em, 1.5em);
    \draw[black, thick, densely dashed]  (7.em, 1.5em) -- (6em, 1.5em);
    \draw[black, thick, densely dashed]  (2.em, -1.5em) -- (3em, -1.5em);
    \draw[black, thick, densely dashed]  (7.em, -1.5em) -- (6em, -1.5em);
    \draw[black, thick] (3.em, 1.5em) -- (6.em, 1.5em) -- (6.em, -1.5em) --(3.em, -1.5em) -- cycle;
    \draw[black, thick, densely dashed]  (4em, 1.5em) arc (180:270:2em);
    \draw[black, thick, fill=white] (4.586em, .086em) circle (.5em);
    \node at (6em, -1.5em)[circle,fill,inner sep=.1em]{};
    \node at (6em, 1.5em)[circle,fill,inner sep=.1em]{};
    \node at (3em, -1.5em)[circle,fill,inner sep=.1em]{};
    \node at (3em, 1.5em)[circle,fill,inner sep=.1em]{};
    \node at (4em, 1.5em)[circle,fill,inner sep=.1em]{};
    \node at (6em,- 0.5em)[circle,fill,inner sep=.1em]{};
    \node at (4.3em, 0.55em)[circle,fill,inner sep=.1em]{};
    \node at (5.05em, -0.2em)[circle,fill,inner sep=.1em]{};
    \end{tikzpicture} 
    &
    \begin{tikzpicture}
    \draw[black, thick, densely dashed]  (2.em, 1.5em) -- (3em, 1.5em);
    \draw[black, thick, densely dashed]  (7.em, 1.5em) -- (6em, 1.5em);
    \draw[black, thick, densely dashed]  (2.em, -1.5em) -- (3em, -1.5em);
    \draw[black, thick, densely dashed]  (7.em, -1.5em) -- (6em, -1.5em);
    \draw[black, thick] (3.em, 1.5em) -- (6.em, 1.5em) -- (6.em, -1.5em) --(3.em, -1.5em) -- cycle;
    \draw[black, thick, densely dashed]  (3em, 0em) -- (6em, 0em);
    \draw[black, thick, fill=white] (4.5em, 0em) circle (.5em);
    \node at (6em, -1.5em)[circle,fill,inner sep=.1em]{};
    \node at (6em, 1.5em)[circle,fill,inner sep=.1em]{};
    \node at (3em, -1.5em)[circle,fill,inner sep=.1em]{};
    \node at (3em, 1.5em)[circle,fill,inner sep=.1em]{};
    \node at (4em, 0em)[circle,fill,inner sep=.1em]{};
    \node at (5em, 0em)[circle,fill,inner sep=.1em]{};
    \node at (3em, 0em)[circle,fill,inner sep=.1em]{};
    \node at (6em, 0em)[circle,fill,inner sep=.1em]{};
    \end{tikzpicture}
          \\ $3k$ & $3l$ &  $3m$ & $3n$ & $3o$ 
   \end{tabular}
  \caption{Three  loop diagrams involving fermion bubble contributions to the scalar $\beta$-function. }
  \label{fig:quartB}
\end{figure} 
\FloatBarrier
\noindent
The corresponding contributions are then
\begin{align}
   {\tilde \beta}^{abcd}_{\lambda{\hskip 0.5pt}B} = {}& 
   \beta_{\lambda{\hskip 0.5pt}3g} \; \S_3 \,\lambda^{abef} \lambda^{cdgh}   \,\tr(y^{eg}) \,\tr(y^{fh})
+\beta_{\lambda{\hskip 0.5pt}3h} \; \S_3 \,\lambda^{abef} \lambda^{cdeg}  \,\tr(y^{fh})\,\tr(y^{hg}) \nn \\
&{} + \beta_{\lambda{\hskip 0.5pt}3i} \;\S_6\,  \lambda^{abef} \lambda^{cegh} \lambda^{dfgi} \, \tr(y^{hi})
   +  \beta_{\lambda{\hskip 0.5pt}3j} \; \S_{12} \, \lambda^{abef} \lambda^{cegh} \lambda^{dghi} \, \tr(y^{fi}) \nn  \\
 &{} +\beta_{\lambda{\hskip 0.5pt}3k} \; \S_{12}  \, \lambda^{abef} \tr(y^{fg}) \,\tr(y^{gecd})
+  \beta_{\lambda{\hskip 0.5pt}3l} \; \S_{6}  \,   \lambda^{abef}\tr(y^{fg}) \,\tr(y^{gced}) \nn \\
 &{}  + \S_{12} \big ( \beta_{\lambda{\hskip 0.5pt}3m} \;  \tr(y^{abcdef}) + \beta_{\lambda{\hskip 0.5pt}3n}\; \tr(y^{abcedf}) \big ) \tr(y^{ef}) \nn \\
&{}  +  \beta_{\lambda{\hskip 0.5pt}3o} \;  \S_6 \, \tr(y^{abecdf}) \,  \tr(y^{ef}) \, ,
  \end{align}
  with
  \begin{align}
&  \beta_{\lambda{\hskip 0.5pt}3g} = - \tfrac14  \, , && \beta_{\lambda{\hskip 0.5pt}3h} = - \tfrac12  \, , &&
   \beta_{\lambda{\hskip 0.5pt}3i} = 2  \, , && \beta_{\lambda{\hskip 0.5pt}3j} = -\tfrac12  \, , && \beta_{\lambda{\hskip 0.5pt}3k} = 3  \, , 
   \quad \beta_{\lambda{\hskip 0.5pt}3l} = 2  \, , \nn \\
 &  \beta_{\lambda{\hskip 0.5pt}3m} = -\tfrac{25}{8}   \, , && \beta_{\lambda{\hskip 0.5pt}3n} = - 4  \, , && \beta_{\lambda{\hskip 0.5pt}3o} = -3  \, .
   \end{align}

There are further  ${\rm O}(\lambda^2)$ diagrams which are
\FloatBarrier
\begin{figure}[hbt!]
  \centering
  \begin{tabular}{ccccc}
  \begin{tikzpicture}
    \draw[black, thick, densely dashed]  (1.em, 1em) -- (2em, 0em) -- (1em,-1em);
    \draw[black, thick, densely dashed]  (6.6em, 1em) -- (5.6em, 0em) -- (6.6em,-1em);
    \draw[black, thick, densely dashed]  (3.8em,0em) circle (1.8em);
    \draw[black, thick, fill=white] (3.8em, -1.25em) circle (1em);
    \draw[black, thick, densely dashed] (2.87em, -0.95em) to [bend right](4.73em, -0.95em);
    \node at (2em, 0em)[circle,fill,inner sep=.1em]{};
    \node at (5.6em, 0em)[circle,fill,inner sep=.1em]{};
    \node at (2.87em, -1.5em)[circle,fill,inner sep=.1em]{};
    \node at (4.73em, -1.5em)[circle,fill,inner sep=.1em]{};
    \node at (2.87em, -0.95em)[circle,fill,inner sep=.1em]{};
    \node at (4.73em, -0.95em)[circle,fill,inner sep=.1em]{};
    \end{tikzpicture}
    &
    \begin{tikzpicture}
    \draw[black, thick, densely dashed]  (1.em, 1em) -- (2em, 0em) -- (1em,-1em);
    \draw[black, thick, densely dashed]  (6.6em, 1em) -- (5.6em, 0em) -- (6.6em,-1em);
    \draw[black, thick, densely dashed]  (3.8em,0em) circle (1.8em);
    \draw[black, thick, fill=white] (3.8em, -1.25em) circle (1em);
    \draw[black, thick, densely dashed] (3.8em, -0.3em) -- (3.8em, -2.2em);
    \node at (2em, 0em)[circle,fill,inner sep=.1em]{};
    \node at (5.6em, 0em)[circle,fill,inner sep=.1em]{};
    \node at (2.87em, -1.5em)[circle,fill,inner sep=.1em]{};
    \node at (4.73em, -1.5em)[circle,fill,inner sep=.1em]{};
    \node at (3.8em, -0.25em)[circle,fill,inner sep=.1em]{};
    \node at (3.8em, -2.2em)[circle,fill,inner sep=.1em]{};
    \end{tikzpicture}
    &
         \begin{tikzpicture}
    \draw[black, thick, densely dashed]  (1.em, 1em) -- (2em, 0em) -- (1em,-1em);
    \draw[black, thick, densely dashed]  (3.8em,0em) circle (1.8em);
    \draw [white]  (3.8em,0em) --+ (45:1.8em)  coordinate (n1);
     \draw  [white]  (3.8em,0em) --+ (-45:1.7em)  coordinate (n2);
    \draw[black, thick, densely dashed]  (n1) --+ (30: 2.2em);
     \draw[black, thick, fill=white] (n2) circle (1em);
     \draw  [white]  (n2) --+ (-25:1em)  coordinate (n3);
      \draw [white]   (n2) --+ (55:1em)  coordinate (n4);
       \draw  [white]  (n2) --+ (130:1em)  coordinate (n5);
        \draw [white]  (n2) --+ (215:1em)  coordinate (n6);
     \draw[black, thick, densely dashed]  (n1) to [bend right=45](n5);
    \draw[black, thick, densely dashed]  (n3) --+ (330:1.3em);
    \node at (2em, 0em)[circle,fill,inner sep=.1em]{};
    \node at (n1)[circle,fill,inner sep=.1em]{};
    \node at (n5)[circle,fill,inner sep=.1em]{};
    \node at (n3)[circle,fill,inner sep=.1em]{};
    \node at (n4)[circle,fill,inner sep=.1em]{};
    \node at (n6)[circle,fill,inner sep=.1em]{};
    \end{tikzpicture}
    &
         \begin{tikzpicture}
    \draw[black, thick, densely dashed]  (1.em, 1em) -- (2em, 0em) -- (1em,-1em);
    \draw[black, thick, densely dashed]  (3.8em,0em) circle (1.8em);
    \draw [white]  (3.8em,0em) --+ (45:1.8em)  coordinate (n1);
     \draw  [white]  (3.8em,0em) --+ (-45:1.7em)  coordinate (n2);
    \draw[black, thick, densely dashed]  (n1) --+ (30: 2.2em);
     \fill [white] (n2) circle (1em);
     \draw  [white]  (n2) --+ (-10:1em)  coordinate (n3);
      \draw [white]   (n2) --+ (55:1em)  coordinate (n4);
       \draw  [white]  (n2) --+ (270:1em)  coordinate (n5);
        \draw [white]  (n2) --+ (215:1em)  coordinate (n6);
         \draw [white]  (n2) --+ (133:1em)  coordinate (n7);
     \draw[black, thick, densely dashed]  (n1) to [bend right=45](n5);
     \fill [white] (n7) circle (0.3em);
       \draw[black, thick] (n2) circle (1em);
    \draw[black, thick, densely dashed]  (n3) --+ (330:1.3em);
    \node at (2em, 0em)[circle,fill,inner sep=.1em]{};
    \node at (n1)[circle,fill,inner sep=.1em]{};
    \node at (n5)[circle,fill,inner sep=.1em]{};
    \node at (n3)[circle,fill,inner sep=.1em]{};
    \node at (n4)[circle,fill,inner sep=.1em]{};
    \node at (n6)[circle,fill,inner sep=.1em]{};
    \end{tikzpicture}
&
       \\ $3p$ & $3q$  & $3r$  & $3s$   \\ [0.5em]
    \begin{tikzpicture}
    \draw[black, thick, densely dashed]  (1.em, 1em) -- (2em, 0em) -- (1em,-1em);
    \draw[black, thick, densely dashed]  (8.em, 1em) -- (7em, 0em) -- (8em,-1em);
    \draw[black, thick, densely dashed] (3.em, 1.5em) -- (2em, 0em) -- (3.em,-1.5em);
    \draw[black, thick, densely dashed] (6.em, 1.5em) -- (7em, 0em) -- (6.em,-1.5em);
    \draw[black, thick] (3.em, 1.5em) -- (6.em, 1.5em) -- (6.em, -1.5em) --(3.em, -1.5em) -- cycle;
    \draw[transparent, thick] (4em, -2em) circle (.5em);
    \node at (2em, 0em)[circle,fill,inner sep=.1em]{};
    \node at (7em, 0em)[circle,fill,inner sep=.1em]{};
    \node at (6em, -1.5em)[circle,fill,inner sep=.1em]{};
    \node at (6em, 1.5em)[circle,fill,inner sep=.1em]{};
    \node at (3em, -1.5em)[circle,fill,inner sep=.1em]{};
    \node at (3em, 1.5em)[circle,fill,inner sep=.1em]{};
    \end{tikzpicture}
    &
    \begin{tikzpicture}
    \draw[black, thick, densely dashed]  (1.em, 1em) -- (2em, 0em) -- (1em,-1em);
    \draw[black, thick, densely dashed]  (8.em, 1em) -- (7em, 0em) -- (8em,-1em);
    \draw[black, thick, densely dashed] (3.em, 1.5em) -- (2em, 0em) -- (3.em,-1.5em);
    \draw[black, thick, densely dashed] (6.em, 1.5em) -- (7em, 0em) -- (6.em,-1.5em);
\draw[black, thick]  (3em, 1.5em) --  (6em, -1.5em) ;
   \draw[black, thick]  (3em, 1.5em)  --(6em, 1.5em);
 \fill [white]  (4.5em, 0em) circle (0.3em);
 \draw[black, thick]  (6em, 1.5em) --  (3em, -1.5em) ;
 \draw[black, thick]  (6em,- 1.5em) --  (3em, -1.5em) ;
      \draw[transparent, thick] (4em, -2em) circle (.5em);
    \node at (2em, 0em)[circle,fill,inner sep=.1em]{};
    \node at (7em, 0em)[circle,fill,inner sep=.1em]{};
    \node at (6em, -1.5em)[circle,fill,inner sep=.1em]{};
    \node at (6em, 1.5em)[circle,fill,inner sep=.1em]{};
    \node at (3em, -1.5em)[circle,fill,inner sep=.1em]{};
    \node at (3em, 1.5em)[circle,fill,inner sep=.1em]{};
    \end{tikzpicture}
    &
    \begin{tikzpicture}
    \draw[black, thick, densely dashed]  (-.5em, 1.5em) --  (2em, 1.5em);
    \draw[black, thick, densely dashed]  (-.5em, -1.5em) --  (2em, -1.5em);
    \draw[black, thick]  (2em, 1.5em) --  (2em, -1.5em) --  (5em, -1.5em) --(5em, 1.5em) -- cycle;
    \draw[black, thick, densely dashed] (5em, 1.5em) -- (6em, 1.5em);
    \draw[black, thick, densely dashed] (5em, -1.5em) -- (6em, -1.5em);
    \draw[black, thick, densely dashed] (.5em, 1.5em) to [bend right=60 ] (.5em,-1.5em) 
    to [bend right=60] (.5em, 1.5em);
    \draw[transparent, thick] (4em, -2em) circle (.5em);
    \node at (.5em, 1.5em)[circle,fill,inner sep=.1em]{};
    \node at (.5em, -1.5em)[circle,fill,inner sep=.1em]{};
    \node at (5em, 1.5em)[circle,fill,inner sep=.1em]{};
    \node at (5em, -1.5em)[circle,fill,inner sep=.1em]{};
    \node at (2em, 1.5em)[circle,fill,inner sep=.1em]{};
    \node at (2em, -1.5em)[circle,fill,inner sep=.1em]{};
    \end{tikzpicture}
    &
    \begin{tikzpicture}
    \draw[black, thick, densely dashed]  (-.5em, 1.5em) --  (2em, 1.5em);
    \draw[black, thick, densely dashed]  (-.5em, -1.5em) --  (2em, -1.5em);
    \draw[black, thick]  (2em, 1.5em) --  (5em, -1.5em) ;
   \draw[black, thick]  (2em, 1.5em)  --(5em, 1.5em);
 \fill [white]  (3.5em, 0em) circle (0.3em);
 \draw[black, thick]  (5em, 1.5em) --  (2em, -1.5em) ;
   \draw[black, thick]  (2em, -1.5em)  --(5em, -1.5em);
    \draw[black, thick, densely dashed] (5em, 1.5em) -- (6em, 1.5em);
    \draw[black, thick, densely dashed] (5em, -1.5em) -- (6em, -1.5em);
    \draw[black, thick, densely dashed] (.5em, 1.5em) to [bend right=60] (.5em,-1.5em)
    to [bend right=60] (.5em, 1.5em);
    \draw[transparent, thick] (4em, -2em) circle (.5em);
    \node at (.5em, 1.5em)[circle,fill,inner sep=.1em]{};
    \node at (.5em, -1.5em)[circle,fill,inner sep=.1em]{};
    \node at (5em, 1.5em)[circle,fill,inner sep=.1em]{};
    \node at (5em, -1.5em)[circle,fill,inner sep=.1em]{};
    \node at (2em, 1.5em)[circle,fill,inner sep=.1em]{};
    \node at (2em, -1.5em)[circle,fill,inner sep=.1em]{};
    \end{tikzpicture}
    \\ $3t$  & $3u$  & $3v$  & $3w$ \\
              \end{tabular}
              \vskip- 12pt
  \caption{Three-loop ${\rm O}(\lambda^2)$ diagrams contributing to the scalar $\beta$-function. }
  \label{fig:quartC}
\end{figure}
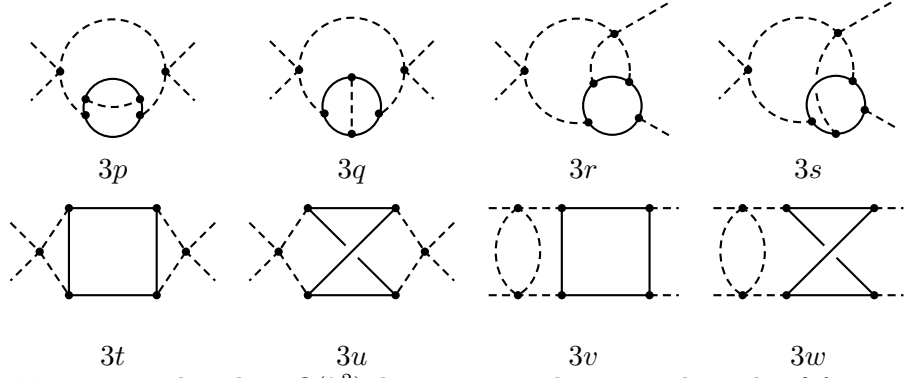

\noindent
\FloatBarrier
which give
  \begin{align}
   {\tilde \beta}^{abcd}_{\lambda{\hskip 0.5pt}C} = {}& \S_3\,  \lambda^{abef} \lambda^{cdeg} 
\big ( \beta_{\lambda{\hskip 0.5pt}3p} \,  \tr(y^{fghh}) + \beta_{\lambda{\hskip 0.5pt}3q} \, \tr(y^{fhgh}) \big  ) \nn  \\
\noalign{\vskip 2pt}
&{} +  \S_{12}  \, \lambda^{abef} \lambda^{cegh} \big ( \beta_{\lambda{\hskip 0.5pt}3r} \,  \tr(y^{dfgh})
+ \beta_{\lambda{\hskip 0.5pt}3s}  \,\tr(y^{dgfh}) \big  )  \nn \\
\noalign{\vskip 2pt}
& {} + \S_{3}\,  \lambda^{abef} \lambda^{cdgh}
\big  (  \beta_{\lambda{\hskip 0.5pt}3t}  \, \tr(y^{efgh})  +  \beta_{\lambda{\hskip 0.5pt}3u}  \, \tr(y^{egfh})  \big  ) \nn \\
\noalign{\vskip 2pt}
     &{} +\S_{12} \,  \lambda^{aefg} \lambda^{befh}  \beta_{\lambda{\hskip 0.5pt}3v}\,  \tr(y^{cdgh}) + 
     \S_{6} \, \lambda^{aefg} \lambda^{befh}  
     \beta_{\lambda{\hskip 0.5pt}3w}\,\tr(y^{cgdh}  \big ) \, .
  \end{align}

  Diagrams with a single scalar vertex are
  \FloatBarrier 
    \begin{figure}[ht!]
  \centering
  \begin{tabular}{ccccc}
   \begin{tikzpicture}
    \draw[black, thick, densely dashed]  (6em, 1.5em) -- (7.5em, 1.5em);
    \draw[black, thick, densely dashed]  (6em, -1.5em) -- (7.5em, -1.5em);
    \draw[black, thick, densely dashed]  (6em, 0em) -- (7.5em, 0em);
    \draw[black, thick, densely dashed] (3.em, 1.5em) -- (2em, 0em) -- (3.em,-1.5em);
    \draw[black, thick, densely dashed] (1em, 0em) -- (3.em, 0em);
    \draw[black, thick] (3.em, 1.5em) -- (6.em, 1.5em) -- (6.em, -1.5em) --(3.em, -1.5em) -- cycle;
    \node at (2em, 0em)[circle,fill,inner sep=.1em]{};
    \node at (3em, 0em)[circle,fill,inner sep=.1em]{};
    \node at (6em, 0em)[circle,fill,inner sep=.1em]{};
    \node at (6em, -1.5em)[circle,fill,inner sep=.1em]{};
    \node at (6em, 1.5em)[circle,fill,inner sep=.1em]{};
    \node at (3em, -1.5em)[circle,fill,inner sep=.1em]{};
    \node at (3em, 1.5em)[circle,fill,inner sep=.1em]{};
    \end{tikzpicture}
        &
    \begin{tikzpicture}
    \draw[black, thick, densely dashed]  (1.em, 1em) -- (2em, 0em) -- (1em,-1em);
    \draw[black   , thick, densely dashed]  (6em, 1.5em) -- (7.5em, 1.5em);
    \draw[black, thick, densely dashed]  (6em, -1.5em) -- (7.5em, -1.5em);
    \draw[black, thick, densely dashed] (3.em, 1.5em) -- (2em, 0em) -- (3.em,-1.5em);
    \draw[black, thick] (3.em, 1.5em) -- (6.em, 1.5em) -- (6.em, -1.5em) --(3.em, -1.5em) -- cycle;
    \draw[black, thick, densely dashed] (3em, .75em) arc (90:-90:.75em);
    \node at (2em, 0em)[circle,fill,inner sep=.1em]{};
    \node at (6em, -1.5em)[circle,fill,inner sep=.1em]{};
    \node at (6em, 1.5em)[circle,fill,inner sep=.1em]{};
    \node at (3em, -1.5em)[circle,fill,inner sep=.1em]{};
    \node at (3em, 1.5em)[circle,fill,inner sep=.1em]{};
    \node at (3em, .75em)[circle,fill,inner sep=.1em]{};
    \node at (3em, -.75em)[circle,fill,inner sep=.1em]{};
    \end{tikzpicture}
    &    
    \begin{tikzpicture}
    \draw[black, thick, densely dashed]  (1.em, 1em) -- (2em, 0em) -- (1em,-1em);
    \draw[black, thick, densely dashed]  (6em, 1.5em) -- (7.5em, 1.5em);
    \draw[black, thick, densely dashed]  (6em, -1.5em) -- (7.5em, -1.5em);
    \draw[black, thick, densely dashed] (3.em, 1.5em) -- (2em, 0em) -- (3.em,-1.5em);
    \draw[black, thick] (3.em, 1.5em) -- (6.em, 1.5em) -- (6.em, -1.5em) --(3.em, -1.5em) -- cycle;
    \draw[black, thick, densely dashed] (6em, .75em) arc (90:270:.75em);
    \node at (2em, 0em)[circle,fill,inner sep=.1em]{};
    \node at (6em, -1.5em)[circle,fill,inner sep=.1em]{};
    \node at (6em, 1.5em)[circle,fill,inner sep=.1em]{};
    \node at (3em, -1.5em)[circle,fill,inner sep=.1em]{};
    \node at (3em, 1.5em)[circle,fill,inner sep=.1em]{};
    \node at (6em, .75em)[circle,fill,inner sep=.1em]{};
    \node at (6em, -.75em)[circle,fill,inner sep=.1em]{};
    \end{tikzpicture}
      &
        \begin{tikzpicture}
    \draw[black, thick, densely dashed]  (1.em, 1em) -- (2em, 0em) -- (1em,-1em);
    \draw[black, thick, densely dashed]  (6em, 1.5em) -- (7.5em, 1.5em);
    \draw[black, thick, densely dashed]  (6em, -1.5em) -- (7.5em, -1.5em);
    \draw[black, thick, densely dashed] (3.em, 1.5em) -- (2em, 0em) -- (3.em,-1.5em);
    \draw[black, thick] (3.em, 1.5em) -- (6.em, 1.5em) -- (6.em, -1.5em) --(3.em, -1.5em) -- cycle;
    \draw[black, thick, densely dashed] (3.75em, -1.5em) arc (180:0:.75em);
    \node at (2em, 0em)[circle,fill,inner sep=.1em]{};
    \node at (6em, -1.5em)[circle,fill,inner sep=.1em]{};
    \node at (6em, 1.5em)[circle,fill,inner sep=.1em]{};
    \node at (3em, -1.5em)[circle,fill,inner sep=.1em]{};
    \node at (3em, 1.5em)[circle,fill,inner sep=.1em]{};
    \node at (5.25em, -1.5em)[circle,fill,inner sep=.1em]{};
    \node at (3.75em, -1.5em)[circle,fill,inner sep=.1em]{};
    \end{tikzpicture}
        \\ $3x$ &  $3y$ & $3z$  & $3\ta$ \\ [0.5em]
            \begin{tikzpicture}
    \draw[black, thick, densely dashed]  (1.em, 1em) -- (2em, 0em) -- (1em,-1em);
    \draw[black, thick, densely dashed]  (6em, 1.5em) -- (7.5em, 1.5em);
    \draw[black, thick, densely dashed]  (6em, -1.5em) -- (7.5em, -1.5em);
    \draw[black, thick, densely dashed] (3.em, 1.5em) -- (2em, 0em) -- (3.em,-1.5em);
    \draw[black, thick] (3.em, 1.5em) -- (6.em, 1.5em) -- (6.em, -1.5em) --(3.em, -1.5em) -- cycle;
    \draw[black, thick, densely dashed] (3em, 0em) arc (90:0:1.5em);
    \node at (2em, 0em)[circle,fill,inner sep=.1em]{};
    \node at (6em, -1.5em)[circle,fill,inner sep=.1em]{};
    \node at (6em, 1.5em)[circle,fill,inner sep=.1em]{};
    \node at (3em, -1.5em)[circle,fill,inner sep=.1em]{};
    \node at (3em, 1.5em)[circle,fill,inner sep=.1em]{};
    \node at (3em, 0em)[circle,fill,inner sep=.1em]{};
    \node at (4.5em, -1.5em)[circle,fill,inner sep=.1em]{};
    \end{tikzpicture}
    &
    \begin{tikzpicture}
    \draw[black, thick, densely dashed]  (1.em, 1em) -- (2em, 0em) -- (1em,-1em);
    \draw[black, thick, densely dashed]  (6em, 1.5em) -- (7.5em, 1.5em);
    \draw[black, thick, densely dashed]  (6em, -1.5em) -- (7.5em, -1.5em);
    \draw[black, thick, densely dashed] (3.em, 1.5em) -- (2em, 0em) -- (3.em,-1.5em);
    \draw[black, thick] (3.em, 1.5em) -- (6.em, 1.5em) -- (6.em, -1.5em) --(3.em, -1.5em) -- cycle;
    \draw[black, thick, densely dashed] (6em, 0em) arc (90:180:1.5em);
    \node at (2em, 0em)[circle,fill,inner sep=.1em]{};
    \node at (6em, -1.5em)[circle,fill,inner sep=.1em]{};
    \node at (6em, 1.5em)[circle,fill,inner sep=.1em]{};
    \node at (3em, -1.5em)[circle,fill,inner sep=.1em]{};
    \node at (3em, 1.5em)[circle,fill,inner sep=.1em]{};
    \node at (6em, 0em)[circle,fill,inner sep=.1em]{};
    \node at (4.5em, -1.5em)[circle,fill,inner sep=.1em]{};
    \end{tikzpicture}
    &
    \begin{tikzpicture}
    \draw[black, thick, densely dashed]  (1.em, 1em) -- (2em, 0em) -- (1em,-1em);
    \draw[black, thick, densely dashed]  (6em, 1.5em) -- (7.5em, 1.5em);
    \draw[black, thick, densely dashed]  (6em, -1.5em) -- (7.5em, -1.5em);
    \draw[black, thick, densely dashed] (3.em, 1.5em) -- (2em, 0em) -- (3.em,-1.5em);
    \draw[black, thick] (3.em, 1.5em) -- (6.em, -1.5em);
     \draw[black, thick] (3.em, 1.5em) -- (6.em, 1.5em);
     \draw[black, thick] (3.em, -1.5em) -- (6.em, -1.5em);
     \fill [white] (4.5em, 0em) circle (0.3em);
   \draw[black, thick] (3.em, -1.5em) -- (6.em, 1.5em);
    \draw[black, thick, densely dashed] (4em, 1.5em) arc (180:360:.5em);
    \node at (2e m, 0em)[circle,fill,inner sep=.1em]{};
    \node at (6em, -1.5em)[circle,fill,inner sep=.1em]{};
    \node at (6em, 1.5em)[circle,fill,inner sep=.1em]{};
    \node at (3em, -1.5em)[circle,fill,inner sep=.1em]{};
    \node at (3em, 1.5em)[circle,fill,inner sep=.1em]{};
    \node at (4em, 1.5em) [circle,fill,inner sep=.1em]{};
    \node at (5em, 1.5em) [circle,fill,inner sep=.1em]{};
    \end{tikzpicture}
    &
    \begin{tikzpicture}
    \draw[black, thick, densely dashed]  (1.em, 1em) -- (2em, 0em) -- (1em,-1em);
    \draw[black, thick, densely dashed]  (6em, 1.5em) -- (7.5em, 1.5em);
    \draw[black, thick, densely dashed]  (6em, -1.5em) -- (7.5em, -1.5em);
    \draw[black, thick, densely dashed] (3.em, 1.5em) -- (2em, 0em) -- (3.em,-1.5em);
    \draw[black, thick] (3.em, 1.5em) -- (6.em, -1.5em);
     \draw[black, thick] (3.em, 1.5em) -- (6.em, 1.5em);
     \draw[black, thick] (3.em, -1.5em) -- (6.em, -1.5em);
     \fill [white] (4.5em, 0em) circle (0.3em);
     \draw[black, thick] (3.em, -1.5em) -- (6.em, 1.5em);
    \draw[black, thick, densely dashed] (4.5em, 1.5em) to [bend right] (5.25em,.75em);
    \node at (2em, 0em)[circle,fill,inner sep=.1em]{};
    \node at (6em, -1.5em)[circle,fill,inner sep=.1em]{};
    \node at (6em, 1.5em)[circle,fill,inner sep=.1em]{};
    \node at (3em, -1.5em)[circle,fill,inner sep=.1em]{};
    \node at (3em, 1.5em)[circle,fill,inner sep=.1em]{};
    \node at (4.5em, 1.5em) [circle,fill,inner sep=.1em]{};
    \node at (5.25em, .75em) [circle,fill,inner sep=.1em]{};
    \end{tikzpicture}
        &
    \begin{tikzpicture}
    \draw[black, thick, densely dashed]  (1.em, 1em) -- (2em, 0em) -- (1em,-1em);
    \draw[black, thick, densely dashed]  (6em, 1.5em) -- (7.5em, 1.5em);
    \draw[black, thick, densely dashed]  (6em, -1.5em) -- (7.5em, -1.5em);
    \draw[black, thick, densely dashed] (3.em, 1.5em) -- (2em, 0em) -- (3.em,-1.5em);
    \draw[black, thick] (3.em, 1.5em) -- (6.em, -1.5em);
     \draw[black, thick] (3.em, 1.5em) -- (6.em, 1.5em);
     \draw[black, thick] (3.em, -1.5em) -- (6.em, -1.5em);
     \fill [white] (4.5em, 0em) circle (0.3em);
   \draw[black, thick] (3.em, -1.5em) -- (6.em, 1.5em);
    \draw[black, thick, densely dashed] (4.5em, 1.5em) to [bend left] (3.75em,.75em);
    \node at (2em, 0em)[circle,fill,inner sep=.1em]{};
    \node at (6em, -1.5em)[circle,fill,inner sep=.1em]{};
    \node at (6em, 1.5em)[circle,fill,inner sep=.1em]{};
    \node at (3em, -1.5em)[circle,fill,inner sep=.1em]{};
    \node at (3em, 1.5em)[circle,fill,inner sep=.1em]{};
    \node at (4.5em, 1.5em) [circle,fill,inner sep=.1em]{};
    \node at (3.75em, .75em) [circle,fill,inner sep=.1em]{};
     \end{tikzpicture}
            \\  $3\tb$ &  $3\tc$ &  $3\td $ &$3\te $ & $ 3\tf $ \\ [0.5em]
               \begin{tikzpicture}
    \draw[black, thick, densely dashed]  (1.em, 1em) -- (2em, 0em) -- (1em,-1em);
    \draw[black, thick, densely dashed]  (6em, 1.5em) -- (7.5em, 1.5em);
    \draw[black, thick, densely dashed]  (6em, -1.5em) -- (7.5em, -1.5em);
    \draw[black, thick, densely dashed] (3.em, 1.5em) -- (2em, 0em) -- (3.em,-1.5em);
    \draw[black, thick] (3.em, 1.5em) -- (6.em, -1.5em);
     \draw[black, thick] (3.em, 1.5em) -- (6.em, 1.5em);
     \draw[black, thick] (3.em, -1.5em) -- (6.em, -1.5em);
     \fill [white] (4.5em, 0em) circle (0.3em);
   \draw[black, thick] (3.em, -1.5em) -- (6.em, 1.5em);
    \draw[black, thick, densely dashed] (5.25em, .75em) to (3.75em, .75em);
    \node at (2em, 0em)[circle,fill,inner sep=.1em]{};
    \node at (6em, -1.5em)[circle,fill,inner sep=.1em]{};
    \node at (6em, 1.5em)[circle,fill,inner sep=.1em]{};
    \node at (3em, -1.5em)[circle,fill,inner sep=.1em]{};
    \node at (3em, 1.5em)[circle,fill,inner sep=.1em]{};
    \node at (5.25em, .75em) [circle,fill,inner sep=.1em]{};
    \node at (3.75em, .75em) [circle,fill,inner sep=.1em]{};
    \end{tikzpicture}
    &
        \begin{tikzpicture}
    \draw[black, thick, densely dashed]  (1.em, 1em) -- (2em, 0em) -- (1em,-1em);
    \draw[black, thick, densely dashed]  (6em, 1.5em) -- (7.5em, 1.5em);
    \draw[black, thick, densely dashed]  (6em, -1.5em) -- (7.5em, -1.5em);
    \draw[black, thick, densely dashed] (3.em, 1.5em) -- (2em, 0em) -- (3.em,-1.5em);
    \draw[black, thick] (3.em, 1.5em) -- (6.em, 1.5em) -- (6.em, -1.5em) --(3.em, -1.5em) -- cycle;
    \draw[black, thick, densely dashed] (4.5em, 1.5em) -- (4.5em, -1.5em);
    \node at (2em, 0em)[circle,fill,inner sep=.1em]{};
    \node at (6em, -1.5em)[circle,fill,inner sep=.1em]{};
    \node at (6em, 1.5em)[circle,fill,inner sep=.1em]{};
    \node at (3em, -1.5em)[circle,fill,inner sep=.1em]{};
    \node at (3em, 1.5em)[circle,fill,inner sep=.1em]{};
    \node at (4.5em, 1.5em)[circle,fill,inner sep=.1em]{};
    \node at (4.5em, -1.5em)[circle,fill,inner sep=.1em]{};
    \end{tikzpicture}
    &
    \begin{tikzpicture}
    \draw[black, thick, densely dashed]  (1.em, 1em) -- (2em, 0em) -- (1em,-1em);
    \draw[black, thick, densely dashed]  (6em, 1.5em) -- (7.5em, 1.5em);
    \draw[black, thick, densely dashed]  (6em, -1.5em) -- (7.5em, -1.5em);
    \draw[black, thick, densely dashed] (3.em, 1.5em) -- (2em, 0em) -- (3.em,-1.5em);
    \draw[black, thick] (3.em, 1.5em) -- (6.em, 1.5em) -- (6.em, -1.5em) --(3.em, -1.5em) -- cycle;
    \draw[black, thick, densely dashed] (3em, 0em) -- (6em, 0em);
    \node at (2em, 0em)[circle,fill,inner sep=.1em]{};
    \node at (6em, -1.5em)[circle,fill,inner sep=.1em]{};
    \node at (6em, 1.5em)[circle,fill,inner sep=.1em]{};
    \node at (3em, -1.5em)[circle,fill,inner sep=.1em]{};
    \node at (3em, 1.5em)[circle,fill,inner sep=.1em]{};
    \node at (3em, 0em)[circle,fill,inner sep=.1em]{};
    \node at (6em, 0em)[circle,fill,inner sep=.1em]{};
    \end{tikzpicture}
    &
        \begin{tikzpicture}
    \draw[black, thick, densely dashed]  (6em, 1.5em) -- (7.8em, 1.5em);
    \draw[black, thick, densely dashed]  (6em, -1.5em) -- (7.8em, -1.5em);
    \draw[black, thick, densely dashed]  (6em, 0em) -- (7.8em, 0em);
    \draw[black, thick, densely dashed] (3.em, 1.5em) -- (1.4 em, 0. em) -- (3.em,-1.5em);
    \draw[black, thick, densely dashed] (0em, 0em) -- (3.em, 0.em);
    \draw[black, thick] (6.em, 1.5em)  --  (3.em, 1.5em)  ;
    \draw[black, thick,name path=a ]  (3.em, 1.5em) -- (6.em, 0em)  ;
    \draw[white, thick, name path=b]  (3.em,0em)-- (6.em, 1.5em)  ;
      \path [name intersections={of=a and b,by=in}];
            \node[fill=white, inner sep=3pt, rotate=45] at (in) {};
      \draw[black, thick]  (3.em,0em)-- (6.em, 1.5em)  ;
    \draw[white, thick,name path=c ]  (3.em,-1.5em)-- (6.em, 0.em)  ;
    \draw[black, thick]  (3.em,-1.5em)-- (6.em, -1.5em);
    \draw[black, thick,name path=d ]  (6.em, -1.5em) -- (3.em, 0em) ;
     \path [name intersections={of=c and d,by=in2}];
            \node[fill=white, inner sep=3pt, rotate=45] at (in2) {};
     \draw[black, thick]  (3.em,-1.5em)-- (6.em, 0.em)  ;
    \node at (1.4em, 0em)[circle,fill,inner sep=.1em]{};
    \node at (3em, 0em)[circle,fill,inner sep=.1em]{};
    \node at (6em, 0em)[circle,fill,inner sep=.1em]{};
    \node at (6em, -1.5em)[circle,fill,inner sep=.1em]{};
    \node at (6em, 1.5em)[circle,fill,inner sep=.1em]{};
    \node at (3em, -1.5em)[circle,fill,inner sep=.1em]{};
    \node at (3em, 1.5em)[circle,fill,inner sep=.1em]{};
    \end{tikzpicture}       
    &
     \begin{tikzpicture}
    \draw[black, thick, densely dashed]  (6em, 1.5em) -- (7.5em, 1.5em);
    \draw[black, thick, densely dashed]  (6em, -1.5em) -- (7.5em, -1.5em);
    \draw[black, thick, densely dashed]  (6em, 0em) -- (7.5em, 0em);
    \draw[black, thick, densely dashed] (3.em, 1.5em) -- (2em, 0em) -- (3.em,-1.5em);
    \draw[black, thick, densely dashed] (0.5em, 0em) -- (3.em, 0em);
     \draw[black, thick] (3.em, 1.5em)--(6.em, -1.5em) ;
      \filldraw [white] (4.5em,0em) circle [radius=2.5pt]; 
     \draw[black, thick] (3.em, 1.5em)--(6.em, 1.5em)  -- (6.em, 0em) --  (3.em, 0em)--(3.em, -1.5em) -- (6.em,-1.5em);
    \node at (2em, 0em)[circle,fill,inner sep=.1em]{};
    \node at (3em, 0em)[circle,fill,inner sep=.1em]{};
    \node at (6em, 0em)[circle,fill,inner sep=.1em]{};
    \node at (6em, -1.5em)[circle,fill,inner sep=.1em]{};
    \node at (6em, 1.5em)[circle,fill,inner sep=.1em]{};
    \node at (3em, -1.5em)[circle,fill,inner sep=.1em]{};
    \node at (3em, 1.5em)[circle,fill,inner sep=.1em]{};
    \end{tikzpicture}
    \\  $3\tg$&  $3\thh$ &  $3\ti$ & $3\tj$ & $3\tk$
      \end{tabular}
  \caption{Three-loop diagrams containing one scalar vertex and one fermion loop.}
  \label{fig:quartD}
\end{figure}
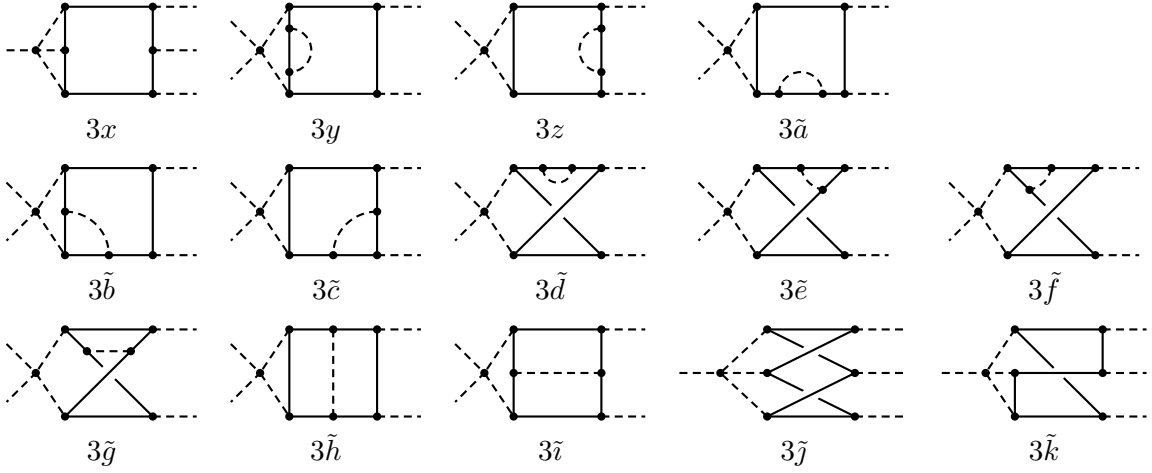
\FloatBarrier 
\noindent
These correspond to
  \begin{align}
   {\tilde  \beta}^{abcd}_{\lambda{\hskip 0.5pt}D} = 
   {}& \beta_{\lambda{\hskip 0.5pt}3x} \; \S_{12} \,\lambda^{aefg}  \,\tr(y^{bcdefg}) +
   \S_6 \, \lambda^{abef} \big (  \beta_{\lambda{\hskip 0.5pt}3y} \,\tr(y^{cdeggf}) + \beta_{\lambda{\hskip 0.5pt}3z} \, \tr(y^{cggdef}) \big ) \nn  \\
   &{}+ \S_{12} \,\lambda^{abef}  \big (  \beta_{\lambda{\hskip 0.5pt}3\ta} \; \tr(y^{cdggef}) +  \beta_{\lambda{\hskip 0.5pt}3\tb} \; \tr(y^{cdgegf}) 
   +  \beta_{\lambda{\hskip 0.5pt}3\tc} \; \tr(y^{efgcgd}) \big )\nn  \\
    & + \S_{12}\,   \lambda^{abef}\, \big (  \beta  \raisebox{-1.5 pt}{$\scriptstyle \lambda{\hskip 0.5pt}3\td$} \;  \tr(y^{cedggf}) 
    + \beta_{\lambda{\hskip 0.5pt}3\te} \; \tr(y^{cegdgf}) \big )  
    +   \beta_{\lambda{\hskip 0.5pt}3\tf} \; \S_6 \, \lambda^{abef} \,\tr(y^{cgegdf})\nn  \\
     &{}+ \S_{6} \,\lambda^{abef}  \big (  \beta_{\lambda{\hskip 0.5pt}3\tg} \; \tr(y^{cegdfg}) +  \beta_{\lambda{\hskip 0.5pt}3\thh} \; \tr(y^{cdgefg}) 
   +  \beta_{\lambda{\hskip 0.5pt}3\ti} \; \tr(y^{egfcgd}) \big )\nn  \\
     &{} +  \beta_{\lambda{\hskip 0.5pt}3\tj} \;  \S_4 \, \lambda^{aefg}  \,\tr(y^{becfdg}) 
     +  \beta  \raisebox{-1.5 pt}{$\scriptstyle \lambda{\hskip 0.5pt}3\tk$} \;  \S_{24} \, \lambda^{aefg}  \,\tr(y^{bcedfg}) \, .
     \end{align}

The remaining diagrams have no quartic scalar vertex. Those which involve two fermion loops are just
\FloatBarrier
  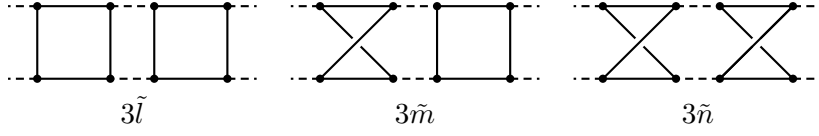
\begin{figure}[ht]
  \centering
\begin{tabular}{cccccc}
      \begin{tikzpicture}
    \draw[black, thick, densely dashed] (1em, 2.2em) -- (0em, 2.2em);
    \draw[black, thick, densely dashed] (1em, -0.3em) -- (0em, -0.3em);
    \draw[black, thick, densely dashed] (3.5em, 2.2em) -- (5em, 2.2em);
    \draw[black, thick, densely dashed] (5em, -0.3em) -- (3.5em, -0.3em);
    \draw[black, thick, densely dashed] (7.5em, 2.2em) -- (8.5em, 2.2em);
    \draw[black, thick, densely dashed] (7.5em, -0.3em) -- (8.5em, -0.3em);
     \draw[black, thick] (1em, 2.2em) -- (3.5em, 2.2em) -- (3.5em, -0.3em) -- (1em,-0.3em) -- cycle;
     \draw[black, thick] (7.5em, 2.2em) -- (7.5em, -0.3em) -- (5em, -0.3em) -- (5em,2.2em) -- cycle;
    \node at (1em, -0.3em)[circle,fill,inner sep=.1em]{};
    \node at (1em, 2.2em)[circle,fill,inner sep=.1em]{};
    \node at (3.5em, -0.3em)[circle,fill,inner sep=.1em]{};
    \node at (3.5em, 2.2em)[circle,fill,inner sep=.1em]{};
    \node at (5em, 2.2em)[circle,fill,inner sep=.1em]{};
    \node at (5em, -0.3em)[circle,fill,inner sep=.1em]{};
    \node at (7.5em, 2.2em)[circle,fill,inner sep=.1em]{};
    \node at (7.5em, -0.3em)[circle,fill,inner sep=.1em]{};
    \end{tikzpicture}
    &
      \begin{tikzpicture}
    \draw[black, thick, densely dashed] (1em, 2.2em) -- (0em, 2.2em);
    \draw[black, thick, densely dashed] (1em, -0.3em) -- (0em, -0.3em);
    \draw[black, thick, densely dashed] (3.5em, 2.2em) -- (5em, 2.2em);
    \draw[black, thick, densely dashed] (5em, -0.3em) -- (3.5em, -0.3em);
    \draw[black, thick, densely dashed] (7.5em, 2.2em) -- (8.5em, 2.2em);
    \draw[black, thick, densely dashed] (7.5em, -0.3em) -- (8.5em, -0.3em);
     \draw[black, thick] (1em, 2.2em) -- (3.5em, 2.2em);
    \draw[black, thick] (1em, -0.3em) -- (3.5em, -0.3em);
     \draw[black, thick] (1em, 2.2em) -- (3.5em, -0.3em);
      \fill [white] (2.25em,0.95em) circle (0.2em);
       \draw[black, thick] (3.5em, 2.2em) -- (1em, -0.3em);
     \draw[black, thick] (7.5em, 2.2em) -- (7.5em, -0.3em) -- (5em, -0.3em) -- (5em,2.2em) -- cycle;
    \node at (1em, -0.3em)[circle,fill,inner sep=.1em]{};
    \node at (1em, 2.2em)[circle,fill,inner sep=.1em]{};
    \node at (3.5em, -0.3em)[circle,fill,inner sep=.1em]{};
    \node at (3.5em, 2.2em)[circle,fill,inner sep=.1em]{};
    \node at (5em, 2.2em)[circle,fill,inner sep=.1em]{};
    \node at (5em, -0.3em)[circle,fill,inner sep=.1em]{};
    \node at (7.5em, 2.2em)[circle,fill,inner sep=.1em]{};
    \node at (7.5em, -0.3em)[circle,fill,inner sep=.1em]{};
    \end{tikzpicture}
    &
       \begin{tikzpicture}
    \draw[black, thick, densely dashed] (1em, 2.2em) -- (0em, 2.2em);
    \draw[black, thick, densely dashed] (1em, -0.3em) -- (0em, -0.3em);
    \draw[black, thick, densely dashed] (3.5em, 2.2em) -- (5em, 2.2em);
    \draw[black, thick, densely dashed] (5em, -0.3em) -- (3.5em, -0.3em);
    \draw[black, thick, densely dashed] (7.5em, 2.2em) -- (8.5em, 2.2em);
    \draw[black, thick, densely dashed] (7.5em, -0.3em) -- (8.5em, -0.3em);
     \draw[black, thick] (1em, 2.2em) -- (3.5em, 2.2em);
    \draw[black, thick] (1em, -0.3em) -- (3.5em, -0.3em);
     \draw[black, thick] (1em, 2.2em) -- (3.5em, -0.3em);
      \fill [white] (2.25em,0.95em) circle (0.2em);
       \draw[black, thick] (3.5em, 2.2em) -- (1em, -0.3em);
       \draw[black, thick] (7.5em, 2.2em) -- (5em, -0.3em);
          \draw[black, thick] (5em, 2.2em) -- (7.5em, 2.2em);
    \draw[black, thick] (5em, -0.3em) -- (7.5em, -0.3em);
     \draw[black, thick] (5em, 2.2em) -- (7.5em, -0.3em);
      \fill [white] (6.25em,0.95em) circle (0.2em);
       \draw[black, thick] (7.5em, 2.2em) -- (5em, -0.3em);
    \node at (1em, -0.3em)[circle,fill,inner sep=.1em]{};
    \node at (1em, 2.2em)[circle,fill,inner sep=.1em]{};
    \node at (3.5em, -0.3em)[circle,fill,inner sep=.1em]{};
    \node at (3.5em, 2.2em)[circle,fill,inner sep=.1em]{};
    \node at (5em, 2.2em)[circle,fill,inner sep=.1em]{};
    \node at (5em, -0.3em)[circle,fill,inner sep=.1em]{};
    \node at (7.5em, 2.2em)[circle,fill,inner sep=.1em]{};
    \node at (7.5em, -0.3em)[circle,fill,inner sep=.1em]{};
    \end{tikzpicture}
        \\
\end{tabular}
  \caption{Double fermion loop diagrams without scalar vertex.}
  \label{fig:quartE}
  \end{figure}
  \FloatBarrier
  \be
    {\tilde  \beta}^{abcd}_{\lambda{\hskip 0.5pt}E} = \S_6 \big ( \beta  \raisebox{-1.5 pt}{$\scriptstyle \lambda{\hskip 0.5pt}3\tl$} \; \tr(y^{abef}) + 
     \beta_{\lambda{\hskip 0.5pt}3\tm} \; \tr( y^{aebf} ) \big ) \,\tr(y^{cdef}) 
    +   \beta_{\lambda{\hskip 0.5pt}3\tn} \, \S_3\, \tr(y^{aebf}) \,\tr(y^{cedf}) \, .
\ee

The final set of  diagrams is then
\FloatBarrier
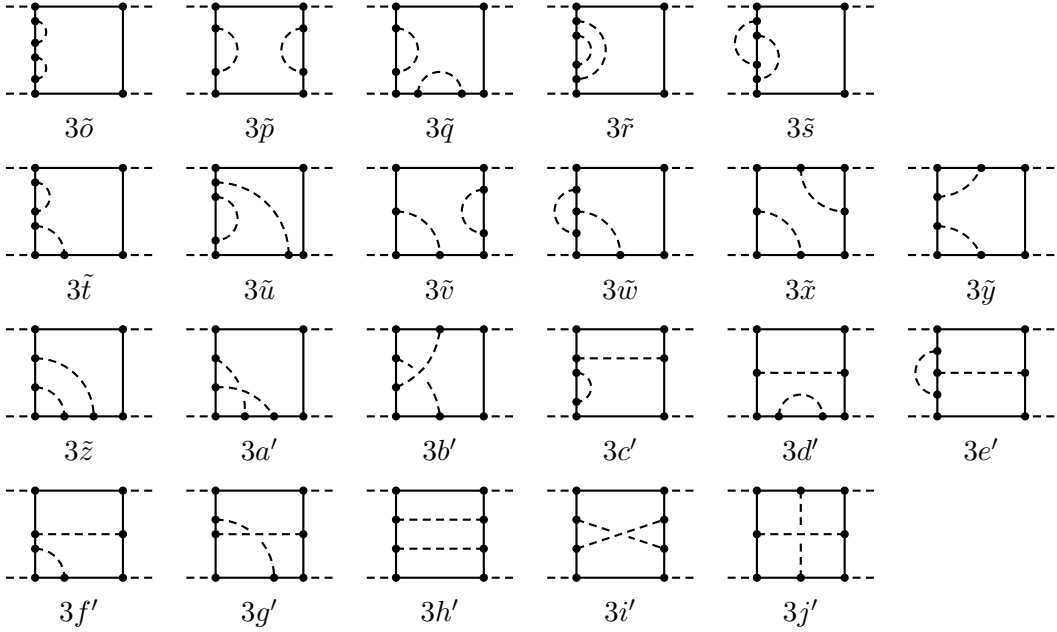
\begin{figure}[ht]
  \centering
  \begin{tabular}{cccccc}
        \begin{tikzpicture}
    \draw[black, thick, densely dashed]  (2.em, 1.5em) -- (3em, 1.5em);
    \draw[black, thick, densely dashed]  (7.em, 1.5em) -- (6em, 1.5em);
    \draw[black, thick, densely dashed]  (2.em, -1.5em) -- (3em, -1.5em);
    \draw[black, thick, densely dashed]  (7.em, -1.5em) -- (6em, -1.5em);
    \draw[black, thick] (3.em, 1.5em) -- (6.em, 1.5em) -- (6.em, -1.5em) --(3.em, -1.5em) -- cycle;
    \draw[black, thick, densely dashed] (3em, 1.em) arc (90:-90:.375em);
    \draw[black, thick, densely dashed] (3em, -1.em) arc (-90:90:.375em);
    \node at (6em, -1.5em)[circle,fill,inner sep=.1em]{};
    \node at (6em, 1.5em)[circle,fill,inner sep=.1em]{};
    \node at (3em, -1.5em)[circle,fill,inner sep=.1em]{};
    \node at (3em, 1.5em)[circle,fill,inner sep=.1em]{};
    \node at (3em, 1.em)[circle,fill,inner sep=.1em]{};
    \node at (3em, .25em)[circle,fill,inner sep=.1em]{};
    \node at (3em, -1.em)[circle,fill,inner sep=.1em]{};
    \node at (3em, -.25em)[circle,fill,inner sep=.1em]{};
    \end{tikzpicture}
    &
    \begin{tikzpicture}
    \draw[black, thick, densely dashed]  (2.em, 1.5em) -- (3em, 1.5em);
    \draw[black, thick, densely dashed]  (7.em, 1.5em) -- (6em, 1.5em);
    \draw[black, thick, densely dashed]  (2.em, -1.5em) -- (3em, -1.5em);
    \draw[black, thick, densely dashed]  (7.em, -1.5em) -- (6em, -1.5em);
    \draw[black, thick] (3.em, 1.5em) -- (6.em, 1.5em) -- (6.em, -1.5em) --(3.em, -1.5em) -- cycle;
    \draw[black, thick, densely dashed] (3em, .75em) arc (90:-90:.75em);
    \draw[black, thick, densely dashed] (6em, .75em) arc (90:270:.75em);
    \node at (6em, -1.5em)[circle,fill,inner sep=.1em]{};
    \node at (6em, 1.5em)[circle,fill,inner sep=.1em]{};
    \node at (3em, -1.5em)[circle,fill,inner sep=.1em]{};
    \node at (3em, 1.5em)[circle,fill,inner sep=.1em]{};
    \node at (3em, 0.75em)[circle,fill,inner sep=.1em]{};
    \node at (3em, -.75em)[circle,fill,inner sep=.1em]{};
    \node at (6em, 0.75em)[circle,fill,inner sep=.1em]{};
    \node at (6em, -.75em)[circle,fill,inner sep=.1em]{};
    \end{tikzpicture}
    &
    \begin{tikzpicture}
    \draw[black, thick, densely dashed]  (2.em, 1.5em) -- (3em, 1.5em);
    \draw[black, thick, densely dashed]  (7.em, 1.5em) -- (6em, 1.5em);
    \draw[black, thick, densely dashed]  (2.em, -1.5em) -- (3em, -1.5em);
    \draw[black, thick, densely dashed]  (7.em, -1.5em) -- (6em, -1.5em);
    \draw[black, thick] (3.em, 1.5em) -- (6.em, 1.5em) -- (6.em, -1.5em) --(3.em, -1.5em) -- cycle;
    \draw[black, thick, densely dashed] (3em, .75em) arc (90:-90:.75em);
    \draw[black, thick, densely dashed] (3.75em, -1.5em) arc (180:0:.75em);
    \node at (6em, -1.5em)[circle,fill,inner sep=.1em]{};
    \node at (6em, 1.5em)[circle,fill,inner sep=.1em]{};
    \node at (3em, -1.5em)[circle,fill,inner sep=.1em]{};
    \node at (3em, 1.5em)[circle,fill,inner sep=.1em]{};
    \node at (3em, 0.75em)[circle,fill,inner sep=.1em]{};
    \node at (3em, -.75em)[circle,fill,inner sep=.1em]{};
    \node at (3.75em, -1.5em)[circle,fill,inner sep=.1em]{};
    \node at (5.25em, -1.5em)[circle,fill,inner sep=.1em]{};
    \end{tikzpicture}
    &
    \begin{tikzpicture}
    \draw[black, thick, densely dashed]  (2.em, 1.5em) -- (3em, 1.5em);
    \draw[black, thick, densely dashed]  (7.em, 1.5em) -- (6em, 1.5em);
    \draw[black, thick, densely dashed]  (2.em, -1.5em) -- (3em, -1.5em);
    \draw[black, thick, densely dashed]  (7.em, -1.5em) -- (6em, -1.5em);
    \draw[black, thick] (3.em, 1.5em) -- (6.em, 1.5em) -- (6.em, -1.5em) --(3.em, -1.5em) -- cycle;
    \draw[black, thick, densely dashed] (3em, 1.em) arc (90:-90:1.em);
    \draw[black, thick, densely dashed] (3em, .5em) arc (90:-90:.5em);
    \node at (6em, -1.5em)[circle,fill,inner sep=.1em]{};
    \node at (6em, 1.5em)[circle,fill,inner sep=.1em]{};
    \node at (3em, -1.5em)[circle,fill,inner sep=.1em]{};
    \node at (3em, 1.5em)[circle,fill,inner sep=.1em]{};
    \node at (3em, 1.em)[circle,fill,inner sep=.1em]{};
    \node at (3em, .5em)[circle,fill,inner sep=.1em]{};
    \node at (3em, -1.em)[circle,fill,inner sep=.1em]{};
    \node at (3em, -.5em)[circle,fill,inner sep=.1em]{};
    \end{tikzpicture}
    &
    \begin{tikzpicture}
    \draw[black, thick, densely dashed]  (2.em, 1.5em) -- (3em, 1.5em);
    \draw[black, thick, densely dashed]  (7.em, 1.5em) -- (6em, 1.5em);
    \draw[black, thick, densely dashed]  (2.em, -1.5em) -- (3em, -1.5em);
    \draw[black, thick, densely dashed]  (7.em, -1.5em) -- (6em, -1.5em);
    \draw[black, thick] (3.em, 1.5em) -- (6.em, 1.5em) -- (6.em, -1.5em) --(3.em, -1.5em) -- cycle;
    \draw[black, thick, densely dashed] (3em, 1.em) arc (90:270:.75em);
    \draw[black, thick, densely dashed] (3em, .5em) arc (90:-90:.75em);
    \node at (6em, -1.5em)[circle,fill,inner sep=.1em]{};
    \node at (6em, 1.5em)[circle,fill,inner sep=.1em]{};
    \node at (3em, -1.5em)[circle,fill,inner sep=.1em]{};
    \node at (3em, 1.5em)[circle,fill,inner sep=.1em]{};
    \node at (3em, 1.em)[circle,fill,inner sep=.1em]{};
    \node at (3em, .5em)[circle,fill,inner sep=.1em]{};
    \node at (3em, -1.em)[circle,fill,inner sep=.1em]{};
    \node at (3em, -.5em)[circle,fill,inner sep=.1em]{};
    \end{tikzpicture}
    \\
\end{tabular}
  \caption{Three-loop diagrams  containing only Yukawa couplings with a single fermion loop.}
  \label{fig:quartF}
\end{figure}
\FloatBarrier
  \begin{align}
     {\tilde  \beta}^{abcd}_{\lambda{\hskip 0.5pt}F}  =  {}&  \beta_{\lambda{\hskip 0.5pt}3\too} \;  \S_{12} \, \tr(y^{abcdee f\! f}) +
       \beta_{\lambda{\hskip 0.5pt}3\tp} \;  \S_{6} \, \tr(y^{abeecd\hskip0.1pt f\! f}) + \beta_{\lambda{\hskip 0.5pt}3\tq} \;  \S_{12} \,   \tr(y^{aeebf\!fcd})\nn \\
    &{} + \S_{12} \, \big ( \beta_{\lambda{\hskip 0.5pt}3\trr} \;  \tr(y^{abcdef\!fe}) + \beta_{\lambda{\hskip 0.5pt}3\ts} \; \tr(y^{abcdefef}) \big ) \nn \\
 &{}  +  \S_{24} \, \big ( \beta_{\lambda{\hskip 0.5pt}3\ttt} \;  \tr(y^{af \! febecd}) + \beta_{\lambda{\hskip 0.5pt}3\tu} \; \tr(y^{aef\! fbecd}) 
 +  \beta_{\lambda{\hskip 0.5pt}3\tv} \; \tr(y^{aebecf \! fd}) \big ) \nn  \\
    &{} + \beta_{\lambda{\hskip 0.5pt}3\tw} \;  \S_{24} \, \tr(y^{aefebfcd})  +  \beta_{\lambda{\hskip 0.5pt}3\tx} \; \S_6 \, \tr(y^{aebecfdf}) 
    +  \beta_{\lambda{\hskip 0.5pt}3\ty} \; \S_{12} \, \tr(y^{abecefdf} ) \nn \\
 &   {} + \S_{12} \,  \big ( \beta_{\lambda{\hskip 0.5pt}3\tz} \;  \tr(y^{aefbfecd})  + \beta_{\lambda{\hskip 0.5pt}3a'} \;  \tr(y^{aefbefcd}) + 
 \beta_{\lambda{\hskip 0.5pt}3b'} \;  \tr(y^{abecfedf})  \big )   \nn  \\
 & {}+ \beta_{\lambda{\hskip 0.5pt}3c'} \;  \S_{24}\,  \tr(y^{af\! febced}) +   
 \S_{12}\,\big ( \beta_{\lambda{\hskip 0.5pt}3d'} \; tr(y^{abecf\!fde} ) +   \beta_{\lambda{\hskip 0.5pt}3e'} \;  \tr(y^{aefebcfd}) \big )   \nn  \\
       & {}+  \S_{24}\,  \big ( \beta_{\lambda{\hskip 0.5pt}3f'} \;  \tr(y^{aefbfced})  + \beta_{\lambda{\hskip 0.5pt}3g'} \; \tr(y^{aefbecfd}) \big ) \nn \\
 &    {}+  \S_{6}\,  \big ( \beta_{\lambda{\hskip 0.5pt}3h'} \; \tr(y^{aefbcfed})  
 +  \beta_{\lambda{\hskip 0.5pt}3i'} \; \tr(y^{aefbcefd}) \big )
  +   \beta_{\lambda{\hskip 0.5pt}3j'} \; \S_3\, \tr(y^{aebfcedf}) \, .
  \end{align}
  Of the quartic scalar diagrams $2g, 3f, 3\thh, 3\ti, 3\tj, 3\tk, 3b', 3g',3h',3i' ,3j'$ are primitive. 
  The diagrams $3f, 3l, 3s, 3u, 3w, 3\td, 3\te, 3\tf, 3\tg, 3\tj, 3\tk, 3a', 3b', 3g', 3i', 3j'$ are non planar.

 As explained subsequently the 62 individual coefficients can be determined so that
 \begin{align}\label{quartic3}
& \beta_{\lambda{\hskip 0.5pt}3a}  =  - \tfrac38\, ,  &&  \beta_{\lambda{\hskip 0.5pt}3b}  =  2 \, , &&
 \beta_{\lambda{\hskip 0.5pt}3c}  =   \tfrac12 \, , &&  \beta_{\lambda{\hskip 0.5pt}3d}  =  - \tfrac12\, , \nn \\ 
 &\beta_{\lambda{\hskip 0.5pt}3e}  =  -  \tfrac12 \, , 
 && \beta_{\lambda{\hskip 0.5pt}3f}  =  12 \zeta_3  \, , &&   \beta_{\lambda{\hskip 0.5pt}3g}  =  - \tfrac14  \, , &&   
 \beta_{\lambda{\hskip 0.5pt}3h}  =  - \tfrac12 \, ,  \nn \\  
 & \beta_{\lambda{\hskip 0.5pt}3i}  =   2\, , && \beta_{\lambda{\hskip 0.5pt}3j}  =   - \tfrac12\, ,  
 && \beta_{\lambda{\hskip 0.5pt}3k}  =   3 \, , && \beta_{\lambda{\hskip 0.5pt}3l}  =  2 \, , \nn \\ 
 & \beta_{\lambda{\hskip 0.5pt}3m}  =   -\tfrac{25}{8}   \, , &&   \beta_{\lambda{\hskip 0.5pt}3n}  =   -4 \, , && \beta_{\lambda{\hskip 0.5pt}3o}  =  -3 \, , 
 &&  \beta_{\lambda{\hskip 0.5pt}3p}  = \tfrac{25}{8}   \, , \nn \\ 
 & \beta_{\lambda{\hskip 0.5pt}3q}  =  \tfrac54\, , &&
 \beta_{\lambda{\hskip 0.5pt}3r}  =   2   \, , &&   \beta_{\lambda{\hskip 0.5pt}3s}  =   2 (3\zeta_3 -2) \, , &&\beta_{\lambda{\hskip 0.5pt}3t}  =   2 \, , \nn \\
&  \beta_{\lambda{\hskip 0.5pt}3u}  =   3\zeta_3 -1  \, , &&   \beta_{\lambda{\hskip 0.5pt}3v}  =   3 \, , &&   \beta_{\lambda{\hskip 0.5pt}3w}  =   2 \, , 
&&   \beta_{\lambda{\hskip 0.5pt}3x}  =  -10  \, ,  \nn \\ 
& \beta_{\lambda{\hskip 0.5pt}3y}  =   - 1 \, ,  && \beta_{\lambda{\hskip 0.5pt}3z}  =   - 3 \, , && \beta_{\lambda{\hskip 0.5pt}3\ta}  =   -3 \, , &&
\beta_{\lambda{\hskip 0.5pt}3\tb}  =  -2 \, ,  \nn \\
& \beta_{\lambda{\hskip 0.5pt}3\tc}  =  -6 \, , && 
\beta  \raisebox{-1.5 pt}{$\scriptstyle \lambda{\hskip 0.5pt}3\td$}  =  -4 \, , &&  \beta_{\lambda{\hskip 0.5pt}3\te}  =  - 4 \, , 
&& \beta  \raisebox{-1.5 pt}{$\scriptstyle \lambda{\hskip 0.5pt}3\tf$}   =  4(3\zeta_3-2) \, , \nn \\
& \beta_{\lambda{\hskip 0.5pt}3\tg}=  2(6 \zeta_3 -5)  \, ,  && \beta_{\lambda{\hskip 0.5pt}3\thh} =  - 2  \, , && 
\beta_{\lambda{\hskip 0.5pt}3\ti} =  - 10  \, , && \beta_{\lambda{\hskip 0.5pt}3\tj} =  -24\zeta_3 \, , \nn \\
& \beta  \raisebox{-1.5 pt}{$\scriptstyle \lambda{\hskip 0.5pt}3\tk$} = -12\zeta_3\, , && 
\beta  \raisebox{-1.5 pt}{$\scriptstyle \lambda{\hskip 0.5pt}3\tl$} =  -8 \, ,  && \beta_{\lambda{\hskip 0.5pt}3\tm} =  -12 \, , &&
 \beta_{\lambda{\hskip 0.5pt}3\tn}  =  -4  \,  \nn \\
& \beta_{\lambda{\hskip 0.5pt}3\too}  =  1 \, ,  && \beta_{\lambda{\hskip 0.5pt}3\tp}  =  1 \, , && \beta_{\lambda{\hskip 0.5pt}3\tq}  =  1   \, ,  
&& \beta_{\lambda{\hskip 0.5pt}3\trr}  =  -\tfrac78    \, ,  \nn \\
&  \beta_{\lambda{\hskip 0.5pt}3\ts}  =  2 \, ,  &&
   \beta_{\lambda{\hskip 0.5pt}3\ttt}  =  2  \, , & &  \beta_{\lambda{\hskip 0.5pt}3\tu}  =  - 4  \, ,  &&  \beta_{\lambda{\hskip 0.5pt}3\tv}  =  2  \, , \nn \\
&  \beta_{\lambda{\hskip 0.5pt}3\tw}  =  2    \, ,  && \beta_{\lambda{\hskip 0.5pt}3\tx}  =  4   \, , 
&&  \beta_{\lambda{\hskip 0.5pt}3\ty}  =  4\, ,  &&  \beta_{y{\hskip 0.5pt}3\tz}  =  -8  \, , \nn \\
& \beta_{\lambda{\hskip 0.5pt}3a'}  =  - 2(6\zeta_3-5) \, , &&  \beta_{\lambda{\hskip 0.5pt}3b'}  =  0 \, , &&
 \beta_{\lambda{\hskip 0.5pt}3c'}  =   -1 \, ,  &&\beta_{\lambda{\hskip 0.5pt}3d'}  =  - 1 \, , \nn \\
 &  \beta_{\lambda{\hskip 0.5pt}3e'}  =  -  6 \, , 
 &&  \beta_{\lambda{\hskip 0.5pt}3f'}  =  -2   \, , &&   \beta_{\lambda{\hskip 0.5pt}3g'}  =  - 12\zeta_3  \, , &&   
 \beta_{\lambda{\hskip 0.5pt}3h'}  =  - 4  \, ,   \nn \\
 &   \beta_{\lambda{\hskip 0.5pt}3i'}  =  - 12\zeta_3 \, , &&  \beta_{\lambda{\hskip 0.5pt}3j'}  =   - 24\zeta_3\, .
\end{align}
This completes the expressions given in~\cite{Steudtner:2021fzs}, which have been obtained using $\mathcal{N}=1$ SUSY relations as well as explicit literature results for the SM~\cite{Chetyrkin:2012rz,Bednyakov:2013eba,Chetyrkin:2013wya,Bednyakov:2013cpa,Bednyakov:2014pia} and Gross-Neveu type models \cite{Zerf:2017zqi,Mihaila:2017ble}. In this paper $\mathcal{N}=\tfrac12$ SUSY conditions are also considered, which are not sufficient to obtain \eqref{quartic3}, but overcomplete the conditions in \cite{Steudtner:2021fzs} without inconsistencies. Hence, literature results \cite{Zerf:2017zqi,Mihaila:2017ble} are cross-checked by the SUSY relations and explicit SM computations.

 \section{Reduction to \texorpdfstring{$U(1)$}{U(1)} Symmetry}
 
 For complex fields with a $U(1)$ symmetry the number of diagrams is significantly reduced\footnote{This example
 was considered in \cite{Jack:2013sha}}. 
 This restriction is achieved by taking $\phi^a= ( \vphi_i, \bphi^i)$, so that $ \phi^a \phi'{}^a
 = \vphi_i \, \bphi'{}^i + \bphi^i \, \vphi'{\!}_i$, and 
 \be
 \tfrac{1}{24}\, \lambda^{abcd}\, \phi^a\phi^b\phi^c\phi^d \to \tfrac14\, \lambda_{ij}{}^{kl}\, \bphi^i\bphi^j\vphi_k\vphi_l \, ,
\quad  \phi^a y^a \to \vphi_i\,  y^i \left ( \begin{smallmatrix} 0 & 1\\ 0 & 0 \end{smallmatrix} \right ) + 
  \bphi^i \, \by_i \left ( \begin{smallmatrix} 0 & 0\\ 1  & 0 \end{smallmatrix} \right ) \, .
 \ee
 The scalar and fermion lines on each diagram then have arrows with the basic vertices for the Yukawa
 couplings $y^i, \, \by_i$ represented by 
 $ \raisebox{-6pt}{
{\begin{tikzpicture}
\begin{scope}[very thick,decoration={
    markings,
    mark=at position 0.6 with {\arrow{>}}}
    ] 
  \draw   [postaction={decorate}] [black, thick, densely dashed]  (0,0.3)--(0.3,0);
\draw  [postaction={decorate}] [black, thick]   (0,-0.3)--(0.3,0);
 \draw  [postaction={decorate}] [black, thick]    (0.7,0)--(0.3,0);
          \node at (0.3, 0)[circle,fill,inner sep=.1em]{};  
  \end{scope}
   \end{tikzpicture} }}
$\, , 
$
\raisebox{-6pt}{
{\begin{tikzpicture}
\begin{scope}[very thick,decoration={
    markings,
    mark=at position 0.7 with {\arrow{>}}}
    ] 
  \draw  [postaction={decorate}] [black, thick] (0.3,0)--(-0.1,0) ;
 \draw  [postaction={decorate}] [black, thick]  (0.3,0)--(0.6,-0.3);
  \draw  [postaction={decorate}] [black, thick, densely dashed]    (0.3,0)--(0.6,0.3);  
   \node at (0.3, 0)[circle,fill,inner sep=.1em]{}; 
   \end{scope}
   \end{tikzpicture} }}    
$
and for the scalar quartic coupling $\lambda_{ij}{}^{kl}$ by
$
\raisebox{-6pt}{
{\begin{tikzpicture}
\begin{scope}[very thick,decoration={
    markings,
    mark=at position 0.5 with {\arrow{<}}}
    ] 
  \draw [postaction={decorate}] [black, thick, densely dashed]  (0.3,0) -- (-0.1,0.3);
  \draw  [postaction={decorate}] [black, thick, densely dashed] (0.3,0)  --  (-0.1,-0.3) ;
  \draw [postaction={decorate}] [black, thick, densely dashed]     (0.7,0.3)--(0.3,0) ;
   \draw[postaction={decorate}]  [black, thick, densely dashed]      (0.7,-0.3)--(0.3,0)  ;
          \node at (0.3, 0)[circle,fill,inner sep=.1em]{};  
 \end{scope}
   \end{tikzpicture}}
   } 
   $. 
 The triangle graphs present for real couplings are no longer allowed. With this prescription then
 for traces over the Yukawa couplings 
 \be
 \tr ( y^{a_1 a_2 a_3\dots a_{2n}})  \to \begin{cases}
 \tfrac12  \big ( \tr(\by_{i_1}y^{i_2} \by_{i_3} \dots y^{i_{2n}})   + \tr( y^{i_{2n}} \dots y^{i_2} \by_{i_1}) \big )  \, , \\ 
\tfrac12  \big (  \tr ( y^{i_1} \by_{i_2}y^{i_3} \dots \by_{i_{2n}} ) + \tr(\by_{i_{2n}}\dots \by_{i_2} y^{i_1})  \big )  \, .
\end{cases}
\ee
 
 In general, for the anomalous  dimensions 
 \be
 \gamma_\phi{}^{ab} = \big ( \gamma_{\vphi\, i}{}^j, \, \gamma_{\vphi\, j}{}^i \big ) \, , \qquad
 \gamma_\psi = \begin{pmatrix} \gamma_\psi & 0 \\ 0 & ~ {\bar \gamma}_\psi \end{pmatrix} \, ,
 \ee 
 and for the $\beta$-functions from \eqref{betay} and \eqref{betalambda}
 \begin{align}
 & \beta_{y}{}^{i} =  {\tilde \beta}_{y}{}^{ i} +  \gamma_{\psi}\, y^i + y^i \, {\bar \gamma}_{\psi} 
 + y^j \gamma_{\vphi\, j}{}^i \, , \quad
 \beta_{\by\hskip 0.5pt i} ={\tilde \beta}_{\by \hskip 0.5pt i} + {\bar \gamma}_{\psi}\, \by_i + 
 \by_i \, \gamma_{\psi}  +  \gamma_{\vphi\, i}{}^j \, \by_j \, ,  \nn \\
 & \beta_{\lambda \, ij}{}^{kl } =  {\tilde \beta}_{\lambda \, ij}{}^{kl } + \gamma_{\vphi\, i}{}^m \lambda_{mj}{}^{kl}
 +  \gamma_{\vphi\, i}{}^m \lambda_{im }{}^{kl} +   \lambda_{j}{}^{ml}\, \gamma_{\vphi\, m}{}^k
 +   \lambda_{j}{}^{km}\, \gamma_{\vphi\, m}{}^l  \, ,
 \end{align}
 where $\gamma_\psi \to {\bar \gamma}_\psi{\hskip 0.5pt} \ \beta_{y}{}^{i} \to \beta_{\by\hskip 0.5pt i} $ by taking 
 $y^i\leftrightarrow \by_i, \ \lambda_{ij}{}^{kl} \to \lambda_{kl}{}^{ij}$ in each contribution.
  For $\by_i = (y^i)^\dagger, \, \lambda_{ \, kl}{}^{ij} = (\lambda_{ \, ij}{}^{kl})^* $ then ${\gamma}_\psi = 
 \gamma_\psi{}^\dagger$ and  $ \gamma_{\vphi\, j}{}^i = ( \gamma_{\vphi\, i}{}^j)^*$.
At one loop
\begin{align}
\gamma_{\vphi\, i}{}^j {}^{(1)}  = {}& \gamma_{\phi{\hskip 0.5pt}1} \, \tr( \by_i\,  y^j) \, , \qquad \gamma_{\psi} {}^{(1)} 
= \gamma_{\psi{\hskip 0.5pt}1} \, y^i \by_i \, , \qquad  {\tilde \beta}_{y}{}^{ i}  {}^{(1)} = 0 \, , \nn \\
 {\tilde \beta}_{\lambda \, ij}{}^{kl }{}^{(1)} = {}& \beta_{\lambda{\hskip 0.5pt}1a} \big ( \lambda_{ij}{}^{mn} \lambda_{mn}{}^{kl} 
 +2\,  \S_2 \,  \lambda_{im}{}^{kn} \lambda_{jn}{}^{ml} \big )   + 
 \tfrac12\,  \beta_{\lambda{\hskip 0.5pt}1b} \, \S_2\,  \tr ( \by_i \, y^k \by_j \, y^l) \, ,
\end{align}
where $\S_2 \, X_{ij}{}^{kl} =  X_{ij}{}^{kl} +  X_{ji}{}^{kl} =  X_{ij}{}^{kl} +  X_{ij}{}^{lk} $.
At two loops
\begin{align}
\gamma_{\vphi\, i}{}^j {}^{(2)}  = {}& 3\, \gamma_{\phi{\hskip 0.5pt}2a} \, \lambda_{ik}{}^{mn} \lambda_{mn}{}^{kj} 
+ \tfrac12\,  \gamma_{\phi{\hskip 0.5pt}2b} \,  \big (\tr(\by_i\, y^k \by_k\, y^j ) + \tr( \by_i \, y^j \by_k\, y^k )\big ) \, , \nn \\
\gamma_{\psi} {}^{(2)} = {}& \gamma_{\psi{\hskip 0.5pt}2a} \, y^i \by_j \,  \tr (\by_i \, y^j ) + 
\gamma_{\psi{\hskip 0.5pt}2b} \,  y^i \by_j \, y^j \by_i \, , \nn \\
{\tilde \beta}_{y}{}^{ i}  {}^{(2)} = {}& \beta_{y{\hskip 0.5pt}2a} \, \lambda_{jk}{}^{li}\,  y^j \by_l\, y^k + 
\beta_{y{\hskip 0.5pt}2f} \,  y^j \by_k\, y^i \by_j\, y^k \, , \nn \\
{\tilde \beta}_{\lambda \, ij}{}^{kl }{}^{(2)} = {}& \beta_{\lambda{\hskip 0.5pt}2a} \big (2\, \lambda_{ij}{}^{mn} \lambda_{mp}{}^{qk} 
\lambda_{nq}{}^{pl} +  2\, \lambda_{ip}{}^{mq} \lambda_{jq}{}^{np} \lambda_{mn}{}^{kl} 
 +  \S_4 \,  \lambda_{ip}{}^{qk} \lambda_{jq}{}^{mn} \lambda_{mn}{}^{pl}   \nn \\
 \noalign{\vskip -2pt}
 & \hskip 1.5cm + 2\, \S_4 \,  \lambda_{im}{}^{pn} \lambda_{jn}{}^{qk} \lambda_{pq}{}^{ml} \big )  \nn \\
 &{}   +  \beta_{\lambda{\hskip 0.5pt}2b}  \big ( \lambda_{ij}{}^{mp} \lambda_{mq}{}^{kl} 
 +  \S_4 \, \lambda_{im}{}^{kp} \lambda_{jq}{}^{lm} \big ) \, \tr( \by_p\, y^q) \nn \\
&{}   +  \tfrac12 \, \beta_{\lambda{\hskip 0.5pt}2c} \, \S_4 \,  \lambda_{im}{}^{kn} \big ( \tr( \by_j\, y^l \by_n \, y^m ) + 
 \tr( \by_j\, y^m \by_n \, y^l ) \big ) \nn \\
 &{}   +  \beta_{\lambda{\hskip 0.5pt}2d}  \, \big (  \lambda_{ij}{}^{mn} \, \tr( \by_m\, y^k \by_n \, y^l ) + 
 \lambda_{mn}{}^{kl} \, \tr( \by_i\, y^m \by_j \, y^n ) \big ) \nn \\
&{} +  \tfrac12\,  \beta_{\lambda{\hskip 0.5pt}2e} \, \S_4\,  \big ( \tr ( \by_m\, y^k \by_i \, y^l \by_j \, y^m ) 
+ \tr ( \by_m\, y^m \by_i \, y^k \by_j \, y^l ) \big )  \nn \\
&{} +  \tfrac12\,  \beta_{\lambda{\hskip 0.5pt}2g} \, \S_4\, \tr ( \by_i\, y^k \by_m \, y^l \by_j \, y^m ) \, ,
\end{align}
with $\S_4 \, X_{ij}{}^{kl} =  X_{ij}{}^{kl} +  X_{ji}{}^{kl} + X_{ij}{}^{lk} +  X_{ji}{}^{lk} $.

At three loops the results here reduce to 8 contributions for $\gamma_\phi$
\begin{align}
\gamma_{\vphi\, i}{}^j {}^{(3)}  = {}& \gamma_{\phi{\hskip 0.5pt}3a} \,
\big (  \lambda_{ik}{}^{mn} \lambda_{mn}{}^{pq} \lambda_{pq}{}^{kj} 
+ 4 \, \lambda_{ik}{}^{mn} \lambda_{ml}{}^{kp} \lambda_{np}{}^{lj} \big ) \nn \\
&{} + \gamma_{\phi{\hskip 0.5pt}3b} \, \big (  \lambda_{ik}{}^{mn} \lambda_{mn}{}^{lj} + 
2 \, \lambda_{im}{}^{ln} \lambda_{k n }{}^{m j} \big ) \, \tr(\by_l\, y^k)\nn \\
&{}  +  \gamma_{\phi{\hskip 0.5pt}3c} \,\big ( \lambda_{ik}{}^{lm} \, \tr (\by_l \, y^k \,\by_m \, y^j ) 
+ \tr(\by_i\,y^l \by_k\, y^m ) \, \lambda_{lm}{}^{kj} \big )\nn  \\
&{}+ \tfrac12 \, \gamma_{\phi{\hskip 0.5pt}3d}  \, \big ( \tr(\by_i\, y^k \by_l \, y^j )+  \tr( \by_i \, y^ j \by_l\, y^k\, ) \big )  \, \tr(\by_k\, y^l)
\nn \\
&{} + \tfrac12 \, \gamma_{\phi{\hskip 0.5pt}3f}  \,  \big ( \tr(\by_i \, y^k \by_k \, y^l \, \by_l \, y^j) +
\tr(\by_i \, y^j \by_k \, y^k \by_l \, y^l \, )  \big ) \nn \\
&{}  + \tfrac12 \, \gamma_{\phi{\hskip 0.5pt}3g}  \,  \big ( \tr(\by_i\, y^k \by_l \, y^l \, \by_k \, y^j )  +
\tr( \by_i \, y^j \by_k \, y^l \by_l \, y^k  ) \big )\nn  \\
&{} + \gamma_{\phi{\hskip 0.5pt}3h}  \,  \tr(\by_k \, y^k \by_i \, y^l\, \by_l \, y^j )  + \gamma_{\phi{\hskip 0.5pt}3m}  \, 
\tr(\by_k \, y^l \by_i \, y^k \by_l \, y^j )  \, .
\end{align}
and 9 for $\gamma_\psi$
\begin{align}
\gamma_{\psi} {}^{(3)}  = {}& 3\, \gamma_{\psi{\hskip 0.5pt}3a} \, y^i \by_j \, \lambda_{im}{}^{kl} \lambda_{kl}{}^{mj} 
+  \gamma_{\psi{\hskip 0.5pt}3b} \, y^i \by_k\, y^j \by_l \, \lambda_{ij}{}^{kl} \nn \\
&{}+  \gamma_{\psi{\hskip 0.5pt}3c}\, y^i \by_j \, \tr (y^j \by_k) \,  \tr (y^k \by_i) 
 +\big (  \gamma_{\psi{\hskip 0.5pt}3d}\,  y^i \by_k\, y^k \by_j  + \gamma_{\psi{\hskip 0.5pt}3e}\, y^k \by_j\, y^i \by_k  \big )
  \, \tr (y^j \by_i) \nn  \\ 
  &{} + \tfrac12 \, \gamma_{\psi{\hskip 0.5pt}3g}\,  y^i \by_j  \, \big ( \tr (y^j \by_k \, y^k  \by_i) +  \tr (y^k \by_k \, y^j  \by_i) \big ) \nn \\
&{} +  \gamma_{\psi{\hskip 0.5pt}3i}\,   y^i \by_j \, y^j \by_k\, y^k \by_i 
+ \gamma_{\psi{\hskip 0.5pt}3j}\,  y^i \by_j \, y^k \by_k\, y^j \by_i + 
 \gamma_{\psi{\hskip 0.5pt}3p} \, y^i \by_j \, y^k \by_i \, y^j \by_k \, ,
\end{align}
and 12 for $ {\tilde \beta}_{y}$
\begin{align} \hskip - 0.4cm
 {\tilde \beta}_{y}{}^{ i}  {}^{(3)} \! = {}& 
 2\, \beta_{y{\hskip 0.5pt}3b}  \,  y^j\by_k\, y^l \big ( \lambda_{jm}{}^{ni}\lambda_{nl}{}^{km}  +  
\lambda_{jm}{}^{kn}\lambda_{nl}{}^{mi} \big )  
+  \beta_{y{\hskip 0.5pt}3c} \, y^j\by_k \, y^l \, \lambda_{jl}{}^{mn}\lambda_{mn}{}^{ki} \nn \\
&{}+ \beta_{y{\hskip 0.5pt}3d} \, y^j \by_m\, y^k \lambda_{jk}{}^{li} \, \tr(y^m \by_l )  + 
\beta_{y{\hskip 0.5pt}3e} \, \big (  y^k \by_l\, y^m +   y^m \by_l\,  y^k \big ) \lambda_{jk}{}^{li} \, \tr(y^j \by_m ) \,
\nn  \\
&{}+  \beta_{y{\hskip 0.5pt}3f}  \, y^k \by_m \, y^i \by_n \, y^l \, \lambda_{kl}{}^{mn} 
+ \beta_{y{\hskip 0.5pt}3j} \, \big ( y^k \by_m\, y^m \by_j\, y^l +  y^k \by_j \, y^m\by_m y^l \big )\,  \lambda_{kl}{}^{ji}  \nn \\
&{} + \beta_{y{\hskip 0.5pt}3l}  \, \big ( y^m \by_j \, y^k \by_m \, y^l + y^k \by_m \, y^l\by_j \, y^m \big )\, \lambda_{kl}{}^{ji} \nn \\ 
&{}+  \beta_{y{\hskip 0.5pt}3s}  \, \big ( y^j \by_l \, y^i \by_k \, y^l  + y^l \by_k \, y^i \by_l \, y^j \big ) \,  \tr(y^k \by_j ) 
+  \tfrac12\,  \beta_{y{\hskip 0.5pt}3w}   \, \big ( y^j \by_k \, y^l + y^l \by_k \, y^j \big ) \,  \tr ( \by_j \, y^k \by_l \, y^i ) \nn  \\
&{} +  \beta_{y{\hskip 0.5pt}3\too}  \, \big ( y^k \by_j \, y^j \by_l \, y^i \by_k \, y^l + y^k \by_l \, y^i \by_k \, y^j \by_j \, y^l \big ) \nn \\
&{} +  \beta_{y{\hskip 0.5pt}3\tp}  \, \big ( y^k \by_l \, y^j \by_j \, y^i \by_k \, y^l + y^k \by_l \, y^i \by_j \, y^j \by_k \, y^l \big )\nn  \\
&{} +  \beta_{y{\hskip 0.5pt}3\tz} \, \big ( y^j \by_k \, y^l \by_j \, y^i \by_l \, y^k +y^k \by_l \, y^i \by_j \, y^l \by_k \, y^j \big ) \, . 
\end{align}

For the scalar quartic $\beta$-function at three loops the 1PI contributions are restricted to 43 diagrams as
 $3n, \, 3q, \, 3\tb, \, 3\tc, \, 3\te, \, 3\tf, \, 3\ts, \, 3\ttt, \, 3\tu, \, 3\tv, \, 3\tw, \, 3\tx, \, 3\ty{\hskip 0.5pt}, \, 3\tz, \, 3e', \, 3f', \,
3g', \,3i',  3j'$ are no longer present. There remain 7 primitive 3 loop diagrams.

There is one possible antisymmetric term at three loops
\be
\upsilon_{\vphi\, i}{}^j {}^{(3)}  = \upsilon_{\phi{\hskip 0.5pt}3c}\, \big ( \lambda_{ik}{}^{lm} \, \tr( \by_l \, y^k \by_m \, y^j) 
- \tr ( \by_i\, y^l \by_k \, y^m) \, \lambda_{lm}{}^{kj} \big ) \, .  
\label{U1anti}
\ee

\section{Supersymmetry Relations}

Supersymmetry of course relates bosons and fermions. Imposing symmetry on the scalar
fermion theory leads to linear relations between the anomalous dimension and $\beta$-function
coefficients which we describe below.

\subsection{\texorpdfstring{$\N=1$}{N=1} Supersymmetry}
\label{susyred} 

The Wess Zumino theory for scalars and fermions is a special case which can be obtained by restricting
the couplings of the theory with $U(1)$ symmetry so that
\be
y^i \to Y^{ijk} = Y^{(ijk)} \, , \qquad \by_i \to {\bar Y}_{ijk} = {\bar Y}_{(ijk)} \, , \qquad 
\lambda_{ij}{}^{kl} \to {\bar Y}_{ijm} Y^{mkl} \, .
\label{U1susy}
\ee
The usual non renormalisation theorems require
\be
\gamma_\psi{}^i{}_j = \gamma_{\vphi}{}_j{}^i = \gamma^i{}_j \, , \qquad 
 {\tilde \beta}_Y{\!}^{ijk} = 0 \, , \qquad {\tilde \beta}_{\lambda\, ij}{}^{kl} = 
 2 \,  {\bar Y}_{ijm} \, \gamma^m{}_n Y^{nkl} \, .
 \label{conN1}
\ee

At one loop this just imposes
\be
 \gamma^S_1 =\gamma_{\psi{\hskip 0.5pt}1} = \gamma_{\phi{\hskip 0.5pt}1} \, , \qquad 4 \,  \beta_{\lambda{\hskip 0.5pt}1a} + \beta_{\lambda{\hskip 0.5pt}1b} = 0 \, , \qquad  \beta_{\lambda{\hskip 0.5pt}1a} = 2\,  \gamma^S_1 \, .
\ee
At two loops the necessary conditions are
\begin{align}
& \gamma^S_2= \gamma_{\psi{\hskip 0.5pt}2a} + \gamma_{\psi{\hskip 0.5pt}2b} = 3\, \gamma_{\phi{\hskip 0.5pt}2a} + \gamma_{\phi{\hskip 0.5pt}2b} \, , \qquad
\beta_{y{\hskip 0.5pt}2a} + \beta_{y{\hskip 0.5pt}2f} = 0 \, , \nn \\
& 2\, \beta_{\lambda{\hskip 0.5pt}2a} +  \beta_{\lambda{\hskip 0.5pt}2d} = 0 \, , \quad 
4\, \beta_{\lambda{\hskip 0.5pt}2a} + 2\, \beta_{\lambda{\hskip 0.5pt}2c} +  \beta_{\lambda{\hskip 0.5pt}2g} = 0 \, ,\quad
\beta_{\lambda{\hskip 0.5pt}2a} +  \beta_{\lambda{\hskip 0.5pt}2b} +  \beta_{\lambda{\hskip 0.5pt}2e} = 0 , \,  \nn \\
& \beta_{\lambda{\hskip 0.5pt}2b}  = 2 \,  \gamma^S_2  \, .
\end{align} 
At three loops the conditions on the anomalous dimensions and Yukawa couplings are then
\begin{align}
& \gamma^S_{3A} = \gamma_{\psi{\hskip 0.5pt}3c} + \gamma_{\psi{\hskip 0.5pt}3i} = \gamma_{\phi{\hskip 0.5pt}3a} + \gamma_{\phi{\hskip 0.5pt}3f}   \, , \qquad
 \gamma^S_{3B} = \gamma_{\psi{\hskip 0.5pt}3d} =  \gamma_{\phi{\hskip 0.5pt}3b} + \gamma_{\phi{\hskip 0.5pt}3h} \, ,  \nn \\
&   \gamma^S_{3C} = 3 \, \gamma_{\psi{\hskip 0.5pt}3a} + \gamma_{\psi{\hskip 0.5pt}3e} + \gamma_{\psi{\hskip 0.5pt}3g}  + \gamma_{\psi{\hskip 0.5pt}3j} 
 = 2\, \gamma_{\phi{\hskip 0.5pt}3b}  + \gamma_{\phi{\hskip 0.5pt}3d}  + \gamma_{\phi{\hskip 0.5pt}3g}   \, , \nn \\ 
&  \gamma^S_{3D}=  \gamma_{\psi{\hskip 0.5pt}3b} +  \gamma_{\psi{\hskip 0.5pt}3p} = 
4\, \gamma_{\phi{\hskip 0.5pt}3a} + 2\,\gamma_{\phi{\hskip 0.5pt}3c} + \gamma_{\phi{\hskip 0.5pt}3m}  \, , \nn \\
& 
 4 \,  \beta_{y{\hskip 0.5pt}3b }  +  \beta_{y{\hskip 0.5pt}3f} +  2\, \beta_{y{\hskip 0.5pt}3l} +  \beta_{y{\hskip 0.5pt}3w} + 2\,  \beta_{y{\hskip 0.5pt}3\tz} = 0 \, , \nn \\
&  \beta_{y{\hskip 0.5pt}3e} + \beta_{y{\hskip 0.5pt}3j} +   \beta_{y{\hskip 0.5pt}3s} +  \beta_{y{\hskip 0.5pt}3\too}  =0\, , \qquad
\beta_{y{\hskip 0.5pt}3c} +  \beta_{y{\hskip 0.5pt}3d}  + 2\,  \beta_{y{\hskip 0.5pt}3\tp} = 0 \, ,
 \end{align}
 so that
  \begin{align}
\gamma^{S(1)}_\phi = {}& \gamma^S_1 \, 
\tikz[baseline=(vert_cent.base)]{
 \node (vert_cent) {\hspace{-13pt}$\phantom{-}$};
 \begin{scope}[very thick,decoration={markings,
    mark=at position 0.55 with {\arrow{>}}}] 
    \draw  [postaction={decorate}, line width = 0.7pt]  (-0.4,0)--(0.1,0);
    \draw  [postaction={decorate}, line width = 0.7pt]      (0.7,0) ++(0:0.6cm) arc (0:180:0.6cm and 0.4cm);
     \draw  [postaction={decorate}, line width = 0.7pt]     (0.7,0) ++(0:0.6cm) arc (360:180:0.6cm and 0.4cm);
   \draw  [postaction={decorate}, line width = 0.7pt]       (1.3,0)--(1.8,0); 
        \end{scope}
        \filldraw [black] (0.1,0) circle [radius=1.5pt];
         \filldraw [black] (1.3,0) circle [radius=1.5pt];
} \, , \qquad
\gamma^{S(2)}_\phi  =  \gamma^S_2  \, 
\tikz[baseline=(vert_cent.base)]{
\begin{scope}[very thick,decoration={markings,
    mark=at position 0.6 with {\arrow{>}}}] 
  \draw  [postaction={decorate}, line width = 0.7pt]  (-0.4,0)--(0.1,0);
  \draw   [postaction={decorate}, line width = 0.7pt]       (0.7,0) ++(0:0.6cm and 0.4cm) arc (0:50:0.6cm and 0.4cm) 
  node (n1)   {} ;
  \draw   [white]     (0.7,0) ++(50:0.6cm and 0.4cm) arc (50:130:0.6cm and 0.4cm)   node (n2)   {} ;
   \draw   [postaction={decorate}, line width = 0.7pt]     (0.7,0) ++(130:0.6cm and 0.4cm) arc (130:50:0.6cm and 0.4cm) ;
 \draw  [postaction={decorate}, line width = 0.7pt]      (0.7,0) ++(130:0.6cm and 0.4cm) arc (130:180:0.6cm and 0.4cm) ;
   \draw     [postaction={decorate}, line width = 0.7pt]        (n2.base) to[out=305,in=235] (n1.base) ;
\draw   [postaction={decorate}, line width = 0.7pt]    (1.3,0)--(1.8,0);
    \end{scope}
    \begin{scope}[very thick,decoration={markings,
    mark=at position 0.55 with {\arrow{>}}}] 
    \draw  [postaction={decorate}, line width = 0.7pt]     (0.7,0) ++(0:0.6cm) arc (360:180:0.6cm and 0.4cm);
    \end{scope}
                  \filldraw [black] (n1) circle [radius=1.5pt]; 
                  \filldraw [black] (n2)  circle [radius=1.5pt]; 
          \filldraw [black]  (0.1,0) circle [radius=1.5pt]; 
           \filldraw [black] (1.3,0) circle [radius=1.5pt];
          } \, , 
    \nn\\  
     \gamma^{S(3)}_\phi = {}& \gamma^S_{3A}   \,
           \tikz[baseline=(vert_cent.base)]{
\begin{scope}[very thick,decoration={markings,
    mark=at position 0.65 with {\arrow{>}}}] 
  \draw  [postaction={decorate}, line width = 0.7pt]  (-0.4,0)--(0.1,0);
  \draw   [postaction={decorate}, line width = 0.7pt]       (0.7,0) ++(0:0.6cm and 0.4cm) arc (0:35:0.6cm and 0.4cm) 
  node (n1)   {} ; 
    \draw   [white]     (0.7,0) ++(145:0.6cm and 0.4cm) arc (145:105:0.6cm and 0.4cm)   node (n2)   {} ;
   \draw   [white]     (0.7,0) ++(35:0.6cm and 0.4cm) arc (35:145:0.6cm and 0.4cm)   node (n4)   {} ;
    \draw   [white]     (0.7,0) ++(35:0.6cm and 0.4cm) arc (35:75:0.6cm and 0.4cm)   node (n3)   {} ;  
     \draw   [postaction={decorate}, line width = 0.7pt]       (0.7,0) ++(75:0.6cm and 0.4cm) arc (75:40:0.6cm and 0.4cm) ;
  \draw   [postaction={decorate}, line width = 0.7pt]       (0.7,0) ++(145:0.6cm and 0.4cm) arc (145:105:0.6cm and 0.4cm) ;  
  \draw   [postaction={decorate}, line width = 0.7pt]       (0.7,0) ++(145:0.6cm and 0.4cm) arc (145:180:0.6cm and 0.4cm) ;
 \draw  [postaction={decorate}, line width = 0.7pt]      (0.7,0) ++(75:0.6cm and 0.4cm) arc (75:105:0.6cm and 0.4cm) ;
 \draw   [postaction={decorate}, line width = 0.7pt]    (1.3,0)--(1.8,0); 
     \end{scope} 
      \begin{scope}[very thick,decoration={markings,
    mark=at position 0.55 with {\arrow{>}}}] 
    \draw  [postaction={decorate}, line width = 0.7pt]     (0.7,0) ++(0:0.6cm) arc (360:180:0.6cm and 0.4cm);
      \draw     [postaction={decorate}, line width = 0.7pt]        (n3.base) to[out=220,in=280] (n1.base) ;
       \draw     [postaction={decorate}, line width = 0.7pt]        (n4.base) to[out=300,in=330] (n2.base) ;
    \end{scope}
     \filldraw [black] (n1) circle [radius=1.5pt]; 
     \filldraw [black] (n2)  circle [radius=1.5pt]; 
     \filldraw [black] (n3) circle [radius=1.5pt]; 
           \filldraw [black] (n4)  circle [radius=1.5pt];
            \filldraw [black]  (0.1,0) circle [radius=1.5pt]; 
           \filldraw [black] (1.3,0) circle [radius=1.5pt];
     }  
                   +  \gamma^S_{3B}  \,
\tikz[baseline=(vert_cent.base)]{
\begin{scope}[very thick,decoration={markings,
    mark=at position 0.6 with {\arrow{>}}}] 
  \draw  [postaction={decorate}, line width = 0.7pt]  (-0.4,0)--(0.1,0);
  \draw   [postaction={decorate}, line width = 0.7pt]       (0.7,0) ++(0:0.6cm and 0.4cm) arc (0:50:0.6cm and 0.4cm) 
  node (n1)   {} ;
  \draw   [white]     (0.7,0) ++(50:0.6cm and 0.4cm) arc (50:130:0.6cm and 0.4cm)   node (n2)   {} ;
   \draw   [postaction={decorate}, line width = 0.7pt]     (0.7,0) ++(130:0.6cm and 0.4cm) arc (130:50:0.6cm and 0.4cm) ;
 \draw  [postaction={decorate}, line width = 0.7pt]      (0.7,0) ++(130:0.6cm and 0.4cm) arc (130:180:0.6cm and 0.4cm) ;
   \draw     [postaction={decorate}, line width = 0.7pt]        (n2.base) to[out=305,in=235] (n1.base) ;
\draw   [postaction={decorate}, line width = 0.7pt]    (1.3,0)--(1.8,0);
\draw    [postaction={decorate}, line width = 0.7pt]         (0.7,0) ++(235:0.6cm and 0.4cm) arc (235:305:0.6cm and 0.4cm)  node(n4) {} ;
    \end{scope}
    \begin{scope}[very thick,decoration={markings,
    mark=at position 0.45 with {\arrow{<}}}] 
\draw    [postaction={decorate}, line width = 0.7pt]    (0.7,0) ++(180:0.6cm and 0.4cm) arc (180:230:0.6cm and 0.4cm) node(n3){} ;
\draw    [postaction={decorate}, line width = 0.7pt]        (0.7,0) ++(310:0.6cm and 0.4cm) arc (310:360:0.6cm and 0.4cm) ;
 \draw      [postaction={decorate}, line width = 0.7pt]     (n4.base) to[out=-235,in=-305] (n3.base) ;
    \end{scope}
                  \filldraw [black] (n1) circle [radius=1.5pt]; 
                  \filldraw [black] (n2)  circle [radius=1.5pt]; 
                  \filldraw [black] (n3) circle [radius=1.5pt]; 
                  \filldraw [black] (n4)  circle [radius=1.5pt]; 
          \filldraw [black]  (0.1,0) circle [radius=1.5pt]; 
           \filldraw [black] (1.3,0) circle [radius=1.5pt];
          }
          +  \gamma^S_{3C}   \,
           \tikz[baseline=(vert_cent.base)]{
\begin{scope}[very thick,decoration={markings,
    mark=at position 0.6 with {\arrow{>}}}] 
  \draw  [postaction={decorate}, line width = 0.7pt]  (-0.4,0)--(0.1,0);
  \draw   [postaction={decorate}, line width = 0.7pt]       (0.7,0) ++(0:0.6cm and 0.4cm) arc (0:40:0.6cm and 0.4cm) 
  node (n1)   {} ; 
    \draw   [white]     (0.7,0) ++(140:0.6cm and 0.4cm) arc (140:110:0.6cm and 0.4cm)   node (n2)   {} ;
   \draw   [white]     (0.7,0) ++(40:0.6cm and 0.4cm) arc (40:140:0.6cm and 0.4cm)   node (n4)   {} ;
    \draw   [white]     (0.7,0) ++(40:0.6cm and 0.4cm) arc (40:70:0.6cm and 0.4cm)   node (n3)   {} ;  
     \draw   [postaction={decorate}, line width = 0.7pt]       (0.7,0) ++(70:0.6cm and 0.4cm) arc (70:40:0.6cm and 0.4cm) ;
  \draw   [postaction={decorate}, line width = 0.7pt]       (0.7,0) ++(140:0.6cm and 0.4cm) arc (140:110:0.6cm and 0.4cm) ;  
  \draw   [postaction={decorate}, line width = 0.7pt]       (0.7,0) ++(140:0.6cm and 0.4cm) arc (140:180:0.6cm and 0.4cm) ;
 \draw  [postaction={decorate}, line width = 0.7pt]      (0.7,0) ++(70:0.6cm and 0.4cm) arc (70:110:0.6cm and 0.4cm) ;
 \draw   [postaction={decorate}, line width = 0.7pt]    (1.3,0)--(1.8,0); 
     \end{scope} 
      \begin{scope}[very thick,decoration={markings,
    mark=at position 0.55 with {\arrow{>}}}] 
    \draw  [postaction={decorate}, line width = 0.7pt]     (0.7,0) ++(0:0.6cm) arc (360:180:0.6cm and 0.4cm);
      \draw     [postaction={decorate}, line width = 0.7pt]        (n4.base) to[out=290,in=250] (n1.base) ;
       \draw     [postaction={decorate}, line width = 0.7pt]        (n3.base) to[out=270,in=270] (n2.base) ;
    \end{scope}
     \filldraw [black] (n1) circle [radius=1.5pt]; 
     \filldraw [black] (n2)  circle [radius=1.5pt]; 
     \filldraw [black] (n3) circle [radius=1.5pt]; 
           \filldraw [black] (n4)  circle [radius=1.5pt];
            \filldraw [black]  (0.1,0) circle [radius=1.5pt]; 
           \filldraw [black] (1.3,0) circle [radius=1.5pt];
     }
          +  \gamma^S_{3D}   \,
\tikz[baseline=(vert_cent.base)]{
\begin{scope}[very thick,decoration={markings,
    mark=at position 0.6 with {\arrow{>}}}] 
  \draw  [postaction={decorate}, line width = 0.7pt]  (-0.4,0)--(0.1,0);
  \draw   [postaction={decorate}, line width = 0.7pt]       (0.7,0) ++(0:0.6cm and 0.4cm) arc (0:50:0.6cm and 0.4cm) 
  node (n1)   {} ;
  \draw   [white]     (0.7,0) ++(50:0.6cm and 0.4cm) arc (50:130:0.6cm and 0.4cm)   node (n2)   {} ;
   \draw   [postaction={decorate}, line width = 0.7pt]     (0.7,0) ++(130:0.6cm and 0.4cm) arc (130:50:0.6cm and 0.4cm) ;
 \draw  [postaction={decorate}, line width = 0.7pt]      (0.7,0) ++(130:0.6cm and 0.4cm) arc (130:180:0.6cm and 0.4cm) ;
\draw   [postaction={decorate}, line width = 0.7pt]    (1.3,0)--(1.8,0);
\draw    [postaction={decorate}, line width = 0.7pt]         (0.7,0) ++(230:0.6cm and 0.4cm) arc (230:310:0.6cm and 0.4cm)  node(n4) {} ;
    \end{scope}
    \begin{scope}[very thick,decoration={markings,
    mark=at position 0.45 with {\arrow{<}}}] 
\draw    [postaction={decorate}, line width = 0.7pt]    (0.7,0) ++(180:0.6cm and 0.4cm) arc (180:230:0.6cm and 0.4cm) node(n3){} ;
\draw    [postaction={decorate}, line width = 0.7pt]        (0.7,0) ++(310:0.6cm and 0.4cm) arc (310:360:0.6cm and 0.4cm) ;
    \end{scope}
                  \filldraw [black] (n1) circle [radius=1.5pt]; 
                  \filldraw [black] (n2)  circle [radius=1.5pt]; 
                  \filldraw [black] (n3) circle [radius=1.5pt]; 
                  \filldraw [black] (n4)  circle [radius=1.5pt]; 
          \filldraw [black]  (0.1,0) circle [radius=1.5pt]; 
           \filldraw [black] (1.3,0) circle [radius=1.5pt];
          \begin{scope}[very thick,decoration={markings,
    mark=at position 0.3 with {\arrow{<}}}] 
\draw    [postaction={decorate}, line width = 0.7pt]        (n1.base) -- (n3.base) ;
\fill [white] (0.7,0) circle[ radius = 3pt] ;
\draw    [postaction={decorate}, line width = 0.7pt]        (n4.base) -- (n2.base) ;
    \end{scope} 
        } \, ,
         \end{align}
with
\be
\gamma^S_1= \tfrac12 \, , \quad  \gamma^S_2= - \tfrac12 \, , \quad  \gamma^S_{3A}= - \tfrac14 \, , \quad 
 \gamma^S_{3B} = - \tfrac18 \, , \quad  \gamma^S_{3C} = 1 \, , \quad  \gamma^S_{3D}= \tfrac32  \zeta_3 \, .
\ee
The contribution \eqref{U1anti} to $\upsilon$ vanishes on reduction to  supersymmetry as in \eqref{U1susy}.

The constraints arising from \eqref{conN1}  at three loops leads to
 4 conditions  relating ${\tilde \beta}_\lambda$ to the anomalous dimensions
\be
 \beta_{\lambda{\hskip 0.5pt}3h} = 2\, \gamma^S_{3A} \, , \quad   \beta_{\lambda{\hskip 0.5pt}3g} = 2\, \gamma^S_{3B}\, , \quad 
 3\, \beta_{\lambda{\hskip 0.5pt}3a} +  \beta_{\lambda{\hskip 0.5pt}3p}= 2\, \gamma^S_{3C}  \, , \quad
 2\, \beta_{\lambda{\hskip 0.5pt}3c} +  \beta_{\lambda{\hskip 0.5pt}3u}= 2\, \gamma^S_{3D}  \, ,
 \ee
 and  14 linear homogeneous relations for ${\tilde \beta}_\lambda$
\begin{align}
&3\, \beta_{\lambda{\hskip 0.5pt}3a} + \beta_{\lambda{\hskip 0.5pt}3i}  +  \beta_{\lambda{\hskip 0.5pt}3m} +  \beta_{\lambda{\hskip 0.5pt}3p}  
+  \beta_{\lambda{\hskip 0.5pt}3\trr}  = 0 \, , \nn \\
& 4\, \beta_{\lambda{\hskip 0.5pt}3b} + 4\, \beta_{\lambda{\hskip 0.5pt}3d}  + 2\, \beta_{\lambda{\hskip 0.5pt}3r} +  \beta_{\lambda{\hskip 0.5pt}3\ti} = 0 \, , \quad
 \beta_{\lambda{\hskip 0.5pt}3b} + 2\, \beta_{\lambda{\hskip 0.5pt}3j}  + \beta_{\lambda{\hskip 0.5pt}3k} + \beta_{\lambda{\hskip 0.5pt}3\ta}  
+ \beta_{\lambda{\hskip 0.5pt}3c'} = 0  \, , \nn \\
&4\, \beta_{\lambda{\hskip 0.5pt}3c} + 4\, \beta_{\lambda{\hskip 0.5pt}3e}  + 2\, \beta_{\lambda{\hskip 0.5pt}3t} +4\,  \beta_{\lambda{\hskip 0.5pt}3v}  
+ 2\, \beta_{\lambda{\hskip 0.5pt}3\thh} + \beta  \raisebox{-1.5 pt}{$\scriptstyle \lambda{\hskip 0.5pt}3\tl$} +  \beta_{\lambda{\hskip 0.5pt}3h'}  =0 \, , \qquad \nn \\
& 2\, \beta_{\lambda{\hskip 0.5pt}3c} + 2\, \beta_{\lambda{\hskip 0.5pt}3i}  + \beta_{\lambda{\hskip 0.5pt}3o} + 2\,  \beta_{\lambda{\hskip 0.5pt}3y} = 0 \, , \quad
\beta_{\lambda{\hskip 0.5pt}3d} + \, \beta_{\lambda{\hskip 0.5pt}3h}  +  \beta_{\lambda{\hskip 0.5pt}3\too}= 0 \, , \quad
2\, \beta_{\lambda{\hskip 0.5pt}3j} + \beta_{\lambda{\hskip 0.5pt}3\tq} =0  \, , 
\nn \\
& \beta_{\lambda{\hskip 0.5pt}3e} +  2 \, \beta_{\lambda{\hskip 0.5pt}3g}  +  \beta_{\lambda{\hskip 0.5pt}3\tp}= 0 \, , \quad
2\, \beta_{\lambda{\hskip 0.5pt}3e} +   \beta_{\lambda{\hskip 0.5pt}3i}  +   \beta_{\lambda{\hskip 0.5pt}3v} +   
\beta_{\lambda{\hskip 0.5pt}3z}  +  \beta_{\lambda{\hskip 0.5pt}3d'}= 0 \, ,  \nn \\
& \beta_{\lambda{\hskip 0.5pt}3b} +  \beta_{\lambda{\hskip 0.5pt}3i}  +  \beta  \raisebox{-1.5 pt}{$\scriptstyle \lambda{\hskip 0.5pt}3\td$} = 0\, , \quad 
8\, \beta_{\lambda{\hskip 0.5pt}3e} +  4\,  \beta_{\lambda{\hskip 0.5pt}3w}  +  \beta_{\lambda{\hskip 0.5pt}3\tn}= 0 \, ,  \quad 
\quad  4\, \beta_{\lambda{\hskip 0.5pt}3j}  +  \beta_{\lambda{\hskip 0.5pt}3l} = 0 \, , \nn \\
& 4\, \beta_{\lambda{\hskip 0.5pt}3b} + 2\, \beta_{\lambda{\hskip 0.5pt}3s}  + \beta_{\lambda{\hskip 0.5pt}3x} + \beta_{\lambda{\hskip 0.5pt}3a'}  = 0 \, , \quad
 \nn \\
 & 4\, \beta_{\lambda{\hskip 0.5pt}3b} + 4\, \beta_{\lambda{\hskip 0.5pt}3d}   + \beta_{\lambda{\hskip 0.5pt}3f} + 2\, \beta_{\lambda{\hskip 0.5pt}3r} 
 +   \beta_{\lambda{\hskip 0.5pt}3\tg} +  2\, \beta  \raisebox{-1.5 pt}{$\scriptstyle \lambda{\hskip 0.5pt}3\tk$} +   \beta_{\lambda{\hskip 0.5pt}3b'}  = 0 \, .
 \label{N1consis}
\end{align}
 The last 5 relations involve contributions arising from non planar diagrams.

\vskip 2cm

 \subsection{\texorpdfstring{$\N=\tfrac12$}{N=1/2} Supersymmetry} 
 \label{susyhalf}
 
 This is a special case of the general scalar fermion theory where of course the number of real 
 scalars matches the number of fermions with $y^a \to Y^{abc}$ a symmetric real tensor 
 and also $\lambda^{abcd} \to \S_3\, Y^{abe} Y^{cde}$.  For this theory $\phi, \psi$ can be
 combined as a real superfield $\Phi$  and in  a perturbative expansion the diagrams 
 reduce to those  of a simple $\phi^3$ theory. 
 Under renormalisation as a four
 dimensional theory there is a $\beta$-function $\beta_Y{}^{abc} $ 
 and anomalous dimension $\gamma_{\Phi }{}^{ab} $. For  this case, unlike for $\N=1$ supersymmetry, 
 there are non trivial
divergent vertex graphs so that $\beta_Y$ and $\gamma_\Phi$ are independent although the scalar
$\beta$-function $\beta_\lambda$ is determined in terms of these.

At lowest one loop order this gives
 \be
  \gamma_{\Phi\hskip 0.5pt 1}= \gamma_{\phi{\hskip 0.5pt}1} = \gamma_{\psi{\hskip 0.5pt}1} \, , \qquad 
\beta_{ Y \hskip 0.2pt 1} = 2\, \beta_{\lambda{\hskip 0.5pt}1a} = \beta_{y{\hskip 0.5pt}1}  =  4 \, \gamma_{\phi{\hskip 0.5pt}1} \, , \qquad
4 \,  \beta_{\lambda{\hskip 0.5pt}1a} + \beta_{\lambda{\hskip 0.5pt}1b} = 0 \, ,
\ee
where
\be
 \gamma_\Phi{\!}^{(1)}=   \gamma_{\Phi \hskip 0.5pt1} \,
\tikz[baseline=(vert_cent.base)]{
  \node (vert_cent) {\hspace{-13pt}$\phantom{-}$};
  \draw (0,0)--(0.3,0)
        (0.7,0) ++(0:0.4cm) arc (0:360:0.4cm and 0.3cm)
        (1.1,0)--(1.4,0);
}\,  , \qquad
{\tilde \beta}_Y{\!}^{(1)} =  \beta_{ Y \hskip 0.2pt 1}
\tikz[baseline=(vert_cent.base)]{
  \node (vert_cent) {\hspace{-13pt}$\phantom{-}$};
  \draw (0.7,0) ++(0:0.35cm) arc (0:135:0.35cm) node (n1) {}
        (0.7,0) ++(135:0.35cm) arc (135:225:0.35cm) node (n2) {}
        (0.7,0) ++(225:0.35cm) arc (225:360:0.35cm) node (n3) {}
        (n1.base)--+(145:0.5cm)
        (n2.base)--+(215:0.5cm)
        (n3.base)--+(0:0.5cm);
 }  \, .
\ee
The equality $\gamma_{\phi{\hskip 0.5pt}1} = \gamma_{\psi{\hskip 0.5pt}1}$ is a reflection of the choice
of normalisation of fermion traces in the main body of results. Each fermion trace gives the contribution of
a two component real fermion propagating round the loop.

At two loops equality of $\gamma_\phi{\hskip 0.5pt}, \ \gamma_\psi$ and symmetry  of $\beta_y$ requires
\begin{align}
&   \gamma_{\Phi\hskip 0.5pt 2A} = 3\,  \gamma_{\phi{\hskip 0.5pt}2a} + \gamma_{\phi{\hskip 0.5pt}2b} =  \gamma_{\psi{\hskip 0.5pt}2a} + \gamma_{\psi{\hskip 0.5pt}2b} \, , \qquad 
  \gamma_{\Phi\hskip 0.5pt 2B} =  6\,  \gamma_{\phi{\hskip 0.5pt}2a} + \gamma_{\phi{\hskip 0.5pt}2c} = \gamma_{\psi{\hskip 0.5pt}2c}  \, , \nn \\
&  \beta_{Y\hskip 0.5pt 2A}=\beta_{y{\hskip 0.5pt}2b} = \beta_{y{\hskip 0.5pt}2c}  \, , \quad  \beta_{Y\hskip 0.5pt 2B}= \beta_{y{\hskip 0.5pt}2a} + \beta_{y{\hskip 0.5pt}2d} = \beta_{y{\hskip 0.5pt}2e}  \, , 
\quad    \beta_{Y\hskip 0.5pt 2C} = \beta_{y{\hskip 0.5pt}2a} +  \beta_{y{\hskip 0.5pt}2f}  \, ,
\end{align}
where
\begin{align}
\gamma_\Phi{\!}^{(2)}= {}& \gamma_{\Phi\hskip 0.5pt 2A} 
\tikz[baseline=(vert_cent.base)]{
  \node (vert_cent) {\hspace{-13pt}$\phantom{-}$};
  \draw (0,0)--(0.3,0)
        (0.7,0) ++(0:0.4cm and 0.3cm) arc (0:50:0.4cm and 0.3cm) node (n1)
        {}
        (0.7,0) ++(50:0.4cm and 0.3cm) arc (50:130:0.4cm and 0.3cm) node
        (n2) {}
        (0.7,0) ++(130:0.4cm and 0.3cm) arc (130:360:0.4cm and 0.3cm)
        (n1.base) to[out=215,in=325] (n2.base)
        (1.1,0)--(1.4,0); 
}
+ \gamma_{\Phi\hskip 0.5pt 2B}  \tikz[baseline=(vert_cent.base)]{
  \node (vert_cent) {\hspace{-13pt}$\phantom{-}$};
  \draw (0,0)--(0.3,0)
        (0.7,0) ++(0:0.4cm) arc (0:360:0.4cm and 0.3cm)
        (0.7,0.3)--(0.7,-0.3)
        (1.1,0)--(1.4,0);
}   
 \, , \nn \\
{\tilde \beta}_Y{\!}^{(2)} = {}&   {\mathcal S}_3 \bigg (
 \beta_{Y\hskip 0.5pt 2A}
\tikz[baseline=(vert_cent.base)]{
  \node (vert_cent) {\hspace{-13pt}$\phantom{-}$};
  \draw (0.7,0) ++(0:0.35cm) arc (0:30:0.35cm) node (n1) {}
        (0.7,0) ++(30:0.35cm) arc (30:105:0.35cm) node (n2) {}
        (0.7,0) ++(105:0.35cm) arc (105:135:0.35cm) node (n3) {}
        (0.7,0) ++(135:0.35cm) arc (135:225:0.35cm) node (n4) {}
        (0.7,0) ++(225:0.35cm) arc (225:360:0.35cm) node (n5) {}
        (n1.base) to[out=200,in=300] (n2.base)
        (n3.base)--+(145:0.5cm)
        (n4.base)--+(215:0.5cm)
        (n5.base)--+(0:0.5cm);
}
 +  \beta_{Y\hskip 0.5pt 2B}
\tikz[baseline=(vert_cent.base)]{
  \node (vert_cent) {\hspace{-13pt}$\phantom{-}$};
  \draw (0.7,0) ++(0:0.35cm) arc (0:45:0.35cm) node (n1) {}
        (0.7,0) ++(45:0.35cm) arc (45:135:0.35cm) node (n2) {}
        (0.7,0) ++(135:0.35cm) arc (135:225:0.35cm) node (n3) {}
        (0.7,0) ++(225:0.35cm) arc (225:315:0.35cm) node (n4) {}
        (0.7,0) ++(315:0.35cm) arc (315:360:0.35cm) node (n5) {}
        (n1.base) to [out=220,in=140] (n4.base)
        (n2.base)--+(145:0.5cm)
        (n3.base)--+(215:0.5cm)
        (n5.base)--+(0:0.5cm);
}
 \bigg )
+  \beta_{Y\hskip 0.5pt 2C}
\tikz[baseline=(vert_cent.base)]{
  \node (vert_cent) {\hspace{-13pt}$\phantom{-}$};
  \draw (0.7,0) ++(0:0.35cm) arc (0:45:0.35cm) node (n1) {}
        (0.7,0) ++(45:0.35cm) arc (45:135:0.35cm) node (n2) {};
  \draw[white] (0.7,0) ++(135:0.35cm) arc (135:225:0.35cm) node (n3) {};
  \draw (0.7,0) ++(225:0.35cm) arc (225:315:0.35cm) node (n4) {}
        (0.7,0) ++(315:0.35cm) arc (315:360:0.35cm) node (n5) {}
        (n2.base)--+(145:0.5cm)
        (n3.base)--+(215:0.5cm)
        (n5.base)--+(0:0.5cm);
  \draw[name path=a] (n1.base)--(n3.base);
  \draw[white, name path=b] (n2.base)--(n4.base);
  \path[name intersections={of=a and b,by=i}];
  \node[fill=white, inner sep=2pt, rotate=45] at (i) {};
  \draw (n2.base)--(n4.base);
}
\, .
\label{betaS2}
\end{align}
Determining $\beta_\lambda$ in terms of $\beta_Y$ and $\gamma_\phi$ imposes the restrictions
\begin{align}
& \beta_{\lambda{\hskip 0.5pt}2a} =  2\,  \gamma_{\Phi\hskip 0.5pt 2A} =   \beta_{Y\hskip 0.5pt 2A} = \tfrac12 \,   \beta_{Y\hskip 0.5pt 2B}\, ,    \qquad
0 =  \gamma_{\Phi\hskip 0.5pt 2B} =   \beta_{Y\hskip 0.5pt 2C} \, ,  \nn \\
&4\,  \beta_{\lambda{\hskip 0.5pt}2a}  =  4\,  \beta_{\lambda{\hskip 0.5pt}2b}= - 2\,  \beta_{\lambda{\hskip 0.5pt}2d} = - 2\,  \beta_{\lambda{\hskip 0.5pt}2e} = -  \beta_{\lambda{\hskip 0.5pt}2f}  =  -  \beta_{\lambda{\hskip 0.5pt}2g}  \, , \qquad  \beta_{\lambda{\hskip 0.5pt}2c} = 0 \, ,  
\end{align}
which implies further constraints on the Yukawa  $\beta$-functions
\be
 \beta_{y{\hskip 0.5pt}2a} =  -  \beta_{y{\hskip 0.5pt}2f} =  2\,  \beta_{y{\hskip 0.5pt}2b} -  \beta_{y{\hskip 0.5pt}2d} \, , \qquad 
 \beta_{y{\hskip 0.5pt}2b} = 2 ( \gamma_{\psi{\hskip 0.5pt}2a} + \gamma_{\psi{\hskip 0.5pt}2b})\, .
\ee

At three loops  there are 9  propagator diagrams
 \begin{align}
\gamma_\Phi{}^{(3)}
 = {}& \gamma_{\Phi\hskip 0.5pt 3A}
\tikz[baseline=(vert_cent.base)]{
  \node (vert_cent) {\hspace{-13pt}$\phantom{-}$};
  \draw (0,0)--(0.3,0)
        (0.7,0) ++(0:0.4cm and 0.3cm) arc (0:20:0.4cm and 0.3cm) node (n1)
        {}
        (0.7,0) ++(20:0.4cm and 0.3cm) arc (20:70:0.4cm and 0.3cm) node (n2)
        {}
        (0.7,0) ++(70:0.4cm and 0.3cm) arc (70:110:0.4cm and 0.3cm) node
        (n3) {}
        (0.7,0) ++(110:0.4cm and 0.3cm) arc (110:160:0.4cm and 0.3cm) node
        (n4) {}
        (0.7,0) ++(20:0.4cm and 0.3cm) (n1.base) to[out=205,in=250]
        (n2.base)
        (0.7,0) ++(160:0.4cm and 0.3cm) arc (160:360:0.4cm and 0.3cm)
        (n3.base) to[out=290,in=335] (n4.base)
        (1.1,0)--(1.4,0);
}
+ \gamma_{\Phi\hskip 0.5pt 3B}
\tikz[baseline=(vert_cent.base)]{
  \node (vert_cent) {\hspace{-13pt}$\phantom{-}$};
  \draw (0,0)--(0.3,0)
        (0.7,0) ++(0:0.4cm and 0.3cm) arc (0:50:0.4cm and 0.3cm) node (n1)
        {}
        (0.7,0) ++(50:0.4cm and 0.3cm) arc (50:130:0.4cm and 0.3cm) node
        (n2) {}
        (0.7,0) ++(130:0.4cm and 0.3cm) arc (130:230:0.4cm and 0.3cm)
        (n1.base) to[out=215,in=325] (n2.base)
        (0.7,0) ++(230:0.4cm and 0.3cm) node (n3) {}
        (0.7,0) ++(230:0.4cm and 0.3cm) arc (230:310:0.4cm and 0.3cm) node
        (n4) {}
        (0.7,0) ++(310:0.4cm and 0.3cm) arc (310:360:0.4cm and 0.3cm)
        (n3.base) to[out=35,in=145] (n4.base)
        (1.1,0)--(1.4,0);
}
+ \gamma_{\Phi\hskip 0.5pt 3C}
\tikz[baseline=(vert_cent.base)]{
  \node (vert_cent) {\hspace{-13pt}$\phantom{-}$};
  \draw (0,0)--(0.3,0)
        (0.7,0) ++(0:0.4cm and 0.3cm) arc (0:30:0.4cm and 0.3cm) node (n1)
        {}
        (0.7,0) ++(30:0.4cm and 0.3cm) arc (30:55:0.4cm and 0.3cm) node
        (n2) {}
        (0.7,0) ++(55:0.4cm and 0.3cm) arc (55:125:0.4cm and 0.3cm) node
        (n3) {}
        (0.7,0) ++(125:0.4cm and 0.3cm) arc (125:150:0.4cm and 0.3cm) node
        (n4) {}
        (0.7,0) ++(150:0.4cm and 0.3cm) arc (150:360:0.4cm and 0.3cm)
        (n1.base) to[out=215,in=325] (n4.base)
        (0.7,0) ++(60:0.4cm and 0.3cm) (n2.base) to[out=215,in=325]
        (n3.base)
        (1.1,0)--(1.4,0);
}
+ \gamma_{\Phi\hskip 0.5pt 3D} \; \S_2 \,   \tikz[baseline=(vert_cent.base)]{
  \node (vert_cent) {\hspace{-13pt}$\phantom{-}$};
  \draw (0,0)--(0.3,0)
        (0.7,0) ++(0:0.4cm and 0.3cm) arc (0:110:0.4cm and 0.3cm) node (n1)
        {}
        (0.7,0) ++(110:0.4cm and 0.3cm) arc (110:160:0.4cm and 0.3cm) node
        (n2) {}
        (0.7,0) ++(160:0.4cm and 0.3cm) arc (160:360:0.4cm and 0.3cm)
        (n1.base) to[out=290,in=335] (n2.base)
        (1.1,0)--(1.4,0)
        (0.7,0.3)--(0.7,-0.3);
}\nn \\
&\hspace{0.01cm} {}+ \gamma_{\Phi\hskip 0.5pt 3E}
\tikz[baseline=(vert_cent.base)]{
  \node (vert_cent) {\hspace{-13pt}$\phantom{-}$};
  \draw (0,0)--(0.3,0)
        (0.7,0) ++(0:0.4cm) arc (0:360:0.4cm and 0.3cm)
        (0.7,0.3)--(0.7,0.2)
        (0.7,0) ++(0:0.1cm) arc (0:360:0.1cm and 0.2cm)
        (0.7,-0.2)--(0.7,-0.3)
        (1.1,0)--(1.4,0);
}
+ \gamma_{\Phi\hskip 0.5pt 3F}
\tikz[baseline=(vert_cent.base)]{
  \node (vert_cent) {\hspace{-13pt}$\phantom{-}$};
  \draw (0,0)--(0.3,0)
        (0.7,0) ++(0:0.4cm and 0.3cm) arc (0:50:0.4cm and 0.3cm) node (n1)
        {}
        (0.7,0) ++(50:0.4cm and 0.3cm) arc (50:130:0.4cm and 0.3cm) node
        (n2) {}
        (0.7,0) ++(130:0.4cm and 0.3cm) arc (130:360:0.4cm and 0.3cm)
        (n1.base) to[out=215,in=325] (n2.base)
        (1.1,0)--(1.4,0)
        (0.7,0.3)--(0.7,0.15);
}
+ \gamma_{\Phi\hskip 0.5pt 3G}
\tikz[baseline=(vert_cent.base)]{
  \node (vert_cent) {\hspace{-13pt}$\phantom{-}$};
  \draw (0,0)--(0.3,0);
  \draw (1.1,0)--(1.4,0);
  \draw[clip] (0.7,0) ++(0:0.4cm) arc (0:360:0.4cm and 0.3cm);
  \draw (0.45,0.25)--(0.7,0);
  \draw (0.95,0.25)--(0.7,0);
  \draw (0.7,0)--(0.7,-0.3);
}
+ \gamma_{\Phi\hskip 0.5pt 3H}  \tikz[baseline=(vert_cent.base)]{
  \node (vert_cent) {\hspace{-13pt}$\phantom{-}$};
  \draw (0,0)--(0.3,0);
  \draw (1.1,0)--(1.4,0);
  \draw[clip] (0.7,0) ++(0:0.4cm) arc (0:360:0.4cm and 0.3cm);
  \draw (0.56,0.3)--(0.56,-0.3);
  \draw (0.82,0.3)--(0.82,-0.3);
} \nn \\
&{} + \gamma_{\Phi\hskip 0.5pt 3I}
\tikz[baseline=(vert_cent.base)]{
  \node (vert_cent) {\hspace{-13pt}$\phantom{-}$};
  \draw (0,0)--(0.3,0);
  \draw (1.1,0)--(1.4,0);
  \draw[clip] (0.7,0) ++(0:0.4cm) arc (0:360:0.4cm and 0.3cm);
  \draw[name path=a] (0.45,0.3)--(0.95,-0.3);
  \draw[white, name path=b] (0.95,0.3)--(0.45,-0.3);
  \path[name intersections={of=a and b,by=i}];
  \node[fill=white, inner sep=1.5pt, rotate=45] at (i) {};
  \draw (0.95,0.3)--(0.45,-0.3);
} \, ,
\label{gammaS}
\end{align}
with $
\S_2 \, \tikz[baseline=(vert_cent.base)]{
  \node (vert_cent) {\hspace{-13pt}$\phantom{-}$};
  \draw (0,0)--(0.3,0)
        (0.7,0) ++(0:0.4cm and 0.3cm) arc (0:110:0.4cm and 0.3cm) node (n1)
        {}
        (0.7,0) ++(110:0.4cm and 0.3cm) arc (110:160:0.4cm and 0.3cm) node
        (n2) {}
        (0.7,0) ++(160:0.4cm and 0.3cm) arc (160:360:0.4cm and 0.3cm)
        (n1.base) to[out=290,in=335] (n2.base)
        (1.1,0)--(1.4,0)
        (0.7,0.3)--(0.7,-0.3);
}
= \tikz[baseline=(vert_cent.base)]{
  \node (vert_cent) {\hspace{-13pt}$\phantom{-}$};
  \draw (0,0)--(0.3,0)
        (0.7,0) ++(0:0.4cm and 0.3cm) arc (0:110:0.4cm and 0.3cm) node (n1)
        {}
        (0.7,0) ++(110:0.4cm and 0.3cm) arc (110:160:0.4cm and 0.3cm) node
        (n2) {}
        (0.7,0) ++(160:0.4cm and 0.3cm) arc (160:360:0.4cm and 0.3cm)
        (n1.base) to[out=290,in=335] (n2.base)
        (1.1,0)--(1.4,0)
        (0.7,0.3)--(0.7,-0.3);
}
+
 \tikz[baseline=(vert_cent.base)]{
  \node (vert_cent) {\hspace{-13pt}$\phantom{-}$};
  \draw (0,0)--(0.3,0)
        (0.7,0) ++(0:0.4cm and 0.3cm) arc (0:20:0.4cm and 0.3cm) node (n1)
        {}
        (0.7,0) ++(20:0.4cm and 0.3cm) arc (20:70:0.4cm and 0.3cm) node
        (n2) {}
        (0.7,0) ++(70:0.4cm and 0.3cm) arc (70:360:0.4cm and 0.3cm)
        (n1.base) to[out=205,in=250] (n2.base)
        (1.1,0)--(1.4,0)
        (0.7,0.3)--(0.7,-0.3);
}
$.
Reducing general results requires
\begin{align}
  \gamma_{\Phi\hskip 0.5pt 3A}={} & \gamma_{\phi{\hskip 0.5pt}3a} + \gamma_{\phi{\hskip 0.5pt}3f}  =   \gamma_{\psi{\hskip 0.5pt}3c} + \gamma_{\psi{\hskip 0.5pt}3i}  \, , \nn \\
\gamma_{\Phi\hskip 0.5pt 3B} = {} &  \gamma_{\phi{\hskip 0.5pt}3b} +  \gamma_{\phi{\hskip 0.5pt}3h} =  \gamma_{\psi{\hskip 0.5pt}3d}  \, ,  \nn \\
 \gamma_{\Phi\hskip 0.5pt 3C}  = {}& 2\,  \gamma_{\phi{\hskip 0.5pt}3b} + \gamma_{\phi{\hskip 0.5pt}3d} +  \gamma_{\phi{\hskip 0.5pt}3g} =  
 3 \, \gamma_{\psi{\hskip 0.5pt}3a} + \gamma_{\psi{\hskip 0.5pt}3e} + \gamma_{\psi{\hskip 0.5pt}3g}+ \gamma_{\psi{\hskip 0.5pt}3j}    \, , \nn \\
\gamma_{\Phi\hskip 0.5pt 3D}={}&  2 \,  \gamma_{\phi{\hskip 0.5pt}3a} + 2 \gamma_{\phi{\hskip 0.5pt}3b} + \gamma_{\phi{\hskip 0.5pt}3j} 
=  \gamma_{\psi{\hskip 0.5pt}3f} + \gamma_{\psi{\hskip 0.5pt}3n}    \, , \nn \\
\gamma_{\Phi\hskip 0.5pt 3E}  = {}& 2 \, \gamma_{\phi{\hskip 0.5pt}3b} +  \gamma_{\phi{\hskip 0.5pt}3e} =  \gamma_{\psi{\hskip 0.5pt}3o}   \, , \nn \\
\gamma_{\Phi\hskip 0.5pt 3F}  = {} & 2\,  \gamma_{\phi{\hskip 0.5pt}3a} + \gamma_{\phi{\hskip 0.5pt}3i}  =  
6 \, \gamma_{\psi{\hskip 0.5pt}3a} + \gamma_{\psi{\hskip 0.5pt}3h} + \gamma_{\psi{\hskip 0.5pt}3l}  \, , \nn \\
\gamma_{\Phi\hskip 0.5pt 3G}  ={} &  12 \,  \gamma_{\phi{\hskip 0.5pt}3a} + 4 \gamma_{\phi{\hskip 0.5pt}3c} + \gamma_{\phi{\hskip 0.5pt}3l} =  
\gamma_{\psi{\hskip 0.5pt}3b} + 2 \, \gamma_{\psi{\hskip 0.5pt}3m}  \, ,  \nn \\
 \gamma_{\Phi\hskip 0.5pt 3H}  = {} &  4\,  \gamma_{\phi{\hskip 0.5pt}3a} + \gamma_{\phi{\hskip 0.5pt}3k}  
 =   \gamma_{\psi{\hskip 0.5pt}3b} + \gamma_{\psi{\hskip 0.5pt}3k} \, , \nn \\
\gamma_{\Phi\hskip 0.5pt 3I}  = {}&  4 \,  \gamma_{\phi{\hskip 0.5pt}3a} + 2 \gamma_{\phi{\hskip 0.5pt}3c} + \gamma_{\phi{\hskip 0.5pt}3m} =  \gamma_{\psi{\hskip 0.5pt}3b} + \gamma_{\psi{\hskip 0.5pt}3p} \, .
 \end{align}
 
 For an antisymmetric contribution
 \be
\upsilon_\Phi{}^{(3)} = \upsilon_{3D} \; \A_2 \,   \tikz[baseline=(vert_cent.base)]{
  \node (vert_cent) {\hspace{-13pt}$\phantom{-}$};
  \draw (0,0)--(0.3,0)
        (0.7,0) ++(0:0.4cm and 0.3cm) arc (0:110:0.4cm and 0.3cm) node (n1)
        {}
        (0.7,0) ++(110:0.4cm and 0.3cm) arc (110:160:0.4cm and 0.3cm) node
        (n2) {}
        (0.7,0) ++(160:0.4cm and 0.3cm) arc (160:360:0.4cm and 0.3cm)
        (n1.base) to[out=290,in=335] (n2.base)
        (1.1,0)--(1.4,0)
        (0.7,0.3)--(0.7,-0.3);
}
=  \upsilon_{\Phi \hskip 0.5pt 3D} \, \Big (
 \tikz[baseline=(vert_cent.base)]{
  \node (vert_cent) {\hspace{-13pt}$\phantom{-}$};
  \draw (0,0)--(0.3,0)
        (0.7,0) ++(0:0.4cm and 0.3cm) arc (0:110:0.4cm and 0.3cm) node (n1)
        {}
        (0.7,0) ++(110:0.4cm and 0.3cm) arc (110:160:0.4cm and 0.3cm) node
        (n2) {}
        (0.7,0) ++(160:0.4cm and 0.3cm) arc (160:360:0.4cm and 0.3cm)
        (n1.base) to[out=290,in=335] (n2.base)
        (1.1,0)--(1.4,0)
        (0.7,0.3)--(0.7,-0.3);
}
-
 \tikz[baseline=(vert_cent.base)]{
  \node (vert_cent) {\hspace{-13pt}$\phantom{-}$};
  \draw (0,0)--(0.3,0)
        (0.7,0) ++(0:0.4cm and 0.3cm) arc (0:20:0.4cm and 0.3cm) node (n1)
        {}
        (0.7,0) ++(20:0.4cm and 0.3cm) arc (20:70:0.4cm and 0.3cm) node
        (n2) {}
        (0.7,0) ++(70:0.4cm and 0.3cm) arc (70:360:0.4cm and 0.3cm)
        (n1.base) to[out=205,in=250] (n2.base)
        (1.1,0)--(1.4,0)
        (0.7,0.3)--(0.7,-0.3);
}\Big )  \ ,
\ee
where from \eqref{upphi3} and \eqref{uppsi3}
\be
\upsilon_{3D} = \upsilon_{\phi{\hskip 0.5pt}3j} = -  \upsilon_{\psi{\hskip 0.5pt}3f} + \upsilon_{\psi{\hskip 0.5pt}3n}  \, .
\label{susyup}
\ee
 
 At three loops there are  17 1PI contributions to the symmetric $\beta$-function which are expressible
 diagrammatically as
 \begin{align}
{\tilde \beta}_Y{}^{(3)}
= {} & 
{\mathcal S}_3 \bigg( \beta_{Y\hskip 0.5pt 3A}
\tikz[baseline=(vert_cent.base)]{
  \node (vert_cent) {\hspace{-13pt}$\phantom{-}$};
  \draw (0.7,0) ++(0:0.35cm) arc (0:30:0.35cm) node (n1) {}
        (0.7,0) ++(30:0.35cm) arc (30:105:0.35cm) node (n2) {}
        (0.7,0) ++(105:0.35cm) arc (105:135:0.35cm) node (n3) {}
        (0.7,0) ++(135:0.35cm) arc (135:225:0.35cm) node (n4) {}
        (0.7,0) ++(225:0.35cm) arc (225:255:0.35cm) node (n5) {}
        (0.7,0) ++(255:0.35cm) arc (255:330:0.35cm) node (n6) {}
        (0.7,0) ++(330:0.35cm) arc (330:360:0.35cm) node (n7) {}
        (n1.base) to[out=200,in=300] (n2.base)
        (n5.base) to[out=70,in=160] (n6.base)
        (n3.base)--+(145:0.5cm)
        (n4.base)--+(215:0.5cm)
        (n7.base)--+(0:0.5cm);
}
+ \beta_{Y\hskip 0.5pt 3B}
\tikz[baseline=(vert_cent.base)]{
  \node (vert_cent) {\hspace{-13pt}$\phantom{-}$};
  \draw (0.7,0) ++(0:0.35cm) arc (0:15:0.35cm) node (n1) {}
        (0.7,0) ++(15:0.35cm) arc (15:60:0.35cm) node (n6) {}
        (0.7,0) ++(60:0.35cm) arc (60:75:0.35cm) node (n7) {}
        (0.7,0) ++(75:0.35cm) arc (75:120:0.35cm) node (n8) {}
        (0.7,0) ++(120:0.35cm) arc (120:135:0.35cm) node (n3) {}
        (0.7,0) ++(135:0.35cm) arc (135:225:0.35cm) node (n4) {}
        (0.7,0) ++(225:0.35cm) arc (225:360:0.35cm) node (n5) {}
        (n1.base) to[out=170,in=260] (n6.base)
        (n7.base) to[out=235,in=320] (n8.base)
        (n3.base)--+(145:0.5cm)
        (n4.base)--+(215:0.5cm)
        (n5.base)--+(0:0.5cm);
}
+ \beta_{Y\hskip 0.5pt 3C}
\tikz[baseline=(vert_cent.base)]{
  \node (vert_cent) {\hspace{-13pt}$\phantom{-}$};
  \draw (0.7,0) ++(0:0.35cm) arc (0:20:0.35cm) node (n1) {}
        (0.7,0) ++(20:0.35cm) arc (20:40:0.35cm) node (n6) {}
        (0.7,0) ++(40:0.35cm) arc (40:95:0.35cm) node (n7) {}
        (0.7,0) ++(95:0.35cm) arc (95:115:0.35cm) node (n2) {}
        (0.7,0) ++(115:0.35cm) arc (115:135:0.35cm) node (n3) {}
        (0.7,0) ++(135:0.35cm) arc (135:225:0.35cm) node (n4) {}
        (0.7,0) ++(225:0.35cm) arc (225:360:0.35cm) node (n5) {}
        (n1.base) to[out=200,in=295] (n2.base)
        (n6.base) to[out=200,in=295] (n7.base)
        (n3.base)--+(145:0.5cm)
        (n4.base)--+(215:0.5cm)
        (n5.base)--+(0:0.5cm);
}
+ \beta_{Y\hskip 0.5pt 3D}
\tikz[baseline=(vert_cent.base)]{
  \node (vert_cent) {\hspace{-13pt}$\phantom{-}$};
  \draw (0.7,0) ++(0:0.35cm) arc (0:90:0.35cm) node (n1) {}
        (0.7,0) ++(90:0.35cm) arc (90:135:0.35cm) node (n2) {}
        (0.7,0) ++(135:0.35cm) arc (135:150:0.35cm) node (n6) {}
        (0.7,0) ++(150:0.35cm) arc (150:210:0.35cm) node (n7) {}
        (0.7,0) ++(210:0.35cm) arc (210:225:0.35cm) node (n3) {}
        (0.7,0) ++(225:0.35cm) arc (225:270:0.35cm) node (n4) {}
        (0.7,0) ++(270:0.35cm) arc (270:360:0.35cm) node (n5) {}
        (n1.base)--(n4.base)
        (n6.base) to[out=300,in=60] (n7.base)
        (n2.base)--+(145:0.5cm)
        (n3.base)--+(215:0.5cm)
        (n5.base)--+(0:0.5cm);
}
\nn \\
\noalign {\vskip - 4pt}
& \hskip 9.5cm {}+ \beta_{Y\hskip 0.5pt 3E} 
\tikz[baseline=(vert_cent.base)]{
  \node (vert_cent) {\hspace{-13pt}$\phantom{-}$};
  \draw (0.7,0) ++(0:0.35cm) arc (0:90:0.35cm) node (n1) {}
        (0.7,0) ++(90:0.35cm) arc (90:135:0.35cm) node (n2) {}
        (0.7,0) ++(135:0.35cm) arc (135:225:0.35cm) node (n3) {}
        (0.7,0) ++(225:0.35cm) arc (225:270:0.35cm) node (n4) {}
        (0.7,0) ++(270:0.35cm) arc (270:360:0.35cm) node (n5) {}
        (n1.base) node[midway] (n6) {} to (n4.base)
        (n2.base)--+(145:0.5cm)
        (n3.base)--+(215:0.5cm)
        (n5.base)--+(0:0.5cm);
  \draw[fill=white] (n6.base) ++(-0.15cm,0.25cm) arc (0:360:0.1cm and
  0.2cm);
} \bigg ) \nn \\
&{} + \S_6 \bigg ( \beta_{Y\hskip 0.5pt 3F}
\tikz[baseline=(vert_cent.base)]{
  \node (vert_cent) {\hspace{-13pt}$\phantom{-}$};
  \draw (0.7,0) ++(0:0.35cm) arc (0:15:0.35cm) node (n6) {}
        (0.7,0) ++(15:0.35cm) arc (15:75:0.35cm) node (n7) {}
        (0.7,0) ++(75:0.35cm) arc (75:90:0.35cm) node (n1) {}
        (0.7,0) ++(90:0.35cm) arc (90:135:0.35cm) node (n2) {}
        (0.7,0) ++(135:0.35cm) arc (135:225:0.35cm) node (n3) {}
        (0.7,0) ++(225:0.35cm) arc (225:270:0.35cm) node (n4) {}
        (0.7,0) ++(270:0.35cm) arc (270:360:0.35cm) node (n5) {}
        (n1.base)--(n4.base)
        (n6.base) to[out=160,in=290] (n7.base)
        (n2.base)--+(145:0.5cm)
        (n3.base)--+(215:0.5cm)
        (n5.base)--+(0:0.5cm);
}
+ \beta_{Y\hskip 0.5pt 3G} \, 
\tikz[baseline=(vert_cent.base)]{
  \node (vert_cent) {\hspace{-13pt}$\phantom{-}$};
  \draw (0.7,0) ++(0:0.35cm) arc (0:45:0.35cm) node (n1) {}
        (0.7,0) ++(45:0.35cm) arc (45:65:0.35cm) node (n2) {}
        (0.7,0) ++(65:0.35cm) arc (65:115:0.35cm) node (n5) {}
        (0.7,0) ++(115:0.35cm) arc (115:135:0.35cm) node (n3) {}
        (0.7,0) ++(135:0.35cm) arc (135:225:0.35cm) node (n4) {}
        (0.7,0) ++(225:0.35cm) arc (225:315:0.35cm) node (n6) {}
        (0.7,0) ++(315:0.35cm) arc (315:360:0.35cm) node (n7) {}
        (n1.base) to[out=240,in=120] (n6.base)
        (n2.base) to[out=220,in=320] (n5.base)
        (n3.base)--+(145:0.5cm)
        (n4.base)--+(215:0.5cm)
        (n7.base)--+(0:0.5cm);
}
 + \beta_{Y\hskip 0.5pt 3H} 
\tikz[baseline=(vert_cent.base)]{
  \node (vert_cent) {\hspace{-13pt}$\phantom{-}$};
  \draw (0.7,0) ++(0:0.35cm) arc (0:45:0.35cm) node (n6) {}
        (0.7,0) ++(45:0.35cm) arc (45:90:0.35cm) node (n1) {}
        (0.7,0) ++(90:0.35cm) arc (90:135:0.35cm) node (n2) {}
        (0.7,0) ++(135:0.35cm) arc (135:225:0.35cm) node (n3) {}
        (0.7,0) ++(225:0.35cm) arc (225:270:0.35cm) node (n4) {}
        (0.7,0) ++(270:0.35cm) arc (270:360:0.35cm) node (n5) {}
        (n1.base)--(n4.base) node[midway] (n7) {}
        (n2.base)--+(145:0.5cm)
        (n3.base)--+(215:0.5cm)
        (n5.base)--+(0:0.5cm)
        (n6.base)--(n7.base);
} \bigg ) \nn \\
&{}+  {\mathcal S}_3 \bigg(  \beta_{Y\hskip 0.5pt 3I}
\tikz[baseline=(vert_cent.base)]{
  \node (vert_cent) {\hspace{-13pt}$\phantom{-}$};
  \draw (0.7,0) ++(0:0.35cm) arc (0:105:0.35cm) node (n6) {}
        (0.7,0) ++(105:0.35cm) arc (105:135:0.35cm) node (n1) {}
        (0.7,0) ++(135:0.35cm) arc (135:165:0.35cm) node (n7) {}
        (0.7,0) ++(165:0.35cm) arc (165:195:0.35cm) node (n2) {}
        (0.7,0) ++(195:0.35cm) arc (195:225:0.35cm) node (n3) {}
        (0.7,0) ++(225:0.35cm) arc (225:255:0.35cm) node (n8) {}
        (0.7,0) ++(255:0.35cm) arc (255:360:0.35cm) node (n5) {}
        (n1.base)--+(145:0.5cm)
        (n3.base)--+(215:0.5cm)
        (n5.base)--+(0:0.5cm)
        (n6.base) to[out=250,in=20] (n7.base)
        (n2.base) to[out=340,in=110] (n8.base);
}
+  \beta_{Y\hskip 0.5pt 3J}
\tikz[baseline=(vert_cent.base)]{
  \node (vert_cent) {\hspace{-13pt}$\phantom{-}$};
  \draw (0.7,0) ++(0:0.35cm) arc (0:45:0.35cm) node (n1) {}
        (0.7,0) ++(45:0.35cm) arc (45:75:0.35cm) node (n2) {}
        (0.7,0) ++(75:0.35cm) arc (75:135:0.35cm) node (n3) {}
        (0.7,0) ++(135:0.35cm) arc (135:225:0.35cm) node (n4) {}
        (0.7,0) ++(225:0.35cm) arc (225:285:0.35cm) node (n5) {}
        (0.7,0) ++(285:0.35cm) arc (285:315:0.35cm) node (n6) {}
        (0.7,0) ++(315:0.35cm) arc (315:360:0.35cm) node (n7) {}
        (n1.base) to[out=240,in=120] (n6.base)
        (n2.base) to[out=240,in=120] (n5.base)
        (n3.base)--+(145:0.5cm)
        (n4.base)--+(215:0.5cm)
        (n7.base)--+(0:0.5cm);
}
+ \beta_{Y\hskip 0.5pt 3K}
\tikz[baseline=(vert_cent.base)]{
  \node (vert_cent) {\hspace{-13pt}$\phantom{-}$};
  \draw (0.7,0) ++(0:0.35cm) arc (0:30:0.35cm) node (n1) {}
        (0.7,0) ++(30:0.35cm) arc (30:105:0.35cm) node (n2) {}
        (0.7,0) ++(105:0.35cm) arc (105:135:0.35cm) node (n3) {}
        (0.7,0) ++(135:0.35cm) arc (135:225:0.35cm) node (n4) {}
        (0.7,0) ++(225:0.35cm) arc (225:360:0.35cm) node (n5) {}
        (n1.base) to[out=200,in=300] coordinate[midway] (n6) (n2.base)
        (n3.base)--+(145:0.5cm)
        (n4.base)--+(215:0.5cm)
        (n5.base)--+(0:0.5cm)
        (0.7,0) ++(67.5:0.35cm) node (n7) {}
        (n7.base)--(n6.base);
} \bigg ) \nn \\
&  {}+ {\mathcal S}_3 \bigg( \beta_{Y\hskip 0.5pt 3L}
\tikz[baseline=(vert_cent.base)]{
  \node (vert_cent) {\hspace{-13pt}$\phantom{-}$};
  \draw (0.7,0) ++(0:0.35cm) arc (0:15:0.35cm) node (n6) {}
        (0.7,0) ++(15:0.35cm) arc (15:65:0.35cm) node (n7) {}
        (0.7,0) ++(65:0.35cm) arc (65:80:0.35cm) node (n1) {}
        (0.7,0) ++(80:0.35cm) arc (80:135:0.35cm) node (n2) {};
  \draw[white] (0.7,0) ++(135:0.35cm) arc (135:225:0.35cm) node (n3) {};
  \draw (0.7,0) ++(225:0.35cm) arc (225:280:0.35cm) node (n4) {}
        (0.7,0) ++(280:0.35cm) arc (280:360:0.35cm) node (n5) {}
        (n2.base)--+(145:0.5cm)
        (n3.base)--+(215:0.5cm)
        (n5.base)--+(0:0.5cm);
  \draw[name path=a] (n1.base)--(n3.base);
  \draw[white, name path=b] (n2.base)--(n4.base);
  \path[name intersections={of=a and b,by=i}];
  \node[fill=white, inner sep=1.5pt, rotate=45] at (i) {};
  \draw (n2.base)--(n4.base);
  \draw (n6.base) to[out=170,in=270] (n7.base);
}
+ \beta_{Y\hskip 0.5pt 3M} 
\tikz[baseline=(vert_cent.base)]{
  \node (vert_cent) {\hspace{-13pt}$\phantom{-}$};
  \draw (0.7,0) ++(0:0.35cm) arc (0:30:0.35cm) node (n6) {}
        (0.7,0) ++(30:0.35cm) arc (30:45:0.35cm) node (n1) {}
        (0.7,0) ++(45:0.35cm) arc (45:135:0.35cm) node (n2) {};
  \draw[white] (0.7,0) ++(135:0.35cm) arc (135:225:0.35cm) node (n3) {};
  \draw (0.7,0) ++(225:0.35cm) arc (225:315:0.35cm) node (n4) {}
        (0.7,0) ++(315:0.35cm) arc (315:330:0.35cm) node (n7) {}
        (0.7,0) ++(330:0.35cm) arc (330:360:0.35cm) node (n5) {}
        (n2.base)--+(145:0.5cm)
        (n3.base)--+(215:0.5cm)
        (n5.base)--+(0:0.5cm)
        (n6.base) to[out=240,in=120] (n7.base);
  \draw[name path=a] (n1.base)--(n3.base);
  \draw[white, name path=b] (n2.base)--(n4.base);
  \path[name intersections={of=a and b,by=i}];
  \node[fill=white, inner sep=1.5pt, rotate=45] at (i) {};
  \draw (n2.base)--(n4.base);
}  \bigg ) 
+ \beta_{Y\hskip 0.5pt 3N} \hskip  - 0.3cm
\tikz[baseline=(vert_cent.base)]{
  \node (vert_cent) {\hspace{-13pt}$\phantom{-}$};
  \draw (0.7,0) ++(0:0.35cm) arc (0:22.5:0.35cm) node (n6) {}
        (0.7,0) ++(22.5:0.35cm) arc (22.5:45:0.35cm) node (n1) {}
        (0.7,0) ++(45:0.35cm) arc (45:67.5:0.35cm) node (n8) {};
\draw  (0.7,0) ++(67.5:0.35cm) arc (67.5:135:0.35cm) node (n2){};
  \draw[white] (0.7,0) ++(135:0.35cm) arc (135:225:0.35cm) node (n3) {};
  \draw (0.7,0) ++(225:0.35cm) arc (225:315:0.35cm) node (n4) {}
        (0.7,0) ++(315:0.35cm) arc (315:360:0.35cm) node (n5) {}
        (n2.base)--+(145:0.5cm)
        (n3.base)--+(215:0.5cm)
        (n5.base)--+(0:0.5cm);
  \draw[white][name path=a] (n1.base) to node[pos=0.25] (n7) {} (n3.base);
  \draw[white, name path=b] (n2.base)--(n4.base);
  \path[name intersections={of=a and b,by=i}];
   \draw (n6.base)--(n7.base);
    \node[fill=white, inner sep=1pt, rotate=45] at (i) {};
  \draw (n7.base)--(n3.base);
   \node[fill=white, inner sep=1.5pt, rotate=45] at (i) {};
   \draw (n2.base)--(n4.base);
   \draw (n7.base)--(n8.base);
}
 \nn \\
&  {}+ {\mathcal S}_3 \bigg(  \beta_{Y\hskip 0.5pt 3O}
\tikz[baseline=(vert_cent.base)]{
  \node (vert_cent) {\hspace{-13pt}$\phantom{-}$};
  \draw (0.7,0) ++(0:0.35cm) arc (0:45:0.35cm) node (n1) {}
        (0.7,0) ++(45:0.35cm) arc (45:100:0.35cm) node (n6) {}
        (0.7,0) ++(100:0.35cm) arc (100:135:0.35cm) node (n2) {}
        (0.7,0) ++(135:0.35cm) arc (135:225:0.35cm) node (n3) {}
        (0.7,0) ++(225:0.35cm) arc (225:260:0.35cm) node (n7) {}
        (0.7,0) ++(260:0.35cm) arc (260:315:0.35cm) node (n4) {}
        (0.7,0) ++(315:0.35cm) arc (315:360:0.35cm) node (n5) {}
        (n2.base)--+(145:0.5cm)
        (n3.base)--+(215:0.5cm)
        (n5.base)--+(0:0.5cm);
  \draw[name path=a] (n1.base)--(n7.base);
  \draw[white, name path=b] (n6.base)--(n4.base);
  \path[name intersections={of=a and b,by=i}];
  \node[fill=white, inner sep=1.5pt, rotate=45] at (i) {};
  \draw (n6.base)--(n4.base);
}
+ \beta_{Y\hskip 0.5pt 3P}
\tikz[baseline=(vert_cent.base)]{
  \node (vert_cent) {\hspace{-13pt}$\phantom{-}$};
  \draw (0.7,0) ++(0:0.35cm) arc (0:60:0.35cm) node (n1) {}
        (0.7,0) ++(60:0.35cm) arc (60:135:0.35cm) node (n2) {}
        (0.7,0) ++(135:0.35cm) arc (135:155:0.35cm) node (n6) {}
        (0.7,0) ++(155:0.35cm) arc (155:205:0.35cm) node (n7) {}
        (0.7,0) ++(205:0.35cm) arc (205:225:0.35cm) node (n3) {}
        (0.7,0) ++(225:0.35cm) arc (225:300:0.35cm) node (n8) {}
        (0.7,0) ++(300:0.35cm) arc (300:360:0.35cm) node (n5) {}
        (n2.base)--+(145:0.5cm)
        (n3.base)--+(215:0.5cm)
        (n5.base)--+(0:0.5cm);
  \draw[name path=a] (n1.base)--(n7.base);
  \draw[white, name path=b] (n6.base)--(n8.base);
  \path[name intersections={of=a and b,by=i}];
  \node[fill=white, inner sep=1.5pt, rotate=45] at (i) {};
  \draw (n6.base)--(n8.base);
}
\bigg) 
 + \beta_{Y\hskip 0.5pt 3Q} \tikz[baseline=(vert_cent.base)]{
  \node (vert_cent) {\hspace{-13pt}$\phantom{-}$};
  \draw (0.7,0) ++(0:0.35cm) arc (0:60:0.35cm) node (n1) {}
        (0.7,0) ++(60:0.35cm) arc (60:135:0.35cm) node (n2) {}
        (0.7,0) ++(135:0.35cm) arc (135:225:0.35cm) node (n3) {}
        (0.7,0) ++(135:0.35cm) arc (135:180:0.35cm) node (n6) {}
        (0.7,0) ++(180:0.35cm) arc (180:300:0.35cm) node (n4) {}
        (0.7,0) ++(300:0.35cm) arc (300:360:0.35cm) node (n5) {}
        node at (0.7,0) (n7) {}
        (n1.base)--(n7.base) 
        (n4.base)--(n7.base)
        (n2.base)--+(145:0.5cm)
        (n3.base)--+(215:0.5cm)
        (n5.base)--+(0:0.5cm)
        (n6.base)--(n7.base);
} 
\, ,
\label{beta3}
\end{align} 
with $\mathcal S_6$ denoting the sum over six inequivalent permutations.
 Imposing symmetry on the general Yukawa $\beta$-function in this case requires 18 relations
 \begin{align}
 &\beta_{y{\hskip 0.5pt}3p} - \beta_{y{\hskip 0.5pt}3y} =0 \, , \quad   \beta_{y{\hskip 0.5pt}3m} - \beta_{y{\hskip 0.5pt}3z} =0 \, , \quad
 6 \,  \beta_{y{\hskip 0.5pt}3a} + \beta_{y{\hskip 0.5pt}3v} - \beta_{y{\hskip 0.5pt}3\te} =0  \, , \nn \\
 & \beta_{y{\hskip 0.5pt}3d} - \beta_{y{\hskip 0.5pt}3n} + \beta_{y{\hskip 0.5pt}3\tj} =0 \, , \quad
    \beta_{y{\hskip 0.5pt}3e} - \beta_{y{\hskip 0.5pt}3o} + \beta  \raisebox{-1.5 pt}{$\scriptstyle y{\hskip 0.5pt}3\thh$} =0 \, , 
  \quad \beta_{y{\hskip 0.5pt}3e} + \beta_{y{\hskip 0.5pt}3q} - \beta_{y{\hskip 0.5pt}3\tc} =0 \, , \nn \\ 
&  \beta_{y{\hskip 0.5pt}3j} + \beta_{y{\hskip 0.5pt}3r} - \beta\raisebox{-1.5 pt}{$\scriptstyle y{\hskip 0.5pt}3\tb$} =0 \, , \nn \\
&   3\, \beta_{y{\hskip 0.5pt}3a} - \beta_{y{\hskip 0.5pt}3t} + \beta_{y{\hskip 0.5pt}3u} - \beta_{y{\hskip 0.5pt}3\ta}  =0 \, , \quad
 \beta_{y{\hskip 0.5pt}3b} - \beta_{y{\hskip 0.5pt}3j} - \beta_{y{\hskip 0.5pt}3r} + \beta_{y{\hskip 0.5pt}3\ti}  =0  \, ,  \quad
    \nn \\
&   \beta_{y{\hskip 0.5pt}3g} + \beta_{y{\hskip 0.5pt}3k} - \beta  \raisebox{-1.5 pt}{$\scriptstyle y{\hskip 0.5pt}3\tf $} 
+ \beta  \raisebox{-1.5 pt}{$\scriptstyle y{\hskip 0.5pt}3\tl$}  =0 \, , \qquad
  \beta_{y{\hskip 0.5pt}3g} - \beta_{y{\hskip 0.5pt}3h} - \beta_{y{\hskip 0.5pt}3\tg} + \beta_{y{\hskip 0.5pt}3\tv}  =0 \, , \nn \\
  & \beta_{y{\hskip 0.5pt}3h} - \beta_{y{\hskip 0.5pt}3i} + \beta \raisebox{-1.5 pt}{$\scriptstyle y{\hskip 0.5pt}3\td$} - \beta_{y{\hskip 0.5pt}3\tm}  =0  \, , \quad
2\, \beta_{y{\hskip 0.5pt}3b} - \beta_{y{\hskip 0.5pt}3f} - 2\,  \beta_{y{\hskip 0.5pt}3k} + \beta_{y{\hskip 0.5pt}3\tq}  -  \beta  \raisebox{-1.5 pt}{$\scriptstyle y{\hskip 0.5pt}3\ttt$}  =0  \, , \ \nn \\
 & 2\,  \beta_{y{\hskip 0.5pt}3b} + \beta_{y{\hskip 0.5pt}3g} - \beta_{y{\hskip 0.5pt}3h} + \beta_{y{\hskip 0.5pt}3l}- \beta \raisebox{-1.5 pt}{$\scriptstyle y{\hskip 0.5pt}3\td$}  +  \beta_{y{\hskip 0.5pt}3\tn}  =0  \, , \nn \\
 & 2\,  \beta_{y{\hskip 0.5pt}3e} -  \beta_{y{\hskip 0.5pt}3j} + 2 \, \beta_{y{\hskip 0.5pt}3q} - \beta_{y{\hskip 0.5pt}3\tc} - \beta  \raisebox{-1.5 pt}{$\scriptstyle y{\hskip 0.5pt}3\tk$}  =0  \, , \nn \\
 & 2\, \beta_{y{\hskip 0.5pt}3c} - \beta_{y{\hskip 0.5pt}3k} - \beta_{y{\hskip 0.5pt}3\ts} + \beta_{y{\hskip 0.5pt}3\tu}  =0 \, ,
 \qquad \beta_{y{\hskip 0.5pt}3g} - \beta_{y{\hskip 0.5pt}3h} - \beta_{y{\hskip 0.5pt}3\tg} + \beta_{y{\hskip 0.5pt}3\tv}  =0 \, , 
 \nn \\
 &  \beta_{y{\hskip 0.5pt}3c} +  \beta_{y{\hskip 0.5pt}3d} - \beta_{y{\hskip 0.5pt}3e}- \beta_{y{\hskip 0.5pt}3j} -\beta_{y{\hskip 0.5pt}3s} -\beta_{y{\hskip 0.5pt}3\too} +2\, \beta_{y{\hskip 0.5pt}3\tp}  =0  \, , \nn \\
 & 2\, \beta_{y{\hskip 0.5pt}3b} +  2\,\beta_{y{\hskip 0.5pt}3c} - \beta_{y{\hskip 0.5pt}3f}+ \beta_{y{\hskip 0.5pt}3l} + \beta_{y{\hskip 0.5pt}3x} + \beta_{y{\hskip 0.5pt}3\tw} -
 2\, \beta_{y{\hskip 0.5pt}3\tx} -   \beta_{y{\hskip 0.5pt}3\ty}=0  \, .
 \label{3loopYconsis}
 \end{align} 
  Subject to \eqref{3loopYconsis}
\begin{align}
&\beta_{Y\hskip 0.5pt3A} = \beta_{y{\hskip 0.5pt}3p}  \, , \quad \beta_{Y\hskip 0.5pt 3B} = 
\beta_{y{\hskip 0.5pt}3m}  \, , \quad 
\beta_{Y\hskip 0.5pt 3C} = 3\, \beta_{y{\hskip 0.5pt}3a} + \beta_{y{\hskip 0.5pt}3u}   \, , \quad   \beta_{Y\hskip 0.5pt 3D} =  \beta_{y{\hskip 0.5pt}3o} \, , 
\quad \beta_{Y\hskip 0.5pt 3E} = \beta_{y{\hskip 0.5pt}3n}  \, , \nn \\
&\beta_{Y\hskip 0.5pt 3F} = \beta_{y{\hskip 0.5pt}3e} +   \beta_{y{\hskip 0.5pt}3q} \, , \quad
\beta_{Y\hskip 0.5pt 3G} = \beta_{y{\hskip 0.5pt}3j} + \beta_{y{\hskip 0.5pt}3r} \, , \quad 
\beta_{Y\hskip 0.5pt 3H} = \beta_{y{\hskip 0.5pt}3h} +  \beta\raisebox{-1.5 pt}{$\scriptstyle y{\hskip 0.5pt}3\td$} \, ,  \nn \\
&\beta_{Y\hskip 0.5pt3I} = 2\,  \beta_{y{\hskip 0.5pt}3b} + \beta_{y{\hskip 0.5pt}3\tq} \, , \quad \beta_{Y\hskip 0.5pt 3J} = 
 \beta\raisebox{-1.5 pt}{$\scriptstyle y{\hskip 0.5pt}3\tf$}   \, , \quad
\beta_{Y\hskip 0.5pt 3K} = 6\, \beta_{y{\hskip 0.5pt}3a} + \beta_{y{\hskip 0.5pt}3v} \, , \nn \\
& \beta_{Y\hskip 0.5pt 3L} =  \beta_{y{\hskip 0.5pt}3e} + \beta_{y{\hskip 0.5pt}3j} + \beta_{y{\hskip 0.5pt}3s} + \beta_{y{\hskip 0.5pt}3\too} \, , \qquad
 \beta_{Y\hskip 0.5pt 3M} =  \beta_{y{\hskip 0.5pt}3k} + \beta_{y{\hskip 0.5pt}3\ts} \, ,  \nn \\
 & \beta_{Y\hskip 0.5pt 3N} =  2\, \beta_{y{\hskip 0.5pt}3c} + \beta_{y{\hskip 0.5pt}3i} + 2 \, \beta_{y{\hskip 0.5pt}3\trr}  \, , \qquad
 \beta_{Y\hskip 0.5pt 3O} =  \beta_{y{\hskip 0.5pt}3h} + \beta_{y{\hskip 0.5pt}3\tg} \, , \nn \\
 & \beta_{Y\hskip 0.5pt 3P} =   \beta_{y{\hskip 0.5pt}3f} + 2\, \beta_{y{\hskip 0.5pt}3\tx}  + \beta_{y{\hskip 0.5pt}3\ty} \, , \quad
 \beta_{Y\hskip 0.5pt 3Q} =   4\, \beta_{y{\hskip 0.5pt}3b} +  \beta_{y{\hskip 0.5pt}3f}  + 2 \,  \beta_{y{\hskip 0.5pt}3l} +  \beta_{y{\hskip 0.5pt}3w} + 2\,  \beta_{y{\hskip 0.5pt}3\tz} \, .
 \label{Yuk3S}
\end{align}

Explicit  results for this $\N=\tfrac12$ theory in the  ${\overline {M\!S}}$ scheme are then
\begin{align} 
\hskip -1cm
& \gamma_{\Phi\hskip 0.5pt1}  = \tfrac12 \, ,  && \gamma_{\Phi\hskip 0.5pt 2A}  = - \tfrac12 \, , &&  
\gamma_{\Phi\hskip 0.5pt 2B} =0\, ,  &&
\gamma_{\Phi\hskip 0.5pt 3A}  = - \tfrac14 \, , && \gamma_{\Phi\hskip 0.5pt 3B} = - \tfrac18 \, , && 
\gamma_{\Phi\hskip 0.5pt 3C} = 1 \, , \nn \\
& \gamma_{\Phi\hskip 0.5pt 3D}  = 0 \, , &&
 \gamma_{\Phi\hskip 0.5pt 3E} =  \tfrac14 \, , && \gamma_{\Phi\hskip 0.5pt 3F} = - \tfrac12 \, , && 
 \gamma_{\Phi\hskip 0.5pt 3G} = 1\, , &&
 \gamma_{\Phi\hskip 0.5pt 3H}  = \tfrac32 \, , &&  \gamma_{\Phi\hskip 0.5pt 3I}  = \tfrac32 \zeta_3 \, ,
 \label{dres}
 \end{align}
with $\upsilon_{\Phi\hskip 0.5pt 3D}  = - \tfrac34$ and 
\begin{align}
& \beta_{Y \skip 0.5pt1} = 2 \, , &&  \beta_{Y\hskip 0.5pt 2A}= - 1 \, , &&  \beta_{Y\hskip 0.5pt 2B}= - 2 \, , &&  
\beta_{Y\hskip 0.5pt 2C} =0\, , \nn \\
& \beta_{3A} = - \tfrac12 \, , &&  \beta_{Y\hskip 0.5pt 3B} = -\tfrac12  \, ,  
&&  \beta_{Y\hskip 0.5pt 3C} = 2 \, , 
&& \beta_{Y\hskip 0.5pt 3D}=  -1 \, ,  \nn \\
& \beta_{Y\hskip 0.5pt 3E} =  2  \, , && \beta_{Y\hskip 0.5pt 3F} = 2 \, , &&  \beta_{Y\hskip 0.5pt 3G}= -1 \, , &&   \beta_{3H} = 4 \, , \nn \\
& \beta_{Y\hskip 0.5pt 3I} = -2\, ,  &&  \beta_{Y\hskip 0.5pt 3J} = 4\, , 
&& \beta_{Y\hskip 0.5pt 3K} = - 1 \, ,  \nn \\
 & \beta_{Y\hskip 0.5pt 3N} = 12 \zeta_3\, , 
&&  \beta_{Y\hskip 0.5pt 3O} = 6\zeta_3 \, , &&
\beta_{3P} = 12 \zeta_3 \, , && 
\beta_{Y\hskip 0.5pt 3L} =  \beta_{Y\hskip 0.5pt 3M} =  \beta_{Y\hskip 0.5pt 3Q} = 0\, .
\label{cres}
 \end{align}
 These results can be obtained directly from superspace calculations \cite{JackU}.

 Reducing the three loop scalar $\beta$-function to the $\N= \tfrac12$ theory requires large numbers
 of relations. For the anomalous dimension and the symmetric $\beta$ function
 \begin{align}
 & \beta_{\lambda{\hskip 0.5pt}3h} = 2\, \gamma_{\Phi\hskip 0.5pt 3A} \, , \quad  \beta_{\lambda{\hskip 0.5pt}3g} = 2\, \gamma_{\Phi\hskip 0.5pt 3B}  \, , \quad
3\,   \beta_{\lambda{\hskip 0.5pt}3a} +  \beta_{\lambda{\hskip 0.5pt}3p} = 2\, \gamma_{\Phi\hskip 0.5pt 3C}  \, , \quad  0 =  \gamma_{\Phi\hskip 0.5pt 3D}  \, , \nn \\ 
& \beta_{\lambda{\hskip 0.5pt}3c} = 2\, \gamma_{\Phi\hskip 0.5pt 3E} =  \tfrac12 \, \gamma_{\Phi\hskip 0.5pt 3G} 
 \, , \quad  6\,   \beta_{\lambda{\hskip 0.5pt}3a} +  \beta_{\lambda{\hskip 0.5pt}3q} = 2\, \gamma_{\Phi\hskip 0.5pt 3F}   \, , \nn  \\
 & 2\, \beta_{\lambda{\hskip 0.5pt}3c} +  \beta_{\lambda{\hskip 0.5pt}3t} = 2\, \gamma_{\Phi\hskip 0.5pt 3H} \, , \quad
 2\, \beta_{\lambda{\hskip 0.5pt}3c} +  \beta_{\lambda{\hskip 0.5pt}3u} = 2\, \gamma_{\Phi\hskip 0.5pt 3I} \, , \nn \\
 \noalign{\vskip 4pt}
 &  \beta_{\lambda{\hskip 0.5pt}3d} = 2\,  \beta_{\lambda{\hskip 0.5pt}3g} =   \beta_{\lambda{\hskip 0.5pt}3h} 
=  \beta_{\lambda{\hskip 0.5pt}3j} =
  \tfrac12( \beta_{\lambda{\hskip 0.5pt}3i} +  \beta_{\lambda{\hskip 0.5pt}3z} ) =   \tfrac12( \beta_{\lambda{\hskip 0.5pt}3b} +  \beta_{\lambda{\hskip 0.5pt}3\ta} )  
  =   \tfrac14(2\, \beta_{\lambda{\hskip 0.5pt}3b} +  \beta_{\lambda{\hskip 0.5pt}3\tc} ) \nn \\
&{} =  \tfrac12(6\,\beta_{\lambda{\hskip 0.5pt}3a} +  \beta_{\lambda{\hskip 0.5pt}3q} )
 = \beta_{Y\hskip 0.5pt 3A} =\beta_{3B} = \tfrac12 \, \beta_{Y\hskip 0.5pt 3D} 
 = \tfrac12 \,  \beta_{Y\hskip 0.5pt 3G}   = \tfrac14 \,  \beta_{Y\hskip 0.5pt 3I} = \tfrac12 \,  \beta_{Y\hskip 0.5pt 3K} \ \, , \nn \\
  \noalign{\vskip 4pt}
& 3\, \beta_{\lambda{\hskip 0.5pt}3a}  +  \beta_{\lambda{\hskip 0.5pt}3p} =  \beta_{\lambda{\hskip 0.5pt}3b} = \beta_{\lambda{\hskip 0.5pt}3i} =
2\, \beta_{\lambda{\hskip 0.5pt}3j}  +  \beta_{\lambda{\hskip 0.5pt}3k} = 2\, \beta_{\lambda{\hskip 0.5pt}3c}  +  \beta_{\lambda{\hskip 0.5pt}3t}  
+  \tfrac12\, \beta\raisebox{-1.5 pt}{$\scriptstyle \lambda{\hskip 0.5pt}3\thh$}  \nn \\
&{} =  \beta_{Y\hskip 0.5pt 3C} =  \beta_{Y\hskip 0.5pt 3E} =  \beta_{Y\hskip 0.5pt 3F} = \tfrac12 \,  \beta_{Y\hskip 0.5pt 3H} = \tfrac12 \,  \beta_{Y\hskip 0.5pt 3J} \, , 
\nn \\
& \hskip 4cm   2\, \beta_{\lambda{\hskip 0.5pt}3c}  +  \beta_{\lambda{\hskip 0.5pt}3y} =  4\, \beta_{\lambda{\hskip 0.5pt}3c}  + 
 \beta\raisebox{-1.5 pt}{$\scriptstyle \lambda{\hskip 0.5pt}3\tb$} = 4\, \beta_{\lambda{\hskip 0.5pt}3d} +   \beta_{\lambda{\hskip 0.5pt}3r}  = 0 \, , \nn \\
   \noalign{\vskip 4pt}
&   \beta_{\lambda{\hskip 0.5pt}3b} +  \beta_{\lambda{\hskip 0.5pt}3i} +    \beta\raisebox{-1.5 pt}{$\scriptstyle \lambda{\hskip 0.5pt}3\td$} =   
4\, \beta_{\lambda{\hskip 0.5pt}3j}  +  \beta_{\lambda{\hskip 0.5pt}3l} =  \beta_{Y\hskip 0.5pt 3L} \, , \qquad
  0=  2\, \beta_{\lambda{\hskip 0.5pt}3b} +  \beta_{\lambda{\hskip 0.5pt}3\te} =  \beta_{Y\hskip 0.5pt 3M} \, , \nn \\
   \noalign{\vskip 4pt}
  & 4\, \beta_{\lambda{\hskip 0.5pt}3b} +  \beta\raisebox{-1.5 pt}{$\scriptstyle \lambda{\hskip 0.5pt}3\tf$}
  =  \beta_{Y\hskip 0.5pt 3N} \, , \nn \\
   \noalign{\vskip 4pt}
 &  2\, \beta_{\lambda{\hskip 0.5pt}3b} +  \beta_{\lambda{\hskip 0.5pt}3s} =  4\, \beta_{\lambda{\hskip 0.5pt}3c} + 2\, \beta_{\lambda{\hskip 0.5pt}3u} 
  =  2\, \beta_{\lambda{\hskip 0.5pt}3b} +  2\, \beta_{\lambda{\hskip 0.5pt}3d} +   \beta_{\lambda{\hskip 0.5pt}3r} +  \tfrac12\, \beta_{\lambda{\hskip 0.5pt}3\tg}
  =  \beta_{Y\hskip 0.5pt 3O} = \tfrac12 \,  \beta_{Y\hskip 0.5pt 3P} \, , \nn \\
   \noalign{\vskip 4pt}
  & 4\, \beta_{\lambda{\hskip 0.5pt}3b} +  4\, \beta_{\lambda{\hskip 0.5pt}3d} +  2\,  \beta_{\lambda{\hskip 0.5pt}3r} + \beta_{\lambda{\hskip 0.5pt}3\ti}
  =  \beta_{Y\hskip 0.5pt 3Q} \, .
  \label{Rel3beta}
 \end{align}
 There are here 18 linear relations on the 3 loop $\beta_\lambda$ coefficients. 
 There are also 33 additional consistency equations. For those involving contributions from planar
 diagrams
  \begin{align}
 &
  4\, \beta_{\lambda{\hskip 0.5pt}3b} +   \beta_{\lambda{\hskip 0.5pt}3\tz}  = 0 \, , \quad \beta_{\lambda{\hskip 0.5pt}3b} +   \beta_{\lambda{\hskip 0.5pt}3i} +    \beta_{\lambda{\hskip 0.5pt}3\tu } = 0 \, , \quad 2\, \beta_{\lambda{\hskip 0.5pt}3i} +  \beta_{\lambda{\hskip 0.5pt}3n}  = 0 \, ,  \quad 
  2\, \beta_{\lambda{\hskip 0.5pt}3j} +  \beta_{\lambda{\hskip 0.5pt}3\tq}  = 0 \, , 
  \nn \\
&  2\, \beta_{\lambda{\hskip 0.5pt}3d} +  2\, \beta_{\lambda{\hskip 0.5pt}3j} +  \beta\raisebox{-1.5 pt}{$\scriptstyle  \lambda{\hskip 0.5pt}3\ttt$}  = 0\, , \quad  \
  4\, \beta_{\lambda{\hskip 0.5pt}3d} +  4\, \beta_{\lambda{\hskip 0.5pt}3e} +  \beta_{\lambda{\hskip 0.5pt}3\ty}  = 0\, , \quad
   \beta_{\lambda{\hskip 0.5pt}3d} +   \beta_{\lambda{\hskip 0.5pt}3h} +  \beta_{\lambda{\hskip 0.5pt}3\too}  = 0\, , 
  \nn \\
&  8\, \beta_{\lambda{\hskip 0.5pt}3e} +  \beta_{\lambda{\hskip 0.5pt}3\tx}  = 0 \, , \quad  
 2\, \beta_{\lambda{\hskip 0.5pt}3e} +  2\, \beta_{\lambda{\hskip 0.5pt}3j} +  \beta_{\lambda{\hskip 0.5pt}3\tv}  = 0\, , \quad
  \beta_{\lambda{\hskip 0.5pt}3e} +  2\, \beta_{\lambda{\hskip 0.5pt}3g} +  \beta_{\lambda{\hskip 0.5pt}3\tp}  = 0\, ,
 \nn \\
 & 6\,  \beta_{\lambda{\hskip 0.5pt}3a} +  2\,  \beta_{\lambda{\hskip 0.5pt}3d} +   \beta_{\lambda{\hskip 0.5pt}3q} + \beta_{\lambda{\hskip 0.5pt}3\ts} =0 \, , \quad
  3\,  \beta_{\lambda{\hskip 0.5pt}3a} +  \beta_{\lambda{\hskip 0.5pt}3i} + \beta_{\lambda{\hskip 0.5pt}3m} +  \beta_{\lambda{\hskip 0.5pt}3p} 
+ \beta_{\lambda{\hskip 0.5pt}3\trr}  =0 \, , 
  \nn \\
& 4\,  \beta_{\lambda{\hskip 0.5pt}3b} +  4\,  \beta_{\lambda{\hskip 0.5pt}3d} + 2\, \beta_{\lambda{\hskip 0.5pt}3r} +  \beta_{\lambda{\hskip 0.5pt}3\ti} =0 \, ,  \
 \qquad 4\,  \beta_{\lambda{\hskip 0.5pt}3b} +  4\,  \beta_{\lambda{\hskip 0.5pt}3c} +   
 2\,  \beta\raisebox{-1.5 pt}{$\scriptstyle  \lambda{\hskip 0.5pt}3\tb$} + \beta_{\lambda{\hskip 0.5pt}3e'} =0 \, , 
 \nn \\
 &  4\,  \beta_{\lambda{\hskip 0.5pt}3b} +  4\, \beta_{\lambda{\hskip 0.5pt}3d} + \beta_{\lambda{\hskip 0.5pt}3r} +  \beta_{\lambda{\hskip 0.5pt}3x} + \beta_{\lambda{\hskip 0.5pt}3\tw}  = 0 \, , \quad
 2\,  \beta_{\lambda{\hskip 0.5pt}3b} +  4\, \beta_{\lambda{\hskip 0.5pt}3e} + 2\, \beta_{\lambda{\hskip 0.5pt}3v} +  \beta_{\lambda{\hskip 0.5pt}3\tc} 
+ \beta_{\lambda{\hskip 0.5pt}3f'}  =0 \, ,  \nn \\
& \beta_{\lambda{\hskip 0.5pt}3b} +   2\, \beta_{\lambda{\hskip 0.5pt}3j} +  \beta_{\lambda{\hskip 0.5pt}3k} +  \beta_{\lambda{\hskip 0.5pt}3\ta} 
+ \beta_{\lambda{\hskip 0.5pt}3c'}  =0 \, ,  \quad
 2\,  \beta_{\lambda{\hskip 0.5pt}3c} +  2\,  \beta_{\lambda{\hskip 0.5pt}3i}   +  \beta_{\lambda{\hskip 0.5pt}3o} + 2\,  \beta_{\lambda{\hskip 0.5pt}3y} =0 \, , 
\nn \\
 &   2\,  \beta_{\lambda{\hskip 0.5pt}3e} +  \beta_{\lambda{\hskip 0.5pt}3i} + \beta_{\lambda{\hskip 0.5pt}3v} +  \beta_{\lambda{\hskip 0.5pt}3z} 
+ \beta_{\lambda{\hskip 0.5pt}3d'}  =0 \, ,  \quad 4\, \beta_{\lambda{\hskip 0.5pt}3v}  + \beta_{\lambda{\hskip 0.5pt}3\tm} =0 \, , 
  \nn \\
& 4\,  \beta_{\lambda{\hskip 0.5pt}3c} +   4\, \beta_{\lambda{\hskip 0.5pt}3e} + 2\,  \beta_{\lambda{\hskip 0.5pt}3t} +  4\, \beta_{\lambda{\hskip 0.5pt}3v} 
+  2\, \beta\raisebox{-1.5 pt}{$\scriptstyle  \lambda{\hskip 0.5pt}3\thh$} +  \beta\raisebox{-1.5 pt}{$\scriptstyle  \lambda{\hskip 0.5pt}3\tl$} 
+ \beta_{\lambda{\hskip 0.5pt}3h'}  =0 \, .
\label{Consis3l} 
 \end{align}
 For the relations which involve contributions from the 
 non planar diagrams for the quartic $\beta$-function, 
 \begin{align}
 &  \beta_{\lambda{\hskip 0.5pt}3b} +   \beta_{\lambda{\hskip 0.5pt}3i} +  \beta\raisebox{-1.5 pt}{$\scriptstyle  \lambda{\hskip 0.5pt}3\td$}= 0 \, , \qquad
 2\, \beta_{\lambda{\hskip 0.5pt}3b} +   \beta_{\lambda{\hskip 0.5pt}3\te}  = 0 \, ,  \qquad
 4\, \beta_{\lambda{\hskip 0.5pt}3j} +  \beta_{\lambda{\hskip 0.5pt}3l}  = 0 \, , \nn \\
 & 4\, \beta_{\lambda{\hskip 0.5pt}3e} +  \beta_{\lambda{\hskip 0.5pt}3w}  = 0 \, ,  
 \qquad 8\,  \beta_{\lambda{\hskip 0.5pt}3e} +  4\,  \beta_{\lambda{\hskip 0.5pt}3w} +    \beta_{\lambda{\hskip 0.5pt}3\tn } = 0 \, , \nn \\
 \noalign{\vskip 4pt} 
 & 2\, \beta_{\lambda{\hskip 0.5pt}3f} +  \beta_{\lambda{\hskip 0.5pt}3\tj}  = 0 \, , \qquad  
  4\,  \beta_{\lambda{\hskip 0.5pt}3b} +  \beta_{\lambda{\hskip 0.5pt}3f} +   \beta\raisebox{-1.5 pt}{$\scriptstyle  \lambda{\hskip 0.5pt}3\tf$} + 
  2\, \beta\raisebox{-1.5 pt}{$\scriptstyle  \lambda{\hskip 0.5pt}3\tk$} =0 \, , \nn \\
 & 4\,  \beta_{\lambda{\hskip 0.5pt}3b} +  2\,  \beta_{\lambda{\hskip 0.5pt}3s} +   \beta_{\lambda{\hskip 0.5pt}3x} + \beta_{\lambda{\hskip 0.5pt}3a'} =0 \, , \nn \\
 &4\,  \beta_{\lambda{\hskip 0.5pt}3b} +   4\, \beta_{\lambda{\hskip 0.5pt}3d} +  \beta_{\lambda{\hskip 0.5pt}3f} +  2\, \beta_{\lambda{\hskip 0.5pt}3r} 
+  \beta_{\lambda{\hskip 0.5pt}3\tg}
+  2\, \beta\raisebox{-1.5 pt}{$\scriptstyle  \lambda{\hskip 0.5pt}3\tk$} + \beta_{\lambda{\hskip 0.5pt}3b'}  =0 \, ,  \nn \\
& 4\,  \beta_{\lambda{\hskip 0.5pt}3b} +   \beta_{\lambda{\hskip 0.5pt}3f} +  2\,  \beta_{\lambda{\hskip 0.5pt}3s} +  
 \beta\raisebox{-1.5 pt}{$\scriptstyle  \lambda{\hskip 0.5pt}3\tk$} + \beta_{\lambda{\hskip 0.5pt}3g'}  =0 \, , \nn \\
 & 8\,  \beta_{\lambda{\hskip 0.5pt}3c} +   \beta_{\lambda{\hskip 0.5pt}3f} +  4\,  \beta_{\lambda{\hskip 0.5pt}3u} + 2\, \beta_{\lambda{\hskip 0.5pt}3i'} =0 \, , \qquad
 2\, \beta_{\lambda{\hskip 0.5pt}3f} +  \beta_{\lambda{\hskip 0.5pt}3j'}  = 0 \, .
 \label{Consis3l2} 
 \end{align}
 The 14 homogeneous relations in \eqref{N1consis} are contained in \eqref{Consis3l},  \eqref{Consis3l2}.
 Combining \eqref{Consis3l},  \eqref{Consis3l2} 
 with \eqref{Rel3beta} would apparently generate 51 conditions but 2 are redundant. Two of the conditions in  
 \eqref{Consis3l} imply $\beta_{Y{\hskip 0.5pt}3L}=\beta_{Y{\hskip 0.5pt}3Q} =0$ and the relations in \eqref{Rel3beta}
 $ 0=  2\, \beta_{\lambda{\hskip 0.5pt}3b} +  \beta_{\lambda{\hskip 0.5pt}3\te} $ and
 $\beta_{\lambda{\hskip 0.5pt}3b} +  \beta_{\lambda{\hskip 0.5pt}3i} +    \beta\raisebox{-1.5 pt}{$\scriptstyle \lambda{\hskip 0.5pt}3\td$} =   
4\, \beta_{\lambda{\hskip 0.5pt}3j}  +  \beta_{\lambda{\hskip 0.5pt}3l} $ can be omitted since they are all zero in \eqref{Consis3l}.
There remain 49 independent equations.
 
 The conditions $ \beta_{Y\hskip 0.5pt 3A} =\beta_{3B} = \tfrac12 \, \beta_{Y\hskip 0.5pt 3D} 
 = \tfrac12 \,  \beta_{Y\hskip 0.5pt 3G}   = \tfrac14 \,  \beta_{Y\hskip 0.5pt 3I} = \tfrac12 \,  \beta_{Y\hskip 0.5pt 3K} $, 
$\beta_{3C} =  \beta_{Y\hskip 0.5pt 3E} =  \beta_{Y\hskip 0.5pt 3F} = \tfrac12 \,  \beta_{Y\hskip 0.5pt 3H} = \tfrac12 \,  \beta_{Y\hskip 0.5pt 3J}$, 
$ \beta_{Y\hskip 0.5pt 3O} = \tfrac12 \,  \beta_{Y\hskip 0.5pt 3P}$ and $\beta_{3L} = \beta_{Y\hskip 0.5pt 3M} = \beta_{Y\hskip 0.5pt 3Q}=  0$  impose 
13 further relations on the Yukawa $\beta$-functions from \eqref{Yuk3S}.

\section{Special Cases and Fixed Points}

To analyse the RG flow in scalar fermion theories, and potentially find fixed points in an $\vep=4-d$ expansion, 
it is generally necessary to restrict to cases where the RG flow is constrained to a small number of 
couplings. Here we describe various examples where symmetries are imposed so that  the RG flow is
 reduced to two scalar couplings and one Yukawa coupling. Of course with minimal subtraction $\vep$ only 
 appears at zeroth order in a loop expansion so that various perturbative results listed here can easily be
 used in the hunt for fixed points. Possible fixed points are first determined by using the 
 one loop contributions  to the $\beta$-functions and are described  in  this section. Corresponding 
 two and three formulae which give $\vep^2,\vep^3$ contributions are 
obtained by restriction of the general  results and are presented in supplementary material. 

At one loop the results  obtained here for the general case give
\begin{align}
{\beta}_{y}{\!}^{(1)a}  = {}& 2\, y^b y^a y^b + \tfrac12 \big ( y^b y^b  \, y^a + y^a \, y^b  y^b \big ) 
+ \tfrac12\, y^b \,  \tr ( y^b  y^a ) \, , \nn \\
{\beta}_{\lambda}{\!}^{(1)abcd}  = {}& 3\, \lambda^{ef(ab} \lambda^{cd)ef}
 + 2 \, \lambda^{e(abc}\, \tr \big (  y^{d)}  y^e \big )  - 12 \,  \tr \big ( y^{(a} y^b y^c y^{d)}\big  )  \, , \nn \\
   \gamma_{\phi}{\!}^{(1)ab} = {}& \tfrac12   \tr (y^a y^b ) \, , \qquad \gamma_{\psi}{\!}^{(1)}  = \tfrac12 \, y^a y^a \, ,
   \label{oneloop}
\end{align}
with $\{y^a\}$ symmetric and real.

For $n_s$ real scalars and $n_f$ pseudo real Majorana fermions  the reduction to three couplings is achieved
by assuming
\be
y^a \to y\, t^a \, {\mathds 1}_{m } \, , \quad
\lambda^{abcd} = \lambda \big ( \delta^{ab}  \delta^{cd} +  \delta^{ac}  \delta^{bd}+  \delta^{ad}  \delta^{bc}\big  )
+ g\, h^{abcd} \, ,
\label{ylh}
\ee
with $\{ t^a \} $ a set of real traceless $n\times n$ symmetric matrices and $m$ essentially arbitrary.
We assume\footnote{The 
corresponding scalar potential should be bounded below. The constraints on $\lambda,g$
may be determined by the inequalities for any hermitian traceless $n\times n$ $t$
$$
k_+  \tr(t^2)^2 \ge \tr (t^4 ) \ge k_-   \tr(t^2)^2 \, , \qquad k_+ =  \frac{(n-1)^3 +1}{n^2(n-1)} \, , \quad
k_- = \begin{cases} \frac1n \, , & n\ \mbox{even} \\
\frac{n^2+3}{n(n^2-1)} \, ,  & n\  \mbox{odd} \end{cases} \, .
$$
For the potential to be bounded below it is possible for either $g$ or $\lambda$ to be negative
 so long as $\lambda>0$,  $3\,  \lambda + k_+  n\alpha \, g>0$
or $g>0$, $ 3\, \lambda + k_- n\alpha \, g>0$. 
}
\be
\tr \big  (t^{(a}\hskip 0.5pt t^ b \hskip 0.5pt  t^c \hskip 0.5pt  t^{d)}\big )  =  n\, \alpha\, h^{abcd} \, . 
\label{hdef}
\ee
In general the symmetry group is given by 
\be
R^{-1} y^a \, R  = {\tilde R}^{ab} y^b \quad \mbox{for} \quad  R \in H_f \subset O(n_f) \, , \ \  
[{\tilde R}^{ab}] \in  {H_s} \subset O(n_s) \, .
\label{inv}
\ee
and then $h^{abcd}$ is an $H_s$ invariant $O(n_s)$ symmetric tensor. Assuming \eqref{ylh}
$H_f \simeq H\times O(m)$ with $H\subset O(n)$.  For simplicity we take $t^a$  to be  traceless and
 $H_s \simeq  H/ {\mathbb Z}_2$ with $H$ simple.

At one loop a consistent RG flow is achieved by 
requiring for the Yukawa $\beta$-function the conditions
\be
 t^a {\hskip 0.5pt} t^{a }  = n_s \, \alpha \, {\mathds 1}_{n}  \, , \qquad 
\tr (t^a  {\hskip 0.5pt}  t^b) =  n \, \alpha \, \delta^{ab}  \, , \qquad
t^b {\hskip 0.5pt} t^a{\hskip 0.5pt} t^b  = \beta\,  t^a  \, ,
\label{tid}
\ee
where  $n_f = n \, m $ and $\alpha>0$ depends on a choice of scale for $\{t^a\}$.
For the scalar coupling  it is also necessary that
\be
 h^{ef(ab} h^{cd)ef} = A  \big ( \delta^{ab}  \delta^{cd} +  \delta^{ac}  \delta^{bd}+  \delta^{ad}  \delta^{bc}\big  )
 + B \, h^{abcd} \, .
\label{hid}
\ee
 The tensor $h^{abcd}$ may be further decomposed as 
\be
h^{abcd} = d^{abcd}  + r \big ( \delta^{ab}  \delta^{cd} +  \delta^{ac}  \delta^{bd}+  \delta^{ad}  \delta^{bc}\big  )
\, , \qquad r=  \frac{2\, n_s \, \alpha + \beta }{3(n_s+2)}\, ,
\label{hdrel}
\ee
for $d^{abcd}$ symmetric and traceless and \eqref{ylh} is alternatively expressed as
\be
\lambda^{abcd} ={\hat  \lambda} \big ( \delta^{ab}  \delta^{cd} +  \delta^{ac}  \delta^{bd}+  \delta^{ad}  \delta^{bc}\big  )
+ g\, d^{abcd}  \, , \qquad {\hat  \lambda} = \lambda + r\, g \, .
\ee
With this definition \eqref{hid} is equivalent to
\be
 d^{ef(ab} d^{cd)ef} = a\, \delta^{(ab}  \delta^{cd)}  + b \, d^{abcd} \, ,
\label{did}
\ee
where $b= B - 4 \,r, \ a = 3 ( A + r B )-  (n_s+8) r^2$.

At higher loops the necessary constraints are such that the $\beta$-functions are reduced to $\beta_y, \,
\beta_\lambda,\, \beta_g$ and the anomalous dimension matrices have the form $\gamma_\phi \, \delta^{ab},
\ \gamma_\psi \,{\mathds 1}_m \times {\mathds 1}_n$.

Using
\be
h^{ab cc} = \tfrac13  ( 2\,  n_s \, \alpha + \beta ) \, \delta^{ab} \, ,
\ee
the lowest order results \eqref{oneloop} are consistent with this form and  give,
\begin{align}
\beta_y{\!}^{(1)} ={}& ( n_s \, \alpha + 2\,  \beta +   \tn_f ) y^3 \, , \nn \\
\beta_\lambda{\!}^{(1)} ={}&  (n_s + 8 ) \lambda^2 
+ \tfrac23 ( 2 \, n_s \, \alpha + \beta ) \lambda \, g  + 4 \, \tn_f \, \lambda \, y^2+ 3\, A\,  g^2 \, , \nn \\
\beta_g{\!}^{(1)} ={}&   12 \, \lambda \, g + 3 \, B \, g^2 +  4\, \tn_f \, g \, y^2 - 24\,  \tn_f  \,  y^4 \, , \nn \\
\gamma_\phi{\!}^{(1)} ={}&    \tn_f \, y^2 \, , \qquad 
\gamma_\psi{\!}^{(1)} = \tfrac12 \, n_s \, \alpha\, y^2  \, , \qquad \tn_f = \tfrac12 \, \alpha \, n_f \, .
\label{oneloop2}
\end{align}
Alternatively
\begin{align}
 {\hat  \lambda}= \lambda  + r\, g \, , \quad  \beta\raisebox{ -1.5pt}{$\scriptstyle{\hat \lambda}$}{}^{(1)} =  {}& 
 (n_s + 8 ){\hat \lambda}^2  +a \, g^2 +  4 \, \tn_f \,{\hat \lambda} \, y^2  -  24\,  \tn_f \, r\,  y^4 \, , \nn \\
 \beta_g{\!}^{(1)} ={}&   12 \, {\hat \lambda} \, g + 3 \, b \, g^2 +  4\, \tn_f \, g \, y^2 - 24\,  \tn_f  \,  y^4 \, . 
\label{oneloop3} 
\end{align}

For quadratic scalar operators then at lowest order the anomalous dimension for the singlet $\sigma=\phi^2$
and the corresponding matrix for $\rho^{ab} =\phi^a\phi^b - \frac{1}{n_s} \delta^{ab} \phi^2$ are just
\begin{align}
\gamma_\sigma{\!}^{(1)} = {}& (n_s+ 2){\hat \lambda}  + 2\,  \tn_f \, y^2 \, , \nn \\
\gamma_\rho{\!}^{(1)ab,cd} = {}& 2 ({\hat \lambda}+ \tn_f \, y^2) 
\big (\tfrac12(\delta^{ac}  \delta^{bd}+  \delta^{ad}  \delta^{bc}
- \tfrac{1}{n_s}  \delta^{ab}  \delta^{cd} \big  )+ g\, d^{abcd} \, .
\end{align}

At higher orders, besides the symmetric traceless tensor $d^{abcd}$, it is necessary to take into account the
mixed symmetry tensor $w^{abcd}$ defined by
\be
 2 \, \tr \big ( t^{(a}t^{b)} t^{(c}t^{d)} \big ) -  \tr \big ( t^{a}t^{c} t^{b} t^{d}\big  )  - 
\tr\big  ( t^{a}t^{d} t^{b} t^{c}\big  )
= n\, \alpha \big (  w^{abcd} -   s ( \delta^{ac}  \delta^{bd}+  \delta^{ad}  \delta^{bc}
- 2\, \delta^{ab}  \delta^{cd})   \big  ) \, , 
\label{wdef}
\ee
which with
\be
s = \frac{ n_s \, \alpha - \beta}{n_s - 1} \, ,
\ee
satisfies $w^{abcd} = w^{(ab)(cd)} = w^{cdab}, \, w^{a(bcd)}=0$ and is traceless on contraction of any pair of indices.
This contributes to $\gamma_\rho{\!}^{ab,cd}$ at two and higher loops. The anomalous dimensions 
are then dictated by the eigenvalues of $d^{abcd}$ and $w^{abcd}$ as $\frac12(n-1)(n+2)\times
\frac12(n-1)(n+2)$ symmetric matrices. There are discussed in appendix \ref{appdw}. In general
there are three eigenvalues as  symmetric traceless tensors decompose into components belonging
to representation spaces of the reduced symmetry group $H$.

If $a=0$ then in  \eqref{hdrel} $d^{abcd} =0 $ and the $g$ coupling is redundant. The scalar $\beta$-function
at one loop is given just by $\beta\raisebox{ -1.5pt}{$\scriptstyle{\hat \lambda}$}{}^{(1)}$ in \eqref{oneloop3} with $g=0$. 
 This restriction necessarily holds for $n=2,3$ since, for any traceless $t$, 
$\tr( t^4) = \frac12 \tr(t^2)^2$. This translates into the condition $\frac12(n_s+2) n\, \alpha = 2 \, n_s \, \alpha + \beta$.
In this case $H=O(n_s)$ and there are just two anomalous dimensions for quadratic scalars $\gamma_\sigma, \,
\gamma_\rho$ with
\be
\gamma_\sigma{\!}^{(1)} = (n_s+ 2){\hat \lambda} + 2\,  \tn_f \, y^2 \, , \qquad
\gamma_\rho{\!}^{(1)} = 2 ( {\hat \lambda} +  \tn_f \, y^2 )  \, .
\ee

Two extreme examples of matrices satisfying \eqref{tid} are given by

\noindent  1, Symmetric. $t^a  \to s^a $ where $\{ s^a \}$ are a basis for 
symmetric traceless $n\times n$ real matrices with $n\ge 2$ satisfying the completeness condition
$(s^a)_{\alpha\beta} (s^a)_{\gamma \delta} =  \delta_{\alpha\gamma} \,\delta_{\beta \delta} + 
 \delta_{\alpha\delta} \, \delta_{\beta \gamma} - \frac{2}{n} \, 
 \delta_{\alpha\beta }\,\delta_{\gamma\delta}$, and $\tr(s^a s ^b) = 2\, \delta^{ab}$. $\{ s^a \}$ are the generators
 corresponding to the coset $Sl(n,{\mathbb R})/SO(n)$.
 
 \noindent  2, Diagonal.  $t^a$ are traceless  diagonal  $n\times n$ real matrices with $n\ge 2$. A basis is 
 obtained by taking $(t^a)_{\alpha\beta}  = e^a_\alpha \, \delta_{\alpha\beta}$ where $e^a_\alpha, \ \alpha=1,\dots,n$
 form the $n$ vertices of a $n-1$ dimensional hypertetrahedron and satisfy $\sum_\alpha  e^a_\alpha =0, \,
 \sum_\alpha e^a_\alpha e^b_\alpha = 2\,\delta^{ab} $ with $e^a_\alpha e^a_\beta =2\, \delta_{\alpha\beta} -  \frac2n$.
 In this example the tensor $w^{abcd}$ in \eqref{wdef} vanishes.

\noindent For these cases we have
\begin{align}
\text{
 \begin{tabular}{  c   c   c  c  c  c  c c }
$y^a$ &  $n_s$ & $n_f$ &  $~~~ \alpha $ & $ \beta  $ & $ A $ & $B$ & $H$  \\
\noalign {\vskip 3pt}
\hline
\noalign {\vskip 4pt}
1. & $ \tfrac12(n-1)(n+2)  $ & $ n\, m $& $~~~ \tfrac 2n$ & $\tfrac1n  (n-2)$&  $ \tfrac{2}{9n^2} (n^2+6)$ & $ \tfrac{1}{9 n}(2n^2 + 9n -36) $ & $O(n) $
 \nn \\
 \noalign {\vskip 4pt}
2. & $ n- 1$ & $ n\, m $& $~~~ \tfrac 2n$ & $\tfrac2n  (n-1)$&  $ \tfrac{4}{3n^2}$ & $ \tfrac{2}{n}(n-2) $ & $\S_{n} $
\end{tabular} 
} \nn \\
\noalign{\vskip -6pt}
\label{tableS}
\end{align}
For purely scalar theories these examples were described long ago in \cite{Priest}.

In general defining
\begin{align}
S_{\alpha\beta\gamma\delta} = {}&  (t^a)_{(\alpha\beta}(t^a)_{\gamma\delta)} - \tfrac{2\, n_s \, \alpha} { 3(n+2)}\big  (
\delta_{\alpha\beta} \delta_{\gamma\delta} + \delta_{\alpha\gamma} \delta_{\beta\delta} + 
\delta_{\alpha\delta} \delta_{\beta\gamma} \big ) \, , \nn \\
W_{\alpha\beta\gamma\delta} = {}&  (t^a)_{\alpha\beta}(t^a)_{\gamma\delta} 
-\tfrac12\big ((t^a)_{\alpha\gamma}(t^a)_{\beta\delta}  + (t^a)_{\alpha\delta}(t^a)_{\beta\gamma}  \big ) \nn \\
\noalign{\vskip -2pt} 
& \hskip 2cm {}+ \tfrac{n_s \, \alpha} {n- 1}\big  (
\delta_{\alpha\beta} \delta_{\gamma\delta} - \tfrac12( \delta_{\alpha\gamma} \delta_{\beta\delta} + 
\delta_{\alpha\delta} \delta_{\beta\gamma} ) \big ) \, ,
\label{SW}
\end{align}
then positivity of $S_{\alpha\beta\gamma\delta} S_{\alpha\beta\gamma\delta} $ and  
$W_{\alpha\beta\gamma\delta} W_{\alpha\beta\gamma\delta} $ give the bounds
\be
 \big ( \tfrac{2 \, n_s}{n+2} -  \tfrac12 \, n\big )   \alpha   \le \beta \le  \big ( n  - \tfrac{n_s}{n-1} \big ) \alpha \, ,
 \label{bone}
\ee
which entails $n_s\le \tfrac12 (n-1)(n+2)$. This is of course saturated in case 1 and the upper bound on $\beta$
is saturated in case 2.
With $O^{ab} = t^a\, t^b - \tfrac{1}{n_s} \, \delta^{ab} \, t^c t^c$ then since  $| \tr(O^{ab} \, O^{ab} )| \le  \tr(O^{ba} \, O^{ab} )$ we must have also the bounds
\be
- (n_s-2) \alpha  \le \beta \le n_s \, \alpha  \, .
\label{btwo}
\ee

General results for fermion scalar theories can be restricted to  $n_s$  real scalars and $n_f$ Dirac fermions  
by taking
\be
y^a \to \begin{pmatrix} 0 &y\,  t^a \\ \by \, \bt^a & 0 \end{pmatrix} {\mathds 1}_{m} \, , \qquad \bt^a = (t^a)^\dagger \, .
\label{yDirac}
\ee
with $\{t^a \}$ $n\times n$ matrices so that $n_f = n\, m$. Assuming the  Yukawa interaction satisfies
\be
U^{-1} y^a \, U  = {\tilde R}^{ab} y^b \quad \mbox{for} \quad  U  \in H_f \subset U(n_f)\times U(n_f) \, , \ \  
[{\tilde R}^{ab}] \in  {H_s} \subset O(n_s) \, ,
\label{inv2}
\ee
to preserve the form  \eqref{yDirac} $U =\left( \begin{smallmatrix} U_- & 0 \\ 0 & U_+ \end{smallmatrix}\right )$ with 
$U_\pm \in H \times U(m) $ so that the symmetry groups become
$H_f= H\times H \times U(m)$ with ${H_s}= H/U(1)$.
Here we require as previously that $\{t^a\}$ are 
traceless so that it is necessary to take $U_\pm=U$. 

For each fermion trace the reduction of general results is obtained by taking
\begin{align}
\tr ( y^{a_1} y^{a_2}  \dots y^{a_{2p} } ) \to{}& m\, (y\by)^n \big (  \tr ( t^{a_1} \bt^{a_2}  \dots \bt^{a_{2p} } )+ 
\tr ( \bt^{a_1} t^{a_2}  \dots t^{a_{2p} } )\nn \\
\noalign{\vskip - 3pt}  
&{} \hskip 1.6cm  +  \tr ( \bt^{a_{2p}}  \dots \bt^{a_2} t^{a_1} )+  \tr ( t^{a_{2p}}   \dots  t^{a_2} \bt^{a_1 } ) \big ) \, , \nn \\
\tr ( y^{a_1} y^{a_2}  \dots y^{a_{2p+1 } } ) \to{}& 0  \, .
\label{trace}
\end{align}
The identities in \eqref{tid} and also \eqref{hdef} become
\be
t^a {\hskip 0.5pt}{\bar t}^{\hskip 0.8pt a }  = n_s \, \alpha \, {\mathds 1}_{n}  \, , \quad 
\tr (t^a{\hskip 0.5pt} {\bar t} {\hskip 0.8pt}^b) =  n \, \alpha \, \delta^{ab}  \, , \quad
t^b {\hskip 0.5pt} {\bar t}{\hskip 0.5pt}^a{\hskip 0.5pt} t^b  = \beta\,  t^a  \, , 
\quad \tr \big  (t^{\hskip 0.5pt(a} {\bar t}^{\hskip 0.8pt b}  t^{\hskip 0.5pt c}  {\bar t^{\hskip 0.8pt d)}} \big )  = 
n \, \alpha\, h^{abcd} \, .
\label{tid2}
\ee
As before $\alpha>0$ but the bounds  in \eqref{bone} are no longer valid although, as previously with 
$O^{ab} = t^a\,  \bt^b - \tfrac{1}{n_s} \, \delta^{ab} \, t^c \,\bt^c$, \eqref{btwo} remains. The results in \eqref{oneloop}
and \eqref{oneloop2} then remain valid after taking $n_f \to 4\, n_f$.

Various examples of matrices $\{t^a\}$ satisfying \eqref{tid2}  are obtained from the generators
in the fundamental representation of classical Lie groups

\noindent 3, Unitary.  $t^a, \, {\bar t}{\hskip 0.8pt}^a  \to \lambda^a $ where $\{ \lambda^a \}$ are
hermitian traceless $n\times n$ matrices, $n\ge 2$,  forming  generators for $SU(n)$, 
satisfying the completeness condition $(\lambda^a)_\alpha{}^\beta (\lambda^a)_\gamma{}^\delta = 
2( \delta_\alpha{}^\delta \delta_\gamma {}^\beta - \frac{1}{n} \, 
 \delta_\alpha{}^\beta \delta_\gamma{}^\delta)$, and $\tr(\lambda^a \lambda^b) = 2\, \delta^{ab}$.

\noindent  4, Antisymmetric.  $t^a, \,- {\bar t}{\hskip 0.8pt}^a  \to a^a $ where $\{ a^a \}$ are
antisymmetric $n\times n$ real matrices, $n\ge 2$, forming generators for $SO(n)$, 
satisfying the completeness condition
$(a^a)_{\alpha\beta} (a^a)_{\gamma \delta} =  \delta_{\alpha\gamma}\, \delta_{\beta \delta} - 
 \delta_{\alpha\delta}\, \delta_{\beta \gamma} $, and $\tr(a^a a^b) = - 2\, \delta^{ab}$.

\noindent 5, Symplectic.   $t^a, \, {\bar t}{\hskip 0.8pt}^a  \to \sigma^a $ where $\{ \sigma^a \}$ are
hermitian  $n\times n$ matrices, $n=2p, \ p \ge 1$, which are generators of $Sp(n)$ so that 
for  $J_{\alpha\beta}, (J^{-1})^{\alpha\beta}$ antisymmetric matrices then
$J^{-1} \sigma^a \hskip 0.5pt J = - (\sigma^a)^T$ or $(\sigma^a J)^T= \sigma^a J$. The assumed 
completeness relation is then
$(\sigma^a)_\alpha{}^\beta (\sigma^a)_\gamma{}^\delta =  \delta_\alpha{}^\delta \delta_\gamma {}^\beta  + J_{\alpha\gamma}\, (J^{-1})^{\beta\delta}$ and $\tr(\sigma^a \sigma^b) =  2\, \delta^{ab}$.

\noindent 6, Symplectic${}^\prime$.   $t^a, \, {\bar t}{\hskip 0.8pt}^a  \to \tsig^a $ where $\{ \tsig^a \}$ are
hermitian traceless $n\times n$ matrices, $n=2p, \ p \ge 2$, corresponding to  generators belonging to the coset
$SU(n)/Sp(n)$. With $J, \, J^{-1}$ antisymmetric matrices as in case 4  
$J^{-1} \tsig^a \hskip 0.5pt J = (\tsig^a)^T$ or $(\tsig^a J)^T= - \tsig^a J$ and the completeness relation becomes
$(\tsig^a)_\alpha{}^\beta (\tsig^a)_\gamma{}^\delta =  \delta_\alpha{}^\delta \delta_\gamma {}^\beta  
- J_{\alpha\gamma}\, (J^{-1})^{\beta\delta}- \frac{2}{n} \, 
 \delta_\alpha{}^\beta \delta_\gamma{}^\delta$ and $\tr(\tsig^a \tsig^b) =  2\, \delta^{ab}$.

\noindent For the different cases we have
\begin{align}
\hskip -0.5cm
{\text{
 \begin{tabular}{  c   c   c  c  c  c  c c }
$y^a$ &  $n_s$ & $n_f$ &  $~~~ \alpha $ & $ \beta  $ & $ A $ & $B$ & $~~ H$   \\
\noalign {\vskip 3pt}
\hline
\noalign {\vskip 4pt}
3. & $ n^2 - 1 $ & $  n\,  m $  & $~~~\tfrac2n $&   $-\tfrac 2n$  & $\tfrac{4}{9n^2} (n^2 + 3)$ & $\tfrac{4}{9n}(n^2 -9)$ 
& $~~ SU(n)$ \\ 
\noalign {\vskip 4pt}
4. & $ \tfrac12 n (n-1)  $ & $ n\, m  $ & $~~~ \tfrac 2n $ & $~ 1$ & $\tfrac29  $ & $ \tfrac19 (2n-1) $ & $ ~~SO(n)$  \\
\noalign {\vskip 4pt} 
5.  & $\tfrac12 n (n+1) $  & $  n\, m  $& $~~~ \tfrac 2 n$ & $- 1 $ & $ \tfrac29 $ & $ \tfrac{1}{9} (2n+1)$ &
$~~ Sp(n)$ \\
\noalign {\vskip 4pt} 
6.  & $\tfrac12 (n-2) (n+1) $  & $  n\, m  $& $~~~ \tfrac 2 n$ & $- \tfrac1n(n+2) $ & $ \tfrac{2}{9n^2}(n^2+6) $ & 
$ \tfrac{1}{9n} (2n^2-9n-36)$ &
$~~ Sp(n)$
\end{tabular} 
}} \nn \\
\noalign{\vskip -2pt}
\label{table}
\end{align}
For case 4 and $n=2$, $H$ reduces to ${\mathbb Z}_2$.

Beyond one loop there are further conditions necessary on $t^a, \, \bt^a$ for each primitive diagram (which are those 
with no subdivergences).
At two loops it is sufficient to require
\begin{align}
 t^b \,  \bt^{\hskip 0.5pt c} \,  t^a \, \bt^{\hskip 0.5pt b} \, t^c = {}&
\gamma \, t^a \, , \nn \\
 \tr \big ( t^e  \, \bt^{\hskip 0.5pt (a} \,  t^b \, \bt^{\hskip 0.5pt |e|} \, t^c  \, \bt^{\hskip 0.5pt d)} \big )
= {}& n\alpha\big ( \delta \,  h^{abcd} + \tfrac13 \alpha( n_s\alpha-\delta)
\big ( \delta^{ab}  \delta^{cd} +  \delta^{ac}  \delta^{bd}+  \delta^{ad}  \delta^{bc}\big  )\big ) \, .
\end{align}
In general
\be
\gamma = n_s \, \alpha(2\alpha+\delta) - (2\alpha -\beta)\delta -\beta^2\, .
\ee
For the different examples considered here results for $a,b$ in \eqref{did} and also $\gamma,\delta$ are then
\begin{align}
\text{
 \begin{tabular}{  c   c   c  c   c}
$y^a$ &  $a $ & $ b $ & $\gamma$  &  $\delta $ \\
\noalign {\vskip 3pt}
\hline
\noalign {\vskip 4pt}
1. &  $\tfrac{(n-3)(n-2)(n+1)(n+4)(n+6)}{6\, n (n_s+2)^2}$  & $ \tfrac{ 2n^4 + 11n^3 - 71n^2- 90n + 72}{18\, n(n_s+2)} $ 
& $\tfrac{1}{n^2}({\scriptstyle  n^3 + 3n^2 - 4n +4} )$  & $1-\tfrac2n$ \\ 
\noalign {\vskip 4pt}
2. &  $\tfrac{8(n-3)(n-2)}{ n (n_s+2)^2}$  & $2\, \tfrac{n^2-5n + 2}{ n(n_s+2)} $ 
& $\tfrac{4}{n^2}\scriptstyle{ ( n-1)^2 } $  & $2- \tfrac{2}{n}$ \\ 
\noalign {\vskip 4pt}
3. & $ \tfrac{4\,(n_s-3)(n_s-8)}{3(n_s+2)^2}  $ & $ \tfrac{ 4(n^4 -20n^2 + 9) }{9\, n(n_s+2)} $ & 
$  \tfrac{4}{n^2} \scriptstyle{(n^2+1)}$  & $-\tfrac2n $   \\ 
\noalign {\vskip 4pt}
4. & $ \tfrac{8\, (n_s-1)(n_s-3)}{3(n_s+2)^2}   $ & $ \tfrac{ (n-5)(n+4)(2n-1)}{9(n_s+2)}  $ & $ \scriptstyle{3-n}$   
& $-1$ \\
\noalign {\vskip 4pt} 
5.  & $  \tfrac{2(n_s-1)(n_s-3)}{3(n_s+2)^2}   $  & $ \tfrac{ (n-4)(n+5)(2n+1)}{18(n_s+2)}  $ &  $ \scriptstyle{3+n}$ 
&$1$\\
\noalign {\vskip 4pt}
6. &  $\tfrac{(n-6)(n-4)(n-1)(n+2)(n+3)}{6\, n (n_s+2)^2}$  & $\tfrac{ 2n^4 - 11n^3 - 71n^2 + 90n + 72}{18\, n(n_s+2)}  $ &
$\tfrac{1}{n^2}({\scriptstyle  - n^3 +3n^2 + 4n + 4 } )$  & $-1-\tfrac2n$ 
\end{tabular} 
} \nn \\
\noalign{\vskip -6pt}
\label{table3}
\end{align}
The results for $d^{abcd} d^{abcd} = \tfrac12 n_s (n_s+2) a$ correspond in cases 3,4,5 
to the evaluation of the quartic Casimir for  $SU(n), \, SO(n), \, Sp(n)$ \cite{4loopbeta}. 
In general for $a>0$ it is necessary to restrict $n>3$ except for case 6 when  $n>6$ is required.

\subsection{Further Algebraic Relations}

For the characterisation of the different possibilities we may further define for cases 1,2,3,6
additional invariant tensors 
\be
 \tr \big  (t^{(a}\hskip 0.5pt t^{\hskip 0.8pt b} \hskip 0.5pt  t^{c)}\big )  =  n\, \alpha\, d^{abc}  \, , \qquad
 \tikz[baseline=(vert_cent.base)]{
    \draw  [line width = 1pt]  (5.4,0.2)--(5.4,0.45);
       \draw  (5.4,0.2) -- (5.15,-0.15);
     \draw  (5.4,0.2) -- (5.65,-0.15) ;
      \draw  (5.15,-0.15) -- (5.65,-0.15) ;
      \draw [line width = 1pt]  (5.65,-0.15) -- (5.85,-0.35) ; 
    \draw [line width =  1pt]   (5.15,-0.15) -- (4.95,-0.35)  ;
     }  
     =  n \alpha
     \tikz[baseline=(vert_cent.base)]{
    \draw  [line width = 0.8pt]  (6.4,0)--(6.4,0.35);
       \draw  [line width = 0.8pt]  (6.4,0) -- (6.1,-0.3);
     \draw  [line width = 0.8pt] (6.4,0.0) -- (6.7,-0.3) ;
 \filldraw [black] (6.4,0) circle [radius=1.5pt];
 } \ ,
 \label{defd}
\ee
where $ d^{abc}$ is symmetric  and traceless, 
These three index $d$-tensors are constrained by
the one and two loop identities
\be
d^{acd} d^{bcd}  = \alpha_d \, \delta^{ab} \, , \quad
d^{ade} d^{bef} d^{cfd} = \beta_d \, d^{abc} \, , \quad 
d^{dbf} d^{efg} d^{agh} d^{dhi}d^{eic} = \gamma_d \, d^{abc} \, ,
\ee
or diagrammatically
\begin{align}
\tikz[baseline=(vert_cent.base)]{
 \node (vert_cent) {\hspace{-13pt}$\phantom{-}$};
    \draw  [line width = 0.8pt]  (-0.4,0)--(0.0,0);
    \draw  [line width = 0.8pt]      (0.45,0) ++(0:0.45cm) arc (0:180:0.45cm and 0.3cm);
     \draw  [line width = 0.8pt]     (0.45,0) ++(0:0.45cm) arc (360:180:0.45cm and 0.3cm);
   \draw  [line width = 0.8pt]       (0.9,0)--(1.3,0);
        \filldraw [black] (0,0) circle [radius=1.5pt];
         \filldraw [black] (0.9,0) circle [radius=1.5pt];
} 
= \alpha_d \ \tikz[baseline=(vert_cent.base)]{
    \draw  [line width = 0.8pt] (1.4,0)--(2.1,0);} \, , \qquad
   \tikz[baseline=(vert_cent.base)]{
    \draw  [line width = 0.8pt]  (5.4,0.1)--(5.4,0.55);
       \draw  [line width = 0.8pt]  (5.4,0.1) -- (4.85,-0.45);
     \draw  [line width = 0.8pt] (5.4,0.1) -- (5.95,-0.45) ;
      \draw  [line width = 0.8pt] (5.05,-0.25) -- (5.75,-0.25) ;
 \filldraw [black] (5.4,0.1) circle [radius=1.5pt];
 \filldraw [black] (5.05,-0.25) circle [radius=1.5pt];
 \filldraw [black] (5.75,-0.25) circle [radius=1.5pt];
     } = \beta_d \ 
   \tikz[baseline=(vert_cent.base)]{
    \draw  [line width = 0.8pt]  (6.4,0)--(6.4,0.5);
       \draw  [line width = 0.8pt]  (6.4,0) -- (6,-0.4);
     \draw  [line width = 0.8pt] (6.4,0.0) -- (6.8,-0.4) ;
 \filldraw [black] (6.4,0) circle [radius=1.5pt];
 } \, , \qquad
   \tikz[baseline=(vert_cent.base)]{
    \draw  [line width = 0.8pt]  (8.4,0.1)--(8.4,0.55);
       \draw  [line width = 0.8pt]  (8.4,0.1) -- (8.95,-0.45);
     \draw  [line width = 0.8pt] (8.4,0.1) -- (7.85,-0.45) ;
      \draw  [line width = 0.8pt] (8.25,-0.05) -- (8.8,-0.3) ;
 \filldraw [black] (8.4,0.1) circle [radius=1.5pt];
  \filldraw [black] (8.25,-0.05) circle [radius=1.5pt];
   \filldraw [black] (8.8,-0.3) circle [radius=1.5pt];
    \filldraw [white] (8.4,-0.12) circle [radius=2pt];
    \draw  [line width = 0.8pt] (8.0,-0.3) -- (8.55,-0.05) ;
      \filldraw [black] (8.0,-0.3) circle [radius=1.5pt];
   \filldraw [black] (8.55,-0.05) circle [radius=1.5pt];
  } 
 = \gamma_d \ 
\tikz[baseline=(vert_cent.base)]{
    \draw  [line width = 0.8pt]  (9.4,0)--(9.4,0.5);
       \draw  [line width = 0.8pt]  (9.4,0) -- (9,-0.4);
     \draw  [line width = 0.8pt] (9.4,0) -- (9.8,-0.4) ;
 \filldraw [black] (9.4,0) circle [radius=1.5pt];
 } \ ,
 \label{abcd}
\end{align}
and at three loops there are two primitive diagrams and it is then necessary that
\be   
\tikz[baseline=(vert_cent.base)]{
    \draw  [line width = 0.8pt]  (5.4,0.4)--(5.4,0.9);
       \draw  [line width = 0.8pt]  (5.4,0.4) -- (4.4,-0.8);
     \draw  [line width = 0.8pt] (5.4,0.4) -- (6.4,-0.8) ;
      \draw  [line width = 0.8pt] (4.7,-0.44) -- (6.1,-0.44) ;
        \draw  [line width = 0.8pt] (5.4,-0.15) -- (5.4,-0.44) ;
         \draw  [line width = 0.8pt] (5.4,-0.15) -- (5.1,0.04) ;
          \draw  [line width = 0.8pt] (5.4,-0.15) -- (5.7,0.04) ;
 \filldraw [black] (5.4,0.4) circle [radius=1.5pt];
 \filldraw [black] (4.7,-0.44) circle [radius=1.5pt];
 \filldraw [black] (6.1,-0.44) circle [radius=1.5pt];
 \filldraw [black] (5.4,-0.15) circle [radius=1.5pt];
  \filldraw [black] (5.1,0.04) circle [radius=1.5pt];
   \filldraw [black] (5.7,0.04) circle [radius=1.5pt];
     \filldraw [black] (5.4,-0.44) circle [radius=1.5pt];
     } = \delta_d \ 
   \tikz[baseline=(vert_cent.base)]{
    \draw  [line width = 0.8pt]  (6.4,0)--(6.4,0.5);
       \draw  [line width = 0.8pt]  (6.4,0) -- (6,-0.4);
     \draw  [line width = 0.8pt] (6.4,0.0) -- (6.8,-0.4) ;
 \filldraw [black] (6.4,0) circle [radius=1.5pt];
 } \, ,\qquad
 \tikz[baseline=(vert_cent.base)]{
    \draw  [line width = 0.8pt]  (5.4,0.4)--(5.4,0.9);
       \draw  [line width = 0.8pt]  (5.4,0.4) -- (4.4,-0.8);
     \draw  [line width = 0.8pt] (5.4,0.4) -- (6.4,-0.8) ;
      \draw  [line width = 0.8pt] (4.7,-0.44) -- (6.1,-0.44) ;
         \draw  [line width = 0.8pt] (5.15,0.1) -- (5.7,-0.44) ;
          \filldraw [white] (5.4,-0.15) circle [radius=2pt];
          \draw  [line width = 0.8pt] (5.65,0.1) -- (5.1,-0.44) ;
 \filldraw [black] (5.4,0.4) circle [radius=1.5pt];
 \filldraw [black] (4.7,-0.44) circle [radius=1.5pt];
 \filldraw [black] (6.1,-0.44) circle [radius=1.5pt];
 \filldraw [black] (5.7,-0.44) circle [radius=1.5pt];
  \filldraw [black] (5.1,-0.44) circle [radius=1.5pt];
  \filldraw [black] (5.15,0.1) circle [radius=1.5pt];
   \filldraw [black] (5.65,0.1) circle [radius=1.5pt];
     } = \epsilon_d \ 
   \tikz[baseline=(vert_cent.base)]{
    \draw  [line width = 0.8pt]  (6.4,0)--(6.4,0.5);
       \draw  [line width = 0.8pt]  (6.4,0) -- (6,-0.4);
     \draw  [line width = 0.8pt] (6.4,0.0) -- (6.8,-0.4) ;
 \filldraw [black] (6.4,0) circle [radius=1.5pt];
 } \, .
 \label{ded}
 \ee

More general versions of these equations with more than one $d$-tensor were discussed for various $n$ in \cite{Liendo}.

For the particular cases considered here
\begin{align}
\text{
 \begin{tabular}{  c   c   c  c   c}
 & $ \alpha_d $ & $\beta_d$  & $\gamma_d$  \\
\noalign {\vskip 3pt}
\hline
\noalign {\vskip 4pt}
1. & $ \tfrac{1}{2n} \, (n-2) (n+4) $ &
$\tfrac{1}{4n} ( n^2+ 4n- 24)$ & $\tfrac{1}{8n^2}(n-4) (3n^2+4n-80) $ \\ 
\noalign {\vskip 4pt}
2. & $ \tfrac{2}{n} \, (n-2)  $ &
$\tfrac{2}{n} ( n-3)$ & $\tfrac{4}{n^2}(n^2-6n+10) $ \\ 
\noalign {\vskip 4pt}
3. &  $  \tfrac{1}{n}\, (n^2-4)  $ & $\tfrac{1}{2n}(n^2-12)$ & $-\tfrac{4}{n^2} ( n^2- 10)  $  \\ 
\noalign {\vskip 4pt}
6. & $ \tfrac{1}{2n} \, (n-4) (n+2) $ &
$\tfrac{1}{4n} ( n^2- 4n- 24)$ & $ -\tfrac{1}{8n^2}(n+4) (3n^2-4n-80)   $ 
\end{tabular} 
} \nn \\
\noalign{\vskip -14 pt} \ ,
\label{table4}
\end{align}
and
\begin{align}
\text{
 \begin{tabular}{  c   c   c  c  }
 & $ \delta_d $ & $\epsilon_d$  \\
\noalign {\vskip 3pt}
\hline
\noalign {\vskip 4pt}
1. & $ \tfrac{1}{n^3}\big ({\scriptstyle{ \frac{1}{64} n^6+ \frac{9}{64} n^5+ \frac{1}{8}n^4 -\frac{11}{2}n^3 +2  n^2+ 116 n - 256}} \big  )$ &
$\tfrac{1}{n^3}\big ({\scriptstyle{  \frac{5}{64} n^5+ \frac{7}{16}n^4 -\frac{19}{4}n^3 - n^2+ 116 n - 264} } \big )$ \\ 
\noalign {\vskip 4pt}
2. & $ \tfrac{8}{n^3} \, (n^3-9n^2 + 29 n -32)  $ &
$\tfrac{8}{n^3}  (n^3-9n^2 + 29 n -33) $  \\ 
\noalign {\vskip 4pt}
3. &  $  \tfrac{1}{8n^3}\, (n^2 -8) (n^4  - 8 n^2 + 256) $ & $-\tfrac{1}{2n^3}(n^4- 68n^2 + 528)$ \\ 
\noalign {\vskip 4pt}
6. & $ \tfrac{1}{n^3}\big ({\scriptstyle{ \frac{1}{64} n^6- \frac{9}{64} n^5+ \frac{1}{8}n^4 +\frac{11}{2}n^3 +2  n^2 -116 n - 256}} \big  )$  &
$-\tfrac{1}{n^3}\big  ({\scriptstyle\frac{5}{64} n^5 - \frac{7}{16}n^4 -\frac{19}{4}n^3 + n^2+ 116 n + 264}\big )$ 
\end{tabular} 
} \nn \\
\noalign{\vskip -14 pt} \ .
\label{table4a}
\end{align}
The results for cases 2 and 3 were given previously in \cite{McKane0,McKane} and \cite{Gracey6d}. 

For further applications
 \be
{\rm Sym} \hskip -0.2cm \raisebox{-0.5 cm}{
\begin{tikzpicture}[scale=1]
\draw[line width = 0.8pt] (0,-0.25) to (0,0.25);
\draw[line width = 0.8pt] (0.5,-0.25) to (0.5,0.25);
\draw[line width = 0.8pt] (0,-0.25) to (0.5,-0.25);
\draw[line width = 0.8pt] (0,0.25) to (0.5,0.25);
\draw [line width = 1pt] (0.5,0.25) --+ (40:0.5cm);
\draw[line width = 1pt] (0,0.25) --+ (140:0.5cm);
\draw [line width = 1pt]  (0,-0.25) --+ (220:0.5cm);
\draw [line width = 1pt](0.5,-0.25) --+ (320:0.5cm);
\fill  (0,0.25) circle [radius=1.5pt];
\fill  (0,-0.25) circle [radius=1.5pt];
\fill  (0.5,0.25) circle [radius=1.5pt];
\fill  (0.5,-0.25) circle [radius=1.5pt];
\end{tikzpicture}
}
\ = \
A_d \ \Big ( 
\raisebox{-0.15cm}{ 
\begin{tikzpicture}[scale=1]
\draw[line width = 1pt] (0,0.23) to (0.5,0.23);
\draw [line width = 1pt](0,-0.23) to (0.5,-0.23);
\end{tikzpicture}
}
\ + \ 
\raisebox{-0.15cm}{
\begin{tikzpicture}[scale=1]
\draw [line width = 1pt](0,0.25) to (0,-0.25);
\draw[line width = 1pt] (0.5,0.25) to (0.5,-0.25);
\end{tikzpicture}
}
\ + \ 
\raisebox{-0.15cm}{
\begin{tikzpicture}[scale=1]
\draw[line width = 1pt] (0,0.25) to (0.5,-0.25);
\fill [white] (0.25,0) circle [radius=2pt];
\draw[line width = 1pt] (0,-0.25) to (0.5,0.25);
\end{tikzpicture}
}  \Big ) 
\ \ + \ \  B_d/n \alpha \ {\rm Sym}  \hskip -0.2cm
\raisebox{-0.5 cm}{
\begin{tikzpicture}[scale=1]
\draw (0,-0.25) to (0,0.25);
\draw (0.5,-0.25) to (0.5,0.25);
\draw (0,-0.25) to (0.5,-0.25);
\draw  (0,0.25) to (0.5,0.25);
\draw [line width = 1pt] (0.5,0.25) --+ (40:0.5cm);
\draw[line width = 1pt] (0,0.25) --+ (140:0.5cm);
\draw [line width = 1pt]  (0,-0.25) --+ (220:0.5cm);
\draw [line width = 1pt](0.5,-0.25) --+ (320:0.5cm);
\end{tikzpicture}
}
\ee
where $(n_s+2)A_d + \tfrac13( 2n_s \,\alpha+ \beta) B_d = \tfrac13(2 \alpha_d + \beta_d) \alpha_d$.
 For the different cases we have
\begin{align}
\text{
 \begin{tabular}{  c   c   c  c  c}
 & $ A_d $ & $B_d$ & $\beta_1$  \\
\noalign {\vskip 3pt}
\hline
\noalign {\vskip 4pt}
1. & $ \tfrac{1}{12n^2} (3 n^2 + 32) $ & $\tfrac{1}{8n} (n^2 + 8n -64) $ & $ \tfrac12( 5n^2 +14 n - 72)$ \\ 
\noalign {\vskip 4pt}
2. & $ \tfrac{8}{3n^2}  $ & $\tfrac{2}{n}  (n-4) $ & $2 (7n-18)$  \\ 
\noalign {\vskip 4pt}
3. &  $  \tfrac{1}{6n^2}\, (3n^2 + 16) $ & $\tfrac{1}{4n}(n^2-  32 )$  &  $5n^2 - 36 $\\ 
\noalign {\vskip 4pt}
6. & $ \tfrac{1}{12n^2} (3n^2+32) $  & $\tfrac{1}{8n} (n^2-8n-64)$  &  $ \tfrac12( 5n^2 -14 n - 72)$ 
\end{tabular} 
} 
\ .
\label{table5a}
\end{align}

The tensors $d^{abc}$ defined as in \eqref{defd} may be used to form symmetric traceless
Yukawa couplings so that the number of fermions $n_f= m\, n_s$. In this case \eqref{oneloop2} becomes
\begin{align}
\beta_y{\!}^{(1)} ={}& ( \alpha_d + 2\,  \beta_d   +   \tfrac12 \, \alpha_d \, m ) y^3 \, , \nn \\
\beta_\lambda{\!}^{(1)} ={}&  (n_s + 8 ) \lambda^2 
+ \tfrac23 ( 2 \, n_s \, \alpha + \beta ) \lambda \, g  + 2 \,  \alpha_d\, m \, \lambda \, y^2+ 3\, A\,  g^2 
- 12\, A_d \, m \, y^4 \, , \nn \\
\beta_g{\!}^{(1)} ={}&   12 \, \lambda \, g + 3 \, B \, g^2 +  2\, \alpha_d \, m \, g \, y^2 - 12\,  B_d\, m \,  y^4  \nn \\
\gamma_\phi{\!}^{(1)} ={}&    \tfrac12 \, \alpha_d \, m \, y^2 \, , \qquad 
\gamma_\psi{\!}^{(1)} = \tfrac12 \, \alpha_d\, y^2   \, .
\label{oneloop3a}
\end{align}
As was discussed in detail in \cite{Emergent} for $m=1, \, n_s=n_f$ there is a reduction to a single component
 $\N=\frac12$ supersymmetric theory with the couplings constrained by
\be
\lambda = \tfrac2n \, y^2 \, , \qquad g= 3\, y^2 \, .
\ee
The results in \eqref{oneloop3a} with this restriction all correspond to
\be
\beta_{y^2}{\!}^{(1)} = \tfrac1n \, \beta_1 \, y^4 \, ,
\ee
with formulae for $\beta_1$ given in \eqref{table5a}. The reduction depends on non trivial relations between
$A, A_d, B, B_d$ which are satisfied in each of the cases listed. Except for the last case $\beta_1>0$, and 
there are fixed points in an $\vep$-expansion, for $n\ge 3$.
For the $\N=\frac12$ supersymmetric theory then taking $Y^{abc}\to y \, d^{abc}$ 
there is a single coupling theory with a $\beta$-function and $\gamma_\Phi$, with three loop coefficients given in 
\eqref{dres}, \eqref{cres}, expanded as
\begin{align}
\beta_y={} &\big ( \tfrac32 \, \alpha_d + 2\, \beta_d  \big) y^3  
- 3  \big( \tfrac12 \, \alpha_d{\!}^2 + \alpha_d \,  \beta_d + 2\, \beta_d{\!}^2  \big ) y^5\nn \\ 
& +  \big( \tfrac{15}{8} \, \alpha_d{\!}^3 + \tfrac94\, \alpha_d{\!}^2 \beta_d + \tfrac{27}{2} \, \alpha_d\, \beta_d{\!}^2  
+30\,   \beta_d{\!}^3  \big ) y^7 \nn \\
&{}+\big ( ( \tfrac92 \, \alpha_d + 30\,  \beta_d ) \gamma_d + 36\, \epsilon_d   
 \big )  \zeta_3\,  y^7 + {\rm O}(y^9) \, , \nn \\
\gamma_\Phi = {}& \tfrac12 \, \alpha_d \, y^2 - \tfrac12\, \alpha_d{\!}^2 y^4 + \big ( \tfrac58 \, \alpha_d{\!}^3
- \tfrac14 \, \alpha_d{\!}^2 \beta_d + \tfrac52 \, \alpha_d\, \beta_d{\!}^2 
+ \tfrac32\, \alpha_d\, \gamma_d \, \zeta_3\big  ) y^6   + {\rm O}(y^8) \, .
\label{susyb}
\end{align}
Of course the lowest order term corresponds to \eqref{oneloop3}. Furthermore
\begin{align}
\gamma_{\Phi^2}  =  {}& 2 \, \alpha_d \, y^2 -( 5\, \alpha_d{\!}^2 + 4 \, \alpha_d \beta_d )y^4 \nn \\
&{}  + \big ( 10\, \alpha_d{\!}^3 +7 \, \alpha_d{\!}^2 \beta_d + 22 \, \alpha_d\, \beta_d{\!}^2 
+ 6  ( \alpha_d{\!}^2 \beta_d + 6 \, \alpha_d\, \beta_d{\!}^2+ 4\, \alpha_d\, \gamma_d ) \zeta_3\big  ) y^6 \nn \\
&{} + 2\, \gamma_\Phi + {\rm O}(y^8) \, .
\end{align}

The bounds \eqref{bone} becomes $ - \frac{n_s-2}{2(n_s+2)}\, \alpha_d \le \beta_d
\le \tfrac{n_s -2}{n_s-1} \,\alpha_d$. The upper bound for $\beta_d$ corresponds to the vanishing of $W_{\alpha\beta\gamma\delta}$ in \eqref{SW} and holds exactly for case 2 in \eqref{table4} for any $n$.
The lower bound for $\beta_d$ corresponds to the vanishing of $S_{\alpha\beta\gamma\delta}$ in \eqref{SW} which becomes in this case the condition
\be
d^{ab e} d^{cd e} + d^{ad e} d^{bc e} + d^{ac e} d^{db e} =
K\big ( \delta^{ab} \delta^{cd} + \delta^{ad} \delta^{bc} + \delta^{ac} \delta^{db}  \big ) \, , \quad 
K = \tfrac{2\, \alpha_d}{n_s+2} \, , 
\label{dF4}
\ee
or diagrammatically
\be
\raisebox{-0.2cm}{
\begin{tikzpicture}[scale=1]
\draw  [line width = 0.9pt](0,0) to (0.7,0);
\draw [line width = 0.9pt](0.6,0) --+ (40:0.5cm);
\draw [line width = 0.9pt](0.6,0) --+ (320:0.5cm);
\draw [line width = 0.9pt](0,0) --+ (140:0.5cm);
\draw [line width = 0.9pt](0.0,0) --+ (220:0.5cm);
\fill  (0,0) circle [radius=1.5pt];
\fill  (0.6,0) circle [radius=1.5pt];
\end{tikzpicture}
}
\ + \
\raisebox{-0.5 cm}{
\begin{tikzpicture}[scale=1]
\draw[line width = 0.9pt] (0,-0.25) to (0,0.25);
\draw (0,0.25) --+ (40:0.5cm);
\draw[line width = 0.9pt] (0,0.25) --+ (140:0.5cm);
\draw [line width = 0.9pt](0,-0.25) --+ (220:0.5cm);
\draw [line width = 0.9pt](0,-0.25) --+ (320:0.5cm);
\fill  (0,0.25) circle [radius=1.5pt];
\fill  (0,-0.25) circle [radius=1.5pt];
\end{tikzpicture}
}
\ + \
\raisebox{-0.5 cm}{
\begin{tikzpicture}[scale=1]
\draw [line width = 0.9pt](0,-0.25) to (0,0.25);
\draw [line width = 0.9pt](0,0.25) --+ (140:0.5cm);
\draw [line width = 0.9pt](0,0.25) --+ (-48:0.95cm);
\draw [line width = 0.9pt](0,-0.25) --+ (220:0.5cm);
\fill [white] (0.22,0) circle [radius=2pt];
\draw [line width = 0.9pt](0,-0.25) --+ (48:0.95cm);
\fill  (0,0.25) circle [radius=1.5pt];
\fill  (0,-0.25) circle [radius=1.5pt];
\end{tikzpicture}
}
\ = \
K \Big ( 
\raisebox{-0.15cm}{
\begin{tikzpicture}[scale=1]
\draw[line width = 0.9pt] (0,0.23) to (0.5,0.23);
\draw [line width = 0.9pt](0,-0.23) to (0.5,-0.23);
\end{tikzpicture}
}
\ + \ 
\raisebox{-0.15cm}{
\begin{tikzpicture}[scale=1]
\draw [line width = 0.9pt](0,0.25) to (0,-0.25);
\draw[line width = 0.9pt] (0.5,0.25) to (0.5,-0.25);
\end{tikzpicture}
}
\ + \ 
\raisebox{-0.15cm}{
\begin{tikzpicture}[scale=1]
\draw[line width = 0.9pt] (0,0.25) to (0.5,-0.25);
\fill [white] (0.25,0) circle [radius=2pt];
\draw[line width = 0.9pt] (0,-0.25) to (0.5,0.25);
\end{tikzpicture}
}
\Big )
\label{dddiag}
\ee
This was analysed in \cite{birdtracks} and related to the $F_4$ family of  Lie groups.

A uniform treatment is obtained by considering hermitian traceless $n\times n$ matrices $\{e_a\}$,
$a=1,\dots , n_s$,  satisfying
\be
\tfrac12(e_a e_b + e_b e_a ) = \tfrac 1n \, \delta_{ab} \, \I_n +  d_{abc} \, e_c \, , \qquad
d_{abc} = d\raisebox{ -1.5pt}{$\scriptstyle (abc)$} \, , \ \ d_{aa c} = 0 \, .
\label{eerel}
\ee
Of course this implies $e_a e_a = \tfrac1n n_s \, \I_n , \ \tr(e_a e_b) = \delta_{ab}$. 
The algebra defined by \eqref{eerel}  is equivalent to the result that hermitian real, complex and quaternionic
matrices form a special real Jordan algebra. Furthermore $3\times 3$ hermitian octonionic matrices also 
form an exceptional real Jordan, or Albert, algebra, with $F_4$ as the automorphism group.
For $3\times 3$ hermitian traceless matrices
\be
x= x_a e_a \, , \qquad x^3 - \tfrac12 \, \tr(x^2) \, x - \tfrac13\, \tr(x^3) \, \I_3 = 0 \, .
\label{CHt}
\ee
This is just the Cayley Hamilton theorem for real or complex matrices  in  these cases. 
The result also extends to hermitian traceless quaternionic and even octonionic matrices \cite{Octonion} where 
the trace is just the sum of the real diagonal elements. \eqref{CHt} is
equivalent to \eqref{dF4} with just $K=\frac16$. For each case it is straightforward to check
\begin{align}
\text{
 \begin{tabular}{  c   c   c  c  c }
& ${\mathbb R} $ & $ {\mathbb C}$ & ${\mathbb H}$ &$ {\mathbb O} $ \\
\noalign {\vskip 3pt}
\hline
\noalign {\vskip 4pt}
$n_s$ &  5& 8& 14& 26
\end{tabular}
} \ \ .
\label{division}
\end{align}
Defining $ {\tilde \alpha}_d,\, {\tilde \beta}_d, \, {\tilde \gamma}_d,\, {\tilde \delta}_d,\, {\tilde \epsilon}_d$
just as in \eqref{abcd} and \eqref{ded}  with normalisation dictated by \eqref{eerel} then applying \eqref{dddiag}
to each diagram gives the relations
\begin{align}
& 2\,  {\tilde \beta}_d + {\tilde \alpha}_d = \tfrac13\, , \quad 
 {\tilde \gamma}_d + 2\, {\tilde \beta}_d{}^2 = \tfrac16 \big (  {\tilde \beta}_d +2\,   {\tilde \alpha}_d \big ) \, , \quad
 2\,  {\tilde \epsilon}_d + {\tilde \beta}_d \, {\tilde \gamma}_d = \tfrac16  \, {\tilde \beta}_d 
 \big ( 2\,  {\tilde \beta}_d + {\tilde \alpha}_d \big ) \, ,\nn \\
&   {\tilde \delta}_d + {\tilde \epsilon}_d + {\tilde \beta}_d{}^3 = \tfrac16  
  \big  (  {\tilde \gamma}_d +  {\tilde \beta}_d{}^2  + {\tilde \alpha}_d{}^2 \big ) \, .
 \end{align}
 Since \eqref{dF4} directly determines ${\tilde \alpha}_d$ the results are then
 \be
 \hskip - 0.8cm
 \text{
 \begin{tabular}{  c   c   c  c  c }
${\tilde \alpha}_d$ & ${\tilde \beta}_d$ & $ {\tilde \gamma}_d$ & ${\tilde \delta}_d$ &$ {\tilde \epsilon}_d$ \\
\noalign {\vskip 3pt}
\hline
\noalign {\vskip 4pt}
$\tfrac{1}{12}({\scriptstyle n_s+2})$ &  $-\tfrac{1}{24}({ \scriptstyle n_s-2})$&  
$-\tfrac{1}{2^5.3^2}({\scriptstyle n_s{\!}^2 -10\, n_s - 16})$& 
 $\tfrac{1}{2^8.3^3}(\scriptstyle{n_s{\!}^3 -3\, n_s{\!}^2 + 80 \, n_s +  100}$)& 
 $-\tfrac{1}{2^9.3^3}{\scriptstyle n_s}({\scriptstyle n_s -2} )({\scriptstyle n_s -10})$
\end{tabular}
} 
 \ee
 For $n_s=5,8,14$ these results are identical to the corresponding results obtained above apart from a change of normalisation. 
 That the only solutions of \eqref{dF4}  are given by \eqref{division} was demonstrated in \cite{birdtracks}  as a 
 consequence of various bounds following  from  \eqref{dF4}.

\subsection{Fixed Points}

 Extending to $4-\vep$ dimensions the interactions become marginally relevant and the $\beta$-functions
 in a $\overline{M\!S}$ scheme take the form
 \be
 {\hat \beta}_{y}{\!}^{a}  = - \tfrac12 \, \vep \, y^a + {\beta}_{y}{\!}^{a} \, , \qquad 
 {\hat \beta}_{\lambda}{\!}^{abcd} = - \vep \, \lambda^{abcd} + {\beta}_{\lambda}{\!}^{abcd} \, .
 \ee
 There are then fixed points which can be analysed in terms of an $\vep$ expansion.
 
 In the restricted theories described previously we consider first the case where 
 there are just two couplings $y,\lambda$. Within the examples discussed here this
 coresponds to requiring $a=0$ as given in \eqref{table3} where the various possiblities
 arise for $n_s=1,2,3,5,8,14$.
 At lowest order from \eqref{oneloop2} and \eqref{oneloop3}, with $\tn_f = \tfrac12 \alpha \, n_f $ from \eqref{oneloop2},
 \begin{align}
 y_*{\!}^2 = {}& \tfrac12 \, Y \, \vep \, , \qquad
\lambda_{*\pm} =  \frac{1}{2(n_s+8) } \big ( 1 -  2 \, Y\, \tn_f  \pm   \sqrt{Z} \, \big ) \vep \, , 
\label{fpone}
 \end{align}
 for
 \be
 Y =  \frac{1 }{n_s\, \alpha + 2\, \beta +  \tn_f }\, , \quad
 Z=  1 + \frac{12}{n_s+2 }\, \big ( n_s ( n_s+10) \alpha + 4\, \beta \big )   \, Y^2\, \tn_f \,   .
 \label{YZr}
 \ee
 For $n_f$ Dirac fermions the results for fixed points remain unchanged except $y^2 \to y\by$ and it is necessary to take
 in \eqref{fpone} and \eqref{YZr} $n_f\to 4 n_f$.
 For $Z=0$ there is a bifurcation point. This requires $\beta$ to be sufficiently negative but
 with just the lower bound in \eqref{btwo} this is impossible and for non zero $n_f$ $Z>1$
 so that $\lambda_{*-}<0$ which leads to an unstable scalar potential.
 For $n_f=0$ the fixed points  are just $\lambda_{*+}= 
 \vep/(n_s+8), \ \lambda_{*-} =0$ which reproduce the fixed points  for the purely scalar $O(n_s)$ theory.
 From \eqref{fpone} and \eqref{YZr} to leading order for large $\tn_f$
 \be
  y_*{\!}^2 \sim \frac{1}{2\, \tn_f} \, \vep \, , \qquad \lambda_{*+} \sim
 6 \, r \,  \frac{1}{\tn_f} \, \vep \, , \qquad \lambda_{*-} \sim
 -  \frac{1}{n_s+8} \, \vep  \, .
  \ee
  Since $\lambda_{*-} <0$ this case is seemingly not relevant.

 The anomalous dimensions at the fixed point to lowest order are then just
 \be
 \gamma_\phi =\tfrac12 \, \tn_f\, Y\, \vep \, , \qquad 
 \gamma_\psi =\tfrac14 \, n_s\, \alpha\, Y\, \vep  \, .
 \ee
 For just two couplings the stability matrix at a fixed point becomes
 \be
 M = \begin{pmatrix} \pr_\lambda {\hat \beta}_\lambda & \pr_\lambda \beta_y  \\  \pr_y \beta_\lambda
  &  \pr_y {\hat \beta}_y \end{pmatrix}\bigg |_{\lambda = \lambda_* , y = y_*} \, .
 \ee
 At lowest order since $\pr_\lambda \beta_y=0$
 the eigenvalues  obtained from \eqref{oneloop} and \eqref{oneloop2} for the fixed points corresponding to
 $\lambda_{*\pm}$ are  then  given by
 \be
 \kappa_{\pm} = \big (1,  \pm \sqrt Z\,  \big )  \vep \, .
 \ee

  The theories described in \eqref{tableS} and \eqref{table} reduce to the case where there
 are just two couplings $\lambda, \, y$ when $n=2,3$. Theories corresponding to $n=2$ are well known.
 For case 2 in \eqref{tableS}  and case 3 in \eqref{table} there is just a single scalar and both correspond
 to the Gross Neveu model, a fermionic generalisation of the Ising model. For case 1 in \eqref{tableS} then $n_s=2$
 when $n=2$ and this is a renormalisable form of Nambu Jona-Lasinio  model which has complex scalars 
 and extends the $XY$ model.
 Case 3 and case 5 are identical for $n=2$, reflecting $SU(2)\simeq Sp(2)$ and this extends the Heisenberg theory
 for $n_s=3$.  Examples corresponding to taking $n=3$ do not seem to have been considered previously.
 The lowest order results taking $\tn_f \to \frac12 N$, are then  given in terms of
 \begin{align}
\label{calarfp}
\text{
\begin{tabular}{  c   c   c   c  c  c} 
results for &  $n $ & $n_s $ & $y_*$ & $\lambda_{*\pm}$  \\
\noalign {\vskip 2pt}
\hline
\noalign {\vskip 4pt}
Antisymmetric & $ 2 $ & $ 1 $  &  $\tfrac{1}{N+6} $ & $ \tfrac{6 -N \pm \sqrt{N^2 + 132 N + 36}}{18(N+6)}$  \\ 
\noalign {\vskip 2pt}
& $ 3 $ & $ 3 $  &  $\tfrac{1}{N+8} $ & $ \tfrac{8 -N \pm \sqrt{N^2 + 160 N + 64}}{22(N+8)}$ \\
\noalign {\vskip 4pt}
Symmetric  &
 $ 2 $ & $ 2 $  &  $\tfrac{1}{N+4} $ & $ \tfrac{4 -N \pm \sqrt{N^2 + 152 N + 16}}{20(N+4)}$ \\
\noalign {\vskip 2pt}
& $ 3 $ & $ 5 $  &  $\tfrac{1}{N+8} $ & $ \tfrac{8 -N \pm \sqrt{N^2 + 192 N + 64}}{26(N+8)}$ \\
\noalign {\vskip 4pt}
Diagonal  & $ 3 $ & $ 2 $  &  $\tfrac{1}{N+8} $ & $ \tfrac{8 -N \pm \sqrt{N^2 + 144 N + 64}}{20(N+8)}$ \\
\noalign {\vskip 4pt}
Unitary  &
 $ 2 $ & $ 3 $  &  $\tfrac{1}{N+2} $ & $ \tfrac{2 -N \pm \sqrt{N^2 + 172 N + 4}}{22(N+2)}$ \\
\noalign {\vskip 2pt}
& $ 3 $ & $ 8 $  &  $\tfrac{1}{N+8} $ & $ \tfrac{8 -N \pm \sqrt{N^2 + 240 N + 64}}{32(N+8)}$ 
\\
\noalign {\vskip 4pt}
Symplectic${}^\prime$  &
 $ 4$ & $ 5 $  &  $\tfrac{1}{N-1} $ & $ \tfrac{-1 -N \pm \sqrt{N^2 + 106 N + 1}}{26(N-1)}$ \\
\noalign {\vskip 2pt}
& $ 6 $ & $ 14 $  &  $\tfrac{1}{N+4} $ & $ \tfrac{4 -N \pm \sqrt{N^2 + 168 N + 16}}{44(N+4)}$ 
\end{tabular} \, .
}
\end{align}
 Of course the $n=2$  results are in accord in special  cases with lowest order results already 
 in the literature \cite{ZinnJustin,Rosenstein,Moshe,Fei2,Zerf1,Zerf2}.\footnote{The results
 in \cite{Zerf2} correspond to those described here by taking $\lambda\to \lambda/4, \, y \to 2y^2, \, 
 n\to \frac12\tn_f$
and $\gamma_\phi/2\to \gamma_\phi, \,  \gamma_\psi/2\to \gamma_\psi, \, \beta_y/4 \to y\beta_y, \, 
4\,\beta_\lambda\to \beta_\lambda$.  The results in \cite{Fei2} also relate to those here by taking
 $g_2/(4\pi)^2 \to 3\lambda, \, g_1/4\pi \to y, \, N\to 2 \tn_f$ while $(4\pi)^2 \beta_{g_2}/3 \to  \beta_\lambda,
 \, 4\pi \beta_{g_1} \to \beta_y$ .}
 Results in \cite{Zerf2}  extend to four loops. 
 There is an extensive literature considering large $n_f$ \cite{Fei2}.   
  For the $XY$ theory and $\tn_f = 1$ then, as is well known~\cite{Thomas,Grover:2013rc,Fei:2016sgs},
 there is a supersymmetric fixed point with, at lowest order
 $  \lambda_{*+} =  y_*{\!}^2 = \frac16 \, \vep $, $\gamma_{\phi *} = \gamma_{\psi *} =  \frac16 \, \vep $,
 $\gamma_{\sigma*} = \vep, \ \gamma_{\rho*} =  \frac23 \, \vep $ and stability matrix eigenvalues $(1,3)\vep$.
  For the Ising case and $\tn_f = \tfrac12$ 
 there is an apparent $\N=1$  supersymmetric fixed point \cite{Thomas,Grover:2013rc,Fei:2016sgs,Gies} 
 with
 $  \lambda_{*+} =  y_*{\!}^2 = \frac17 \, \vep $, $\gamma_{\phi *} = \gamma_{\psi *} =  \frac{1}{14} \, \vep $,
 $\gamma_{\sigma*} = \tfrac47\, \vep  $ and stability matrix eigenvalues $(1,\frac{13}{7})\vep$.

  For three non zero couplings results are more involved. At lowest order from \eqref{oneloop2}
  fixed points  are determined by solving
  \begin{align}
0  ={}& -\tfrac12 \,  \vep \, y + ( n_s \, \alpha + 2\,  \beta +   \tn_f ) y^3 \, , \nn \\
0 ={}&  \big ( {- \vep}  + 4 \, \tn_f \,y^2 \big )  \lambda  + (n_s + 8 ) \lambda^2 
+ \tfrac23 ( 2 \, n_s \, \alpha + \beta ) \lambda \, g  + 3\, A\,  g^2 \, , \nn \\
0  ={}& \big ( {-\vep} +  4\, \tn_f \, y^2 \big )  g +   12 \, \lambda \, g + 3 \, B \, g^2 - 24\,  \tn_f  \,  y^4 \, , 
\label{FP1}
\end{align}
 for the various choices of $\alpha,\,\beta$ and $n_s$.
  The example of hermitian $y^a$ was considered in \cite{Ji} though our results differ in one term in $\beta_\lambda$. 
  A large $N$ analysis for Yukawa couplings given in terms of Lie algebra generators was discussed in 
  \cite{GraceyGN}.  For the purely scalar theories obtained 
  when $n_f = 0$ there are no fixed points with both $\lambda, \, g$ non zero for the theories discussed here 
  for allowed $n$ except in case 4 when $n=4$ and in case 2 for arbitrary $n$. This latter case corresponds to the hypertetrahedral theory discussed in \cite{ZiaW} and more recently in \cite{Seeking}. Besides the Gaussian 
  fixed point with vanishing couplings there is of course always
  the Heisenberg fixed point with, at lowest order,
  \be
  \lambda_{*H} = \frac{1}{n_s+8}\, \vep \, , \quad g_{H*}=0 \, , \quad
  \big ( \kappa_{1H},  \kappa_{2H} \big ) = \Big (1, \, \tfrac{4-n_s}{n_s+8} \Big ) \vep \, ,
  \ee
  which clearly becomes unstable, in this approximation when $n_s>4$.

  For a non zero Yukawa coupling it is trivial to solve \eqref{FP1} to  determine $y_*{\!}^2$ to
  ${\rm O}(\vep)$. Furthermore $\lambda$ appears only linearly in the $\beta_g$ equation
  so  that the fixed point equations reduce to just finding the roots of a quartic polynomial  $f(g)$.
  In consequence there are  generically four possible roots though of course these may be complex.   
 Once $y_*{\!}^2$  is eliminated the equations \eqref{FP1} have a crucial symmetry under
 \be
 \tn_f \to \frac{(n_s \, \alpha + 2 \, \beta)^2}{\tn_f} \, , \qquad \vep \to - \vep \, .
 \label{lmf}
 \ee
 This relates solutions for large and small $\tn_f$. Under this transformation it is easy to verify from 
 \eqref{fpone} and \eqref{YZr} that, since $Y \to \tn_f Y / (n_s \, \alpha + 2\beta )$,
 $\lambda_{*+}\leftrightarrow \lambda_{*-}$.
 
 When $\tn_f$ is small the Yukawa interaction is weakly relevant. The Gaussian fixed point is perturbed
 to give
 \be 
 g_* \sim - \frac{6}{(n_s \, \alpha + 2 \beta)^2} \, \tn_f \, \vep
 +   \frac{108\, B}{(n_s \, \alpha + 2 \beta)^4} \, \tn_f{\!}^2 \, \vep \, , \qquad
 \lambda_* \sim \frac{108 \, A }{(n_s \, \alpha + 2 \beta)^4} \, \tn_f {\!}^2 \, \vep \, ,
 \label{Gfp}
 \ee
 and starting from the $O(n_s)$ symmetric fixed point
  \be 
 g_* \sim  \frac{6(n_s+8)}{(n_s-4)(n_s \, \alpha + 2 \beta)^2} \, \tn_f \, \vep 
 \, , \quad
 \lambda_* \sim \frac{1}{n_s+8} \, \vep + \frac{6 \big ( n_s(n_s+12)\alpha + 8 \beta\big )  }
 {(n_s-4)(n_s+8)(n_s \, \alpha + 2 \beta)^4} \, \tn_f  \, \vep \, .
 \label{Ofp}
 \ee
 If there are additional fixed point solutions for $\tn_f=0$, as in the special cases described above, these disappear
 for very tiny non zero $\tn_f$.

 When $\tn_f$ is large a corresponding pattern related by \eqref{lmf} emerges.
  For $\lambda, \, g$ both of    ${\rm O}(1)$ as $n_f\to \infty$, then since $4\, \tn_f \, y_*{\!}^2 \to 2 \vep$,
  it is easy to see that the fixed point equations are, to leading order, of
  the same form as the purely scalar theory but with $\vep\to - \vep$. For $\vep>0$ this
  leads in general to a scalar potential which is not bounded below and in any event there are no
  solutions except in the cases described above.     For large  $n_f$ there are  solutions with both 
  $g, \, \lambda$ small which have the form
  \be
  y_*{\!}^2 \sim \frac{1}{2\, \tn_f} \, \vep \, , \qquad  g_{*} \sim
 6 \,   \frac{1}{\tn_f} \, \vep  - 108 \, B \, \frac{1}{\tn_f{\!}^2 } \, \vep  \, , \qquad
\lambda_{*} \sim - 108 \, A \, \frac{1}{\tn_f{\!}^2 } \, \vep  \, ,
\label{ygsol}
  \ee
where the scalar  potential is bounded below,  and also for $\lambda = {\rm O}(1)$
  \be
   g_{*} \sim
 6 \,\frac{n_s+8}{n_s-4}    \frac{1}{\tn_f} \, \vep  \, , \qquad
\lambda_{*} \sim  - \frac{1}{n_s+8}  \, \vep  \, .
\label{ygsol2}
 \ee
 The second solution  leads to instabilities so only \eqref{ygsol} remains as a valid possibility. 
 
 For intermediate  $\tn_f$ the possible fixed points depend on $n$.
  For lowish $n\lesssim 7$ the number of solutions drops to zero as $\tn_f$ increases and
 then goes  back to two (for case 2 this happens if $n<5$ and for case 6  if $n\le 14)$. 
 For higher $n$ the number of solutions jumps from 2 to 4  with
 increasing $\tn_f$ and then reverts to two which match on to \eqref{ygsol} and \eqref{ygsol2} for large
 $\tn_f$ (for case 2 and case 4 if $n=4$ there are four solutions for very large $\tn_f$ and very tiny $\tn_f$ 
 as the purely scalar theories have fixed point in these cases). The critical $n$ dividing the two cases 
 is determined by $d(n_c)=0$ for $d(n) = \tfrac49(2 n_s \alpha  + \beta)^2 - 12 (n_s+8)A$. For $d(n)<0$ as happens 
 for $n < n_c$ then $ h(\lambda,g)=(n_s+ 8)\lambda^2 + \tfrac23 (2 n_s \alpha  + \beta) \lambda g+ 3A \, g^2 > 0$. For
 $4\,  \tn_f \, y_*{\!}^2 = \vep$, or $n_f = n_s \alpha + 2 \beta$,  the lowest order fixed point equations require 
 $h(\lambda_*, g_*) = 0$. This ensures that there can be no fixed point solutions for  a finite region
 $n \lesssim n_c$. Conversely for $n \gtrsim n_c$ $h(\lambda,g)$ is no longer positive definite and there are solutions
 with $g$ non zero. 
 In consequence  for $n\approx n_c$  and $\tn_f \approx n_{f,c} =  (  n_s \alpha + 2 \beta )|_{n=n_c}$
 there are either  $0$ or $4$ solutions  
  For the  symmetric case $n_c =6$ and $\tn_{f,c}=8$.\footnote{For $n=6, \, \tn_f = 8$ the 
 fixed point solutions become $y_*{\!}^2 = \frac{1}{32}\vep, \ 
 \lambda_* =\pm  \frac{1}{24}\vep, \ g_* = \mp \frac14 \vep$. In neither case are the conditions for
 a stable potential satisfied.} For the other cases the results for $(n_c, \, n_{f,c})$ are then
 2. $(4.37,2.31)$, 3. $(6.58,12.24)$, 4. $(7.37, 8.37)$, 5. $(6.37,5.37)$, 6. $(14.11, 10.69)$.
 The jumps are associated with bifurcation points which correspond to 
 there being two coincident roots of the polynomial  $f(g)$, or that its discriminant vanishes. The boundaries
 of the regions where there are jumps from 2 to 0 or 2 to 4 correspond to $\tn_f$ linked by \eqref{lmf} though
 the fixed point couplings have opposite signs.
 At the fixed points in general the couplings do not give potentials which are bounded below except for
 one which matches \eqref{ygsol} when $\tn_f$ is large. The positivity condition  remains satisfied
 as $\tn_f$ is reduced until just above the upper bifurcation point.
 
 Diagrams showing the structure of fixed point solutions outlined above are presented in appendix \ref{fpfigures}.
 
 For the stability matrix eigenvalues the absence of $g,\lambda$ contributions  to the 
 Yukawa $\beta$-function at lowest order ensures that one eigenvalue is $\vep$ for any $n,\tn_f$
 and the remaining eigenvalues are obtained from
 \be
 M = \begin{pmatrix} \pr_\lambda {\hat \beta}_\lambda & \pr_g \beta_\lambda \\  \pr_\lambda \beta_g
  &  \pr_g {\hat \beta}_g \end{pmatrix}\bigg |_{\lambda = \lambda_* ,g=g_*, y = y_*} \, .
 \ee
 For $\tn_f$ small from \eqref{Gfp}
 \be
 \kappa_\pm \sim -\vep  -\Big ( 9B+ n_s \alpha - \beta \pm \sqrt{ (9B -2\, n_s \alpha-\beta)^2 + 648 A}\, \Big )
 \frac {2 \, \tn_f }{(n_s \, \alpha + 2\, \beta )^2 } \, 
 \, \vep \, .
  \ee
  Starting from \eqref{Ofp} the eigenvalues are
  \be
  \kappa_1 \sim - \frac{n_s-4}{n_s+8} \, \vep + {\rm O}(\tn_f ) \, , \qquad
  \kappa_2 \sim  \vep + \frac{6\big (n_s(n_s+10)\alpha+ 4\beta \big )} {(n_s+2) ( n_s \alpha + 2\,\beta)^2}\, 
  \tn_f \, \vep \, .
  \ee
 Otherwise for large $\tn_f$ from the fixed point \eqref{ygsol}
 \be
 \kappa_\pm \sim \vep  + \Big ( 9B+ n_s \alpha - \beta \pm \sqrt{ (9B -2 \, n_s \alpha-\beta)^2 + 648 A} \, \Big )
 \frac{2}{\tn_f} \, \vep \, .
  \ee
 
 For the supersymmetric case using \eqref{susyb} there are possible fixed points
 in the $\vep$ expansion if $ 4 \beta_d + 3 \alpha_d >0$. From the lower bound on
 $\beta_d$ this is satisfied whenever $n_s>2$.
 
 \subsection{$U(1)$ Symmetry}
 
 \label{subU1}
 
 A similar reduction is possible for complex scalar fields where there is at least an overall 
 $U(1)$ symmetry.  In this case we  consider chiral fermions $\psi$ 
 and $\chi$ of opposite chirality which need not be equal in number.
 Writing 
 \be 
 y^i = y\,{\mathds 1}_m  t^i \, , \qquad \by_i = \by \, {\mathds 1}_m\bt_ i \, , \qquad \bt_i = (t^i )^\dagger \, ,
 \ee
 then $t^i$, $i=1,\dots, n$,  need not be a square matrix but is assumed to be $r\times s$. 
 In this case we assume
 \be
 U_-{\!\!}^{-1} t^i \, U_+  = R^i{}_j\,  t^j \, , \quad U_- \in U(r) \, , \ U_+ \in U(s) \, , \quad [  R^i{}_j] \in  H \subset U(n) \, .
 \ee

 Corresponding to the previous discussion we assume, with a choice of normalisation for $t^i$,
 \be
 t^i \,\bt_i = s\,  {\mathds 1}_r \, , \qquad  \bt_i \, t^i =  r \, {\mathds 1}_s \, , \qquad
 \tr( t_i \, \bt^j )  =  rs/n \, \delta_i{}^j \, .
 \label{rest}
 \ee
 Defining 
 \be
 \tr\big  ( \bt_{(i}\, t^{k} \, \bt_{j)}\,  t^{l} \big ) =  h_{ij}{}^{kl} \, , \qquad
 h_{ik}{}^{jk} = \tfrac12 \, rs (r+s)/n \, \delta_i{}^j\, , 
 \ee
 the scalar coupling is assumed to have the form\footnote{The necessary conditions for a positive
 potential can be obtained from 
 $$ \big ( \tr ( \bt\, t ) \big ) ^2 \big /\min(r,s)  \le \tr ( \bt\, t \, \bt\, t ) \le  \big ( \tr ( \bt\,  t ) \big ) ^2 \, , $$
 where $t$ is a $r\times s$ complex matrix and $\bt$ its hermitian conjugate.}
 \be
 \lambda_{ij}{}^{kl} = \lambda \big ( \delta_i{}^k \delta_j{}^l + \delta_i{}^l \delta_j{}^k \big ) + g \, h_{ij}{}^{kl} \, .
 \label{lU1}
 \ee
  
 As above we assume conditions are imposed such that $\beta$-functions determining the RG
 flow are reduced to $\beta_y, \, 
 \beta_\lambda, \ \beta_g$ with $\beta_{\bar y} = (\beta_y)^*$ and the anomalous dimension matrices
 become $\gamma_\vphi \, {\mathds 1}_n,  \gamma_\psi  {\mathds 1}_r, \,
 \gamma_\chi {\mathds 1}_s$ with $\gamma_\vphi = \gamma_{\bar \vphi}$. Quadratic operators
 are decomposed as
 \be
 \vphi_i {\bar \vphi}^j = \sigma \, \delta_i{}^j + \rho_i{}^j \, , \ \ \rho_i{}^i = 0 \, , \quad
 \vphi_i \vphi_j = (\vphi^2)_{ij} \, ,
 \ee 
 and the corresponding anomalous dimension matrices are then of the general form
 \begin{align}
& \gamma_\sigma \, \delta_i{}^j \, , \qquad    \gamma_{\rho \hskip 1.5pt i}{}^j{}_k{}^l
 =  \gamma_{\rho}\,  
 \big (  \delta_i {}^l \delta_k{}^j - \tfrac1n \delta_i{}^j \delta_k{}^l \big ) + 
  \gamma^\prime{\!\!}_{\rho} \, d_{ik}{}^{jl} +  \gamma^{\prime\prime}{\!\!}_{\rho} \, w_{ik}{}^{jl} \, , \nn \\
&   \gamma_{\vphi^2} {\hskip 0.5pt}_{ij}{}^{kl} 
 =  \gamma_{\vphi^2} \tfrac12 \big ( \delta_i{}^k \delta_j{}^l + \delta_i{}^l \delta_j{}^k \big ) + 
  \gamma^\prime{\!\!}_{\vphi^2} \, d_{ij}{}^{kl}  \, , 
  \label{phi2}
  \end{align}
  where
  \begin{align}
  h_{ij}{}^{kl} = {}& q  \big ( \delta_i{}^k \delta_j{}^l + \delta_i{}^l \delta_j{}^k \big ) + d_{ij}{}^{kl} \, , 
  \qquad (n+1) q = \tfrac12 \, rs (r+s)/n\, ,  \nn \\
   \tr\big  ( \bt\raisebox{ -1.5pt}{$\scriptstyle [ i $}\, t^{k} \, 
   \bt\raisebox{ -1.5pt}{$\scriptstyle j ]$}\,  t^{l} \big ) =  {}&  p  \big ( \delta_i{}^k \delta_j{}^l - \delta_i{}^l \delta_j{}^k \big )
    + w_{ij}{}^{kl} \, ,  \qquad (n-1) p = \tfrac12 \, rs(r-s)/n\, ,
    \label{defdw}
  \end{align}
  so that both $d_{ij}{}^{kl} $ and $w_{ij}{}^{kl}$ are zero on contraction of up and down indices. 
  Instead of \eqref{lU1} 
  \be
 \lambda_{ij}{}^{kl} = {\hat \lambda} \big ( \delta_i{}^k \delta_j{}^l + \delta_i{}^l \delta_j{}^k \big ) + g \, d_{ij}{}^{kl} \, ,
 \qquad \hat \lambda = \lambda + q\, g \, .
 \label{lU2}
 \ee
  The eigenvalues and corresponding degeneracies for the $d$ and $w$ tensors as they appear in \eqref{phi2}
  are given in appendix \ref{appdw}.
 
 At one loop it is sufficient to impose the conditions
 \begin{align}
 h_{ij}{}^{mn} h_{mn}{}^{kl} ={}& \tA  \big ( \delta_i{}^k \delta_j{}^l + \delta_i{}^l \delta_j{}^k \big ) + 
 \tB  \, h_{ij}{}^{kl} \, , \nn \\
h_{m(i}{}^{n(k} h_{j)n}{}^{l)m} ={}& \tA^\prime \big ( \delta_i{}^k \delta_j{}^l + \delta_i{}^l \delta_j{}^k \big ) + 
 \tB^\prime \, h_{ij}{}^{kl} \, .
 \label{hhAB}
 \end{align}
 Hence
 \begin{align}
 \beta_y{\!}^{(1)} = {}& \tfrac12  \big ( r+s +m\, {rs}/{n} \big ) y^2\by \, , \nn \\
 \beta_\lambda{\!}^{(1)} ={}&  2(n+4) \lambda^2 + 4(n+1) q \, \lambda\,  g + ( \tA+ 4 \tA^\prime) g^2
 + 2 \, m\, rs/n \, \lambda \, y \by \, , \nn \\
 \beta_g {\!}^{(1)} ={} & 12 \, \lambda \, g  + ( \tB+ 4 \tB^\prime) g^2
 + 2 \, m\, rs/n \, g \, y \by - 4\,m\, (y\by)^2 \, . 
 \end{align}
 A necessary condition is 
 \be
 h_{ik}{}^ {mn}h_{mn}{}^{kj} =  (n+1) \big ( \tA + q \, \tB \big )\delta_i{}^j
 = \big ( 2 (n+1)( \tA^\prime + q \, \tB^\prime ) -  (n+1)^2 q^2 \big ) \delta_i{}^j\, .
 \ee
 Defining 
 \be
 {\tilde a} = \tA + q \, \tB - 2 q^2 \, , \qquad d_{ik}{}^{mn} d_{mn}{}^{kj} = (n+1) {\tilde a} \, \delta_i{}^j \, ,
 \label{defat}
 \ee
 then the one loop scalar $\beta$-function can be alternatively expressed as
 \be
 \beta\raisebox{ -1.5pt}{$\scriptstyle{\hat \lambda}$}{}^{(1)} 
 = 2(n+4) \, {\hat \lambda}{}^2 + 3\, {\tilde a}\, g^2 +  m\, rs/n \, {\hat \lambda} \, y \by 
 - 4  \, m \,q \, (y\by)^2 \, .
 \ee 
 For $ {\tilde a} = 0$ $d_{ij}{}^{kl}=0$  and $h_{ij}{}^{kl} =q  ( \delta_i{}^k \delta_j{}^l + \delta_i{}^l \delta_j{}^k)$ 
 so is no longer independent and the $g$ coupling is redundant.

 This framework encompasses a variety of theories discussed in the literature. As an illustration
 we may consider $\{ t^a\}$ to be a basis for $r\times s$ complex matrices where $n=rs$ and which
 satisfy \eqref{rest}. The scalar field symmetry group is $H=U(r) \times U(s)/U(1)$. 
 Positivity of the potential holds if
 \be
 \lambda > 0 \, , \quad 2 \, \lambda + g > 0  \quad \mbox{or} \quad g>0 \, , \quad 2\, \lambda + g/ \min(r,s) >0 \, .
 \ee
 This example requires \eqref{hhAB}
 \be
\tA = \tfrac12\, , \qquad \tB = 0 \, , \qquad
 \tA^\prime = \tfrac14\, , \qquad \tB^\prime  = \tfrac14 (r+s)\, .
 \ee
  In this case from \eqref{defat} and \eqref{defdw}
 \be
 {\tilde a} = \tfrac{(r^2-1)(s^2-1)}{2 (n+1)^2 } \, , \qquad    q = \tfrac{r+s}{2(n+1)} \, .
 \label{defaq}
 \ee
 For the purely scalar theory, without fermions, results for $\beta$-functions have been obtained to 
 five  loops in \cite{Calabrese} and more recently to six loops in 
 \cite{Kompaniets,Bednyakov} and a bootstrap analysis has been undertaken in \cite{StergiouU}.
 
 The lowest order $\beta$-functions and anomalous dimensions are then, with $n=rs$, 
 \begin{align}
 \beta_ y{\!}^{(1)} ={}&  \big ( \tfrac12 ( r+s+m) y \by \big )\, y\, , \quad 
 \beta_\lambda{\!}^{(1)}  = 2( n + 4 )\lambda^2 + 2(r+s) \lambda \, g
 + \tfrac32 \, g^2 + 2m \, \lambda \, y\by \, , \nn \\
 \beta_g{\!}^{(1)}  = {}& 12\, g\, \lambda + (r+s) \, g^2  + 2m \, g\, y\by - 4m\,  (y\by)^2\, , \nn \\
 \gamma_\vphi{\!}^{(1)} ={}&  \tfrac12 m \, y\by \, , \qquad  \gamma_\psi{\!}^{(1)} = \tfrac12r \,  y\by \, , \qquad 
 \gamma_\chi{\!}^{(1)} = \tfrac12 s\, y\by \, ,
 \end{align} 
 and in \eqref{phi2}
 \begin{align}
 & \gamma_\sigma{\!}^{(1)} =  2(n+1)\,  {\hat \lambda} +m\,  y \by \, , \qquad 
  \gamma_\rho{\!}^{(1)} = \gamma_{\vphi^2}{}^{(1)} =  2\, {\hat \lambda} +m\,  y \by \, , \nn \\
&  \gamma^{\prime}{\!\!}_{\vphi^2}   {}^{(1)}  =  \gamma^{\prime}{\!\!}_{\rho}  {}^{(1)} = 2\, g\, , \qquad
  \gamma^{\prime\prime}{\!\!}_{\rho}  {}^{(1)} = 0  \, .
 \end{align} 
 Setting $s=1$, $r=n$,  so that $\tilde a=0$  and the $d$-tensor vanishes, the $\beta$-function reduces to
 \be
  \beta\raisebox{ -1.5pt}{$\scriptstyle{\hat \lambda}$}{}^{(1)} 
 = 2(n+4) \, {\hat \lambda}{}^2 + 2m   \, {\hat \lambda} \, y \by  - 2 m \, (y\by)^2 \, .
 \ee
 For $n=1$ this coincides  with the the Nambu Jona-Lasinio  extended $XY$ model, 
 with $U(1)$ symmetry when $m=1$, so long as $y\by\to2\,  y^2$.
 
 At higher orders further relations corresponding to primitive diagrams are necessary. The 
  primitive Yukawa diagrams $2a, \ 2f$ and scalar diagram $2g$ correspond to
 \begin{align}
  t^j  \bt_k\hskip 0.5pt  t^l \, d_{jl}{}^{ki} = (n+1) {\tilde a} \, t^i \, , \quad
 t^j \bt_k \hskip 0.5pt t^i \bt_j \hskip 0.5pt t^k = t^i \, ,
 \quad \tr\big  (  \bt_{(i} \hskip 0.7pt  t^m \bt_{j)}\hskip 0.7pt t^{(k} \bt_m \hskip 0.7pt  t^{l)}) =  \tfrac12
 \big ( \delta_i{}^k \delta_j{}^l +  \delta_i{}^l \delta_j{}^k \big )\, . 
 \end{align}
 The two loop 1PI contributions to the Yukawa and scalar $\beta$-functions are, with 
 ${\tilde a}, \, q$ as in \eqref{defaq},   then
 \begin{align}
 {\tilde \beta}_y{\!}^{(2)} = {}& \big ({ - 2(r+s)} \, {\hat \lambda} \, y\by  - 2(n+1)\, {\tilde a} \, g \, y\by 
 +  2\, (y \by)^2\big ) \, y  \, , \nn \\
  \beta\raisebox{ -1.5pt}{$\scriptstyle{\hat \lambda}$}{}^{(2)} 
  = {}& - 4 (5n+11) {\hat \lambda}^3 - 6 (n+7)\,{\tilde a} \, g^2 {\hat \lambda}
 - 4(n-5) q \,{\tilde a} \, g^3  - 2 (n+4) m \,  {\hat \lambda}^2 \, y\by 
 - 3\, {\tilde a}  \,m \,  g^2 \, y\by  \nn \\
 \noalign{\vskip - 2pt}
 &{}+ 8\, q \, m \, {\hat \lambda}\, (y\by)^2 +  4\, {\tilde a}\,  m \,  g\,  (y\by)^2 + 4\big ( (r+s)q + 1 \big)  m \,  (y\by)^3 
 \, , \nn \\
 {\tilde \beta}_g{\!}^{(2)} = {}& - 12(n+7)\, g \, {\hat \lambda}^2  -  2(n-5)q\,  g^2 
 \big ( 12\,  {\hat \lambda} + m\,  y\by \big ) 
 -  2\big ( n+4 -18(n-1)q^2  \big ) g^3 \nn \\
 \noalign{\vskip - 2pt}
 &{} -12\,m  \, g\, {\hat \lambda}\, y\by + 8 \,m \, {\hat \lambda}\, (y\by)^2  - 8\, q\, m \, g \,(y\by)^2 
 + 4 (r+s)m \,(y\by)^3  \, .
 \label{resbeta2}
 \end{align}
 At two loops the anomalous dimensions are given by
 \begin{align}
  \gamma_\vphi{\!}^{(2)} ={}&  \tfrac12 (n+1) \big (  {\hat \lambda}^2 + \tfrac12 \, {\tilde a} \, g^2 \big ) 
  -\tfrac38 \, m \, (r+s) \,  (y \by)^2 \, , \nn \\
   \gamma_\psi{\!}^{(2)} ={}& - \tfrac18 \, r ( s+3m) \, (y \by)^2 \, , \qquad
    \gamma_\chi{\!}^{(2)} = - \tfrac18 \, s ( r+3m) \, (y \by)^2 \, ,  \nn \\
  \gamma_\sigma{\!}^{(2)} = {}&  - 6 (n+1) \big (  {\hat \lambda}^2 + \tfrac12 \, {\tilde a} \, g^2 \big ) 
  - 2(n+1) \, m \, {\hat \lambda} \,  y \by  + 2\, \gamma_\vphi{\!}^{(2)}  \, ,  \nn \\
   \gamma_{\vphi^2}{\!}^{(2)} = {}&
   - 2 (n+3) \,   {\hat \lambda}^2 - 2\, {\tilde a} \, g^2  - 2 \, m \, {\hat \lambda} \,  y \by
  + 4\,q\, m \, (y \by)^2 + 2\, \gamma_\vphi{\!}^{(2)}\, , \nn \\
   \gamma^{\prime}{\!\!}_{\vphi^2} {\!}^{(2)}  = {}& 
   - 8 \,   {\hat \lambda}\, g  - 2(n-3)q \, g^2 
  - 2 \, m \, g \,  y \by + 4\, m \,( y \by )^2\, , \nn \\
   \gamma_\rho{\!}^{(2)} = {}& - 2 (n+3) \,   {\hat \lambda}^2 - \tfrac{n-3}{n-1} \, {\tilde a} \, g^2 
  - 2 \, m \, {\hat \lambda} \,  y \by + 2\, \gamma_\vphi{\!}^{(2)}  \, , \nn \\
  \gamma^{\prime}{\!\!}_{\rho}  {}^{(2)} ={}& - 8 \,   {\hat \lambda}\, g  - (n-7)q \, g^2 
  - 2 \, m \, g \,  y \by  \, , \qquad  \gamma^{\prime\prime}{\!\!}_{\rho}  {}^{(2)} = -\tfrac12 \, (r-s) \, g^2 \, .
 \end{align}

 At three loops further relations are necessary. Corresponding to $3{\tilde z}$
 \be
  t^j \bt_k\hskip 0.5pt  t^l \bt_j \hskip 0.5pt t^i \bt_l\hskip 0.5pt  t^k + 
  t^k \bt_l \hskip 0.5pt  t^i \bt_j \hskip 0.5pt t^l \bt_k \hskip 0.5pt  t^j = (r+s)\,  t^i   \, .
 \ee
 For just the quartic scalar coupling at three loops then corresponding to $3f$ there is the relation
 \begin{align}
 &d_{m(i}{}^ {nr}\, d_{j)r}{}^{pq} \, d_{ps}{}^{m(k} \, d_{nq}{}^{l)s} = {\cal A} 
  ( \delta_i{}^k \delta_j{}^l + \delta_i{}^l \delta_j{}^k) +{\tilde c} \, d_{ij}{}^{kl} \, ,\nn \\
 &  \A = \tfrac18 \big ( n+4  - 18\,  q^2 \big) \,{\tilde a} \, , \qquad  
 {\tilde c} = - \tfrac18  \, ( 5\, n+29) \, q + 2 ( 5\,n-1) \, q^3 \, .
 \end{align}
 
 For the purely scalar theory obtained by setting the Yukawa couplings to zero there are non trivial fixed points
 which to lowest order have the form, for $r,s>1$,
 \be
 g_{*\pm} = \frac{n-2}{D_{rs} \pm 3 \sqrt{R_{rs}}}\, \vep\, , \qquad 
 {\hat \lambda}_{*\pm} = \frac{D_{rs} \pm (n+1)\sqrt{R_{rs}}}{4(n+1)(D_{rs} \pm 3 \sqrt{R_{rs}})}\, \vep\, , 
 \label{glfp}
 \ee
for
\be 
D_{rs} = (n-5) (r+s) \, , \qquad R_{rs} = r^2 + s^2 - 10 \, r s + 24 \, .
\ee 
Since $R_{rs} = ( r- n_+(s))(r-n_-(s))$ for  $n_\pm(s) = 5s \pm 2 \sqrt 6 \sqrt{s^2-1}$ 
then $R_{rs}\ge0$ and there are real fixed points if $r \ge n_+(s)$ or $r\le n_-(s)$. For $r>s$
only the first case is relevant.
The corresponding stability matrix eigenvalues  at the fixed points in \eqref{glfp} are then
\be
\big (\kappa_{1\pm}, \, \kappa_{2\pm} \big )=
\Big (  1, \pm \tfrac{ (n-2) \sqrt{R_{rs}}}{D_{rs} \pm 3 \sqrt{R_{rs}}} \Big ) \vep\, .
\label{kapfp}
\ee
Integer solutions for the bifurcation points when $R_{rs}=0$ can be obtained iteratively, for $r>s$, 
by taking $r_i = 10 \, r_{i-1} - s_{i-1} , \ s_i = r_{i-1}$ starting from $r_{1} = 5 ,  \ s_1=1$.
This scalar theory with $H= U(r)\times U(s)/U(1)$ symmetry is an obvious extension of the bifundamental
theory with real scalars and $O(m)\times O(n)/{\mathbb Z}_2$ symmetry discussed recently in
\cite{Seeking,RychkovS} which contains earlier citations. Defining the invariants
\be
|| \lambda ||^2 = \lambda_{ij}{}^{kl} \lambda_{kl}{}^{ij} = 2 n(n+1) \big ( {\hat \lambda}^2 + 
\tfrac12\, {\tilde a} \, g^2 \big ) \, , \qquad
| \lambda_{*\pm} |  =  \lambda_{ij}{}^{ij} = n(n+1) \, {\hat \lambda} \, , 
\ee
then at the fixed point \eqref{glfp}
\be
|| \lambda_{*\pm} ||^2 = \tfrac{1}{24}\, n  \big ( \vep^2  - \kappa_{2\pm}{\!}^2 \big ) \, ,   \qquad
| \lambda_{*\pm} |  = \tfrac{1}{4}\, n  \big ( \vep  + \kappa_{2\pm} \big ) \, .
\label{Kgl}
\ee
Clearly $|| \lambda_{*\pm} ||^2 \le  \tfrac{1}{24}\, n  \, \vep^2 $
in accord, of course, with the bound obtained by Hogervorst and Toldo \cite{Hogervorst} extending
the results in \cite{RychkovS}.  For any $n$ there is also the Heisenberg fixed point with $O(2n)$ symmetry
where
\be
{\hat \lambda}_{*H} = \frac{1}{2(n+4)} \, \vep \, \qquad g_{*H}=0 \, , 
\ee
and 
\be 
\big (\kappa_{1H},\kappa_{2H} \big ) 
= \Big ( 1 , \tfrac{2-n}{n+4} \Big ) \vep \, , \quad || \lambda_{*H} ||^2 = 
 \tfrac{n(n+1)}{2(n+4)^2} \, \vep^2\, , \quad 
 | \lambda_{*H} | = \tfrac{n(n+1)}{2(n+4)} \, \vep \, . 
 \label{Ogl}
\ee
For $n=2$ these results coincide with \eqref{glfp}, \eqref{kapfp} and \eqref{Kgl}.

For non zero Yukawa couplings the  one loop $\beta$ function requires  
\be
(y\by)_* = \vep/ (r+s+m) \, .
\label{yfp}
\ee
After using \eqref{yfp} the equations determining $\lambda_*, \, g_*$ are then invariant under
\be
m \to \frac{(r+s)^2}{m} \, , \qquad \vep \to -\vep \, ,
\ee
relating results for large and small $m$.  At $m=m_c$ for $m_c = r+s$ there are no solutions
if $2(n+4) \lambda^2 + 2(r+s) \lambda g + \frac 32 g^2 >0$ or
\be
2n \le r^2 + s^2 < n + 12 \, .
\label{rsb}
\ee
This is rather restrictive. Taking $r,s\ge 2$ and $r\le s$ 
the only possibilities for $(r,s)$ are just $(2,2), \, (2,3),\, (3,3), \, (2,4)$ (in the last case $r^2+s^2=n+12$). 
In these cases where $R_{rs}<0$ there are two fixed
points for small and large $m$ but none over some interval centred on $r+s$, the interval shrinks to
zero in the $(2,4)$ case where the bound in \eqref{rsb} is saturated.
Otherwise for $R_{rs}<0$ there are again two fixed points for small and large $m$ but four over
a region centred on $r+s$. If $R_{rs} \ge 0$ there are four fixed points for any $m$. 

At large $m$ there is a fixed point which is the counterpart of the Gaussian fixed point for small $m$
and gives rise to a positive semi-definite potential and positive stability matrix eigenvalues
\be
g_* \sim \frac{4}{m} \, \vep - (r+s) \frac{16}{m^2} \, \vep \, , \quad \lambda_* \sim - \frac{24}{m^2}\, \vep  \, , \quad 
\kappa_\pm \sim \vep + \big ( r+s \pm 4 \big ) \frac{6}{m} \, \vep \, .
\ee

For  $s=1$ and $r=n$ there is just one scalar coupling and at lowest order
\be
{\hat \lambda}_{*\pm} = \frac{1}{4(n+4)}\Big ( - \frac{m-n-1}{m+n+1} \pm \sqrt{\tilde Z} \Big )\vep \, , \quad
{\tilde Z} = \frac{(m-n-1)^2+ 16 (n+4) m}{(m+n+1)^2} \, .
\ee
For $n=1$ this is identical to the $XY$ case as given by \eqref{fpone} and \eqref{YZr}.

\section{Consistency Relations}

\label{sec:consis}

The existence of an $a$-function requires consistency relations between the coefficients for individual 
non primitive graphs appearing in the  $\beta$-functions and the anomalous dimensions. The basic equation
has the form, for couplings $\{g^I\}$~\cite{Jack:2013sha},
\be
\pr_I A = T_{IJ} \, B^J \, ,
\label{ATB}
\ee
where $A$ is constructed from 1PI and 1VI vacuum diagrams, $T_{IJ}$ also from 1PI, 1VI vacuum diagrams
with two vertices identified and
\be
B^I = \beta^I - ( \upsilon g)^I \, ,
\label{bup}
\ee
with $\upsilon(g)$ corresponding to an element of the Lie algebra of the maximal symmetry group of
the Lagrangian kinetic term. In general for a vanishing trace of the energy momentum tensor and hence
conformal symmetry the requirement is that $B^I=0$. In \eqref{ATB} $T_{IJ}$ need not be symmetric, 
although any antisymmetric part has further constraints. 
In \eqref{bup}  $\upsilon(g)$ is necessarily present starting at 3 loops.
There is a freedom in \eqref{ATB} where
\be
A \sim A + g_{IJ}B^I B^J \, , \qquad T_{IJ} \sim T_{IJ} + {\mathcal L}_B \, g_{IJ} + \pr_I (g_{JK} B^K)- 
 \pr_J (g_{IK} B^K) \, ,
 \ee
 for arbitrary symmetric $g_{IJ}$. This does not  preserve the symmetry of $T_{IJ}$.

In the present context the lowest order contribution to $T_{IJ}$ is first present at two loops for the Yukawa
couplings, $T_{yy}$,  and at three for the scalar quartic couplings, $T_{\lambda\lambda}$. 
In consequence \eqref{ATB} provides potential relations between the Yukawa $\beta$-functions and fermion, scalar anomalous
dimensions at $\ell$ loops and the scalar coupling $\beta$-function at $\ell-1$ loops where $A$ involves $\ell+2$ loop
diagrams. Eliminating $A$ and $T_{IJ}$ ensures that the relations contain non linear contributions involving
the $\beta$-functions and anomalous dimensions at lower loop order. The elimination of any particular
contribution to $A$ is possible when the relevant diagram is not vertex transitive. If the diagram has $n$ inequivalent
vertices then \eqref{ATB} leads to $n$ independent equations in this case.
For $\ell = 2,3$, and including  also arbitrary gauge couplings, the possible relations
were exhaustively analysed by Poole and Thomsen~\cite{Poole:2019kcm}.  
At this order the conditions relate contributions to the
$\beta$-functions and anomalous dimensions which have insertions of one loop triangles and one loop bubbles. 

For $\ell=2$ there are 11  5-loop vacuum diagrams for $A$ (3 are vertex transitive)
and 9 possible 3 loop $T_{IJ}$, all of which are symmetric,
and \eqref{ATB} gives rise to 21 equations.
Nevertheless there are 4  conditions on the individual $\beta$-function coefficients which reduce to the
vanishing of
\begin{align}
& B_1 =  \gamma_{\phi{\hskip 0.5pt}1} \, \beta_{y{\hskip 0.5pt}2a} - 3 \, \beta_{\lambda{\hskip 0.5pt}1b}\,  
\gamma_{\phi{\hskip 0.5pt}2a}  \, , \qquad  B_2= 
2 \,  \gamma_{\phi{\hskip 0.5pt}1} \,  \beta_{y{\hskip 0.5pt}2c} + 2\,  \gamma_{\psi{\hskip 0.5pt}1}\,  
\gamma_{\phi{\hskip 0.5pt}2c}   - \beta_{y{\hskip 0.5pt}1} \, \gamma_{\phi{\hskip 0.5pt}2b}  \, , 
\nn \\
&B_3=  \beta_{y{\hskip 0.5pt}1}{\!}^2 \, \gamma_{\psi{\hskip 0.5pt}2b} -  \beta_{y{\hskip 0.5pt}1}  \gamma_{\psi{\hskip 0.5pt}1}\, \big ( \beta_{y{\hskip 0.5pt}2c} + \gamma_{\psi{\hskip 0.5pt}2c}\big ) -  \gamma_{\psi{\hskip 0.5pt}1}{\!}^2 \, 
\big (\beta_{y{\hskip 0.5pt}2d} -  \beta_{y{\hskip 0.5pt}2e} \big )   \, , \nn  \\
&  B_4= \beta_{y{\hskip 0.5pt}1}{\!}^2 \, \gamma_{\psi{\hskip 0.5pt}2a}
-  \beta_{y{\hskip 0.5pt}1} (\gamma_{\psi{\hskip 0.5pt}1}\,  \beta_{y{\hskip 0.5pt}2b} +\gamma_{\phi{\hskip 0.5pt}1} \, \gamma_{\psi{\hskip 0.5pt}2c} )
+ \gamma_{\phi{\hskip 0.5pt}1} \gamma_{\psi{\hskip 0.5pt}1}\big (  \beta_{y{\hskip 0.5pt}2d} -  \beta_{y{\hskip 0.5pt}2e} \big )  \, , 
\label{Brel}
\end{align} 
or inserting one loop results this gives the conditions
\begin{align}
& \beta_{y{\hskip 0.5pt}2a}   = -24 \,  \gamma_{\phi{\hskip 0.5pt}2a}  \, , \quad  \beta_{y{\hskip 0.5pt}2c} =   2\, \gamma_{\phi{\hskip 0.5pt}2b} -  \gamma_{\phi{\hskip 0.5pt}2c} \, , \quad
  4 \, \beta_{y{\hskip 0.5pt}2c}  + \beta_{y{\hskip 0.5pt}2d} -  \beta_{y{\hskip 0.5pt}2e} = 16 \,  \gamma_{\psi{\hskip 0.5pt}2a} -4 \, \gamma_{\psi{\hskip 0.5pt}2c}  \, , \nn \\
& \hskip 3cm \beta_{y{\hskip 0.5pt}2b}  + \beta_{y{\hskip 0.5pt}2c} = 4 \, ( \gamma_{\psi{\hskip 0.5pt}2a} + \gamma_{\psi{\hskip 0.5pt}2b} ) - 2\, \gamma_{\psi{\hskip 0.5pt}2c} \, .
\end{align}
The non planar $\beta_{y{\hskip 0.5pt}2f} $ is not present since the associated vacuum graph obtained by joining the
external lines is vertex transitive.

For $\ell=3$ there are, for a general renormalisable fermion scalar theory{\hskip 0.5pt} 49 5 loop diagrams for $A$ (6 are vertex
transitive) and for $T_{yy}$ there are 33 distinct contributions for $T_{yy}$ which are symmetric and 20 with no symmetry. \eqref{ATB} then generates 152 equations which reduce to 42 conditions on the $\beta$-function, 
anomalous dimension coefficients. We consider first relations for non planar contributions 
to the $\beta$-function and anomalous dimension  where there are 7 relations due to the vanishing of 
\begin{align}
&C_1=  \beta_{y{\hskip 0.5pt}3\tw} - \beta_{y{\hskip 0.5pt}3\tx}  \, , \qquad  
C_2=\beta_{y{\hskip 0.5pt}1} \big ( \beta_{y{\hskip 0.5pt}3\too}  - \beta_{y{\hskip 0.5pt}3\tp} \big ) 
-  \gamma_{\psi{\hskip 0.5pt}1} \big ( \beta_{y{\hskip 0.5pt}3\ts} -  \beta_{y{\hskip 0.5pt}3\tu} \big )   \, , \nn \\
 \noalign{\vskip  2pt}
&C_3 =  \beta_{y{\hskip 0.5pt}1} \big ( \beta_{y{\hskip 0.5pt}3\tg}  - \beta_{y{\hskip 0.5pt}3\tv} \big ) + 
\big ( \beta_{y{\hskip 0.5pt}2d} -  \beta_{y{\hskip 0.5pt}2e} \big )  \beta_{y{\hskip 0.5pt}2f}  \, , \nn \\
 \noalign{\vskip  2pt}  
&C_4=  \beta_{y{\hskip 0.5pt}1} \, \gamma_{\psi{\hskip 0.5pt}3p} +  \gamma_{\psi{\hskip 0.5pt}1} 
\big ( \beta_{y{\hskip 0.5pt}3\ts}  - \beta_{y{\hskip 0.5pt}3\tv} \big ) 
 -  \gamma_{\psi{\hskip 0.5pt}2c} \,  \beta_{y{\hskip 0.5pt}2f}  \, , \nn \\
  \noalign{\vskip 2pt}
 & C_5 = \beta_{y{\hskip 0.5pt}1}\,   \beta_{y{\hskip 0.5pt}3\too} + \gamma_{\psi{\hskip 0.5pt}1} 
 \big ( \beta_{y{\hskip 0.5pt}3\tg}  - \beta_{y{\hskip 0.5pt}3\trr} \big ) 
  -    \beta_{y{\hskip 0.5pt}2c}\,  \beta_{y{\hskip 0.5pt}2f} \, , \nn \\
   \noalign{\vskip  2pt}
   & C_6 = \beta_{y{\hskip 0.5pt}1}\,   \gamma_{\phi{\hskip 0.5pt}3m} - \gamma_{\phi{\hskip 0.5pt}1} 
   \big ( \beta_{y{\hskip 0.5pt}3\tg}  - \beta_{y{\hskip 0.5pt}3\tu} \big ) 
   -   \gamma_{\phi{\hskip 0.5pt}2c}\,  \beta_{y{\hskip 0.5pt}2f} \, , \nn \\
   \noalign{\vskip 2pt}
  &  C_7 =\beta_{y{\hskip 0.5pt}1}\,  \beta_{y{\hskip 0.5pt}3s} 
  - \gamma_{\phi{\hskip 0.5pt}1}  \big ( \beta_{y{\hskip 0.5pt}3\tg}+ \beta_{y{\hskip 0.5pt}3\trr}  + 
  \beta_{y{\hskip 0.5pt}3\ts} -\beta_{y{\hskip 0.5pt}3\tu}- 2\, \beta_{y{\hskip 0.5pt}3\tv} \big ) 
    -  \beta_{y{\hskip 0.5pt}2b}\,  \beta_{y{\hskip 0.5pt}2f} \, .
 \label{Crel}
\end{align}
For the planar three loop contributions there are 35 conditions in total. For those corresponding to 
contributions involving $\lambda$ there are 14 corresponding to the vanishing of
\begin{align}
& D_1 = \gamma_{\phi{\hskip 0.5pt}1} \, \beta_{y{\hskip 0.5pt}3f} - 3 \, \gamma_{\phi{\hskip 0.5pt}2a}\, \beta_{\lambda{\hskip 0.5pt}2g} \, , \qquad
\qquad D_2=  \gamma_{\phi{\hskip 0.5pt}1} \, \beta_{y{\hskip 0.5pt}3l} - 3 \, \gamma_{\phi{\hskip 0.5pt}2a}\, \beta_{\lambda{\hskip 0.5pt}2g} \, , \nn \\
&D_3= \gamma_{\psi{\hskip 0.5pt}1} \, \beta_{y{\hskip 0.5pt}3e} + \gamma_{\phi{\hskip 0.5pt}1} \, 
\gamma_{\psi{\hskip 0.5pt}3b} 
- \beta_{y{\hskip 0.5pt}2a} \, \gamma_{\psi{\hskip 0.5pt}2a}  \, ,  \nn \\
& D_4 = \gamma_{\psi{\hskip 0.5pt}1} (  \beta_{y{\hskip 0.5pt}3g}- \beta_{y{\hskip 0.5pt}3k} ) -  
\beta_{y{\hskip 0.5pt}1} \, \gamma_{\psi{\hskip 0.5pt}3b} + 
  \beta_{y{\hskip 0.5pt}2a} \, \gamma_{\psi{\hskip 0.5pt}2c}  \, , 
\nn \\
&D_5=  2\,  \gamma_{\phi{\hskip 0.5pt}1} ( \beta_{y{\hskip 0.5pt}3b}- \beta_{y{\hskip 0.5pt}3c} ) - 
3\,  \gamma_{\phi{\hskip 0.5pt}2a} 
( \beta_{\lambda{\hskip 0.5pt}2c} - 2\,  \beta_{\lambda{\hskip 0.5pt}2d})  \, , \nn \\
&D_6= 2 ( \gamma_{\phi{\hskip 0.5pt}1}\, \gamma_{\phi{\hskip 0.5pt}3a} - \beta_{\lambda{\hskip 0.5pt}1a}\,\gamma_{\phi{\hskip 0.5pt}3b} )
+  3\, \gamma_{\phi{\hskip 0.5pt}2a} \, \beta_{\lambda{\hskip 0.5pt}2b}   \, , 
\nn \\
& D_7= \beta_{y{\hskip 0.5pt}1} ( \beta_{y{\hskip 0.5pt}3g} - \beta_{y{\hskip 0.5pt}3h} )- \beta_{y{\hskip 0.5pt}2a} ( \beta_{y{\hskip 0.5pt}2d} - \beta_{y{\hskip 0.5pt}2e}  )  \, ,
\nn \\
& 
D_8 =\beta_{y{\hskip 0.5pt}1} \,  \beta_{y{\hskip 0.5pt}3d}  + \gamma_{\phi{\hskip 0.5pt}1}  
( 2\,  \beta_{y{\hskip 0.5pt}3g} - \beta_{y{\hskip 0.5pt}3h} - \beta_{y{\hskip 0.5pt}3i} ) 
- \beta_{y{\hskip 0.5pt}2a}\,  \beta_{y{\hskip 0.5pt}2b}    \, , \nn \\
&  
D_9=\beta_{y{\hskip 0.5pt}1} \,  \beta_{y{\hskip 0.5pt}3j}  + \gamma_{\psi{\hskip 0.5pt}1}   ( \beta_{y{\hskip 0.5pt}3h} - \beta_{y{\hskip 0.5pt}3i} ) - \beta_{y{\hskip 0.5pt}2a}\,  \beta_{y{\hskip 0.5pt}2c}   \, , \nn \\
& 
D_{10}=2\,  \beta_{y{\hskip 0.5pt}1} \, \gamma_{\phi{\hskip 0.5pt}3c} - 2\, \gamma_{\phi{\hskip 0.5pt}1} \, 
\beta_{y{\hskip 0.5pt}3h} + 3 \,  \beta_{\lambda{\hskip 0.5pt}1b}\,  \beta_{y{\hskip 0.5pt}3a} 
- \beta_{y{\hskip 0.5pt}2a} ( \beta_{y{\hskip 0.5pt}2b} + 2\, \gamma_{\phi{\hskip 0.5pt}2c} ) + 
6\, \gamma_{\phi{\hskip 0.5pt}2a} \,  \beta_{\lambda{\hskip 0.5pt}2f}  \, , \nn \\
& D_{11}= \gamma_{\phi{\hskip 0.5pt}1} ( \gamma_{\psi{\hskip 0.5pt}1} \, \beta_{y{\hskip 0.5pt}3a} - \beta_{y{\hskip 0.5pt}1} \, \gamma_{\psi{\hskip 0.5pt}3a} ) 
- \gamma_{\phi{\hskip 0.5pt}2a} ( \gamma_{\psi{\hskip 0.5pt}1} \, \beta_{y{\hskip 0.5pt}2b} -  \beta_{y{\hskip 0.5pt}1} \, \gamma_{\psi{\hskip 0.5pt}2a} )  \, , \nn \\
& 
D_{12}=\gamma_{\phi{\hskip 0.5pt}1}  \, \gamma_{\psi{\hskip 0.5pt}1} ( \beta_{y{\hskip 0.5pt}3h} 
+ \beta_{y{\hskip 0.5pt}3k} ) +3\,  \gamma_{\phi{\hskip 0.5pt}2a} \big ( \beta_{y{\hskip 0.5pt}1} \, 
\beta_{\lambda{\hskip 0.5pt}2e} -  \beta_{\lambda{\hskip 0.5pt}1b} \, \beta_{y{\hskip 0.5pt}2c}- 2\,  \gamma_{\psi{\hskip 0.5pt}1} \, \beta_{\lambda{\hskip 0.5pt}2f} \big ) \, ,
\nn \\ 
& D_{13}=2 \,\gamma_{\psi{\hskip 0.5pt}1} ( \gamma_{\phi{\hskip 0.5pt}1} \, \beta_{y{\hskip 0.5pt}3e} - \beta_{\lambda{\hskip 0.5pt}1b}\, \gamma_{\phi{\hskip 0.5pt}3b} 
+ 2\,  \gamma_{\phi{\hskip 0.5pt}1} \,  \gamma_{\phi{\hskip 0.5pt}3c} ) -  \gamma_{\phi{\hskip 0.5pt}1} (\gamma_{\phi{\hskip 0.5pt}2b} \, \beta_{y{\hskip 0.5pt}2a}  - 6\, 
\gamma_{\phi{\hskip 0.5pt}2a} \, \beta_{\lambda{\hskip 0.5pt}2e} )  \, ,  
 \nn \\
&  
D_{14}=\beta_{y{\hskip 0.5pt}1} ( \gamma_{\phi{\hskip 0.5pt}1}  \,  \beta_{y{\hskip 0.5pt}3c}  + 
2 \,  \beta_{\lambda{\hskip 0.5pt}1a}\, \gamma_{\phi{\hskip 0.5pt}3c} -  \beta_{\lambda{\hskip 0.5pt}1b}\, 
\gamma_{\phi{\hskip 0.5pt}3a} )-  \gamma_{\phi{\hskip 0.5pt}1} \, \beta_{\lambda{\hskip 0.5pt}1a}\,  \beta_{y{\hskip 0.5pt}3h}  \nn \\
\noalign{\vskip -2pt}
& \hskip 1.5cm {}- 3\,  \beta_{y{\hskip 0.5pt}1} \, \gamma_{\phi{\hskip 0.5pt}2a}\,  \beta_{\lambda{\hskip 0.5pt}2d} 
-   \beta_{\lambda{\hskip 0.5pt}1a}(  \beta_{y{\hskip 0.5pt}2a}\,  \gamma_{\phi{\hskip 0.5pt}2c} - 3 \, \gamma_{\phi{\hskip 0.5pt}2a}  \,  \beta_{\lambda{\hskip 0.5pt}2f} )  \,  . 
\label{Drela}
 \end{align} 
Of the remaining 21 there is one relation which has total loop order 4 involving the vanishing of
\begin{align}
D_{15}= \beta_{y{\hskip 0.5pt}1} ( \beta_{y{\hskip 0.5pt}3\tj} - \beta\raisebox{-1.5 pt}{$\scriptstyle  y{\hskip 0.5pt}3\tk$}) +  
 \gamma_{\psi{\hskip 0.5pt}1} ( \beta\raisebox{-1.5 pt}{$\scriptstyle y{\hskip 0.5pt}3\tl$}  - \beta_{y{\hskip 0.5pt}3\tn}  )  \, ,
 \label{Drelb}
\end{align}
and otherwise we have, for total loop order 5,   15 relations enforcing the vanishing of
\begin{align}
& 
D_{16}= \beta_{y{\hskip 0.5pt}1}{\!}^2 \gamma_{\phi{\hskip 0.5pt}3d} -  \beta_{y{\hskip 0.5pt}1} \, \gamma_{\phi{\hskip 0.5pt}1} (2\,  \beta_{y{\hskip 0.5pt}3t} + \gamma_{\phi{\hskip 0.5pt}3i}) 
  + 2\,  \gamma_{\phi{\hskip 0.5pt}1}{\!}^2 \,  \beta_{y{\hskip 0.5pt}3\te}  
 - 2 \, \gamma_{\phi{\hskip 0.5pt}2c}  (  \beta_{y{\hskip 0.5pt}1} \, \gamma_{\psi{\hskip 0.5pt}2a} -   \gamma_{\phi{\hskip 0.5pt}1}\, \gamma_{\psi{\hskip 0.5pt}2c} \big )  \, , \nn \\
 & 
D_{17}= \beta_{y{\hskip 0.5pt}1}{\!}^2 \gamma_{\phi{\hskip 0.5pt}3e} -  \beta_{y{\hskip 0.5pt}1} \, \gamma_{\phi{\hskip 0.5pt}1} (\beta_{y{\hskip 0.5pt}3n} - \beta_{y{\hskip 0.5pt}3o}+ \gamma_{\phi{\hskip 0.5pt}3l}) 
  +2\,  \gamma_{\phi{\hskip 0.5pt}1}{\!}^2 ( \beta\raisebox{-1.5 pt}{$\scriptstyle y{\hskip 0.5pt}3\td$} -  \beta_{y{\hskip 0.5pt}3\tq}   ) \nn \\
\noalign{\vskip -2pt}
& \hskip 1.5cm {} 
 -  \gamma_{\phi{\hskip 0.5pt}2c}  (  \beta_{y{\hskip 0.5pt}1} \, \beta_{y{\hskip 0.5pt}2b} -  2\,  \gamma_{\phi{\hskip 0.5pt}1}\, \beta_{y{\hskip 0.5pt}2d} \big )  \, , \nn \\
& 
D_{18}= \beta_{y{\hskip 0.5pt}1}{\!}^2 \gamma_{\phi{\hskip 0.5pt}3g} -  2\, \beta_{y{\hskip 0.5pt}1} \, \gamma_{\phi{\hskip 0.5pt}1} \, \beta_{y{\hskip 0.5pt}3\ta} 
 -  \beta_{y{\hskip 0.5pt}1} \, \gamma_{\psi{\hskip 0.5pt}1} \, \gamma_{\phi{\hskip 0.5pt}3i} + 2 \, \gamma_{\phi{\hskip 0.5pt}1} \, \gamma_{\psi{\hskip 0.5pt}1} \,  \beta_{y{\hskip 0.5pt}3\te}  \nn \\
 \noalign{\vskip -2pt}
& \hskip 1.5cm {} 
 - 2 \, \gamma_{\phi{\hskip 0.5pt}2c}  (  \beta_{y{\hskip 0.5pt}1} \, \gamma_{\psi{\hskip 0.5pt}2b} -   \gamma_{\psi{\hskip 0.5pt}1} \, \gamma_{\psi{\hskip 0.5pt}2c} \big )  \, , \nn \\
 & 
D_{19}= \beta_{y{\hskip 0.5pt}1}{\!}^2 \gamma_{\psi{\hskip 0.5pt}3e} -  \beta_{y{\hskip 0.5pt}1} ( \gamma_{\phi{\hskip 0.5pt}1} \gamma_{\psi{\hskip 0.5pt}3l} + \gamma_{\psi{\hskip 0.5pt}1}\, \beta_{y{\hskip 0.5pt}3t})
  +   \gamma_{\phi{\hskip 0.5pt}1}\, \gamma_{\psi{\hskip 0.5pt}1}  \,  \beta_{y{\hskip 0.5pt}3\te}  \nn \\
\noalign{\vskip -2pt}
& \hskip 1.5cm {} -  \gamma_{\psi{\hskip 0.5pt}1}\, \gamma_{\psi{\hskip 0.5pt}2a} (\beta_{y{\hskip 0.5pt}2d} - \beta_{y{\hskip 0.5pt}2e})
 + \gamma_{\psi{\hskip 0.5pt}2c}
  \big ( \gamma_{\psi{\hskip 0.5pt}1}\, \beta_{y{\hskip 0.5pt}2b} - 2\,  \beta_{y{\hskip 0.5pt}1} \, \gamma_{\psi{\hskip 0.5pt}2a}  
  +  2\, \gamma_{\phi{\hskip 0.5pt}1} \,  \gamma_{\psi{\hskip 0.5pt}2c}\big )  
 \, , \nn \\
 & 
 D_{20}=\beta_{y{\hskip 0.5pt}1}{\!}^2 \gamma_{\psi{\hskip 0.5pt}3g} -  \beta_{y{\hskip 0.5pt}1} \, \gamma_{\psi{\hskip 0.5pt}1} ( \beta_{y{\hskip 0.5pt}3u} + 2\,  \gamma_{\psi{\hskip 0.5pt}3h})
  + 2\,   \gamma_{\psi{\hskip 0.5pt}1}{\!}^2  \,  \beta_{y{\hskip 0.5pt}3v}   
 + 2\, \beta_{y{\hskip 0.5pt}2c}\big ( \gamma_{\psi{\hskip 0.5pt}1}  \beta_{y{\hskip 0.5pt}2b} - \beta_{y{\hskip 0.5pt}1} \, \gamma_{\psi{\hskip 0.5pt}2a} \big )  
 \, , \nn \\
 & 
D_{21}= \beta_{y{\hskip 0.5pt}1}{\!}^2 \gamma_{\psi{\hskip 0.5pt}3j} -  \beta_{y{\hskip 0.5pt}1} \, \gamma_{\psi{\hskip 0.5pt}1} ( \beta_{y{\hskip 0.5pt}3\ta} +  \gamma_{\psi{\hskip 0.5pt}3l})
  +  \gamma_{\psi{\hskip 0.5pt}1}{\!}^2  \,  \beta_{y{\hskip 0.5pt}3\te}   
 -  \gamma_{\psi{\hskip 0.5pt}1}\big (  ( \beta_{y{\hskip 0.5pt}2d} - \beta_{y{\hskip 0.5pt}2e} )  \gamma_{\psi{\hskip 0.5pt}2b} + \beta_{y{\hskip 0.5pt}2c} \, \gamma_{\psi{\hskip 0.5pt}2c} \big )
 \, , \nn \\
 & 
D_{22}= \beta_{y{\hskip 0.5pt}1}{\!}^2 \gamma_{\psi{\hskip 0.5pt}3o} 
+  \beta_{y{\hskip 0.5pt}1} \, \gamma_{\psi{\hskip 0.5pt}1} ( 
 \beta\raisebox{-1.5 pt}{$\scriptstyle y{\hskip 0.5pt}3\thh$}-  \beta_{y{\hskip 0.5pt}3\tj}  - 2\,   \gamma_{\psi{\hskip 0.5pt}3m})
  +  \gamma_{\psi{\hskip 0.5pt}1}{\!}^2  \big (  \beta_{y{\hskip 0.5pt}3\tm}   + \beta_{y{\hskip 0.5pt}3\tn}  
   - \beta_{y{\hskip 0.5pt}3\tq}   - 
  \beta\raisebox{-1.5 pt}{$\scriptstyle y{\hskip 0.5pt}3\ttt$}  \big )   \nn \\
\noalign{\vskip -2pt}
& \hskip 1.5cm {} 
 + \gamma_{\psi{\hskip 0.5pt}2c}\big (  \gamma_{\psi{\hskip 0.5pt}1}( \beta_{y{\hskip 0.5pt}2d} + \beta_{y{\hskip 0.5pt}2e}) - \beta_{y{\hskip 0.5pt}1}\,  \beta_{y{\hskip 0.5pt}2c} \big )
 \, , \nn \\
 & 
D_{23}= \beta_{y{\hskip 0.5pt}1}{\!}^2  \beta_{y{\hskip 0.5pt}3p} - \beta_{y{\hskip 0.5pt}1} \, 
\gamma_{\phi{\hskip 0.5pt}1} (  
  \beta\raisebox{-1.5 pt}{$\scriptstyle y{\hskip 0.5pt}3\thh$} + \beta_{y{\hskip 0.5pt}3\ti}  )   
   + \beta_{y{\hskip 0.5pt}1} \, \gamma_{\psi{\hskip 0.5pt}1}  ( \beta_{y{\hskip 0.5pt}3n} - \beta_{y{\hskip 0.5pt}3o}) \nn \\
  & \hskip 1.5cm {}  -  \gamma_{\phi{\hskip 0.5pt}1} \, \gamma_{\psi{\hskip 0.5pt}1}   
 \big (  2\,  \beta\raisebox{-1.5 pt}{$\scriptstyle y{\hskip 0.5pt}3\td$} + 
 \beta\raisebox{-1.5 pt}{$\scriptstyle y{\hskip 0.5pt}3\tl$}  
 -  \beta_{y{\hskip 0.5pt}3\tn} - 2\,  \beta_{y{\hskip 0.5pt}3\tq} \big )  
-  \beta_{y{\hskip 0.5pt}2c}  \big (  \beta_{y{\hskip 0.5pt}1}  \beta_{y{\hskip 0.5pt}2b} -  2\, \gamma_{\phi{\hskip 0.5pt}1}\, \beta_{y{\hskip 0.5pt}2d} \big )  \, , \nn \\ 
 & 
 D_{24}=\beta_{y{\hskip 0.5pt}1}{\!}^2 ( \beta_{y{\hskip 0.5pt}3n} - \beta_{y{\hskip 0.5pt}3q} )-  \beta_{y{\hskip 0.5pt}1} \, \gamma_{\phi{\hskip 0.5pt}1}
(  2\, \beta\raisebox{-1.5 pt}{$\scriptstyle y{\hskip 0.5pt}3\td$}  -\beta\raisebox{-1.5 pt}{$\scriptstyle y{\hskip 0.5pt}3\tl$}  - \beta_{y{\hskip 0.5pt}3\tn} )  \nn \\\noalign{\vskip -2pt}
& \hskip 1.5cm {} 
+ (  \beta_{y{\hskip 0.5pt}2d} - \beta_{y{\hskip 0.5pt}2e} )\big (  \beta_{y{\hskip 0.5pt}1} \, \beta_{y{\hskip 0.5pt}2b} -  2\, \gamma_{\phi{\hskip 0.5pt}1}\, \beta_{y{\hskip 0.5pt}2d} \big )  \, , \nn \\
& 
 D_{25}=\beta_{y{\hskip 0.5pt}1}{\!}^2 \beta_{y{\hskip 0.5pt}3y}  -  \beta_{y{\hskip 0.5pt}1} \, \gamma_{\psi{\hskip 0.5pt}1}
\big (  2\, \beta\raisebox{-1.5 pt}{$\scriptstyle y{\hskip 0.5pt}3\tb$}  + 2\, \beta\raisebox{-1.5 pt}{$\scriptstyle y{\hskip 0.5pt}3\thh$}  - \beta_{y{\hskip 0.5pt}3\ti}
 - \beta_{y{\hskip 0.5pt}3\tj} \big  )  \nn \\
 \noalign{\vskip -2pt}
& \hskip 1.5cm {} 
  +  \gamma_{\psi{\hskip 0.5pt}1}{\!}^2 \big (  2\, \beta\raisebox{-1.5 pt}{$\scriptstyle y{\hskip 0.5pt}3\td$}  -
  \beta\raisebox{-1.5 pt}{$\scriptstyle y{\hskip 0.5pt}3\tl$}  - \beta_{y{\hskip 0.5pt}3\tm} -2\,  \beta_{y{\hskip 0.5pt}3\tn}  + \beta_{y{\hskip 0.5pt}3\tq}  + 
  \beta\raisebox{-1.5 pt}{$\scriptstyle y{\hskip 0.5pt}3\ttt$}  \big )\nn \\
 \noalign{\vskip -2pt}
& \hskip 1.5cm {} 
-  \beta_{y{\hskip 0.5pt}2c} \big ( \beta_{y{\hskip 0.5pt}1}\,   \beta_{y{\hskip 0.5pt}2c} +  
\gamma_{\psi{\hskip 0.5pt}1}  ( 3\, \beta_{y{\hskip 0.5pt}2d} - \beta_{y{\hskip 0.5pt}2e}) \big )
+ ( 3\, \beta_{y{\hskip 0.5pt}2d} + \beta_{y{\hskip 0.5pt}2e} ) (\beta_{y{\hskip 0.5pt}1}\,  \gamma_{\psi{\hskip 0.5pt}2b} - \gamma_{\psi{\hskip 0.5pt}1}\,  \gamma_{\psi{\hskip 0.5pt}2c} ) \, , \nn \\
& 
D_{26}= \beta_{y{\hskip 0.5pt}1}{\!}^2 ( \beta_{y{\hskip 0.5pt}3\tc} - \beta_{y{\hskip 0.5pt}3\tj} )-  \beta_{y{\hskip 0.5pt}1} \, \gamma_{\psi{\hskip 0.5pt}1}
(  \beta\raisebox{-1.5 pt}{$\scriptstyle y{\hskip 0.5pt}3\td$} +\beta\raisebox{-1.5 pt}{$\scriptstyle y{\hskip 0.5pt}3\tf$} - \beta_{y{\hskip 0.5pt}3\tm} - \beta_{y{\hskip 0.5pt}3\tn} )  \nn \\
\noalign{\vskip -2pt}
& \hskip 1.5cm {}  + (  \beta_{y{\hskip 0.5pt}2d} - \beta_{y{\hskip 0.5pt}2e} )\big (  \beta_{y{\hskip 0.5pt}1} \, \beta_{y{\hskip 0.5pt}2c} - \gamma_{\psi{\hskip 0.5pt}1}
 (  \beta_{y{\hskip 0.5pt}2d} + \beta_{y{\hskip 0.5pt}2e} )\big )  \, , \nn \\
 & 
D_{27}= \beta_{y{\hskip 0.5pt}1}{\!}^2 ( \beta_{y{\hskip 0.5pt}3u} + 2\, \gamma_{\phi{\hskip 0.5pt}3j} )- 2 \, \beta_{y{\hskip 0.5pt}1} \, \gamma_{\psi{\hskip 0.5pt}1}
(  \beta_{y{\hskip 0.5pt}3v} + 2\, \gamma_{\phi{\hskip 0.5pt}3k} + \gamma_{\phi{\hskip 0.5pt}3l}  )  - 4\,   \beta_{y{\hskip 0.5pt}1} \, \gamma_{\phi{\hskip 0.5pt}1}\, \beta_{y{\hskip 0.5pt}3\tj} \nn \\
\noalign{\vskip -2pt}
& \hskip 1.5cm {} 
  - 4\, \gamma_{\phi{\hskip 0.5pt}1} \, \gamma_{\psi{\hskip 0.5pt}1}   
 \big (  \beta\raisebox{-1.5 pt}{$\scriptstyle y{\hskip 0.5pt}3\tf$} - \beta_{y{\hskip 0.5pt}3\tm}   -  \beta_{y{\hskip 0.5pt}3\tn} +  \beta_{y{\hskip 0.5pt}3\tq} \big )  \nn \\
\noalign{\vskip -2pt}
& \hskip 1.5cm {} 
  - 2 \, \beta_{y{\hskip 0.5pt}2c}\big  ( \beta_{y{\hskip 0.5pt}1} ( \beta_{y{\hskip 0.5pt}2b} + 2\, \gamma_{\phi{\hskip 0.5pt}2c})  - 2\, \gamma_{\phi{\hskip 0.5pt}1}\,  \beta_{y{\hskip 0.5pt}2d}  \big ) 
\nn \\
\noalign{\vskip -2pt}
& \hskip 1.5cm {} 
+  4 (  \beta_{y{\hskip 0.5pt}2d} + \beta_{y{\hskip 0.5pt}2e} )\big  (  \beta_{y{\hskip 0.5pt}1} \, \gamma_{\psi{\hskip 0.5pt}2a} - \gamma_{\psi{\hskip 0.5pt}1}\, 
 \beta_{y{\hskip 0.5pt}2b} -   \gamma_{\phi{\hskip 0.5pt}1}\,  \gamma_{\psi{\hskip 0.5pt}2c}  +    \gamma_{\psi{\hskip 0.5pt}1}\,  \gamma_{\phi{\hskip 0.5pt}2c} \big )  \, , \nn \\
 & 
D_{28}=\beta_{y{\hskip 0.5pt}1}{\!}^2 ( \beta_{y{\hskip 0.5pt}3\ta} + 2\, \gamma_{\psi{\hskip 0.5pt}3n} )-  \beta_{y{\hskip 0.5pt}1} \, \gamma_{\psi{\hskip 0.5pt}1}
(  \beta_{y{\hskip 0.5pt}3\te}  + 2\,  \beta_{y{\hskip 0.5pt}3\tj} +  2\, \gamma_{\psi{\hskip 0.5pt}3k} + 2\, \gamma_{\psi{\hskip 0.5pt}3m}  )  \nn \\
\noalign{\vskip -2pt}
& \hskip 1.5cm {} 
  - 2\, \gamma_{\psi{\hskip 0.5pt}1} {\!}^2  
 \big (  \beta\raisebox{-1.5 pt}{$\scriptstyle y{\hskip 0.5pt}3\tl$} - \beta_{y{\hskip 0.5pt}3\tm}   -  \beta_{y{\hskip 0.5pt}3\tn} +  
 \beta\raisebox{-1.5 pt}{$\scriptstyle y{\hskip 0.5pt}3\ttt$}\big )  \nn \\
\noalign{\vskip -2pt} & \hskip 1.5cm {} + \beta_{y{\hskip 0.5pt}1} 
\big ( - \beta_{y{\hskip 0.5pt}2c}{\!}^2 + 2\,  \beta_{y{\hskip 0.5pt}2d} \,  \gamma_{\psi{\hskip 0.5pt}2b} - 2\,  \beta_{y{\hskip 0.5pt}2c} \,  \gamma_{\psi{\hskip 0.5pt}2c}\big )\nn \\
\noalign{\vskip -2pt}
& \hskip 1.5cm {} 
+  \gamma_{\psi{\hskip 0.5pt}1} \big ( - \beta_{y{\hskip 0.5pt}2c}( \beta_{y{\hskip 0.5pt}2d} - \beta_{y{\hskip 0.5pt}2e} ) + 2\,  \beta_{y{\hskip 0.5pt}2e} \,  \gamma_{\psi{\hskip 0.5pt}2c}\big )  \, ,
\nn \\
& D_{29}=\
\beta_{y{\hskip 0.5pt}1}{\!}^2 ( \beta_{y{\hskip 0.5pt}3t} + 2\, \gamma_{\psi{\hskip 0.5pt}3f} )  - 2\, \beta_{y{\hskip 0.5pt}1} \,\gamma_{\psi{\hskip 0.5pt}1} \,\beta_{y{\hskip 0.5pt}3n}  
- \beta_{y{\hskip 0.5pt}1} \,\gamma_{\phi{\hskip 0.5pt}1} \left(\beta_{y{\hskip 0.5pt}3\te} 
+ 2\, \gamma_{\psi{\hskip 0.5pt}3k} + 2\,\gamma_{\psi{\hskip 0.5pt}3m}\right)  \nn \\
\noalign{\vskip -2pt}
& \hskip 1.5cm {} 
+ 2\,\gamma_{\psi{\hskip 0.5pt}1}  
\gamma_{\phi{\hskip 0.5pt}1} (2\,\beta\raisebox{-1.5 pt}{$\scriptstyle y{\hskip 0.5pt}3\td$}
 - \beta\raisebox{-1.5 pt}{$\scriptstyle y{\hskip 0.5pt}3\tl$}  - \beta_{y{\hskip 0.5pt}3\tq})  \nn \\
\noalign{\vskip -2pt}
& \hskip 1.5cm {} 
- \beta_{y{\hskip 0.5pt}1} \big ( \beta_{y{\hskip 0.5pt}2b} ( \beta_{y{\hskip 0.5pt}2c}+ 2 \, \gamma_{\psi{\hskip 0.5pt} 2c}) + 2\,\beta_{y{\hskip 0.5pt}2d} \,\gamma_{\psi{\hskip 0.5pt}2a}\big ) 
+ 2\,\gamma_{\psi{\hskip 0.5pt}1} \, \beta_{y{\hskip 0.5pt}2b} (\beta_{y{\hskip 0.5pt}2d} + \beta_{y{\hskip 0.5pt}2e}) \nn \\
\noalign{\vskip -2pt}
& \hskip 1.5cm {} 
+ \gamma_{\phi{\hskip 0.5pt}1} \big ( 6\,\gamma_{\psi{\hskip 0.5pt}2c}\,\beta_{y{\hskip 0.5pt}2d}  + \beta_{y{\hskip 0.5pt}2c}(\beta_{y{\hskip 0.5pt}2d}-\beta_{y{\hskip 0.5pt}2e})\big )\, , \nn \\
& D_{30}=
  \beta_{y{\hskip 0.5pt}1}\,\gamma_{\phi{\hskip 0.5pt}1} \big (\beta\raisebox{-1.5 pt}{$\scriptstyle y{\hskip 0.5pt}3\tb$} + \beta_{y{\hskip 0.5pt}3\tj} \big ) 
  - \beta_{y{\hskip 0.5pt}1}\, \gamma_{\psi{\hskip 0.5pt}1}\big  (\beta_{y{\hskip 0.5pt}3n} + \beta_{y{\hskip 0.5pt}3r} \big ) \nn \\
 \noalign{\vskip -2pt}
& \hskip 1.5cm {} + \gamma_{\phi{\hskip 0.5pt}1}\, \gamma_{\psi{\hskip 0.5pt}1}\big  (\beta\raisebox{-1.5 pt}{$\scriptstyle y{\hskip 0.5pt}3\td$} 
 - \beta\raisebox{-1.5 pt}{$\scriptstyle y{\hskip 0.5pt}3\tf$} + \beta\raisebox{-1.5 pt}{$\scriptstyle y{\hskip 0.5pt}3\tl$} 
 - \beta_{y{\hskip 0.5pt}3\tn} - \beta_{y{\hskip 0.5pt}3\tq} + \beta\raisebox{-1.5 pt}{$\scriptstyle y{\hskip 0.5pt}3\ttt$} \big  )  \nn \\
 \noalign{\vskip -2pt}
& \hskip 1.5cm {} + 
(\beta_{y{\hskip 0.5pt}2d} + \beta_{y{\hskip 0.5pt}2e}) \big (2\,\beta_{y{\hskip 0.5pt}1}\,\gamma_{\psi{\hskip 0.5pt}2a}  - \gamma_{\psi{\hskip 0.5pt}1}\,\beta_{y{\hskip 0.5pt}2b} - \gamma_{\phi{\hskip 0.5pt}1} (\beta_{y{\hskip 0.5pt}2c} + 2\,\gamma_{\psi{\hskip 0.5pt}2c}) \big ) \, ,
\label{Drelc}
\end{align}
and finally 5 with loop order six
\begin{align}
& 
D_{31}=    \beta_{y{\hskip 0.5pt}1}{\!}^3 \,\gamma_{\psi{\hskip 0.5pt}3c} + \beta_{y{\hskip 0.5pt}1}{\!}^2 
    \big (\gamma_{\phi{\hskip 0.5pt}1} (\beta_{y{\hskip 0.5pt}3t} - \gamma_{\psi{\hskip 0.5pt}3h}) 
    - \gamma_{\psi{\hskip 0.5pt}1}\,\beta_{y{\hskip 0.5pt}3m} \big )
    \nn \\
    & \hskip 1.5cm {}
    - \beta_{y{\hskip 0.5pt}1}\,\gamma_{\psi{\hskip 0.5pt}1}\,\gamma_{\phi{\hskip 0.5pt}1} (2\,\beta_{y{\hskip 0.5pt}3n} 
    - \beta_{y{\hskip 0.5pt}3v}) + \beta_{y{\hskip 0.5pt}1}\,\gamma_{\phi{\hskip 0.5pt}1}{\!}^2
    \big (\beta\raisebox{-1.5 pt}{$\scriptstyle y{\hskip 0.5pt}3\tb$} - \beta_{y{\hskip 0.5pt}3\te} 
    + \beta_{y{\hskip 0.5pt}3\tj}   - \gamma_{\psi{\hskip 0.5pt}3k}\big) \nn \\
     & \hskip 1.5cm {}
    + \gamma_{\phi{\hskip 0.5pt}1}{\!}^2\,\gamma_{\psi{\hskip 0.5pt}1} 
    \big (3\,\beta\raisebox{-1.5 pt}{$\scriptstyle y{\hskip 0.5pt}3\td$} - \beta_{y{\hskip 0.5pt}3\tm} 
    - \beta_{y{\hskip 0.5pt}3\tn} - 2\,\beta_{y{\hskip 0.5pt}3\tq}   + 
    \beta\raisebox{-1.5 pt}{$\scriptstyle y{\hskip 0.5pt}3\ttt$} \big) \nn \\
        & \hskip 1.5cm {} + \beta_{y{\hskip 0.5pt}1}{\!}^2 \, \gamma_{\psi{\hskip 0.5pt}2a}\, \gamma_{\phi{\hskip 0.5pt}2c}
    + \beta_{y{\hskip 0.5pt}1}\,\gamma_{\phi{\hskip 0.5pt}1}
     \big (\gamma_{\psi{\hskip 0.5pt}2a}\,\beta_{y{\hskip 0.5pt}2d} 
     - \beta_{y{\hskip 0.5pt}2b} (\beta_{y{\hskip 0.5pt}2c} + \gamma_{\psi{\hskip 0.5pt}2c}) \big)  
       - \beta_{y{\hskip 0.5pt}1}\,\gamma_{\psi{\hskip 0.5pt}1}\,\gamma_{\phi{\hskip 0.5pt}2c}\,\beta_{y{\hskip 0.5pt}2b}
    \nn \\
       & \hskip 1.5cm {}
    + \gamma_{\phi{\hskip 0.5pt}1}{\!}^2 \big (\gamma_{\psi{\hskip 0.5pt}2c}(\beta_{y{\hskip 0.5pt}2d} 
    - \beta_{y{\hskip 0.5pt}2e}) - 2\, \beta_{y{\hskip 0.5pt}2c}\,\beta_{y{\hskip 0.5pt}2e}\big) 
     + \gamma_{\phi{\hskip 0.5pt}1}\,\gamma_{\psi{\hskip 0.5pt}1}\,\beta_{y{\hskip 0.5pt}2b}\,\beta_{y{\hskip 0.5pt}2e}
     \, , \nn \\
& 		
D_{32}= \beta_{y{\hskip 0.5pt}1}{\!}^3 \gamma_{\phi{\hskip 0.5pt}3h} 
		+ \beta_{y{\hskip 0.5pt}1}{\!}^2 \gamma_{\psi{\hskip 0.5pt}1}\, \beta_{y{\hskip 0.5pt}3u} 
    - \beta_{y{\hskip 0.5pt}1} \,  \gamma_{\phi{\hskip 0.5pt}1} \,\gamma_{\psi{\hskip 0.5pt}1} 
    \big(2\, \beta\raisebox{-1.5 pt}{$\scriptstyle y{\hskip 0.5pt}3\tb$}  
    - \beta_{y{\hskip 0.5pt}3\ti} + \beta_{y{\hskip 0.5pt}3\tj} \big) 
    - 2\,\beta_{y{\hskip 0.5pt}1}\,  \gamma_{\psi{\hskip 0.5pt}1}{\!}^2 
     (\beta_{y{\hskip 0.5pt}3v} + \gamma_{\phi{\hskip 0.5pt}3k}) \nn \\
    & \hskip 1.5cm {}
    +  \gamma_{\phi{\hskip 0.5pt}1} \, \gamma_{\psi{\hskip 0.5pt}1}{\!}^2 
    \big(2\,\beta\raisebox{-1.5 pt}{$\scriptstyle y{\hskip 0.5pt}3\td$} - 
    2\, \beta\raisebox{-1.5 pt}{$\scriptstyle y{\hskip 0.5pt}3\tf$} 
    - \beta\raisebox{-1.5 pt}{$\scriptstyle y{\hskip 0.5pt}3\tl$} + \beta_{y{\hskip 0.5pt}3\tm} - \beta_{y{\hskip 0.5pt}3\tq} 
    + \beta_{y{\hskip 0.5pt}3\ttt} \big) \nn \\
    & \hskip 1.5cm   
    {} - \beta_{y{\hskip 0.5pt}1} \, \gamma_{\phi{\hskip 0.5pt}1} \, \beta_{y{\hskip 0.5pt}2c}{\!}^2 
    - \beta_{y{\hskip 0.5pt}1} \, \gamma_{\psi{\hskip 0.5pt}1} 
		\big(2 \, \beta_{y{\hskip 0.5pt}2c}(\beta_{y{\hskip 0.5pt}2b} + \gamma_{\phi{\hskip 0.5pt}2c}) 
		+ \gamma_{\psi{\hskip 0.5pt}2a} (\beta_{y{\hskip 0.5pt}2d} 
		- \beta_{y{\hskip 0.5pt}2e}) \big)  \nn \\
    & \hskip 1.5cm {}
    + \gamma_{\phi{\hskip 0.5pt}1} \,\gamma_{\psi{\hskip 0.5pt}1}
    \big(\gamma_{\psi{\hskip 0.5pt}2c} (\beta_{y{\hskip 0.5pt}2d} - \beta_{y{\hskip 0.5pt}2e}) 
    + 2\, \beta_{y{\hskip 0.5pt}2c}\, \beta_{y{\hskip 0.5pt}2e} \big) \nn \\
    &\hskip 1.5cm {} + \gamma_{\psi{\hskip 0.5pt}1}{\!}^2 \big(\beta_{y{\hskip 0.5pt}2b} (\beta_{y{\hskip 0.5pt}2d} 
    - \beta_{y{\hskip 0.5pt}2e}) + 2\,\beta_{y{\hskip 0.5pt}2e} \, \gamma_{\phi{\hskip 0.5pt}2c} \big )
    \,, \nn \\
&		
D_{33}=	\beta_{y{\hskip 0.5pt}1}{\!}^3\, \gamma_{\psi{\hskip 0.5pt}3i}  
		+ \beta_{y{\hskip 0.5pt}1}{\!}^2  \gamma_{\psi{\hskip 0.5pt}1}\, (\beta_{y{\hskip 0.5pt}3\ta} 
		- \beta_{y{\hskip 0.5pt}3z} - \gamma_{\psi{\hskip 0.5pt}3l}) 
		+ \beta_{y{\hskip 0.5pt}1}\,\gamma_{\psi{\hskip 0.5pt}1}{\!}^2  
		(\beta\raisebox{-1.5 pt}{$\scriptstyle y{\hskip 0.5pt}3\tb $}
		- \beta_{y{\hskip 0.5pt}3\tj} - \gamma_{\psi{\hskip 0.5pt}3k}) \nn \\
		&\hskip 1.5cm {}  + \gamma_{\psi{\hskip 0.5pt}1}{\!}^3 
		\big( -\beta\raisebox{-1.5 pt}{$\scriptstyle y{\hskip 0.5pt}3\td $}
		+ \beta_{y{\hskip 0.5pt}3\tm} + \beta_{y{\hskip 0.5pt}3\tn}  - 
		\beta\raisebox{-1.5 pt}{$\scriptstyle y{\hskip 0.5pt}3\ttt $} \big )
		 \nn \\
		& \hskip 1.5cm {} 
		- \beta_{y{\hskip 0.5pt}1}\,\gamma_{\psi{\hskip 0.5pt}1} \big (\beta_{y{\hskip 0.5pt}2c}{\!}^2 
		+ \beta_{y{\hskip 0.5pt}2c} \gamma_{{\hskip 0.5pt}2c}
		+ \beta_{y{\hskip 0.5pt}2d} \,\gamma_{\psi{\hskip 0.5pt}2b}
		 - \gamma_{\psi{\hskip 0.5pt}2c}{\!}^2 \big )  
		+ \gamma_{\psi{\hskip 0.5pt}1}{\!}^2
		 \big(\beta_{y{\hskip 0.5pt}2c}\,\beta_{y{\hskip 0.5pt}2e} 
		 + 2\,\beta_{y{\hskip 0.5pt}2d}\,\gamma_{\psi{\hskip 0.5pt}2c}\big) 
 		\, ,  \nn \\
&
D_{34}=	 \beta_{y{\hskip 0.5pt}1}{\!}^3 \,\gamma_{\psi{\hskip 0.5pt}3d} + \beta_{y{\hskip 0.5pt}1}{\!}^2 
	 \big(\gamma_{\psi{\hskip 0.5pt}1}\, \beta_{y{\hskip 0.5pt}3t} 
	 + \gamma_{\phi{\hskip 0.5pt}1}\,\beta_{y{\hskip 0.5pt}3\ta}\big) 
	  - \beta_{y{\hskip 0.5pt}1}\, \gamma_{\phi{\hskip 0.5pt}1}  \, \gamma_{\psi{\hskip 0.5pt}1} 
  	  \big(2\,\beta_{y{\hskip 0.5pt}3\te} + \beta_{y{\hskip 0.5pt}3\ti} + \beta_{y{\hskip 0.5pt}3\tj} 
  	  + 2\,\gamma_{\psi{\hskip 0.5pt}3k}\big) 
	 \nn \\
	& \hskip 1.5cm {} 	 - \gamma_{\phi{\hskip 0.5pt}1} \,\gamma_{\psi{\hskip 0.5pt}1}{\!}^2 
	 \big(3\,\beta\raisebox{-1.5 pt}{$\scriptstyle y{\hskip 0.5pt}3\tl$} - \beta_{y{\hskip 0.5pt}3\tm} 
	 - 2\,\beta_{y{\hskip 0.5pt}3\tn} 
	 - \beta_{y{\hskip 0.5pt}3\tq}   + \beta_{y{\hskip 0.5pt}3\ttt}\big ) \nn \\
      & \hskip 1.5cm {} - \beta_{y{\hskip 0.5pt}1} \, \gamma_{\phi{\hskip 0.5pt}1} \, \beta_{y{\hskip 0.5pt}2c} 
 	\big(\beta_{y{\hskip 0.5pt}2c} + \gamma_{\psi{\hskip 0.5pt}2c}\big)
    -  \beta_{y{\hskip 0.5pt}1} \, \gamma_{\psi{\hskip 0.5pt}1} \big(
    \beta_{y{\hskip 0.5pt}2b} (2\,\beta_{y{\hskip 0.5pt}2c} + \gamma_{\psi{\hskip 0.5pt}2c}) +
    \gamma_{\psi{\hskip 0.5pt}2a}  (3\,\beta_{y{\hskip 0.5pt}2d} - \beta_{y{\hskip 0.5pt}2e}) \big) \nn \\
    & \hskip 1.5cm {}  
           + \gamma_{\phi{\hskip 0.5pt}1}\,\gamma_{\psi{\hskip 0.5pt}1}
    \big(\beta_{y{\hskip 0.5pt}2c} (3\,\beta_{y{\hskip 0.5pt}2d} - \beta_{y{\hskip 0.5pt}2e}) 
    + \gamma_{\psi{\hskip 0.5pt}2c} (5\,\beta_{y{\hskip 0.5pt}2d} - \beta_{y{\hskip 0.5pt}2e}) \big) 
     + 2\, \gamma_{\psi{\hskip 0.5pt}1}{\!}^2\,\beta_{y{\hskip 0.5pt}2b} \,\beta_{y{\hskip 0.5pt}2d} 
	\,, \nn \\
& 
D_{35}=    \beta_{y{\hskip 0.5pt}1}{\!}^3 \,\gamma_{\phi{\hskip 0.5pt}3f} 
    + \beta_{y{\hskip 0.5pt}1}{\!}^2 \big( \gamma_{\psi{\hskip 0.5pt}1} (\beta_{y{\hskip 0.5pt}3u} 
    - \gamma_{\phi{\hskip 0.5pt}3i}) - 2\, \gamma_{\phi{\hskip 0.5pt}1}\,\beta_{y{\hskip 0.5pt}3z}\big) 
    - 2\, \beta_{y{\hskip 0.5pt}1}\, \gamma_{\psi{\hskip 0.5pt}1}{\!}^2 \big(\beta_{y{\hskip 0.5pt}3v} 
    + \gamma_{\phi{\hskip 0.5pt}3k}\big) \nn \\
    & \hskip 1.5cm {}
    + 2\, \beta_{y{\hskip 0.5pt}1}\,\gamma_{\phi{\hskip 0.5pt}1}  \,\gamma_{\psi{\hskip 0.5pt}1} 
    \big(\beta_{y{\hskip 0.5pt}3\te} + \beta_{y{\hskip 0.5pt}3\ti} - \beta_{y{\hskip 0.5pt}3\tj}\big)   
    - 2\, \gamma_{\phi{\hskip 0.5pt}1} \,\gamma_{\psi{\hskip 0.5pt}1}{\!}^2
	\big (\beta_{y{\hskip 0.5pt}3\tf} - \beta_{y{\hskip 0.5pt}3\tl} 
	- \beta_{y{\hskip 0.5pt}3\tm} + \beta_{y{\hskip 0.5pt}3\tq}\big ) \nn \\
	&\hskip 1.5cm {}
    + 2\, \beta_{y{\hskip 0.5pt}1} \,\gamma_{\psi{\hskip 0.5pt}1} 
    \big( \gamma_{\phi{\hskip 0.5pt}2c}\, \gamma_{\psi{\hskip 0.5pt}2c} 
    + \gamma_{\psi{\hskip 0.5pt}2a} (\beta_{y{\hskip 0.5pt}2d} 
    + \beta_{y{\hskip 0.5pt}2e}) - \beta_{y{\hskip 0.5pt}2c} (\beta_{y{\hskip 0.5pt}2b} + \gamma_{\phi{\hskip 0.5pt}2c}) \big) \nn \\
    & \hskip 1.5cm {} - 2\,\gamma_{\phi{\hskip 0.5pt}1}\,\gamma_{\psi{\hskip 0.5pt}1}\,\gamma_{\psi{\hskip 0.5pt}2c}
    \big(\beta_{y{\hskip 0.5pt}2d} + \beta_{y{\hskip 0.5pt}2e} \big)
    + 2\,\gamma_{\psi{\hskip 0.5pt}1}{\!}^2 \big(\gamma_{\phi{\hskip 0.5pt}2c}\,
    \beta_{y{\hskip 0.5pt}2e} - \beta_{y{\hskip 0.5pt}2b}(\beta_{y{\hskip 0.5pt}2d} + \beta_{y{\hskip 0.5pt}2e})\big)
    \, .
\end{align}

To satisfy \eqref{ATB} with three loop $\beta$-functions it is necessary to include
contributions to $\upsilon$ as in in \eqref{upphi3} and \eqref{uppsi3}. The consistency
relations require
\begin{align}
 2\, \beta_{y{\hskip 0.5pt}1} \,  \upsilon_{\phi{\hskip 0.5pt}3c}  = {}& - 3 \, \beta_{\lambda{\hskip 0.5pt}1b}\, 
 \beta_{y{\hskip 0.5pt}3a}  +  \beta_{y{\hskip 0.5pt}2a} \,  \beta_{y{\hskip 0.5pt}2b} \, , \nn \\
 2\, \beta_{y{\hskip 0.5pt}1} \,  \upsilon_{\phi{\hskip 0.5pt}3j}   = {}&  -  \beta_{y{\hskip 0.5pt}1}\, \beta_{y{\hskip 0.5pt}3u} 
 + \gamma_{\psi{\hskip 0.5pt}1}\, \beta_{y{\hskip 0.5pt}3v} 
+  2\,  \beta_{y{\hskip 0.5pt}2b} \,  \beta_{y{\hskip 0.5pt}2c} \, , \nn \\
  2\, \beta_{y{\hskip 0.5pt}1} ( \beta_{y{\hskip 0.5pt}1} \,  \upsilon_{\psi{\hskip 0.5pt}3n}  
  - \gamma_{\psi{\hskip 0.5pt}1} \,  \upsilon_{\psi{\hskip 0.5pt}3m}  ) = {}&
-  \beta_{y{\hskip 0.5pt}1} ( \beta_{y{\hskip 0.5pt}1} \,  \beta_{y{\hskip 0.5pt}3\ta}  
- \gamma_{\psi{\hskip 0.5pt}1} \,  \beta_{y{\hskip 0.5pt}3\te}  )  \nn \\
\noalign{\vskip -2pt}
& {} +  \gamma_{\psi{\hskip 0.5pt}1} \,  \beta_{y{\hskip 0.5pt}2c} (  \beta_{y{\hskip 0.5pt}2d} -  \beta_{y{\hskip 0.5pt}2e} ) 
+ \beta_{y{\hskip 0.5pt}1}\,  \beta_{y{\hskip 0.5pt}2c}{\!}^2 \, , \nn \\
 2\, \beta_{y{\hskip 0.5pt}1} ( \beta_{y{\hskip 0.5pt}1} \,  \upsilon_{\psi{\hskip 0.5pt}3f}  
 - \gamma_{\phi{\hskip 0.5pt}1} \,  \upsilon_{\psi{\hskip 0.5pt}3m}  ) = {}&
 \beta_{y{\hskip 0.5pt}1} ( \beta_{y{\hskip 0.5pt}1} \,  \beta_{y{\hskip 0.5pt}3t}  -
  \gamma_{\phi{\hskip 0.5pt}1} \,  \beta_{y{\hskip 0.5pt}3\te}  )  \nn \\
\noalign{\vskip -2pt}
& {} +  \gamma_{\phi{\hskip 0.5pt}1} \,  \beta_{y{\hskip 0.5pt}2c} (  \beta_{y{\hskip 0.5pt}2d} 
-  \beta_{y{\hskip 0.5pt}2e} ) - \beta_{y{\hskip 0.5pt}1}\, \beta_{y{\hskip 0.5pt}2b}\,  \beta_{y{\hskip 0.5pt}2c} \, .
\label{resup}
\end{align}
Together with \eqref{susyup} these suffice to determine the results in \eqref{Phiup} and \eqref{Psiup}.
 
 \subsection{Supersymmetry Reduction}
 
 \label{Sred}
 
 For the reduction as described in  subsection \ref{susyhalf} the consistency relations reduce to just one at two loops
 given by the vanishing of 
 \be
 S_0 = \beta_{ Y \hskip 0.2pt 1} \, \gamma_{\Phi\hskip 0.5pt 2A} -
  2\,  \gamma_{\Phi \hskip 0.5pt 1} (   \beta_{Y\hskip 0.5pt 2A} +  \gamma_{\Phi\hskip 0.5pt 2B} )  \, ,
 \ee
 and at three loops for planar contributions there are 7 relations obtained by setting to zero
 \begin{align}
 & S_1= \beta_{ Y \hskip 0.2pt 1} ( \beta_{Y\hskip 0.5pt 3E} -  \beta_{Y\hskip 0.5pt 3F} ) -  \gamma_{\Phi\hskip 0.5pt 1} ( \beta_{Y\hskip 0.5pt 3H} -  \beta_{Y\hskip 0.5pt 3J} ) 
 \, , \nn \\
 & S_2= \beta_{ Y \hskip 0.2pt 1} (2\, \beta_{Y\hskip 0.5pt 3A} -  \beta_{Y\hskip 0.5pt 3C}  - 2\, 
 \gamma_{\Phi\hskip 0.5pt 3D} + 4\,  \gamma_{\Phi\hskip 0.5pt 3E}) 
 + 2\,   \gamma_{\Phi\hskip 0.5pt 1} ( \beta_{Y\hskip 0.5pt 3F} -  \beta_{Y\hskip 0.5pt 3G} + \beta_{Y\hskip 0.5pt 3K}  
 -  \gamma_{\Phi\hskip 0.5pt 3G} + 2 \,  \gamma_{\Phi\hskip 0.5pt 3H})  \, , \nn \\
 & S_3= \beta_{Y\hskip 0.5pt1} ( \beta_{Y\hskip 0.5pt 3A} +  \gamma_{\Phi\hskip 0.5pt 3E} ) -  
 \gamma_{\Phi\hskip 0.5pt 1} ( \beta_{Y\hskip 0.5pt 3D} -  \beta_{Y\hskip 0.5pt 3E} 
+  \beta_{Y\hskip 0.5pt 3F} + \beta_{Y\hskip 0.5pt 3G} +  \gamma_{\Phi\hskip 0.5pt 3G}) \nn \\
& \hskip 8cm {} -   \beta_{Y\hskip 0.5pt 2A} (   \beta_{Y\hskip 0.5pt 2A}  +  \gamma_{\Phi\hskip 0.5pt 2B} )
+     \beta_{Y\hskip 0.5pt 2B} \,  \gamma_{\Phi\hskip 0.5pt 2A}  \, , \nn\\
&S_4=  \beta_{ Y \hskip 0.2pt 1} (\gamma_{\Phi\hskip 0.5pt 3A }+  \gamma_{\Phi\hskip 0.5pt 3B} ) + 
  \gamma_{\Phi\hskip 0.5pt 1} ( \beta_{Y\hskip 0.5pt 3A} - 2\, \beta_{Y\hskip 0.5pt 3B} 
+  \beta_{Y\hskip 0.5pt 3C} - 2\, \gamma_{\Phi\hskip 0.5pt 3D} 
+ 2\, \gamma_{\Phi\hskip 0.5pt 3E}-  \gamma_{\Phi\hskip 0.5pt 3F})  \nn \\
& \hskip 8cm {}-  \gamma_{\Phi\hskip 0.5pt 2A} ( 2\,   \beta_{Y\hskip 0.5pt 2A} 
-   \gamma_{\Phi\hskip 0.5pt 2B} )  \, , \nn \\
& S_5= \beta_{ Y \hskip 0.2pt 1}{}^2\, \beta_{Y\hskip 0.5pt 3A} - \beta_{ Y \hskip 0.2pt 1} \gamma_{\Phi\hskip 0.5pt 1} 
(2\, \beta_{Y\hskip 0.5pt 3D} - \beta_{Y\hskip 0.5pt 3F}  + \beta_{Y\hskip 0.5pt 3G} ) 
+ 2\,   \gamma_{\Phi\hskip 0.5pt 1}{}^2
( \beta_{Y\hskip 0.5pt 3I} -   \beta_{Y\hskip 0.5pt 3J} ) \nn \\
& \hskip 8cm {}+  \beta_{Y\hskip 0.5pt 2A} ( 2\,  \gamma_{\Phi\hskip 0.5pt 1}  \beta_{Y\hskip 0.5pt 2B} 
-  \beta_{ Y \hskip 0.2pt 1}   \beta_{Y\hskip 0.5pt 2A} )    \, , \nn \\
&S_6= 2\, \beta_{ Y \hskip 0.2pt 1}{}^2 \gamma_{\Phi\hskip 0.5pt 3B} -  \beta_{ Y \hskip 0.2pt 1} 
\gamma_{\Phi\hskip 0.5pt 1} 
(2\, \beta_{Y\hskip 0.5pt 3A} -  \beta_{Y\hskip 0.5pt 3C} + 2\, \gamma_{\Phi\hskip 0.5pt 3D}) + 2 \,   
\gamma_{\Phi\hskip 0.5pt 1}{}^2 
( 2\, \beta_{Y\hskip 0.5pt 3D}  -  \beta_{Y\hskip 0.5pt 3K}  +  \gamma_{\Phi\hskip 0.5pt 3G} )  
 \nn \\ \noalign{\vskip -2pt} & \hskip 8cm {} 
- 2\,  \gamma_{\Phi\hskip 0.5pt 1} (  \beta_{Y\hskip 0.5pt 2A}{}^2 +   \beta_{Y\hskip 0.5pt 2B}  \, 
\gamma_{\Phi\hskip 0.5pt 2A} ) \, , \nn \\
& S_7= \beta_{ Y \hskip 0.2pt 1}{}^2\, \gamma_{\Phi\hskip 0.5pt 3C} -2\, \beta_{ Y \hskip 0.2pt 1}
 \gamma_{\Phi\hskip 0.5pt 1} 
(\beta_{3C} +  \gamma_{\Phi\hskip 0.5pt 3F}) + 4  \,   \gamma_{\Phi\hskip 0.5pt 1}{}^2 \beta_{Y\hskip 0.5pt 3K} 
 -   4\,  \gamma_{\Phi\hskip 0.5pt 1}  \beta_{Y\hskip 0.5pt 2A} \, \gamma_{\Phi\hskip 0.5pt 2B} =0 \, ,
   \end{align}
   and for the non planar diagrams just 2
\begin{align}
& S_8= \beta_{ Y \hskip 0.2pt 1} \beta_{Y\hskip 0.5pt 3L} -  \gamma_{\Phi\hskip 0.5pt 1}  ( \beta_{Y\hskip 0.5pt 3N} - 
2\,  \beta_{Y\hskip 0.5pt 3O} ) - 2\,  \beta_{Y\hskip 0.5pt 2C} \,  \beta_{Y\hskip 0.5pt 2A}  \, , \nn \\
& S_9= \beta_{ Y \hskip 0.2pt 1} \gamma_{\Phi\hskip 0.5pt 3I} + \gamma_{\Phi\hskip 0.5pt 1}  
( \beta_{Y\hskip 0.5pt 3M} - \beta_{Y\hskip 0.5pt 3O} )  -  \beta_{Y\hskip 0.5pt 2C} \, \gamma_{\Phi\hskip 0.5pt 2B}  \, .
\end{align}

\section{Scheme Variations for Scalar Fermion Theory to Three Loops}

\label{scheme}
 
 The coefficients appearing in the expansions of the $\beta$-functions and anomalous dimensions
 for a general scalar fermion theory are in general dependent on the choice of regularisation scheme.
 At $\ell$ loops possible scheme variations in  $\gamma_\phi{\!}^{(\ell)}, \, \gamma_\psi{\!}^{(\ell)}, \,
{\tilde  \beta}_y{\!}^{(\ell)}, \, {\tilde  \beta}_\lambda{\!}^{(\ell)}$ are determined in terms of arbitrary
parameters related to the expansions of  $\gamma_\phi{\!}^{(\ell-1)}, \, \gamma_\psi{\!}^{(\ell-1)}, \,
{\tilde  \beta}_y{\!}^{(\ell-1)}, \, {\tilde  \beta}_\lambda{\!}^{(\ell-1)}$. This depends on preserving the form of
the functions ${\tilde \beta}, \gamma$ in terms of contributions corresponding to 1PI and 1VI diagrams. 
Labelling the coefficients $\alpha$ in the expansion at $\ell$ loops  by $\gv,  \ell, r$, where here 
$\gv = \phi, \psi ,  y,  \lambda$, 
the general forms of the variations for $\alpha_g\equiv \alpha_{\gv \hskip 0.5pt \ell r}$, with $\alpha \to \gamma$ for 
$\gv = \phi, \, \psi$ and $\alpha \to \beta$ for $\gv \to y, \, \lambda$, are shown in appendix \ref{varG}
to  involve a sum over  contributions
\be
X_{g',g''} = - X_{g'',g'} = \alpha_{g'} \, \epsilon_{g''} -  \epsilon_{g'} \, \alpha_{g''}  \, , \qquad
g' = \gv' \hskip0.8pt\ell' r'  \, , \quad g''= \gv'' \hskip0.8pt \ell'' r'' \, ,
 \label{defX}
 \ee
where  $\{ \epsilon_g  \}$ are arbitrary parameters,  so that
 \be
 \delta \alpha_g = \sum_ {\genfrac{}{}{0pt}{3}{g',g''}{\ell'+\ell''=\ell, \gv'' = \gv}} 
 \hskip - 0.3cm  \N_g{}^{g'g''} \, X_{g',g''} \, , 
 \label{varA}
 \ee
 and  $\N_g{}^{g'g''}= - \N_g{}^{g''g'}$  are  integer coefficients. Of course one
 loop coefficients are scheme invariant and higher loops coefficients corresponding to primitive diagrams,
 which have a different topology and do not lead to integrals which have subdivergences, are also scheme 
 invariant. Any $e^g$ such that for a given $\ell$
 \be
  \sum_ {g, \, \ell'+\ell'' =\ell} \hskip - 0.2cm  e^g\,  \N_g{}^{g'g''} =0 \, , 
 \ee
 gives rise to a linear $\ell$ loop scheme invariant $\sum_g  e^g \alpha_g$.
 
 At two loops there are 10 possible $X$'s but only 7 appear in scheme variations as 
 $X_{\phi1,\lambda1b}$, $X_{\psi1,\lambda1a}$, $X_{y1,\lambda1a}$,  are not present, so we have
 \begin{align}
&  \delta\gamma_{\phi\hskip 0.5pt 2a} = \delta\gamma_{\psi\hskip 0.5pt 2b} = 
 \delta \beta_{y\hskip 0.5pt 2a}=\delta \beta_{y\hskip 0.5pt 2d}=\delta \beta_{y\hskip 0.5pt 2e}=
 \delta \beta_{y\hskip 0.5pt 2f}=
 \delta \beta_{\lambda\hskip 0.5pt 2a}=\delta \beta_{\lambda\hskip 0.5pt 2g}=0 \, , \nn \\
 & \delta\gamma_{\phi\hskip 0.5pt 2b}= 4\, X_{\psi 1,\phi 1} \, ,\quad 
 \delta\gamma_{\phi\hskip 0.5pt 2c}=2\,X_{y1,\phi1} \, , \quad
 \delta\gamma_{\psi\hskip 0.5pt 2a}=2\, X_{\phi 1,\psi1} \, , 
 \quad \delta\gamma_{\psi\hskip 0.5pt 2c}=2\, X_{y1,\psi 1}  \, ,  \nn \\
&  \delta \beta_{y \hskip 0.5pt 2b}= 2\, X_{\phi1,y1} \, , \quad 
\delta \beta_{y\hskip 0.5pt 2c}=2\, X_{\psi 1,y 1} \, , \quad
 \delta \beta_{\lambda\hskip 0.5pt 2b}= 4\,X_{\phi 1,\lambda 1a} \, ,\quad 
\delta \beta_{\lambda\hskip 0.5pt 2c}= 2\, X_{\lambda 1b,\lambda1a} \, ,\nn \\
& \delta \beta_{\lambda\hskip 0.5pt 2d}= X_{\lambda 1b ,\lambda 1a} \, ,\quad 
\delta \beta_{\lambda\hskip 0.5pt 2e}= 2\, X_{\psi 1,\lambda1b } \, ,\quad 
\delta \beta_{\lambda\hskip 0.5pt 2f}= X_{y1 ,\lambda 1b} \, .
\label{scheme2}
 \end{align}
 The cases $y2a, \, y2f, \, \lambda2g$ correspond to primitive diagrams and so the variation is
 necessarily zero. Apart from those coefficients which are individually invariant there are four
 linear scheme invariants. 
 
 At three loops the results for scheme variations separate into different groups. 
 There are six primitive three loop diagrams for the Yukawa $\beta$-function so that
\be
\delta \beta_{y\hskip 0.5pt 3f}= \delta \beta_{y\hskip 0.5pt 3l} =\delta \beta_{y\hskip 0.5pt 3\tilde w}
=\delta \beta_{y\hskip 0.5pt 3\tilde x}=\delta \beta_{y\hskip 0.5pt 3\tilde y}
=\delta \beta_{y\hskip 0.5pt 3\tilde z}= 0 \, ,
\ee
and there are cases where the individual terms are scheme invariant
\be
\delta \gamma_{\psi\hskip 0.5pt 3j}= \delta \beta \raisebox{-1.5 pt}{$\scriptstyle y{\hskip 0.5pt}3\tilde f$} = 
\delta \beta\raisebox{-1.5 pt}{$\scriptstyle y{\hskip 0.5pt}3\tilde l$} =\delta \beta_{y,3\tilde n}=  0 \, . 
\label{scheme0}
\ee
 For those diagrams containing
 the non planar subgraph corresponding to $\beta_{y2f}$ the variations are
 \begin{subequations}
 \begin{align}
&
\delta \beta_{y\hskip 0.5pt 3s} = 2\, X_{\phi 1,y\hskip 0.5pt  2f} \, ,\quad
\delta \beta_{y\hskip 0.5pt 3\tilde o} = \delta \beta_{y\hskip 0.5pt 3\tilde p}= 2\, X_{\psi 1,y \hskip 0.5pt 2f} \, , \nn \\
&   \delta\gamma_{\phi\hskip 0.5pt 3m}= 2\, X_{y 2f,\phi 1} \, ,  \quad 
\delta \gamma_{\psi\hskip 0.5pt 3p}= 2 \, X_{y2f,\psi 1} \, ,  \label{scheme3yaa} \\
&  \delta \beta_{y\hskip 0.5pt 3\tilde g} =\delta \beta_{y\hskip 0.5pt 3\tilde v}= X_{y\hskip 0.5pt  2f ,y1}\, ,
\quad
 \delta \beta_{y\hskip 0.5pt 3\tilde r}=\delta \beta_{y\hskip 0.5pt 3\tilde s}
 =\delta \beta_{y\hskip 0.5pt 3\tilde u}=  X_{y 1,y\hskip 0.5pt  2f}\, . 
 \label{scheme3yab}
\end{align}
 \end{subequations}
 There are evidently 3 linear invariants from \eqref{scheme3yaa} and 4 more from \eqref{scheme3yab}.
 Otherwise for the variations of the anomalous dimension contributions arising from planar diagrams we have
 \begin{subequations}
 \begin{align}
& \delta\gamma_{\phi\hskip 0.5pt 3a}=6\,X_{\lambda 1a ,\phi 2a} \, ,\quad 
\delta\gamma_{\phi\hskip 0.5pt 3b}=6\, X_{\phi 1,\phi 2a} \, ,\quad 
\delta\gamma_{\phi\hskip 0.5pt 3c}=3\, X_{\lambda 1b ,\phi 2a } +X_{y 2a ,\phi 1} \, ,\nn \\
& \delta\gamma_{\phi\hskip 0.5pt 3d}= 2\,  X_{\phi 1 ,\phi 2b} +4\, X_{\psi 2a,\phi 1} \, ,\quad
\delta\gamma_{\phi\hskip 0.5pt 3f}= 4\, X_{\psi 1,\phi 2b} \, ,\quad
 \delta\gamma_{\phi\hskip 0.5pt 3g}= 2 \, X_{\psi 1 ,\phi 2b} +4\, X_{\psi 2b,\phi 1} \, ,\nn \\
& \delta\gamma_{\phi\hskip 0.5pt 3h}= 2 \, X_{\psi 1,\phi 2b} \, , \label{scheme3phia}   \\
& \delta\gamma_{\phi\hskip 0.5pt 3e}= 2 \,X_{\phi 1,\phi 2c} + 2 \, X_{y 2b,\phi 1}\, ,\quad
 \delta\gamma_{\phi\hskip 0.5pt 3i}= 4\, X_{\psi 2c,\phi 1}+2\, X_{y 1,\phi 2b} \, , \nn \\
& \delta\gamma_{\phi\hskip 0.5pt 3j}= 4\, X_{\psi 1,\phi 2c} +X_{y 1 ,\phi 2b}+ 2\, X_{y2c ,\phi 1} \, ,\quad
\delta\gamma_{\phi\hskip 0.5pt 3k}= 2 \, X_{y 1,\phi 2c} + 2\, X_{y 2e ,\phi 1} \,, \nn \\
&\delta\gamma_{\phi\hskip 0.5pt 3l}= 2 \, X_{y1 ,\phi 2c} +4\, X_{y 2d, \phi 1}\,  , 
\label{scheme3phib}
\end{align}
\end{subequations}
and
 \begin{subequations}
\begin{align}
&\delta \gamma_{\psi\hskip 0.5pt 3a}= 2\, X_{\phi2a,\psi 1} \, ,\quad
\delta \gamma_{\psi\hskip 0.5pt 3b}= 2\, X_{y2a ,\psi1} \, ,\quad
\delta \gamma_{\psi\hskip 0.5pt 3c}= 4\,X_{\phi 1,\psi2a} \, ,\nn \\
& \delta \gamma_{\psi\hskip 0.5pt 3d}= 2\,X_{\psi1,\psi2a}+2\, X_{\phi 1,\psi2b} \, ,\quad
\delta \gamma_{\psi\hskip 0.5pt 3e}= 2\, X_{\phi 1,\psi2b} +2\, X_{\psi2a,\psi 1} \, ,\nn \\
& \delta \gamma_{\psi\hskip 0.5pt 3g}= 4\, X_{\psi 1,\psi 2a} + 2\, X_{\phi2b,\psi 1} \, ,\quad
 \delta \gamma_{\psi\hskip 0.5pt 3i}= 4\, X_{\psi 1 ,\psi2b} \, , \label{scheme3psia} \\
&\delta \gamma_{\psi\hskip 0.5pt 3f}= 2\, X_{\phi 1,\psi2c} + X_{y 1,\psi2a} + X_{y2b,\psi 1} \, ,\quad
 \delta \gamma_{\psi\hskip 0.5pt 3h}= 2\, X_{y1,\psi2a} + 2 \, X_{\phi2c ,\psi 1} \, ,\nn \\
& \delta \gamma_{\psi\hskip 0.5pt 3k}= 2\, X_{y 1,\psi2c} +2 \, X_{y2d , \psi 1} \, ,\quad
\delta \gamma_{\psi\hskip 0.5pt 3l}= 2\, X_{\psi2c ,\psi 1} +2\, X_{y 1,\psi 2b} \, , \nn \\
& \delta \gamma_{\psi\hskip 0.5pt 3m}= X_{y1,\psi2c} + X_{y2e,\psi 1} + X_{y2d ,\psi 1} \, ,\quad
 \delta \gamma_{\psi\hskip 0.5pt 3n}= 2 \, X_{\psi 1,\psi2c}+ X_{y1,\psi 2b}  +X_{y2c,\psi 1} \, ,\nn \\
&\delta \gamma_{\psi\hskip 0.5pt 3o}= 2\, X_{\psi 1,\psi2c} +2\, X_{y2c,\psi 1} \, .
\label{scheme3psib}
\end{align}
\end{subequations}
The remaining scheme variations are then
\begin{subequations}
\begin{align}
& \delta \beta_{y\hskip 0.5pt 3b}=\delta \beta_{y\hskip 0.5pt 3c}=X_{\lambda 1a,y2a}\, ,\quad
\delta \beta_{y\hskip 0.5pt 3d}=\delta \beta_{y\hskip 0.5pt 3e}=2\,X_{\phi1,y2a} \, ,\quad
\delta \beta_{y\hskip 0.5pt 3j}=2\,X_{\psi 1,y 2a} \, , \nn \\
&  \delta \beta_{y\hskip 0.5pt 3w} =  \delta \beta_{y\hskip 0.5pt 3x} =X_{\lambda 1b,y2a}\, , \label{scheme3ya} \\
& \delta \beta_{y\hskip 0.5pt 3a}=2\, X_{\phi 2a, y1 }\, ,\quad
\delta \beta_{y\hskip 0.5pt 3g}=\delta \beta_{y\hskip 0.5pt 3h}=X_{y2a,y1} \, ,\quad
\delta \beta_{y\hskip 0.5pt 3i}=\delta \beta_{y\hskip 0.5pt 3k}=X_{y1,y2a} \, ,\nn \\
&  \delta \beta_{y\hskip 0.5pt 3m}=4\,X_{\phi 1,y 2b}\, ,\quad
\delta \beta_{y\hskip 0.5pt 3n}=2\,X_{\phi 1,y2e}+X_{y 2b,y1 } \, ,\quad 
\delta \beta_{y\hskip 0.5pt 3o}=2\, X_{\phi 1,y 2e}+X_{y1 ,y 2b}  \, ,\nn \\
& \delta \beta_{y\hskip 0.5pt 3p}= 2\, X_{\phi 1,y 2c} + 2\,X_{\psi 1,y 2b}\, , \quad
\delta \beta_{y\hskip 0.5pt 3q}=2\, X_{\phi 1,y 2d}+X_{y2b,y 1} \, ,\quad
\delta \beta_{y\hskip 0.5pt 3r}=2\, X_{\phi 1,y 2d}+X_{y1,y2b} \, , \nn \\
& \delta  \beta_{y\hskip 0.5pt 3t}=2\, X_{\phi 1,y 2c}+2\,X_{\psi 2a,y1} \, , \quad
\delta \beta_{y\hskip 0.5pt 3u}=4\, X_{\psi 1,y 2b}+2\, X_{\phi 2b,y1} \, ,\nn \\
& \delta \beta_{y\hskip 0.5pt 3v}=2\, X_{y1,y 2b}+2\, X_{\phi 2c,y1} \, ,\quad
\delta \beta_{y\hskip 0.5pt 3y}=\delta \beta_{y\hskip 0.5pt 3z}=4 \,X_{\psi1 ,y2c} \, , \nn \\
& \delta \beta_{y\hskip 0.5pt 3\tilde a}= 2\, X_{\psi 1,y 2c}+2 \, X_{\psi 2b,y1} \, , \quad 
\delta \beta \raisebox{-1.5 pt}{$\scriptstyle y{\hskip 0.5pt}3\tilde b$} =
2\, X_{\psi 1,y 2e} + X_{y1,y2c}\, ,\quad
\delta \beta_{y\hskip 0.5pt 3\tilde c}=2\, X_{\psi 1,y 2e} +X_{y2c,y1} \, , \nn \\
& \delta \beta \raisebox{-1.5 pt}{$\scriptstyle y{\hskip 0.5pt}3\tilde d$}  = X_{y1,y2e}+X_{y2d,y1} \, ,\quad
\delta \beta_{y\hskip 0.5pt 3\tilde e}=2\, X_{\psi 2c,y1 }+2\, X_{y1,y 2c} \, , \nn \\
&  \delta \beta \raisebox{-1.5 pt}{$\scriptstyle y{\hskip 0.5pt}3\tilde h$} =
\delta \beta \raisebox{-1.5 pt}{$\scriptstyle y{\hskip 0.5pt}3\tilde i$}= 2\, X_{\psi 1,y 2d} + X_{y1,y2c} \, ,\quad
\delta \beta \raisebox{-1.5 pt}{$\scriptstyle y{\hskip 0.5pt}3\tilde j$} 
= \delta \beta \raisebox{-1.5 pt}{$\scriptstyle y{\hskip 0.5pt}3\tilde k$}  =
2\, X_{\psi1,y2d} +X_{y2c,y1}  \, ,\nn \\
&  \delta \beta_{y\hskip 0.5pt 3\tilde m}= X_{y1,y2d}+X_{y2e,y1}\, ,\quad
\delta \beta_{y\hskip 0.5pt 3\tilde q}=X_{y1,y2e}  +X_{y1,y2d}\, , \quad
 \delta \beta\raisebox{-1.5 pt}{$\scriptstyle y{\hskip 0.5pt}3\tilde t$} = 2\,X_{y1,y2d} \, .
\label{scheme3yb}
\end{align}
\end{subequations}
Each variation is necessarily such that the number of fermion loops is conserved.
The variations in \eqref{scheme3yaa},  \eqref{scheme3phia}, \eqref{scheme3psia} and \eqref{scheme3ya}, apart from 
that for $\beta_{y3x}$, correspond to the restriction to the $U(1)$ invariant case. 
The 21 variations in \eqref{scheme3phia}, \eqref{scheme3psia} and \eqref{scheme3ya} involve 14 different $X$'s
and there are 7 linear invariants. The 40 variations in \eqref{scheme3phib}, \eqref{scheme3psib} and \eqref{scheme3yb} involve 23 different $X$'s but there are 18 linear invariants since the equations
are invariant under 
\be
\delta X_{\phi1, \psi2c} = \delta X_{\psi1, y2b} = \rho \, ,  \quad
\delta X_{\phi1, y2c} = \delta X_{\psi1, \phi 2c} = \delta X_{y1, \psi 2a} = - \rho \, ,   \quad
\delta X_{y1, \phi 2b} = 2 \rho \, .
\label{Xrho}
\ee

Individual coefficients in the expansions of $\beta$ or $\gamma$,  besides those corresponding to primitive diagrams,
are scheme invariant when the associated vertex or propagator 
subgraphs are a nested sequence all of the same form. Examples appear in \eqref{scheme0}.
In this case $\gamma_{\psi 1}, \, \gamma_{\psi\hskip 0.5pt 2b}, \, 
 \gamma_{\psi\hskip 0.5pt 3j}$ correspond to rainbow diagrams and $ \beta_{y\hskip 0.5pt 1}, 
 \beta_{y\hskip 0.5pt 2e}, \, \beta \raisebox{-1.5 pt}{$\scriptstyle y{\hskip 0.5pt}3\tilde f$}$ correspond to vertex
 ladder diagrams. For these cases there are exact all orders results 
 \cite{Delbourgo2,Broadhurst2} obtained by solving quadratic equations
 \begin{align}
 \gamma_\psi \big |_{\rm rainbow} ={}&  \sqrt{1+ y^2} - 1 = \tfrac12 \, y^2-  \tfrac18 \, y^4 + 
 \tfrac{1}{16} \, y^6  - \tfrac{5}{128} \, y^8 + \dots \, , \nn \\
\beta_y/y \big |_{\rm ladder} ={}&  \sqrt{1+ 4 \, y^2} - 1 = 2 \, y^2-  2 \, y^4 + 
 4 \, y^6  - 10 \, y^8 + \dots \, .
 \end{align} 
 Further sequences of nested diagrams are  also associated with $  \beta_{y\hskip 0.5pt 1}, \, \beta_{y\hskip 0.5pt 2d}$ 
 and $ \beta\raisebox{-1.5 pt}{$\scriptstyle y{\hskip 0.5pt}3\tilde l$}$ or $ \beta_{y,3\tilde n}$ so these are
 necessarily scheme invariant.
 
 For the restriction to the $U(1)$ theory discussed in section 6, the scheme variations  in 
 \eqref{scheme2}, \eqref{scheme3phia},  \eqref{scheme3psia}, \eqref{scheme3ya}
 consistently restrict as they only involve the $\gamma$ and $\beta$-function coefficients relevant in that case.
 The sequence of nested diagrams for $ \gamma_\psi \big |_{\rm rainbow} $ remains in this case.
 The scheme variations may be restricted to $\N=1$ and $\N=\tfrac12$ supersymmetry. They are 
 consistent with the various constraints obtained earlier so long as all lower order conditions are
 imposed. For the latter case at two loops
 \begin{align}
  \delta \gamma_{\Phi 2B}=2\,  X_{Y 1,\Phi 1} \, , \quad  \delta \beta_{Y 2A}=2\,  X_{\Phi 1 , Y1} \, , \quad
   \delta \gamma_{\Phi 2A }  =  \delta \beta_{Y 2B} = \delta \beta_{Y 2C} =0 \, ,
   \label{Svar1}
 \end{align} 
 and at three loops individual scheme invariant coefficients correspond to
 \be
 \delta\gamma_{\Phi 3C} = \delta\beta_{Y3H}=\delta\beta_{Y 3J}=\delta\beta_{Y3P}=\delta\beta_{Y3Q}= 0 \, ,
 \ee
 with $ \beta_{Y3P}, \ \beta_{Y3Q}$ arising from primitive diagrams while $\gamma_{\Phi 3C}$, along with
 $\gamma_{\Phi 1} , \, \gamma_{\Phi 2A} $,  forms part of a sequence of nested rainbow diagrams.
 For planar contributions
 \begin{align}
& \delta \gamma_{\Phi3A}=2\, \delta\gamma_{\Phi 3B}= 4\, X_{\Phi 1,\Phi 2A} \, , \quad
\delta\gamma_{\Phi 3D}=  4\, X_{\Phi 1,\Phi 2B} +X_{Y1,\Phi 2A}+ 2\, X_{Y2A,\Phi 1} \, ,\nn \\
& \delta\gamma_{\Phi 3E}= 2\, X_{\Phi 1,\Phi 2B} + 2\, X_{Y2A,\Phi 1} \, ,\quad
\delta\gamma_{\Phi 3F}= 4\, X_{\Phi 2B,\Phi 1} + 2 \, X_{Y1,\Phi 2A} \, ,\nn \\
& \delta\gamma_{\Phi 3G}=2\, X_{Y 1,\Phi 2B} + 4\, X_{Y2B,\Phi 1} \, ,\quad
 \delta\gamma_{\Phi 3H}=2\, X_{Y1,\Phi 2B} +2\, X_{Y2B,\Phi 1} \, , \nn \\
& \delta\beta_{Y3A}=\delta\beta_{Y3B}=4\, X_{\Phi 1,Y 2A} \, ,\quad
\delta\beta_{Y3C}= 4\, X_{\Phi 1,Y2A}+ 2\, X_{\Phi 2A ,Y 1} \, ,\nn \\ 
& \delta\beta_{Y3D}=\delta\beta_{Y3G}=2\, X_{\Phi 1,Y2B} +X_{Y1, Y2A} \, ,\quad
\delta\beta_{Y 3E}=\delta\beta_{Y 3F}= 2\, X_{\Phi 1,Y 2B} +X_{Y2A,Y 1} \, ,\nn \\
&  \delta\beta_{Y3I}=2\, X_{Y1, Y2B} \, ,\quad
\delta\beta_{Y3K}=2\, X_{Y 1, Y 2A } + 2\, X_{\Phi 2B, Y1} \, ,\quad
\label{Svar2}
\end{align}
and for non planar
\begin{align}
& \delta\gamma_{\Phi 3I}= 2\, X_{Y2C,\Phi1} \, , \quad
\delta\beta_{Y3L}=4\, X_{\Phi 1,Y 2C}\, , \nn \\
& \delta\beta_{Y3M}= X_{Y1,Y2C} \, , \quad \delta\beta_{Y 3N}= 2\, X_{Y1,Y2C}\, , 
\quad \delta\beta_{Y3O}= X_{Y2C, Y1}   \, .
\label{Svar3}
\end{align}
From \eqref{Svar1} there is one linear invariant, $ \gamma_{\Phi 2B}+ \beta_{Y 2A}$. \eqref{Svar2} 
contains 8 independent $X$'s and  15 equations giving 7  linear invariants
whereas in \eqref{Svar3}  5 equations and 2  $X$'s lead to 3 linear invariants.

These results may be used to verify the invariance of the consistency conditions obtained in section 8.
The variations are either identically zero, using the antisymmetry of $X$, or lead to antisymmetrised
products of three $X$'s, 
\be
Y_{g_1,g_2,g_3} = \alpha_{g_1}\, X_{g_2,g_3} +  \alpha_{g_3}\, X_{g_1,g_2} +  \alpha_{g_2}\, X_{g_3,g_1} \, ,
\label{defY}
\ee
with $g_i = \gv_i\ell_ir_i$. Given the definition of $X$ in \eqref{defX} necessarily $Y_{g_1,g_2,g_3} =0$
and, from the antisymmetry of $ X_{g_1,g_2}$, there is the identity
\be
\alpha_{g_1} \, Y_{g_2,g_3,g_4}  - \alpha_{g_2} \, Y_{g_3,g_4,g_1}  + \alpha_{g_3} \, Y_{g_4,g_1,g_2}
- \alpha_{g_4} \, Y_{g_1,g_2,g_3}  = 0 \, ,
\label{bY}
\ee
and hence
\be
\alpha_{g_1} \, Y_{g_2,g_3,g_4}  - \alpha_{g_2} \, Y_{g_3,g_4,g_1}  \not\supset   X_{g_3,g_4}\, .
\label{Ycube}
\ee
Apart from linear invariants there are also possible quadratic invariants 
\be
Q_\kappa = \kappa^{g_1 g_2} \alpha_{g_1} \alpha_{g_2} \, , \quad  \kappa^{g_1 g_2} = \kappa^{g_2 g_1} \, ,
\ee
if $\kappa^{g_1g_2}$ is such that
\be
{\textstyle \sum_g }\; \kappa^{g_1 g} \N_g{}^{g_2 g_3} = F^{g_1 g_2 g_3} =  F^{[g_1 g_2 g_3]} \, , 
\ee
as then $\delta Q_\kappa = \frac23 \, \sum_{g_1,g_2,g_3}\,  F^{g_1 g_2 g_3} \,  Y_{g_1,g_2,g_3}$.
Higher order invariants are also possible as demonstrated later. 

Applying this for \eqref{scheme2} there are two quadratic invariants obtained from $Y_{\phi1,\psi1, y_1}$ and
$Y_{\psi 1, y1, \lambda1b}$. However as a consequence of \eqref{Ycube} 
\begin{align}
\gamma_{\psi1} &Y_{\phi1, \lambda1a, \lambda1b} - \beta_{\lambda1a} Y_{\phi1, \psi1, \lambda 1b} \nn \\
&{} = \gamma_{\psi1} \big ( \gamma_{\phi1} \,X_{\lambda1a, \lambda1b} 
+  \beta_{\lambda1b}\, X_{\phi1, \lambda1a}\big )
 - \beta_{\lambda1a} \big ( \gamma_{\phi1} \,X_{\psi1, \lambda1b} +  \beta_{\lambda1b} \,X_{\phi1,\psi1}\big ) \, ,
\end{align}
leads to a further cubic invariant. At the next order
from \eqref{scheme3yaa}, \eqref{scheme3yab} there are three possible $Y$'s,
$Y_{\phi1,\psi1,y2f}$, $Y_{\phi1,y1,y2f}$, $Y_{\psi1,y1,y2f}$, which would lead to three potential 
quadratic invariants. Nevertheless these are not independent due to \eqref{bY} and so there remain
two quadratic invariants for the non planar coefficients.
From \eqref{scheme3phia}, \eqref{scheme3psia} and \eqref{scheme3ya} we can construct
\begin{align}
& Y_{\phi1, \psi1, g} \, ,  \qquad g= \phi2a, \, \phi2b, \, \psi2a, \, \psi2b, \, y2a \, , \nn \\
& Y_{\phi1, \lambda1a, g},  \ Y_{\psi1, \lambda1b, g},  \ Y_{\lambda1a, \lambda1b, g} \, , \qquad g= \phi2a, \, y2a \, ,
\end{align}
which would appear to give 11 quadratic invariants.
These are not independent due to the identity
\begin{align}
& \gamma_{\psi1} \big ( \gamma_{\phi1} Y_{\lambda1a, \lambda1b, g} +   \beta_{\lambda1b} \, 
 Y_{\phi1, \lambda1a,  g}\big ) - \beta_{\lambda 1a} \, \big ( \gamma_{\phi1}  Y_{\psi1,\lambda1b, g} 
 + \beta_{\lambda1b} Y_{\phi1, \psi1,g} \big )\nn \\
 &{}= \alpha_g \big ( \gamma_{\psi1} Y_{\phi1,\lambda1a,\lambda1b} - \beta_{\lambda1a} \, Y_{\phi1, \psi1,\lambda1b} 
\big ) \, , \quad g= \phi2a, y2a \, , \
 \alpha_g = \gamma_{\phi2a}, \beta_{y2a} \, ,
\end{align}
so there remain 9 quadratic invariants when $U(1)$ symmetry is imposed. 
In the general case there are
additional invariants flowing from  \eqref{scheme3phib}, \eqref{scheme3psib} and \eqref{scheme3yb}.
 \begin{subequations}
\begin{align}
& Y_{\phi1, \psi 1,g }\, , \ Y_{\phi1, y 1,g }\, , \ Y_{\psi1, y 1,g }\, , \,  \quad 
g= \phi2c, \ \psi2c , \ y2b, \ y2c, \ y2d, y2e \, , \label{Ya}  \\
&  Y_{\phi1, y1,g }\, , \ Y_{\psi1, y 1,g }\, , \qquad g  = \phi2a, \, \phi2b, \, \psi2a, \, \psi2b , \ y2a \, .
\label{Yb} 
\end{align}
\end{subequations} 
These are not independent due to \eqref{bY}. In \eqref{Ya} we may then reduce to two sets of 6 and in \eqref{Yb} to
one set of 5. Possibilities are further restricted by requiring results are independent of $\rho$ in \eqref{Xrho}
which leaves 16. There are then 11 independent $Y$'s  for which there is no constraint, a possible basis is given by
\begin{align} 
&  Y_{\phi1, y1, \phi2a }, \   Y_{\phi1, y1, \phi2c }, \   Y_{\phi1, y1, \psi2b }, \   Y_{\psi1, y1, \psi2c }, \
    Y_{\phi1, y1, y2a }, \   Y_{\phi1, y1, y2b }, \nn \\
&      Y_{\psi1, y1, y2c }, \   Y_{\phi1, \psi1, y2d }, \   Y_{\phi1, y1, y2d }, \
    Y_{\phi1, \psi1, y2e}, \   Y_{\phi1, y1, y2e } \, ,
\end{align}
and 5 involving pairs of $Y$'s formed from $Y_{\phi1, \psi1, g}$ or $Y_{\psi1, y1, g}$, $g= \phi2c, \, y2b$, 
$Y_{\phi1, \psi1, g}$ or $Y_{\phi1, y1, g}$, $g=\psi2c, \, y2c$, 
$Y_{\phi1, y1, g}$ or $Y_{\psi1, y1, g}$, $g=\psi2a, \, \phi2b$
which are $\rho$ invariant. To achieve this \eqref{Xrho} implies
\begin{align}
& - \delta Y_{\phi1, \psi 1, \phi2c} =  \delta Y_{\phi1, \psi 1, y2b}  = - \delta Y_{\phi1, y1, \psi2a} =  
\gamma_{\phi1} \, \rho \, ,  \qquad \delta Y_{\phi1, y1, \phi 2b}  = 2\, \gamma_{\phi1} \, \rho \, , \nn \\
& -  \delta Y_{\phi1, \psi1, \psi2c} =  \delta Y_{\phi1, \psi 1, y2c}  = -  \delta Y_{\psi1, y1, \psi 2a} = 
\gamma_{\psi1} \, \rho \, ,
\quad  \delta Y_{\psi1, y1, \phi 2b} = 2\, \gamma_{\psi1} \, \rho \, , \nn \\
&- \delta Y_{\phi1, y1, \psi2c} =  \delta Y_{\phi1, y1, y2c}  =  \delta Y_{\psi1, y1, \phi2c} = 
- \delta Y_{\psi1, y1, y2b}  =  \beta_{y1} \, \rho  \, .
\end{align}

In \eqref{Svar2} $Y_{\Phi1,Y1,g}$, $g=\Phi2A, \, \Phi2B, \, Y2A, \, Y2B$, lead to four quadratic invariants 
and from \eqref{Svar3} $Y_{\Phi1,Y1,Y2C}$ to one more.

The various consistency conditions must be scheme invariant. We here check this by reducing their
variations to sums of $Y$'s which then show how they can be expressed in terms of quadratic invariants.
At lowest order the variations of \eqref{Brel}, using \eqref{scheme2}, are just
\be 
\delta B_1 = 0\, , \quad \delta B_2 = 4 \, Y_{ \phi 1, \psi 1, y1} \, , \quad 
\delta B_3 = 0\, , \quad \delta B_4 = 2\, \beta_{y1} \, Y_{ \phi 1, \psi 1, y1} \, ,
 \ee
 and for the non planar conditions \eqref{Crel} at the next order from \eqref{scheme3ya}
 \begin{align}
&  \delta C_1 =  \delta C_2 = \delta C_3 = 0  \, , \nn \\
 &  \delta C_4 =  - \delta C_5 = 
 2 \, Y{\!}\raisebox{-1.5 pt}{$\scriptstyle \psi 1, y1, y2f$} \, , \quad
\delta C_6 =  - \delta C_7 = 2 \, Y{\!}\raisebox{-1.5 pt}{$\scriptstyle \phi 1, y1, y2f$} \, ,
\end{align}
and for the variations of \eqref{Drela}, \eqref{Drelb}, \eqref{Drelc}
\begin{align}
& \delta D_1 =  \delta D_2 = \delta D_5 = \delta D_7= 0 \, , \quad \delta D_3 = -2 \,  Y_{ \phi 1, \psi 1, y2a} \, , \quad
\delta D_4 = \delta D_9 =  -2 \,  Y_{ \psi 1, y 1, y2a}  \, , \nn \\
&  \delta D_6 = - 12 \,  Y_{ \phi 1, \phi 2a,  \lambda 1a} \, , \quad
 \delta D_8=  -2 \,  Y_{ \phi 1, y 1, y2a}  \, , \quad  
 \delta D_{10} = 2 \,  Y_{ \phi 1, y 1, y2a} +  6 \,  Y_{ \phi 2a, y 1, \lambda1b} \, , \nn \\
&  \delta D_{11} = 2 \, \gamma_{\psi 1}  Y_{ \phi 1, \phi 2a , y1} - 2  \,  \beta_{y1} Y_{\phi 1, \phi 2a, \psi 1} \, , \quad
  \delta D_{12}  = - 6 \,  \gamma_{\phi 2a} \,  Y_{ \psi 1, y1 ,  \lambda 1b} \, , \nn \\
 & \delta D_{13} = 12 \, \gamma_{\phi 1} \, Y_{ \phi 2a, \psi 1 , \lambda 1b} - 12  \,  \beta_{\lambda1b} \,
  Y_{\phi 1, \phi 2a, \psi 1} + 4 \, B_1 \, X_{ \phi 1, \psi 1} \, , \nn \\
   & \delta D_{14} = \gamma_{\phi 1} \,  Y_{ y1, y2a  , \lambda 1a}  
   -  3\, \gamma_{\phi 2a} \,  Y_{ y1, \lambda1a  , \lambda 1b}  \nn \\
   \noalign{ \vskip -1pt}
   &\hskip 1.3cm {} + 
    6 \, \beta_{y 1} \, Y_{ \phi 2a, \lambda1a , \lambda 1b} 
    - 2  \,  \beta_{y1} \,  Y_{\phi 1, y 2a, \lambda 1a} 
    +  2  \,  \beta_{y2a} \,  Y_{\phi 1, y 1, \lambda 1a} -  B_1 \, X_{ y1, \lambda 1a} \, , \nn \\
&    \delta D_{15} = 0 \, , \nn \\
& \delta D_{16} = 4\,  \gamma_{\phi 1} \big (  Y_{ \phi1, y1  , y{2c}}  +  Y_{ \phi 1, \psi{2c}  , y1}  \big )
  + 2  \,  \beta_{y1} \big (   Y_{\phi 1, \phi  2b, y1}   -  2\, Y_{ \phi 1, \psi{2a}  , y1}  \big ) \nn \\
   \noalign{ \vskip -1pt}
   &\hskip 1.3cm {} -
    4  \, \gamma_{\phi 2c} \, Y_{ \phi 1, \psi 1 , y 1 }   
  -  2 \, B_2 \, X_{  \phi 1, y1 } \, , \nn \\    
  &  \delta D_{17} = -4  \, \gamma_{\phi 1}  Y_{ \phi 1,  y1,y2d} + 2  \,  \beta_{y1} \big ( Y_{\phi 1, \phi 2c,  y1}
  + Y_{\phi 1, y1 ,  y2b}\big )  \, , \nn \\
  &  \delta D_{18} = 4  \, \gamma_{\phi 1}  Y_{ \psi 1,  y1, y2c} + 4\,  \gamma_{\psi 1} Y_{\phi 1, \psi 2c,  y1}
- 2  \, \beta_{y1} \big ( 2\,  Y_{ \phi 1,  \psi 2b, y 1 } + Y_{\phi 2b,  \psi 1 ,  y 1} \big ) 
 - 2\, B_2 \, X_{ \psi 1, y1} \, , \nn \\
 & \delta D_{19} = 2\, \gamma_{\phi 1} \,  Y_{ \psi 1, \psi 2c  , y1 }  
   +  2\, \gamma_{\psi 1} \,  Y_{ \phi 1, y1 , y 2c}  +  2\, \beta_{y 1} \,  Y_{ \phi 1, \psi 2b  , y1 }  \nn \\
   \noalign{ \vskip -1pt}
   &\hskip 1.3cm {}  
        - 2 \big  ( \gamma_{\psi 1}( \beta_{y2d} - \beta_{y2e} ) + 2 \, \beta_{y1} \, \gamma_{\psi 2c} 
        \big )\big / \beta_{y1} \,  Y_{\phi 1, \psi  1, y1} +
        2 \big (  B_3 \, X_{ \phi 1 ,y1} +  B_4 \, X_{ \psi 1,y1} \big ) / \beta_{y1}  \, , \nn \\
&  \delta D_{20} = 4  \, \gamma_{\psi 1} \big ( Y_{ \psi 1,  y1, y2b} -  Y_{\phi 2c, \psi 1,  y1} \big ) 
+  2 \, \beta_{y1} \big ( 2\,  Y_{ \psi 1,  \psi 2a, y 1 } +  Y_{\phi 2b,  \psi 1 ,  y 1}  \big ) \nn \\
\noalign{ \vskip -1pt}
   &\hskip 1.3cm {}   
-  4  \, \beta_{y 2 c}  Y_{ \phi 1,  \psi 1 , y 1 }  +  2\, B_2 \, X_{ \psi 1, y1} \, , \nn \\
&  \delta D_{21} = 2 \, \gamma_{\psi 1} \big ( Y_{ \psi 1, \psi 2c , y1} +Y_{\psi 1, y1, y2c} \big ) \, , \nn \\
&  \delta D_{22} = - 2 \, \gamma_{\psi 1} \big ( Y_{ \psi 1, y1 , y 2d } +Y_{\psi 1, y1, y2e } \big )
+ 2 \, \beta_{y1}  \big ( Y_{ \psi 1, \psi 2c , y1} +Y_{\psi 1, y1, y2c} \big ) \, , \nn \\
 &  \delta D_{23} = 4  \, \gamma_{\phi 1}  Y_{ \psi 1,  y1,y2d} - 2  \,  \beta_{y1} \big ( Y_{\psi 1, y1,  y 2b}
  + Y_{\phi 1, y1 ,  y2c} \big ) \, , \nn \\
&  \delta D_{24} = 2 \, \beta_{y 1} \big ( Y_{ \phi 1, y1 , y2d} -Y_{\phi 1, y1, y2e} \big ) \, , \quad
 \delta D_{25} = 4 \, \gamma_{\psi 1}  Y_{ \psi 1, y1 , y2e} - 4 \, \beta_{y1} Y_{\psi 1, y1, y2c}  \, , \nn \\
&  \delta D_{26} = 2 \, \beta_{y 1} \big ( Y_{ \psi 1, y1 , y2d} -Y_{\psi 1, y1, y2e} \big ) \, , \nn \\
& \delta D_{27} = -8  \, \beta_{y 1} \big (  Y_{ \phi 1,  \psi 1, y2d} + Y_{ \phi 2c,  \psi 1, y1 } \big ) 
-8 \,  \gamma_{\psi 1} Y_{\phi 1, y 1,  y2e}
+  4  \, \beta_{y1}\big (   Y_{ \phi 1,  y1 , y 2c } -   Y_{ \psi 1,  y1 , y 2b } \big ) \nn \\
\noalign{ \vskip -1pt}
   &\hskip 1.3cm {}   + 8 ( 2\, \beta_{y2d} + \beta_{y2e} )\,   Y_{ \phi 1,  \psi 1 , y 1 }  \, , \nn \\
  &  \delta D_{28} = 4  \, \beta_{y 1}  Y_{ \psi 1, \psi 2c ,y1} - 2 \, \gamma_{\psi 1} 
  \big (  Y_{\psi 1, y1,  y 2d} +  Y_{\psi 1, y1,  y 2e} \big ) \, , \nn \\
  & \delta D_{29} = 2  \, \beta_{y 1} \big ( Y_{ \psi 1, y1, y2b} -  Y_{ \phi 1, y1, y2c} + 2 \, Y_{\phi 1, \psi 2c ,y1} \big )
  - 2 \, \gamma_{\phi 1} \big ( 3\, Y_{ \psi 1, y1, y2d}  + Y_{ \psi 1, y1, y2e}  \big ) \nn \\
  \noalign{ \vskip -1pt}
   &\hskip 1.3cm {}+ 4 \, \gamma_{\psi1} \,   Y_{ \phi 1, y1, y2e} - 4  \,  \beta_{y 2d } \, Y_{\phi 1, \psi 1,  y1} \, , \nn \\
   &  \delta D_{30} = 2  \, \beta_{y 1} \big (  Y_{ \phi 1, \psi 1 ,y2d } + Y_{ \phi 1, \phi 1 ,y2e } \big ) 
  +  2 \, (\beta_{y 2d} + \beta_{y2e} ) \,   Y_{\phi 1, \psi 1,  y 1} \, , \nn \\
  &  \delta D_{31} = 4  \, \beta_{y 1}{\!}^2 \,  Y_{ \psi 1, \psi 2b ,y1 } +  
  2\,  \beta_{y 1} \gamma_{\psi 1}  \big ( Y_{ \psi 1, \psi 2c  ,y1 } +  Y_{ \psi 1, y1  ,y2c}  \big ) 
  -  2 \, \gamma_{\psi 1}{\!}^2 \,   Y_{\psi 1, y 1,  y 2e} + 4\, B_3\,  X_{\psi1,y1} \, , \nn \\
   &  \delta D_{32} = -2  \, \beta_{y 1}{\!}^2 \,  Y_{ \phi 2b, \psi 1  ,y1 } -  
  4\,  \beta_{y 1} \gamma_{\psi 1}  \big ( Y_{ \phi 1,  \psi 1 ,y2e } +  Y_{ \psi 1, y1  ,y2b}  \big ) \nn \\
  \noalign{ \vskip -1pt}
   &\hskip 1.3cm {}  
  -  2 \, \gamma_{\psi 1} \, (   \beta_{y 2d } - 3\,  \beta_{y 2e } ) \,Y_{\phi 1,  \psi 1,  y 1} - 2\, B_2\, \beta_{y1}  X_{\psi1,y1} \, , \nn \\
   &  \delta D_{33} = -4  \, \beta_{y 1}{\!}^2 \,  Y_{ \phi 2b, \psi 1  ,y1 } + 
  4\,  \beta_{y 1} \gamma_{\psi 1}  \big ( Y_{ \phi 1,  \psi 2c ,y1 } -  Y_{ \psi 1, y1  ,y2b}  \big ) 
  +8\,  \beta_{y 1} \gamma_{\phi 1}  \, Y_{ \psi 1,  y1 ,y2c }
  \nn \\
  \noalign{ \vskip -1pt}
   &\hskip 1.3cm {}  
-  4 \, \gamma_{\psi 1}{\!}^2 \,   Y_{\phi 1, y 1,  y 2e}  
+ 4  \, \gamma_{\psi 1}\, (   \beta_{y 2d } +  \beta_{y 2e } ) \,Y_{\phi 1,  \psi 1,  y 1}  - 4\, B_2\,  \beta_{y1}\, X_{\psi1,y1} \, , \nn \\
&  \delta D_{34} = 2  \, \beta_{y 1}{\!}^2 \,\big (   Y_{ \phi 1, \psi 2b  ,y1 } +   Y_{ \psi 1, \psi 2a  ,y1 } \big ) 
-2\, \beta_{y 1} \gamma_{\phi 1}  \, Y_{ \psi 1,  y1 ,y2c }
 -  2 \,  \beta_{y 1} \gamma_{\psi 1} \, Y_{ \phi 1,  y1 , y2c } \nn \\
  \noalign{ \vskip -1pt}
   &\hskip 1.3cm {}  
  -  2 \, \gamma_{\psi 1} (  3\, \beta_{y 2d } -   \beta_{y 2e } ) \,Y_{\phi 1,  \psi 1,  y 1} 
  + 2\, B_3\,  X_{\phi1,y1} + 2\, B_4\,   X_{\psi1,y1}\, , \nn \\
  &  \delta D_{35} = 4 \, \beta_{y 1}{\!}^2 \,   Y_{ \phi 1, \psi 2a  ,y1 }  
-   2\, \beta_{y 1} \gamma_{\phi 1}  \, \big (  Y_{ \phi 1,  y1 ,y2c } + Y_{\phi 2c , \psi 1, y1} \big ) 
 +  4 \,  \beta_{y 1} \gamma_{\psi 1} \, Y_{ \phi 1,  y 1 , y2b } \nn \\
  \noalign{ \vskip -1pt}
   &\hskip 1.3cm {}  
-2   \gamma_{\phi 1}{\!}^2 \,  \big ( 2\,  Y_{\psi 1, y 1,  y 2d}  +  Y_{\psi 1, y 1,  y 2e}  \big ) 
+ 4 \,  \gamma_{\phi 1} \gamma_{\psi 1} \, Y_{ \phi 1,  y 1 , y2e }
 \nn \\
  \noalign{ \vskip -1pt}
   &\hskip 1.3cm {}  
  + 2 \, \big (  \gamma_{\phi 1} \, \beta_{y 2d }  +  \beta_{y 1 } \, \gamma_{\phi 2c} \big ) \,Y_{\phi 1,  \psi 1,  y 1} 
  + 2\, B_4\,   X_{\phi1,y1}   \, .
\end{align}

For the constraints obtained in the reduced $\N=\frac12$ theory listed in sub section \ref{Sred}, corresponding to a
$\Phi^3$ interaction, then with \eqref{Svar1}, 
 \eqref{Svar2},  \eqref{Svar3},
 \begin{align}
 & \delta S_0 =  \delta S_1 =   \delta S_2  = 0 \, , \quad \delta S_3 = - 2\, Y_{\Phi 1, Y1, Y2A} 
 \, , \quad \delta S_4 = -  6\, Y_{\Phi 1, Y1, \Phi 2A} \nn \\
 & \delta S_5 = - 4\,  \beta_{Y1}\, Y_{\Phi 1, Y1, Y2A}  + 4\,  \gamma_{\Phi1}\, Y_{\Phi 1, Y1, Y2B}   \, , \nn \\
&   \delta S_6 = - 4\,  \beta_{Y1}\, Y_{\Phi 1,Y1, \Phi 2A}  + 8\,  \gamma_{\Phi1}\, Y_{\Phi 1, Y1,\Phi 2B}   
  + 4\, S_0\, X_{\Phi 1, Y1} \, , \nn \\
  & \delta S_7 = 8 \, \gamma_{\Phi1}\, \big ( Y_{\Phi 1, Y1, Y2A}  -   Y_{\Phi 1,Y1, \Phi 2B}  \big )   \, ,  \nn \\
  & \delta S_8 = -4\, Y_{\Phi 1, Y1 , Y2C}  \, , \quad \delta S_9 = 2\, Y_{\Phi 1, Y1 , Y2C}  \, .
 \end{align}

\section{Conclusion}\label{sec:dis}

The detailed results given here for $\beta$-functions and anomalous dimensions 
correspond to a  $\overline {M\!S}$ regularisation scheme. As is well known
attempting to extend supersymmetric theories away from their natural dimension
is problematic and generally inconsistent and these issues affect any variant of dimensional
regularisation \cite{Siegel,Avdeev,Avdeev2,Stockinger}. For $\N=1$ 
supersymmetry and scalar fermion theories, without gauge fields, 
there are manifestly supersymmetric
regularisation schemes and  potential problems with  $\overline {M\!S}$ 
arise only beyond three loops so long as the normalisation of fermion traces is
chosen appropriately. 
These issues become significantly more severe for what we term $\N = \tfrac12$ symmetry in this
paper. Traces of three or more odd numbers of three dimensional Dirac gamma matrices 
are potentially non zero due to the appearance of the three dimensional   antisymmetric symbol. This is not
relevant for a fermion loop with three external scalars, due to momentum conservation, 
but such contributions are present if a  fermion loop has five external scalar  lines. Of course analogous
problems with $\gamma_5$ are present with perturbative calculations using ${\overline{M\!S}}$ for chiral fermions.
Such problems also  arise  in four dimensional chiral gauge theories for loops with two external vector lines
and two external scalars and such loop diagrams contribute at four loops to 1PI  contributions with two external
vector lines  and also to the Yukawa $\beta$-function \cite{Poole5}. In \cite{Poole5} it was shown how
consistency with the $a$-function helps resolve some analogous $\gamma_5$ issues.
In three dimensions similar potential problems arising for  five vertex fermion loops 
as  sub graphs occur at four loops in the  Yukawa vertex renormalisation where the relevant diagrams are of the form
\be
\begin{tikzpicture}
    \draw[black, thick] (2.em, 0.em) -- (3em, 1em) -- (5em, 1em) -- (6.em,0.em);
    \node at (3.65em, 1.7938em)  [circle,fill,inner sep=.1em] (n1){};
    \node at (4.35em, 1.7938em) [circle,fill,inner sep=.1em] (n2){};
    \draw[black, thick, densely dashed] (4em, 4.em) -- (4em, 3.1em);
    \draw[black, thick, densely dashed] (3.55em, 1.em) -- (n1);
     \draw[black, thick, densely dashed] (4.45em, 1.em) -- (n2);
    \draw[black, thick, densely dashed] (3em, 1.em) -- (3.3em, 2.4em);
    \draw[black, thick, densely dashed] (5em, 1.em) -- (4.7em, 2.4em);
    \draw[black, thick] (4em, 2.4em) circle (.7em);
        \node at (3.55em, 1em)[circle,fill,inner sep=.1em]{};
         \node at (4.45em, 1em)[circle,fill,inner sep=.1em]{};
    \node at (3em, 1em)[circle,fill,inner sep=.1em]{};
    \node at (5em, 1em)[circle,fill,inner sep=.1em]{};
    \node at (3.3em, 2.4em)[circle,fill,inner sep=.1em]{};
    \node at (4.7em, 2.4em)[circle,fill,inner sep=.1em]{};
     \node at (4em, 3.1em)[circle,fill,inner sep=.1em]{};
    \end{tikzpicture} \ ,
\label{anomalous}
\ee
together with various permutations of the internal vertices on the fermion loop.
Such diagrams are primitive since there are no subdivergences when evaluated in four
dimensions with some prescription for the contraction of two three dimensional $\epsilon$ symbols.
A procedure  for obtaining such contributions was described in \cite{Zerf2}. However starting from four dimensional
Dirac or Majorana fermions contributions related to \eqref{anomalous} are absent \cite{Boyack}. The four dimensional
fermion splits into two three dimensional fermions whose Yukawa couplings have the opposite sign, as shown here
in appendix \ref{Majorana}.

In terms of the discussion of scheme changes and forming scheme invariants an alternative though
equivalent approach is obtained within the framework of the Hopf algebra approach
to Feynman diagrams \cite{Connes1,Connes2,Panzer}.
The requirement of scheme invariance is identical with finding linear sums of graphs
such that the Hopf algebra coproduct is cocommutative. A potentially interesting possibility
is whether there is any extension of the Hopf algebraic approach to deriving consistency conditions, such as those
considered here in section 9, which might avoid some of the rather tortuous analysis required
here and in \cite{Poole:2019kcm}. 

The results obtained here suggest that there are potentially many interesting fixed points in scalar fermion
theories once more than three  scalar fields are allowed and the condition that there is just a single 
Yukawa coupling is relaxed. Finding a large $n_f$ expansion for such theories may be tractable.

\section*{Acknowledgements}

We are very grateful to Colin Poole for sharing with us many of the details of his calculations with
Anders Thomsen which appeared in \cite{Poole:2019kcm}.  HO is happy to acknowledge discussions
with Andy Stergiou which helped elucidate many issues. We  also would  like to thank Shabham Sinha for pointing
out various typos in the first version of this paper.

\vskip 1cm

\section{Note Added, Fixed Points with One Scalar or One Fermion Field}

Since this paper was finished Pannell and Stergiou \cite{StergiouF} have investigated in detail possible fixed 
points in fermion scalar theories with low numbers of scalars and fermions. Many possiblities were discovered.
Here we illustrate some results for either one scalar or one fermion which can be obtained quite easily.

For a single scalar field and $n$ two component real fermions the couplings are just $\lambda$ and
$y$ a real symmetric $n\times n$ matrix.  For four dimensional Majorana fermions $n$ should  be even.
At one loop order the $\beta$-functions in $4-\vep$ dimensions reduce,
with the usual rescaling to eliminate factors of $4\pi$, to
\begin{align}
\beta_y = {}& - \tfrac12 \, \vep \, y + 3 \, y^3 + \tfrac12 \, y \, \tr(y^2) \, , \nn \\
\beta_\lambda = {}& - \vep \, \lambda + 3 \, \lambda^2  + 2 \, \lambda \, \tr(y^2) - 12 \, \tr (y^4) \, .
\end{align}
To solve $\beta_y=0, \ \beta_\lambda =0$ we set $\vep=1$ and diagonalise $y$ by an $O(n)$ transformation
so that it has diagonal elements $y_i, \, i=1,\dots, n$. The Yukawa $\beta$-function then gives
\be
y_i = 6 \, y_i {\!}^3 + y_i \, R \, , \ \ R = {\ts \sum_i} \, y_i{\!}^2 \, ,  \quad \Rightarrow \quad 
y_i{\!}^2 = \frac{1}{n+6} \, , \ \ R =  \frac{n}{n+6} \, .
\label{fpy}
\ee
Substituting in $\beta_\lambda=0$ then gives
\be
\lambda = \frac{1}{6(n+6)} \Big ( 6 -n \pm \sqrt{ n^2 + 132\, n + 36} \, \Big ) \, .
\label{fpl}
\ee

The stability matrix becomes
\be
 M = \begin{pmatrix} \pr_\lambda {\beta}_\lambda & \pr_\lambda \beta_{y_j} \\  \pr_{y_i} \beta_\lambda
  &  \pr_{y_i} {\beta}_{y_j} \end{pmatrix} \, ,
  \label{stabM}
 \ee
where $y_i,\lambda$ are determined by \eqref{fpy}, \ \eqref{fpl}. Without any loss of generality we can take
\be
y_i = \frac{1}{\sqrt{n+6}} \, s_i \, , \qquad s_i = 
\begin{cases} 1\, , \quad & i=1,\dots p, \\
  -1 , \,  & i =p+1,\dots , n \end{cases} \, ,  \quad p= 0, \dots, n \, .
  \label{fp2}
\ee
Then $M$ becomes for the two possible solutions for $\lambda$ in \eqref{fpl}
\be
M= \frac{1} {n+6} \begin{pmatrix} \pm  \sqrt{ n^2 + 132\, n + 36} & 0 \\
* & 6\,\delta_{ij} + s_i s_j \
\end{pmatrix} \, .
\ee
The eigenvalues are then
\be
\pm \tfrac{1}{n+6}  \sqrt{ n^2 + 132\, n + 36}  \, , \quad 1 \, , \quad \tfrac{6}{n+6} \ \mbox{degeneracy} \ n-1 \, ,
\ee
where the eigenvectors for the last two cases can be given by
\be
v_i = s_i \, , \quad v_1=1, \, v_i =\begin{cases} -1, \, \ i=2,\dots ,p \\
1, \, \ i=p+1, \dots n \end{cases}  , \   v_j=0 \, , j\ne 1,i \, .
\ee
The fixed points corresponding to \eqref{fp2} are invariant under $O(p)\times O(n-p)$. Each
point on the orbit corresponding to the coset $O(n)/O(p)\times O(n-p)$, of dimension $p(n-p)$,
 generated by  the action of $O(n)$  defines an equivalent theory.

For one fermion and $n_s=m+1$ scalars then by an $O(n_s)$ rotation the Yukawa interaction can be considered to
involve just one scalar $\sigma$ while the remaining $m$ scalars $\vphi_a$ correspond to a purely scalar theory 
formed by quartic polynomial in $\vphi$ together
with interactions involving $\sigma$. If the maximal $O(m)$ symmetry is preserved then there are three couplings
$\lambda_1$  corresponding the $O(m)$ invariant quartic $(\vphi^2)^2$, $\lambda_2$ for $\sigma^4$ and $g$
for a $\vphi^2 \sigma^2$ interaction. 
For $g=0$ there are two decoupled theories. The resulting lowest order $\beta$-functions
take the form
\begin{align}
\beta_{\lambda_1} = {} & - \vep\,  \lambda_1 + (m+8) \lambda_1^2 + g^2 \, , \nn \\
\beta_{\lambda_2} = {} & - \vep\,  \lambda_2 + 9\, \lambda_2^2 +m\,  g^2  + 2 \, y^2 \lambda_2 - 4 \, y^4 \, , \nn \\
\beta_g = {}& - \vep\,  g + \big  ( (m+2) \lambda_1 + 3 \, \lambda_2 \big ) g + 4 \, g^2 + y^2 g \, .
\end{align}
For $y=0$ this is just a biconical theory, for $\lambda_1=\lambda_2= g$ there is an $O(m+1)$ symmetry. Other
fixed points with $g$ non zero are irrational and have two quadratic invariants,
The Yukawa $\beta$-function gives at a fixed point $y^2= \frac17$. For $m=4, \, \vep=1$ there is a rational solution
$\lambda_1 = \frac{1}{21}, \, \lambda_2 =0, \, g= \frac17$, otherwise the $g\ne 0$ solutions are irrational.
For the scalar invariants
\be
S= 3m(m+2) \, \lambda_1{\!}^2 + 9 \, \lambda_2{\!}^2 + m \, g^2 \, , \qquad a_0=  m(m+2)\, \lambda_1+ 3\, \lambda_2 
+ 2\, m \, g \, .
\ee
For the first few values of $m$ these take the values, with $\vep=1$ and $g,y$ non zero,
\begin{align}
\text{
 \begin{tabular}{  c   c   c  c c c c c c c }
$m $ & $S$   &  &   & & &  $a_0 $ \\
\noalign {\vskip 3pt}
\hline
\noalign {\vskip 4pt}
0 & $\tfrac{9}{49}$&   &&&& $\tfrac37$  \\
\noalign {\vskip 2pt}
1 & 0.29635& 0.23347 & & & &0.8081 & 0.6518  \\
2 &  0.42391 & 0.28682 & &&&1.2569 &  0.9066   \\
3 & 0.55581 &  0.34503 &&&& 1.7530 &  1.1996  \\
4 & 0.68552 & 0.41049 & 0.66475 &$\tfrac{32}{49}$ & &  2.2841 & 1.5422 & 2.3582 &
 $\tfrac{16}{7}$
\end{tabular} 
} \nn \\
\noalign{\vskip -12pt}
\label{table2}
\end{align}
Note that $\tfrac{32}{49} \approx 0.65306$.
As a function of $m$, $S$ has the form
\begin{figure}[hbt!]
\centering
 \includegraphics[width=10cm,scale=0.5]{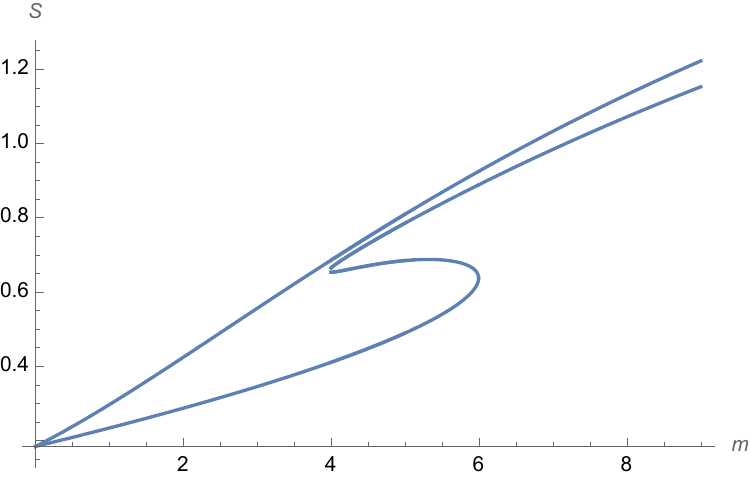}
\end{figure}

\noindent
There are bifurcation points close to $m=4,6$ where two fixed points are created or annihilated.

At lowest order the stability matrix eigenvalues can be determined from the $3\times 3$ 
matrix determined from $\beta_{\lambda_1}, \,  \beta_{\lambda_2}, \beta_g$ analogously to \eqref{stabM}.
For $m=1,2$ the  solution corresponding to the first case in \eqref{table2} has three positive eigenvalues and is 
therefore RG stable.
For $m=4$ the two additional solutions each have a small eigenvalue of opposite sign. 
These both tend to zero as $m$ approaches the bifurcation point, $m\approx 3.965$.

For $m\ge 2$ there are scalar theories with reduced symmetry which should lead to a ranged of
additional fixed points with a Yukawa coupling to a single fermion.

\newpage
\appendix

\section{Majorana Fermions and Their Reduction}

\label{Majorana}

For a spinor field $\Psi$ its conjugate $\bar \Psi$ is defined by
\be
\bar \Psi = \Psi^\dagger A \, , 
\label{psiB}
\ee 
where $A$ satisfies (the choice of both signs is a matter of convention, they are chosen here for later convenience)
\be
A\,  \gamma^\mu A^{-1} = - (\gamma^\mu)^\dagger\, , \qquad 
A^\dagger = -A \, ,
\label{Aprop}
\ee
with, for $d=4$, the  $4\times 4$  Dirac matrices here defined by
\be
\{ \gamma^\mu, \gamma^\nu \} = 2 \, \eta^{\mu\nu} \, \I \, , \qquad \eta^{\mu\nu} = {\rm diag.} (-1,1,1,1) \, .
\ee
Taking  $\gamma_5 = i \gamma^0\gamma^1\gamma^2\gamma^3$, then $A\gamma_5 A^{-1} = - \gamma_5{}^\dagger$
and $\gamma_5{}^2 = \I$.
Under a reflection in the $x^1x^2$ plane, charge conjugation and time reversal
\begin{align}
& \Psi \longrightarrow \hskip -0.6cm{\raisebox{- 6pt}{$\scriptstyle \mathcal R_3$}} \hskip 0.4cm \Psi_{r_3}  = R \, \Psi \big |_{x^3\to -x^3}  
\, ,  &&
{\bar \Psi} \longrightarrow  \hskip -0.6cm{\raisebox{- 6pt}{$\scriptstyle \mathcal R_3$}} \hskip 0.4cm {\bar \Psi}_{r_3}   
=  {\bar \Psi } R^{-1} \big |_{x^3\to -x^3}  \, , \nn \\
& \Psi \longrightarrow \hskip -0.6cm{\raisebox{- 6pt}{$\scriptstyle \mathcal C$}} \hskip 0.5cm \Psi_c  = C \, {\bar \Psi}^T  
\, ,  &&
{\bar \Psi} \longrightarrow  \hskip -0.6cm{\raisebox{- 6pt}{$\scriptstyle \mathcal C$}} \hskip 0.5cm {\bar \Psi}_c   
=-  {\Psi }^T C^{-1}  \, , \nn \\
& \Psi \longrightarrow \hskip -0.6cm{\raisebox{- 6pt}{$\scriptstyle \mathcal T$}} \hskip 0.4cm \Psi_t  = T \, \Psi \big |_{x^0\to -x^0}  
\, , &&
{\bar \Psi} \longrightarrow  \hskip -0.6cm{\raisebox{- 6pt}{$\scriptstyle \mathcal T$}} \hskip 0.4cm {\bar \Psi}_t  
=-  {\bar \Psi } \, T^{-1} \big |_{x^0\to -x^0}  \, , 
\label{PCT}
\end{align}
where $\mathcal T$ is antilinear and
\be
 R^{-1} \gamma^\mu \, R =  \begin{cases} \gamma^\mu , \ &\mu = 0,1,2 \\
 - \gamma^3\, &\mu =3  \end{cases}  ,  \quad
  C^{-1} \gamma^\mu \, C = - (\gamma^\mu)^T \, , \quad
 T^{-1} \gamma^\mu{}^*  \, T   =  \begin{cases} \gamma^\mu , \ &\mu = 1,2,3 \\
 - \gamma^0\, &\mu = 0 \end{cases}   ,
 \ee
 with $ C^{-1} \gamma_5 \, C = \gamma_5{\!}^T, \  T^{-1} \gamma_5{}^* \, T = \gamma_5 $. 
 In general
 \be
 C^T = - C \, , \qquad A^* C^\dagger A = - C^{-1} \, , \qquad A^{-1} T^\dagger A = - T^{-1}   
  \, , \qquad A^{-1} R^\dagger A = R^{-1}  \, . 
 \ee
 Using these results $R,T$ can be given by
\be
R= i\, \gamma^3 \gamma_5\, , \qquad   
T = C^\dagger A \, \gamma^0 \gamma_5  \quad \Rightarrow \quad  T T^* = - \I \, ,
 \ee
 and $\Psi \longrightarrow  \hskip -0.75cm{\raisebox{- 6pt}{$\scriptstyle {\mathcal R}_3{\!}^2$}}\hskip 0.3cm 
 R^2 \Psi = \Psi, \
 \Psi \longrightarrow  \hskip -0.65cm{\raisebox{- 6pt}{$\scriptstyle {\mathcal C}^2$}}\hskip 0.3 cm  
 - C C^{-1}{}^T \Psi = \Psi, \
 \Psi \longrightarrow  \hskip -0.65cm{\raisebox{- 6pt}{$\scriptstyle {\mathcal T}^2$}}\hskip 0.3cm 
 T T^*\Psi = - \Psi$.

For a Majorana fermion
\be
\Psi = \Psi_c \, , 
\ee
and for present purposes we consider the  Lagrangian 
\be
\cL_M  = - i \, \tfrac12 {\bar \Psi}\,  \gamma \cdot \pr \, \Psi - i \, \tfrac12  {\bar \Psi} \, {\mathcal M} \, \Psi 
 - i \, \tfrac12  {\bar \Psi} \, {\mathcal  Y}^a \, \Psi \, \phi^a \, ,
 \label{LSF}
\ee
with both ${\mathcal M} , \, {\mathcal Y^a} $ real, symmetric and $[{\mathcal M} ,{\gamma}^\mu] =
[ {\mathcal Y^a,{ \gamma}^\mu}]=0$. With the conventions \eqref{psiB} and \eqref{Aprop}
$\cL_M{}^\dagger = \cL_M$.

For reduction to three dimensions a convenient basis is obtained by taking $\gamma^\mu \to {\tilde \gamma}^\mu$
with, adapting \cite{Kubota,Boyack,archipelago},
\be
{\tilde \gamma}^\mu = \begin{pmatrix} \tsig^\mu & 0 \\ 0 & -\tsig^\mu \end{pmatrix} \, , \ \mu = 0,1,2 \, , \quad
\tsig^\mu = (i \sigma_2, \sigma_3 , -\sigma_1 ) \, , \quad 
{\tilde \gamma}^3 = \begin{pmatrix} 0 & -i \\ i & 0 \end{pmatrix} \, .
\label{Dirac}
\ee
Here ${\tsig}^\mu, \ - {\tsig}^\mu$ correspond to the two inequivalent two dimensional irreducible
representations for the $d=3$ Dirac algebra. $\tsig^\mu i\sigma_2 = ( - \I_2, \sigma_1, \sigma_3)$ form 
a basis for symmetric $2\times 2$ matrices.
For the $d=4$ representation defined by \eqref{Dirac}
\be
 \tau = {\tilde \gamma}^0{\tilde \gamma}^1 {\tilde \gamma}^2
= \begin{pmatrix} 1 & 0 \\ 0 & -1 \end{pmatrix} \, , \qquad 
{\tilde  \gamma}_5 = i   {\tilde \gamma}^0{\tilde \gamma}^1 {\tilde \gamma}^2 {\tilde \gamma}^3
= \begin{pmatrix} 0 & 1 \\ 1 & 0 \end{pmatrix} \, .
\ee
and
\begin{align}
 & A = {\tilde \gamma}^0 = \begin{pmatrix} i\sigma_2 & 0 \\ 0 & -i \sigma_2 \end{pmatrix}  \, , \hskip 1.4cm
 R = \begin{pmatrix} 1 & 0 \\ 0 & - 1 \end{pmatrix}  \, , \nn \\
&  C = i\, {\tilde \gamma}^0 {\tilde \gamma}^3 {\tilde \gamma_5} = 
 \begin{pmatrix} i\sigma_2 & 0 \\ 0 & i \sigma_2 \end{pmatrix}  \, , \qquad 
  T = \begin{pmatrix}0&  i\sigma_2 \\ i \sigma_2 & 0 \end{pmatrix}  \, .
\end{align}

The representation in \eqref{Dirac} can be related to the more commonplace chiral representation by
\be
U \, {\tilde \gamma}^\mu \, U^{-1} = \begin{pmatrix} 0 & i\, \sigma^\mu \\i\,  {\bar \sigma}^\mu &0 \end{pmatrix}\, , \quad
U \, {\tilde \gamma}_5 \, U^{-1} = \begin{pmatrix} -1 & 0 \\ 0  &1  \end{pmatrix}\, , \quad
\sigma^\mu = ( 1, \, {\boldsymbol \sigma}) \, , \ \ {\bar \sigma^\mu} = ( 1, \, -  {\boldsymbol \sigma}) \, ,
\ee
where
\be
U = \frac12 \begin{pmatrix} \sigma_3 - \sigma_2 & \sigma_2 - \sigma_3 \\
\sigma_3\sigma_2 - 1 &  \sigma_3\sigma_2 - 1\end{pmatrix} \, , \quad
U^{-1} = \frac12 \begin{pmatrix} \sigma_3 - \sigma_2 &  \sigma_2\sigma_3 - 1\\
\sigma_2 - \sigma_3  &  \sigma_2\sigma_3 - 1\end{pmatrix} \, .
\ee

With \eqref{Dirac} the spinor field decomposes into two $d=3$ two-component spinors as 
\be
\Psi = \begin{pmatrix} \psi_1 \\ \psi_2 \end{pmatrix} \, , \quad {\bar \Psi} = ( {\bar \psi_1} , - {\bar \psi}_2 ) \, , \quad
{\bar \psi_a} = \psi_a{\!}^\dagger i \sigma_2 \, , \quad \psi_1 = \psi_1{\!}^*  \, , \ \ \psi_2 = - \psi_2{\!}^* \, .
\label{Psi3}
\ee
 Using the decomposition \eqref{Psi3}, and taking 
${\mathcal M} \to \left( \begin{smallmatrix} m & 0\\0& m \end{smallmatrix}\right ), \ 
{\mathcal Y^a} \to \left( \begin{smallmatrix} y^a & 0\\0& y^a \end{smallmatrix}\right )$,  \eqref{LSF} becomes, 
\begin{align}
\cL_M = {}& - i \, \tfrac12\,  {\textstyle \sum_{a=1,2}}\,  {\bar \psi}_a \,  \tsig \cdot \pr \, \psi_a - 
\tfrac12( {\bar \psi}_1 \hskip 0.8pt \pr_3 \hskip 0.5pt \psi_2
+  {\bar \psi}_2 \hskip 0.8pt  \pr_3\hskip 0.5pt  \psi_1 ) \nn \\
&{} - i\,  \tfrac12  ( {\bar \psi}_1 \hskip 0.5pt  m \hskip 0.5pt \psi_1- {\bar \psi}_2 \hskip 0.5pt  m \hskip 0.5pt\psi_2 ) 
- i\,  \tfrac12 ( {\bar \psi}_1 \, y^a \psi_1- {\bar \psi}_2 \, y^a \psi_2 ) \phi^a \, .
\label{LSF2}
\end{align}
For zero mass, $m=0$, this has a ${\mathbb Z}_2$ symmetry where $\psi_1 \leftrightarrow \psi_2, \, \phi^a \to - \phi^a$.
This ensures the cancellation of fermion loops with odd numbers of Yukawa vertices.
For a symmetry $h^{-1} y^a h = y^a, \ h^{-1} \tsig^\mu h = \tsig^\mu$  with $h\in H_M \subset O(n_f)$ then in general
there is a symmetry $H_M$ but this extends to $(H_M \times H_M) \rtimes {\mathbb Z}_2$ when $d=3$.
For a single scalar this becomes $(O(n_f) \times O(n_f)) \rtimes {\mathbb Z}_2$ when $m=0$.

With conventions from \eqref{PCT}
\begin{align}
& ( \psi_{1r_3}, \, \psi_{2r_3} ) = (\psi_1, - \psi_2) \big |_{x^3\to -x^3}  \ , \qquad 
(\bar  \psi_{1r_3}, \, \bar \psi_{2r_3} ) = ( \bar \psi_1,- \bar \psi_2) \big |_{x^3\to -x^3}  \, , \nn \\
& ( \psi_{1t}, \, \psi_{2t} ) = (i\sigma_2 \psi_2 ,i\sigma_2\psi_1) \big |_{x^0\to -x^0}  \ , \quad 
(\bar  \psi_{1t}, \, \bar \psi_{2t} ) = ( - \bar \psi_2 i \sigma_2 ,- \bar \psi_1 i \sigma_2 )\big |_{x^0\to -x^0}  \, .
\end{align}
Thus \eqref{LSF2} is invariant under ${\mathcal R}_3, {\mathcal T}$
so long as  the time reversal $\mathcal T$ transformation is combined with  $\psi_1 \leftrightarrow \psi_2$, 
$ i \, {\bar \psi}_1\psi_1  \longrightarrow \hskip -0.6cm{\raisebox{- 5pt}{$\scriptstyle \mathcal T$}} \hskip 0.3cm
- i \, {\bar \psi}_{2t} \psi_{2t}$. If $\mathcal T$ is combined with $\psi_1 \leftrightarrow \psi_2$ the sign of the mass term
is reversed.
However for a reflections  ${\mathcal R}_1$ or  ${\mathcal R}_2$, corresponding to  $x^1\to -x^1$ or $x^2\to -x^2$,  
then instead
${\bar \psi}_1\psi_1 \leftrightarrow - {\bar \psi}_2\psi_2 $ and the individual mass terms are not invariant 
by themselves under  either ${\mathcal R}_1$ or  ${\mathcal R}_2$, as expected for three dimensional spinors.
The three dimensional theory with just one two component spinor cannot have mass terms which preserve
${\mathcal T}$ or ${\mathcal R}_1$ invariance \cite{Witten}.

If the Yukawa interaction is modified to 
\be
\cL_Y = \tfrac12 \, {\bar \Psi}\,  {\tilde \gamma}_3 {\tilde \gamma}_5 \, {\mathcal Y}^a  \Psi \, \phi^a  = 
 -i\, \tfrac12  ( {\bar \psi}_1 \, y^a \psi_1+ {\bar \psi}_2 \, y^a \psi_2 ) \phi^a \, ,
 \label{Yukmod}
 \ee
it is then necessary to include the four loop diagrams corresponding to \eqref{anomalous}. 
the symmetry for $d=3$  is enhanced  to a subgroup of $O(2n_f)$, for $n_s=1$ the symmetry is $O(2n_f)$.
This prescription allows for fermion loops with odd numbers of Yukawa vertices but is not relevant up  to three loops.
By including contributions corresponding to diagrams of the form \eqref{Psi3} it was  implicitly followed in  
\cite{Zerf2} in their
four loop calculation. Nevertheless \eqref{Yukmod} breaks Lorentz invariance for $d\ne 3$ though $O(2,1)$
is preserved.  Applying dimensional
regularisation, $d=4-\vep$,  the one loop counterterms necessary when starting from \eqref{LSF} 
or equivalently \eqref{LSF2} are, for a single scalar $\sigma$,
\be
\cL_{\rm ct}^{(1)} = \frac{y^2}{(4\pi)^2\vep} \Big (  i \, \tfrac12 {\bar \Psi}\,  \gamma \cdot \pr \, \Psi 
 - i \, m \,  {\bar \Psi}  \Psi +  (\pr \sigma)^2  + 6 \, m^2 \sigma^2 - i \,  y\,   {\bar \Psi}  \Psi \, \sigma
 + \tfrac{1}{3}\,  y^2 \, \sigma^4 \Big ) \, .
 \ee
For the modified Yukawa interaction \eqref{Yukmod} the result becomes
\begin{align}
\cL_{\rm ct}^{(1)}\big |_{\rm modified}\!\! ={} & \frac{y^2}{(4\pi)^2\vep} \Big (  i \, \tfrac12 {\bar \Psi}\,  \gamma \cdot \pr \, \Psi 
- i \,  {\bar \Psi}\,  \gamma_3\pr_3 \, \Psi - i \, m \,  {\bar \Psi}  \Psi 
+ \tfrac23 \big ((\pr \sigma)^2 - (\pr_3 \sigma)^2\big ) {} +  4 \, m^2 \sigma^2  \nn \\
\noalign{\vskip - 6pt}
& \hskip 3cm{}  +\tfrac12\,   y\,    {\bar \Psi} \gamma_3 \gamma_5 \Psi \,  \sigma 
+ \tfrac{1}{12}\,  y^2 \, \sigma^4 \Big ) \, .
 \end{align}
  Besides breaking Lorentz invariance explicitly in the kinetic terms  the 
 counterterm has  different coefficients for the Yukawa and the quartic scalar terms. 
 For a consistent flow it would be necessary to
 allow for modified kinetic terms for the scalar and fermion fields so that the propagation
 velocity is different in the 3-direction from the 1,2 directions, bringing in two new parameters consistent 
 with the  breaking $O(3,1)$ to $O(2,1)$. Whether the $\vep$-expansion can be applied in this case is unclear.

An alternative possibilty, yet to be explored, is 
to take ${\mathcal M} \to \left( \begin{smallmatrix} m & 0\\0& M \end{smallmatrix}\right )$ and require 
$\Lambda\gg M\gg m$ for $\Lambda$ some cutoff.
This  breaks Lorentz invariance  more softly and  should lead to $\psi_2$ being decoupled so as to generate
 an effective theory for $\psi_1$.
This non Lorentz invariant theory can potentially be extended to $\N=\frac12$ supersymmetry away from $d=3$.

\section{Algebra of $d$ and $w$ Tensors}

\label{appdw}

The tensors defined by \eqref{hdrel} and \eqref{wdef} satisfy identities which allow determination of
eigenvalues,
\begin{align}
& d^{abef} \, d^{cdef} = \tfrac{1}{n_s-1} \, a
\big ( \tfrac12 n_s (\delta^{ac} \delta^{bd}   +  \delta^{ad} \delta^{bc})   -
\delta^{ab} \delta^{cd}  \big ) + e\, w^{abcd} + b \, d^{abcd}  \, , \nn \\
& d^{abef} \, w^{cdef} = w^{abef} \, d^{cdef} =  f\, w^{abcd} + h \, d^{abcd}  \, ,  \nn \\
& w^{abef} \, w^{cdef} = \tfrac{1}{n_s-1} \, a'
\big ( \tfrac12 n_s (\delta^{ac} \delta^{bd}   +  \delta^{ad} \delta^{bc})   -
\delta^{ab} \delta^{cd}  \big ) + e'\, w^{abcd} + b' \, d^{abcd}  \, , 
\label{dwrel}
\end{align}
where $a,b$ are as in \eqref{did}. For consistency
\be
e\,  a'  = h \, a  \, , \quad \ f \, a'  = b'  \, a \, , \quad
 f \,  h = b'  \, e \, , \quad  \tfrac{n_s}{n_s-1} \, a = f^2 + e\, h - b\, f - e\, e'\, .
 \label{consis}
 \ee

 The relevant eigenvalue equations necessary for obtaining the anomalous dimensions $\phi^2$ operators 
 given general perturbative results are
\be
d^{abcd} \, v^{cd} = \mu \, v^{ab} \, ,   \qquad w^{abcd} \, v^{cd} = \nu \, v^{ab} \,\, ,
\label{rhoe2}
\ee
for symmetric traceless $v^{ab}$.  \eqref{dwrel} then requires
\be
\mu^2 = e \, \nu +  b \, \mu +  \tfrac{n_s}{n_s-1} \, a  \, , \quad
\mu \, \nu = f\,  \nu + h\, \mu \, , \quad
\nu^2 = e'\,  \nu +  b'  \, \mu +  \tfrac{n_s}{n_s-1} \, a' \, .
\label{eiggen}
\ee
As a consequence  of \eqref{consis} the last equation is redundant and then 
eliminating $\nu$ leads to a cubic equation for $\mu$ whose solutions determine $\nu$.
There are thus three possibilities  $\mu_i, \, \nu_i$. The associated degeneracies are then determined
by
\be
{\textstyle \sum_i} \, d_i = \tfrac12 (n_s-1) (n_s+2) \, , \qquad 
{\textstyle \sum_i}\,  d_i \, \mu _i  = {\textstyle \sum_i}\,  d_i \, \nu _i = 0 \, .
\label{degen}
\ee

For the examples of interest here the coefficients appearing in \eqref{dwrel}, besides $a,b$ which are listed in
\eqref{table3}, are given by 
\begin{align}
\hskip -1.1cm
{\text{
 \begin{tabular}{  c   c   c  c  c  c  c c }
$y^a$ &  & $e$ &  $f $ & $ h  $ & $ a' $ & $b'$ & $e'$   \\
\noalign {\vskip 3pt}
\hline
\noalign {\vskip 4pt}
1. &  & $  \tfrac{(n-3)(n+6)}{27\, n} $  & 
$ \tfrac{(n-3)(n+6)(n_s-1)}{9\, n(n_s+2)} $&   $\tfrac{2n (n_s+2)}{3(n_s-1)}$  & $ 
\tfrac{3n(n-2)(n+1)(n+4)}{(n_s-1)(n_s+2)}$ & $ 2n $ & 
 $\tfrac{n^3 - 5 n^2 +14 n + 24}{6(n_s-1)} $ \\ 
\noalign {\vskip 4pt}
3. &  & $\tfrac{2(n^2-9)}{27n} $ &
$ \tfrac{2(n^2-9)(n_s-1)}{9\, n(n_s+2)} $ &   $\tfrac{4n (n_s+2)}{3(n_s-1)}$  & $ 
\tfrac{24n^2(n^2-4)}{(n_s-1)(n_s+2)}$ & $ 4n $ & 
 $\tfrac{2n(n_s+11)}{3(n_s-1)} $ \\ 
\noalign {\vskip 4pt}
 4. & & $  \tfrac{2(n+1)}{27} $  & 
$ \tfrac{(n-8)(n+1)^2}{9(n_s+2)} $&   $\tfrac{8 (n_s+2)}{3(n+1)}$  & $ 
\tfrac{24(n-3)(n^2-4)}{(n+1)(n_s+2)}$ & $ 4(n-8) $ & 
 $\tfrac{4(n_s+11)}{3(n+1)} $ \\
 \noalign {\vskip 4pt}
 5. & & $  \tfrac{n-1}{27} $  & 
$ \tfrac{(n-1)^2(n+8)}{18(n_s+2)} $&   $\tfrac{4 (n_s+2)}{3(n-1)}$  & $ 
\tfrac{6 (n^2-4)(n+3)}{(n-1)(n_s+2)}$ & $ 2(n+8) $ & 
 $\tfrac{2(n_s+11)}{3(n-1)} $ 
\\ \noalign {\vskip 4pt} 
6. & & $  \tfrac{(n-6)(n+3)}{27\, n} $  & 
$ \tfrac{(n-6)(n+3)(n_s-1)}{9\, n(n_s+2)} $&   $\tfrac{2n (n_s+2)}{3(n_s-1)}$  & $ 
\tfrac{3n(n-4)(n-1)(n+2)}{(n_s-1)(n_s+2)}$ & $ 2n $ & 
 $\tfrac{n^3 + 5 n^2 +14 n -24}{6(n_s-1)} $ \\ 
\noalign {\vskip 4pt}
\end{tabular} 
}} \nn \\
\label{table5}
\end{align}

Solving \eqref{eiggen} and \eqref{degen} gives
\begin{align}
\hskip -0.5cm
{\text{
 \begin{tabular}{  c   c   c  c  c  c  c }
$y^a$ &  $\mu_1$ & $\nu_1$ &  $\mu_2 $ & $ \nu_2 $ & $ \mu_3 $ & $\nu_3 $  \\
\noalign {\vskip -12pt}\\
& & \hskip - 2.5cm $d_1$ & & \hskip - 2.5 cm $d_2$ & &  \hskip - 2.5cm $d_3$ \\
\noalign {\vskip 3pt}
\hline
\noalign {\vskip 4pt}
1.  & $ \tfrac{(n-3)(n+1)(n+6)}{6(n_s+2)} $ & $ \tfrac{n^2(n+1)}{2(n_s-1)} $ &
$ \tfrac{(n-3)(n^2-4)}{3n(n_s+2)} $&   $-\tfrac{2(n^2-4)}{n_s-1} $ & $-\tfrac{(n-1)(n+4)(n+6)}{6n(n_s+2)} $ & $ \tfrac{(n-1)(n+4)}{n_s-1} $  \\
\noalign {\vskip 2pt}
& & 
  \hskip -2cm $\tfrac{1}{2}\, {\scriptstyle  (n-1) (n+2)} $& &
 \hskip -2cm $\tfrac{1}{24}\, {\scriptstyle n (n-1) (n+1)(n+6)}$  & &
  \hskip -2cm $\tfrac{1}{12}\, {\scriptstyle n (n-3) (n+1)(n+2)}$ 
\\
\noalign {\vskip 6pt}
3. & $\tfrac{2n(n^2-9)}{3(n_s+2)} $ & $ \tfrac{2n^3}{n_s-1} $  & 
$ \tfrac{2(n-3)(n-2)(n+1)}{3n(n_s+2)} $&   $-\tfrac{4(n-2)(n+1)}{n_s-1} $  & $ 
-\tfrac{2(n-1)(n+2)(n+3)}{3n(n_s+2)} $ & $ \tfrac{4(n-1)(n+2)}{n_s-1} $ \\
\noalign {\vskip 2pt}
& & \hskip -2cm $ {\scriptstyle n^2-1}$ & & 
 \hskip -2cm $\tfrac{1}{4}\, {\scriptstyle n^2 (n-1) (n+3)}$ & &
  \hskip -2cm $\tfrac{1}{4}\, {\scriptstyle  n^2(n-3) (n+1)}$
\\
\noalign {\vskip 6pt}
 4.& 
$ \tfrac{n(n-3)(n+1)}{3(n_s+2)} $ & $ \tfrac{2n(n-3)}{n+1} $  & 
$\tfrac{(n-1)(n-8)}{3(n_s+2)} $ & $- \tfrac{4(n-1)}{n+1} $  & 
$- \tfrac{2(n+1)(n+2)}{3(n_s+2)} $&   $\tfrac{8(n+2)}{n + 1} $  \\ 
\noalign {\vskip 2pt}
&&   \hskip -2cm $\tfrac{1}{2}\, {\scriptstyle  (n-1) (n+2)}$ &
& \hskip -2cm $\tfrac{1}{12}\, {\scriptstyle n (n-3) (n+1)(n+2)}$ & & 
 \hskip -2cm $\tfrac{1}{24}\, {\scriptstyle n (n-3) (n-2)(n-1)}$ 
 \\
 \noalign {\vskip 6pt}
 5.&  
$ \tfrac{n(n-1)(n+3)}{6(n_s+2)} $ & $ \tfrac{n(n+3)}{n-1} $  & 
$\tfrac{(n-2)(n-1)}{3(n_s+2)} $ & $- \tfrac{4(n-2)}{n-1} $  & 
$- \tfrac{(n+1)(n+8)}{6(n_s+2)} $&   $\tfrac{2(n+1)}{n - 1} $  \\ 
\noalign {\vskip 2pt}
&&   \hskip -2cm $\tfrac{1}{2}\, {\scriptstyle  (n-2) (n+1)}$ &
& \hskip -2cm $\tfrac{1}{24}\, {\scriptstyle n (n+1)(n+2)(n+3)}$ & & 
 \hskip -2cm $\tfrac{1}{12}\, {\scriptstyle n (n-2) (n-1)(n+3)}$ 
\\  
\noalign {\vskip 6pt}
6.  & $ \tfrac{(n-6)(n-1)(n+3)}{6(n_s+2)} $ & $ \tfrac{n^2(n-1)}{2(n_s-1)} $ &
$ \tfrac{(n-6)(n-4)(n+1)}{6n(n_s+2)} $&   $-\tfrac{(n-4)(n+1)}{n_s-1} $ & $-\tfrac{(n^2-4)(n+3)}{3n(n_s+2)} $ & $ \tfrac{2(n^2-4)}{n_s-1} $  
\\
\noalign {\vskip 2pt}
& & 
  \hskip -2cm $\tfrac{1}{2}\, {\scriptstyle  (n-2) (n+1)} $& &
 \hskip -2cm $\tfrac{1}{12}\, {\scriptstyle n (n-2) (n-1)(n+3)}$  & &
  \hskip -2cm $\tfrac{1}{24}\, {\scriptstyle n (n-6) (n-1)(n+1)}$ 
\\
\end{tabular} 
}} \nn \\
\label{table6}
\end{align}
The results in \eqref{table3} and \eqref{table5} satisfy
\begin{align}
\{ a, b, e, f, h, a',b',e' \}_1 \big |_{n\to -n} = {}& \{ a, - b, - e, - f, - h, a',-b',- e' \}_5 \, , \nn \\
\{ a, b, e, f, h, a',b',e' \}_2 \big |_{n\to -n} ={}&  \{ a, - b, - e, - f, - h, a',-b',- e' \}_2 \, , \nn \\
\{ a, b,  e, f, h, a', b', e' \}_3 \big |_{n\to -n} = {}&  \{ 4 a, - 2 b, - 2 e, - 2 f, - 2 h, 4 a', -2 b',- 2e' \}_4 \, , 
\end{align}
which are a reflection of of $SO(n) \simeq Sp(-n), \ SU(n)\simeq SU(-n)$ \cite{birdtracks}. 
There are corresponding relations for the eigenvalues and degeneracies in \eqref{table5}.

\subsection{Results for $U(1)$ case}

Corresponding to subsection \ref{subU1} a similar analysis can be applied. 
The basic equations relevant in \eqref{phi2} are
\be
d_{ij}{}^{mn} d_{mn}{}^{kl} = {\tilde a}  \big ( \delta_i{}^k \delta_j{}^l + \delta_i{}^l \delta_j{}^k \big ) +{\tilde b}\, 
d_{ij}{}^{kl} \, , \qquad {\tilde a} = \tfrac12 ( 1 - 4 q^2) \, , \ \ {\tilde b}= - 4\, q \, ,
\label{ddrel}
\ee
and, with $n=rs$,
\begin{align}
d_{im}{}^{jn} d_{nk}{}^{ml} ={}& \tfrac{1}{n-1}\, {\tilde a}  \big (n \,  \delta_i{}^l \delta_k{}^j - \delta_i{}^j \delta_k{}^l \big ) 
+ {\hat b} \, d_{ik}{}^{jl}   + {\hat e} \, w_{ik}{}^{jl} \, , \nn \\
d_{im}{}^{jn} w_{nk}{}^{ml} ={}& w_{im}{}^{jn} d_{nk}{}^{ml} =
 {\hat h} \, d_{ik}{}^{jl}   + {\hat f} \, w_{ik}{}^{jl} \, , \nn \\
w_{im}{}^{jn} w_{nk}{}^{ml} ={}& \tfrac{n+1}{(n-1)^2} \, {\tilde a}  
\big (n \,  \delta_i{}^l \delta_k{}^j - \delta_i{}^j \delta_k{}^l \big ) 
+ {\hat b}{}^\prime \, d_{ik}{}^{jl}   + {\hat e}{}^\prime \, w_{ik}{}^{jl} \, , 
\label{dwrel2}
\end{align}
where
\begin{align}
& {\hat b} = \tfrac{n-3}{4(n+1)}(r+s) \, , \quad {\hat e} = \tfrac14(r-s) \, , \quad
{\hat b}^\prime  =  \tfrac14(r+s) \, , \quad {\hat e}^\prime =  \tfrac{n+3}{4(n+1)}(r-s) \, , \nn \\
& {\hat f} = \tfrac{n-1}{4(n+1)}(r+s) \, , \qquad {\hat h} = \tfrac{n+1}{4(n-1)}(r-s) \, .
\label{resbefh}
\end{align}
Eigenvalues and degeneracies are determined as before. From \eqref{ddrel} there are 
two  eigenvalues $\mu_1,\mu_2$ 
corresponding to eigenvectors $v_{ij}= v_{ji}$ which are given, with their associated degeneracies, by
\begin{align}
\hskip -0.5cm
{\text{
 \begin{tabular}{  c   c   c  c  c  c  }
  $\mu_1$ & $d_1$ &  $\mu_2 $ & $ d_2 $  \\
\noalign {\vskip 3pt}
\hline
\noalign {\vskip 4pt}
 $  \tfrac{(r -1)(s - 1)}{n+1 }$ & $ \tfrac14rs(r +  1)(s +1)$ &
$  -\tfrac{(r + 1)(s  + 1)}{n+1}  $&   $\tfrac14rs(r-1)(s-1)$ 
\end{tabular} 
}} \ .
\label{table7}
\end{align}
From \eqref{dwrel2} with \eqref{resbefh} there are three sets of eigenvalues $\mu_u, \nu_u$, $u=1,2,3$, for 
eigenvectors $v_i{}^j, \, v_i{}^i=0$ where $d_{il}{}^{jk} v_k{}^l = \mu \, v_i{}^j, \, w_{il}{}^{jk} v_k{}^l = \nu \, v_i{}^j$
\begin{align}
\hskip -0.5cm
{\text{
 \begin{tabular}{  c   c   c  c  c  c  }
  $\mu_1$ & $\nu_1$ &  $\mu_2 $ & $ \nu_2 $ & $ \mu_3 $ & $\nu_3 $  \\
\noalign {\vskip -18pt}\\
 & \hskip - 1.7 cm $d_1$ & & \hskip - 1.8 cm $d_2$ & &  \hskip - 1.8cm $d_3$ \\
\noalign {\vskip 3pt}
\hline
\noalign {\vskip 4pt}
 $ \tfrac{s(r^2-1)}{2(n+1)} $ & $ \tfrac{s(r^2-1)}{2(n-1)} $ &
$ \tfrac{r(s^2-1)}{2(n+1)} $&   $-\tfrac{r(s^2-1)}{2(n-1)} $ & $-\tfrac{r+s}{2(n+1)} $ & $ \tfrac{r-s}{2(n-1)} $  \\
\noalign {\vskip 4pt}
 & 
  \hskip -1.5cm $s^2-1$& &
 \hskip -1.5cm $r^2-1$  & &
  \hskip -1.5cm $(r^2-1)(s^2-1)$ 
\end{tabular} \ .
}} 
\label{table8}
\end{align}
The degeneracies correspond to expected representations of $SU(r) \times SU(s)$. Related results 
are  given in \cite{StergiouU}.

The results in \eqref{resbeta2} are obtained from
\begin{align}
 {\tilde \beta}_{\hat \lambda}{\!}^{(2)}\big|_{y=0}
  = {}& - 4 (n+11) {\hat \lambda}^3 - 6 (n+7)\,{\tilde a} \, g^2 {\hat \lambda}
 - 2\, \big ( {\tilde b} + 4\,  {\hat b}\big  ) \,{\tilde a} \, g^3 \, , \nn \\
  {\tilde \beta}_g{\!}^{(2)}\big|_{y=0} 
  = {}& - 12(n+7) \, g\,  {\hat \lambda}^2 - 12 \big ( {\tilde b} + 4\,  {\hat b}\big  ) \,  g^2 {\hat \lambda}
  - 8 \, \tfrac{n-2}{n-1}\, {\tilde a} \, g^3 - 8 \big ( {\tilde b} \, {\hat b}+  {\hat b}^2  -{\hat e}\, {\hat h} \big ) \,g^3 \, .
\end{align}

\section{Figures}

\label{fpfigures}

For some of the cases listed in \eqref{tableS} and \eqref{table} we plot the fixed point values of
\be
8\, ||\lambda ||^2 / n_s \, , \quad  |\lambda | /n_s \, ,  \qquad
 ||\lambda ||^2=  \lambda^{abcd}\lambda^{abcd}  \, , \quad |\lambda |= \lambda^{aabb}\, ,
\ee
respectively orange, blue,
as functions of $\log m$ where $m$  gives the number of fermions. The log plots exhibit the symmetry
following from \eqref{lmf}.
For the purely scalar theory, when $m=0$,
$8\,||\lambda ||^2  / n_s\le 1$ \cite{RychkovS} and when this is satisfied $|\lambda | /n_s = \frac12$.
These bounds are clearly violated for a non zero number of fermions.

\begin{figure}[hbt!]
\centering
 \includegraphics[width=\linewidth,scale=0.9]{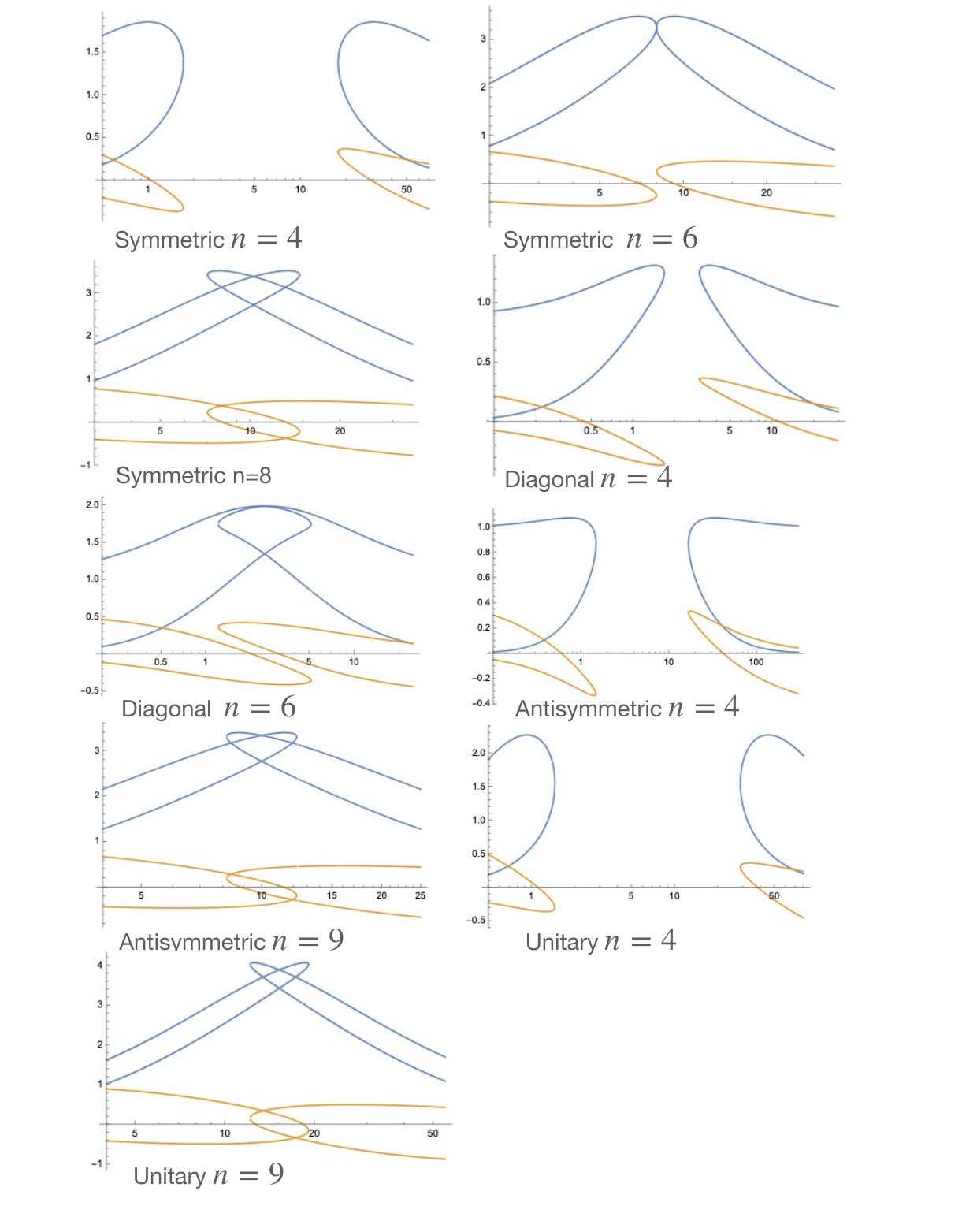}
\end{figure}

At the $O(n_s)$ or Heisenberg fixed point
\be
||\lambda_H|| =  \tfrac{3n_s(n_s+2)}{(n_s+8)^2} \, , \qquad
|\lambda_H| = \tfrac{n_s(n_s+2)}{n_s+8}\, . 
 \ee
 Of course $\lambda^{aabb} <0$ is indicative of an unstable potential.

Corresponding results for $U(r)\times U(s)$ fixed points are given below where plots of $24||\lambda||^2/rs$ and
$|\lambda|/rs$ are given for various representative $r,s$ as functions of $m$ determining the number of
fermions. The intercepts at  $m=0$, and also for $m\to \infty$, are determined by results for the purely
complex scalar $U(r) \times U(s)$ theory and are obtained from \eqref{kapfp}, \eqref{Kgl}, \eqref{Ogl}.
There are two or four fixed points according to whether $R_{rs}<0$ or $R_{rs}>0$. For $r=5, \, s=49$ $R_{rs}=0$
which is a bifurcation point and then $24||\lambda||^2/rs=1$.
\vspace{-0.4cm}
\begin{figure}[hbt!]
\centering
 \includegraphics[width=\linewidth,scale=1]{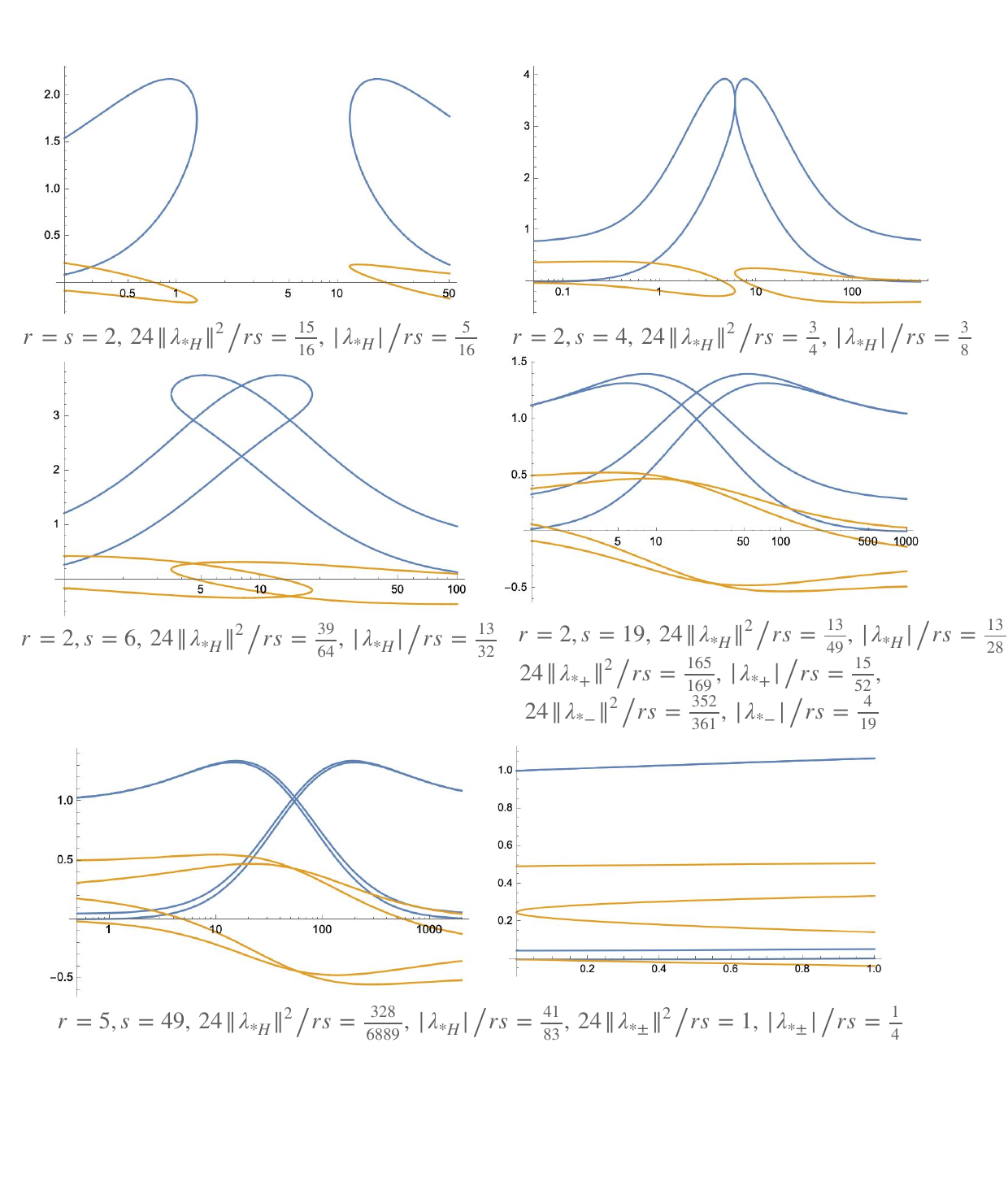}
\end{figure}

\clearpage
For this case the eigenvalues $\kappa$ of the stability matrix, taking $\vep=1$, as functions of $m$ are given by
\begin{figure}[hbt!]
\centering
 \includegraphics[width=\linewidth,scale=1]{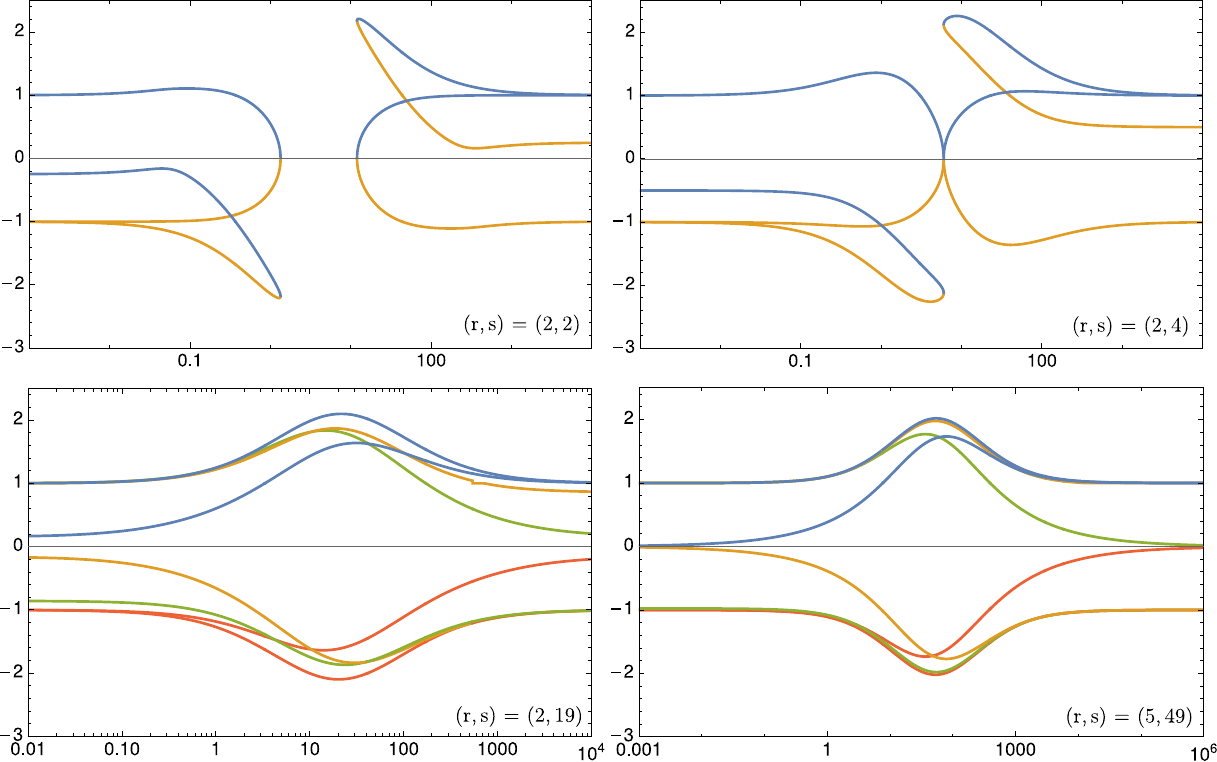}
\end{figure}

\noindent
The intercepts at $m=0$ and $m\to \infty$ are given by \eqref{Ogl} and, when $R_{rs}\ge 0$, \eqref{kapfp}.
For $R_{rs}<0$ there is, as $m\to 0$, one $\kappa\to1$ and two  $\kappa \to -1$, corresponding to the 
Gaussian fixed point. For $R_{rs}>0$ there are two further cases with $\kappa \to 1$.

\newpage
\vskip 2cm
\section{\texorpdfstring{$U(1)$}{U(1)} Scalar Fermion Theory Consistency Equations}

The derivation of consistency relations for the three loop Yukawa couplings can be illustrated by
restricting to case when $U(1)$ symmetry is imposed. The number of couplings is significantly
reduced but there remain non trivial relations which are a subset of the general case\footnote{The consistency
relations in this case were discussed less completely in \cite{Jack:2013sha}}.

At lowest order
\be
T_{IJ}{\!}^{(2)} \, \rmd g^I \rmd ' g^J = t_2 \big ( \tr( \rmd y^i \,\rmd ' \by_i) + 
\tr( \rmd \by_i \, \rmd ' y^i )\big )  \, ,
\ee
and
\be
{  A}^{(3)} =  a_{3a}\, \big ( \tr( \by_i \, y^i \by_j \, y^j ) + \tr ( y^i \by_i \, y^j \by_j ) \big ) 
+ a_{3b} \, \tr ( \by_i \, y^j ) \, \tr ( \by_j \, y^i ) \, .
\label{A3loop}
\ee
The consistency equations are then
\be
 a_{3a} = \tfrac12\,  t_2 \, \gamma_{\psi{\hskip 0.5pt}1} \, , \qquad  a_{3b} = \tfrac12\, t_2 \, \gamma_{\phi{\hskip 0.5pt}1} \, .
\ee

At the next order
\begin{align}
T_{IJ}{\!}^{(3)} \, \rmd g^I \rmd ' g^J \big |_{\rmd \lambda} = {}& 
t_{3\lambda} \, \rmd \lambda_{ij}{}^{kl} \, \rmd ' \lambda_{kl}{}^{ij} \nn \\
T_{IJ}{\!}^{(3)} \, \rmd g^I \rmd ' g^J \big |_{\rmd \by} = {}&  
t_{3a_1} \, \big (  \tr ( \rmd \by_i\,  \rmd ' y^i \, \by_j \, y^j ) + \tr (\rmd \by_i \, y^j  \by_j \, \rmd'  y^i ) \big ) \nn   \\ 
&  {} +t_{3a_2} \, \big (  \tr ( \rmd \by_i\, \rmd ' y^j \, \by_j \, y^i ) + 
 \tr (\rmd \by_i \, y^i \by_j \, \rmd ' y^j ) \big ) \nn  \\
& {} + t_{3a_3} \, \big (   \tr ( \rmd \by_i\, y^i\, \rmd ' \by_j \, y^j ) 
+  \tr (\rmd \by_i \, y^j \, \rmd ' \by_j \, y^i ) \big ) \nn  \\
\noalign{\vskip 1pt}
& {} + t_{3b_1} \, \tr( \rmd \by_i \, \rmd ' y^j ) \, \tr (\by_j \, y^i )
+ t_{3b_2} \, \tr( \rmd \by_i \, y^j ) \, \tr (\by_j \, \rmd ' y^i )\nn \\
\noalign{\vskip 1pt}
& {} + t_{3b_3} \; \tr(\rmd \by_i \, y^j) \, \tr (\rmd ' \by_j \, y^i ) \, ,
\end{align}
with $T_{IJ}{\!}^{(3)} \, \rmd g^I \rmd ' g^J \big |_{\rmd y} $ obtained by conjugation and the notation indicates
which 3 loop contribution in \eqref{A3loop} each term corresponds to. At this order $T_{IJ}{\!}^{(3)} $ is symmetric.
Four loop  vacuum  graphs give 
\begin{align}
{A}^{(4)} = {}& a_{4\lambda}\, \big (
\lambda_{ij}{}^{kl} \lambda_{kl}{}^{mn}\lambda_{mn}{}^{ij} 
+ 4 \, \lambda_{ij}{}^{kl} \lambda_{km}{}^{in}\lambda_{ln}{}^{jm}\big ) \nn \\
&{} +a_{4a} \, \lambda_{ij}{}^{kl}\, \tr( \by_l \, y^m) \, \lambda_{km}{}^{ij} 
+ a_{4b}\, \lambda_{ij}{}^{kl}\,  \tr ( \by_k \, y^i \by_l \, y^j )\nn \\
&{}+  a_{4c} \, \tr ( \by_i \, y^j ) \, \tr(\by_j \, y^k ) \, \tr(\by_k \, y^i )
+a_{4d} \, \big ( \tr ( \by_i \, y^i \, \by_j \, y^j\, \by_k \, y^k) + \tr ( y^i \by_i \, y^j \by_j \, y^k \by_k ) \big )  \nn \\
&{} + a_{4e} \big ( \tr ( \by_i \, y^i \, \by_j\, y^k ) + \tr ( y^i \by_i \, y^k \by_j )  \big ) \, \tr(\by_k \, y^j )
+ a_{4f}\, \tr( \by_i \, y^j \by_k \, y^k \by_j \, y^i )  \nn \\
&{} + a_{4g}\, \tr( \by_i \, y^j \by_k \, y^i \by_j \, y^k ) \, .
\label{A4loop}
 \end{align}
 In this case  there are 11 relations corresponding to the number of inequivalent vertices in \eqref{A4loop}
 \begin{align}
& a_{4\lambda} =\tfrac13\,  t_{3\lambda} \, \beta_{\lambda{\hskip 0.5pt}1a} \, ,  \hskip 1.5cm
 a_{4a} =  3\, t_{2} \, \gamma_{\phi{\hskip 0.5pt}2a} =   2\, t_{3\lambda} \, \gamma_{\phi{\hskip 0.5pt}1} \, ,  \hskip 1.5cm
a_{4b} =  t_{3\lambda} \, \beta_{\lambda{\hskip 0.5pt}1b}   =  \tfrac12\, t_{2} \, \beta_{y{\hskip 0.5pt}2a} \, , \nn \\
\noalign{\vskip 2pt}
&  a_{4c} =  \tfrac13 (t_{3b_1} + t_{3b_2} + t_{3b_3} )  \, \gamma_{\phi{\hskip 0.5pt}1}  \, , \hskip 3cm
 a_{4d} =  \tfrac13 (t_{3a_1} + t_{3a_2} + t_{3a_3} ) \, \gamma_{\psi{\hskip 0.5pt}1}  \, , \nn \\
&   a_{4e} = \tfrac12\,  t_2 \,  \gamma_{\phi{\hskip 0.5pt}2b} + ( t_{3b_2} + t_{3b_3} ) \,  \gamma_{\psi{\hskip 0.5pt}1}  
 =  t_2 \,  \gamma_{\psi{\hskip 0.5pt}2a} +  ( t_{3a_2} + t_{3a_3} ) \,  \gamma_{\phi{\hskip 0.5pt}1}   = 
  t_{3a_1}  \,  \gamma_{\phi{\hskip 0.5pt}1} + t_{3b_1}  \,  \gamma_{\psi{\hskip 0.5pt}1} \, , \nn \\
  \noalign{\vskip 2pt}
&  a_{4f} =   t_2 \,  \gamma_{\psi{\hskip 0.5pt}2b} +  ( t_{3a_2} + t_{3a_3} ) \,  \gamma_{\psi{\hskip 0.5pt}1} = 
 2 \, t_{3a_1} \,  \gamma_{\psi{\hskip 0.5pt}1} \, , \qquad  a_{4g} =  \tfrac13\, t_{2} \, \beta_{y{\hskip 0.5pt}2f} \, .
 \label{Afour}
 \end{align} 
 The ${\rm O}(\beta^2)$ freedom in $A$ at this order corresponds just to the variations
 \be
 \delta a_{4c} =  \epsilon_2 \,  \gamma_{\phi{\hskip 0.5pt}1}{\!}^2 \, , \quad 
  \delta a_{4d} =  \epsilon_2 \,  \gamma_{\psi{\hskip 0.5pt}1}{\!}^2 \, , \quad 
   \delta a_{4e} = 2\, \epsilon_2 \,  \gamma_{\phi{\hskip 0.5pt}1}\gamma_{\psi{\hskip 0.5pt}1} \, , \quad 
    \delta a_{4f} = 2\, \epsilon_2 \,  \gamma_{\psi{\hskip 0.5pt}1}{\!}^2 \, , 
 \label{A4beta}
    \ee
 which is compatible with \eqref{Afour} if
 \be
  \delta t_{3a_1} =  \delta t_{3a_2} = \delta t_{3a3} = \epsilon_2 \,  \gamma_{\psi{\hskip 0.5pt}1}\, , \qquad
   \delta t_{3b_1} =  \delta t_{3b_2} =  \delta t_{3b_3} =  \epsilon_2\,  \gamma_{\phi{\hskip 0.5pt}1}\, .
  \label{vart3}
\ee
There is one non trivial relation at this order from the $a_{4a}, \, a_{4b}$ equations requiring the vanishing of
\be
U_0 = \gamma_{\phi{\hskip 0.5pt}1} \, \beta_{y{\hskip 0.5pt}2a}-  3 \, \beta_{\lambda{\hskip 0.5pt}1b}\,  \gamma_{\phi{\hskip 0.5pt}2a} \, ,
\label{Uscheme0}
\ee
which is identical to $B_1$ in \eqref{Brel}.

At the next order we initially  focus on contributions involving $\lambda$
\begin{align}
T_{IJ}{\!}^{(4)} \, \rmd g^I \rmd ' g^J \big |_{\rmd \lambda} = {}& 
t_{4\lambda} \big (\rmd  \lambda_{ij}{}^{kl}  \, \rmd ' \lambda_{kl}{}^{mn} \,  \lambda_{mn}{}^{ij} + 
\rmd ' \lambda_{ij}{}^{kl}  \, \rmd \lambda_{kl}{}^{mn} \,  \lambda_{mn}{}^{ij} 
+ 8 \,  \rmd  \lambda_{ij}{}^{kl}  \, \rmd ' \lambda_{km}{}^{in} \,  \lambda_{ln}{}^{jm}\, \big ) \nn \\
&{}+ t_{4a_1} \big (  
\rmd  \lambda_{ij}{}^{kl} \, \rmd ' \lambda_{km}{}^{ij} + \rmd  ' \lambda_{ij}{}^{kl}\, \rmd \lambda_{km}{}^{ij} \big ) 
\, \tr(\by_l \, y^m) \nn \\
&{}+ t_{4a_2} \,  \rmd  \lambda_{ij}{}^{kl }\big ( \lambda_{kl}{}^{im} \, 
  \tr(\by_m \, \rmd ' y^j ) +  \tr(\rmd ' \by_l\, y^m )  \lambda_{km}{}^{ij}  \big ) \nn \\
&{}  +   t_{4a_3}\,  \rmd  \lambda_{ij}{}^{kl }\,\big ( \lambda_{kl}{}^{im} \, \tr(\rmd ' \by_m \, y^j ) + \tr(\by_l\, \rmd ' y^m ) \lambda_{km}{}^{ij} \big )  \nn \\
&{}+t_{4b_1}  \,   \rmd  \lambda_{ij}{}^{kl}\, 
\big (  \tr(\by_k \, y^i \by_l \, \rmd ' y^j ) + \tr(\by_k \, y^i \rmd '  \by_l \,  y^j )  \big )  \, ,
\end{align} 
and
\begin{align}
T_{IJ}{\!}^{(4)} \, \rmd g^I \rmd ' g^J \big |_{\rmd \by} = 
{}&  t_{4a_4}\,  \rmd '  \lambda_{ij}{}^{kl }\, \tr(\rmd \by_l\, y^m ) \,  \lambda_{km}{}^{ij} 
+  t_{4a_5} \,  \rmd ' \lambda_{ij}{}^{kl }\, \lambda_{kl}{}^{im} \, \tr(\rmd  \by_m \, y^j )   \nn \\
&{}+ t_{4a_6} \,  \tr( \rmd\by_i\,  \rmd' y^j )\,  \lambda_{jm}{}^{kl }\,  \lambda_{kl}{}^{mi} 
+ t_{4b_2}  \,   \rmd  ' \lambda_{ij}{}^{kl}\, \tr(\rmd \by_k \, y^i  \by_l \,  y^j )\nn \\
&{} +  t_{4b_3} \, \big ( \tr( \rmd \by_i \, \rmd '  y^k \, \by_j \, y^l )  
+ \tr( \rmd \by_i \, y^k \by_j \, \rmd '  y^l ) \big ) \,   \lambda_{kl}{}^{ij } \nn \\
&{}+ t_{4b_4}  \, 
\tr( \rmd \by_i \, y^k \, \rmd ' \by_j \,  y^l )  \,   \lambda_{kl}{}^{ij }\,  ,
\end{align} 
with
\begin{align}
{\ A}^{(5)}\big |_{\lambda}  = {}& \, 
a_{\lambda{\hskip 0.5pt}5a} \big ( \lambda_{ij}{}^{kl} \lambda_{kl}{}^{mn}\lambda_{mn}{}^{pq} \lambda_{pq}{}^{ij} 
+ 8\, \lambda_{ij}{}^{kl} \lambda_{lm}{}^{mn}\lambda_{mn}{}^{pq} \lambda_{pq}{}^{ij} \big ) \nn \\
&{} + a_{\lambda{\hskip 0.5pt}5b} \big ( \lambda_{ij}{}^{kl}   \lambda_{km}{}^{jn}  \lambda_{ln}{}^{pq} \lambda_{pq}{}^{im}
+  \lambda_{ij}{}^{kl} \lambda_{lm}{}^{in}\lambda_{kp}{}^{mq } \lambda_{nq}{}^{jp}  \big ) 
+ a_{\lambda{\hskip 0.5pt}5c}\,  \lambda_{ik}{}^{lm} \lambda_{lm}{}^{kj}\lambda_{jn}{}^{pq} \lambda_{pq}{}^{ni} 
\nn \\
&{} + a_{5a}  \, 
\big ( \lambda_{ij}{}^{mn} \lambda_{mn}{}^{pq}\lambda_{pq}{}^{jk} 
+ 4 \, \lambda_{ij}{}^{mn} \lambda_{mp}{}^{jq}\lambda_{nq}{}^{pk} \big ) \, \tr(\by_k\, y^i )  \nn \\
&{} + a_{5b}\,   \big ( \lambda_{ij}{}^{mn} \lambda_{mn}{}^{kl} 
+ 2\,  \lambda_{(i|m}{}^{nl} \lambda_{j)n}{}^{mk} \big ) \, \tr (\by_k\, y^i ) \,  \tr (\by_l \, y^j ) \nn \\
&{} + a_{5c} \, \lambda_{ik}{}^{lm} \lambda_{lm}{}^{kj} \, \tr (\by_j \, y^n ) \,  \tr (\by_n \, y^i )  
\nn \\
&{}+ a_{5d}\,   \lambda_{ik}{}^{lm} \lambda_{lm}{}^{kj} \, \big (
\tr (\by_j \, y^i \, \by_n\, y^n) + \tr (\by_j \, y^n\by_n \, y^i ) \big ) \nn \\
&{} + a_{5e} \, 
\lambda_{ij}{}^{mn} \lambda_{mn}{}^{kl} \, \tr (\by_k\, y^i \, \by_l \, y^j )
+ a_{5f} \, \lambda_{(i|m}{}^{(k|n} \lambda_{j)n}{}^{l)m} \, \tr (\by_k\, y^i \, \by_l \, y^j ) \nn \\
&{} +a_{5g}  \, \lambda_{ij}{}^{kl }\, \big (  \tr( \by_k \, y^m) \, \tr ( \by_m \, y^i \, \by_l\, y^j )  + 
\tr (\by_k \,y^i \, \by_l\, y^m )\, \tr( \by_m \, y^j ) \big ) 
\nn \\ 
&{}+  a_{5h}  \, \lambda_{ij}{}^{kl }\, \big ( \tr ( \by_m \, y^m \by_k\, y^i\, \by_l \, y^j )  
+ \tr ( \by_m \,y^i\,  \by_k\, y^j\, \by_l\,  y^m  ) \big ) \nn \\
&{} + a_{5i} \,\lambda_{ij}{}^{kl }\, \tr ( \by_k \, y^m \by_l\, y^i\,  \by_m \, y^j ) \, .
\end{align}
If $T_{IJ}$ is symmetric only if
\be
t_{4a_2}= t_{4a_4} \, , \qquad  t_{4a_3}= t_{4a_5} \, , \qquad t_{4b_1}= t_{4b_2} \, .
\label{t4sym}
\ee

The associated consistency equations deriving from $\rmd \lambda$ variations in $A^{(5)}$ are then
\begin{align}
& a_{\lambda{\hskip 0.5pt}5a} = \tfrac12 \, t_{4\lambda} \, \beta_{\lambda{\hskip 0.5pt}1a} \, , \quad a_{\lambda{\hskip 0.5pt}5b} = 
 2\, t_{3\lambda} \,  \beta_{\lambda{\hskip 0.5pt}2a} + 4\, t_{4\lambda} \, \beta_{\lambda{\hskip 0.5pt}1a}\, , \quad 
a_{\lambda{\hskip 0.5pt}5c} =  3\, t_{3\lambda} \, \gamma_{\phi{\hskip 0.5pt}2a} \, , \nn \\
\noalign{\vskip 2pt}
& a_{5a} =   t_{3\lambda}\, \beta_{\lambda{\hskip 0.5pt}2b} + 4 \, t_{4\lambda} \, \gamma_{\phi{\hskip 0.5pt}1} =
 2 \, t_{4\lambda} \, \gamma_{\phi{\hskip 0.5pt}1}  + t_{4a_1}\, \beta_{\lambda{\hskip 0.5pt}1a} \, ,  \qquad \,  a_{5b} =  t_{4a_1} \, \gamma_{\phi{\hskip 0.5pt}1}  \, , \nn \\
 \noalign{\vskip 2pt}
 & a_{5c}= (t_{4a_1} + t_{4a_2} + t_{4a_3} )  \gamma_{\phi{\hskip 0.5pt}1}  \, , \qquad \qquad
 a_{5d}=  t_{3\lambda} \,  \gamma_{\phi{\hskip 0.5pt}2b} + ( t_{4a_2} + t_{4a_3} )  \gamma_{\psi{\hskip 0.5pt}1}  \, , \nn \\
 \noalign{\vskip 2pt}
  & a_{5e}=  t_{3\lambda} \,  \beta_{\lambda{\hskip 0.5pt}2d} + t_{4\lambda} \,   \beta_{\lambda{\hskip 0.5pt}1b}  \, , \quad
a_{5f}=  2\, t_{3\lambda} \,  \beta_{\lambda{\hskip 0.5pt}2c} + 4\, t_{4\lambda} \,   \beta_{\lambda{\hskip 0.5pt}1b}  \, , \quad
 a_{5g} =   t_{4a_1}\, \beta_{\lambda{\hskip 0.5pt}1b} + t_{4b_1} \, \gamma_{\phi{\hskip 0.5pt}1}  \, , \nn \\
 \noalign{\vskip 2pt}
& a_{5h} =  2\, t_{3\lambda} \, \beta_{\lambda{\hskip 0.5pt}2e} + 2\, t_{4b_1} \, \gamma_{\psi{\hskip 0.5pt}1} \, , \hskip 1.8cm
 a_{5i} = 2\,  t_{3\lambda}\, \beta_{\lambda{\hskip 0.5pt}2g} \, ,
 \label{a5lambda}
\end{align}
and from $\rmd \by$ variations 
\begin{align}
& a_{5a} = t_2 \, \gamma_{\phi{\hskip 0.5pt}3a} + ( t_{4a_4} + t_{4a_5} ) \beta_{\lambda{\hskip 0.5pt}1a} \, , \quad
a_{5b} = \tfrac12\, t_2 \, \gamma_{\phi{\hskip 0.5pt}3b} +  \tfrac12( t_{4a_4} + t_{4a_5} ) \gamma_{\phi{\hskip 0.5pt}1} \, ,   \nn \\
\noalign{\vskip 2pt}
& a_{5c} = 3 \, t_{3b_2} \, \gamma_{\phi{\hskip 0.5pt}2a} + ( t_{4a_5} + t_{4a_6} )  \gamma_{\phi{\hskip 0.5pt}1} = 
 3 \,( t_{3b_1} + t_{3b_3})  \, \gamma_{\phi{\hskip 0.5pt}2a} + t_{4a_4}\,  \gamma_{\phi{\hskip 0.5pt}1} \, , \nn \\
 \noalign{\vskip 2pt}
  & a_{5d} = 3 \, t_2  \, \gamma_{\psi{\hskip 0.5pt}3a} + 3( t_{3a_2} + t_{3a_3} )  \gamma_{\phi{\hskip 0.5pt}2a}
 = 3 \, t_{3a_1}  \, \gamma_{\phi{\hskip 0.5pt}2a} +  t_{4a_6} \,  \gamma_{\psi{\hskip 0.5pt}1} \, , \nn \\
 \noalign{\vskip 2pt}
 & a_{5e} = \tfrac12\, t_2 \, \beta_{y{\hskip 0.5pt}3c} + \tfrac12\, t_{4b_2}\, \beta_{\lambda{\hskip 0.5pt}1a} \, , \qquad \qquad
  a_{5f} = 2\, t_2 \, \beta_{y{\hskip 0.5pt}3b} + 2\, t_{4b_2}\, \beta_{\lambda{\hskip 0.5pt}1a} \, , \nn \\
  & a_{5g} = t_2  \, \beta_{y{\hskip 0.5pt}3b} + ( t_{4b_2} + t_{4b_3} )  \gamma_{\phi{\hskip 0.5pt}1}
  =  t_2  \, \beta_{y{\hskip 0.5pt}3d} + ( t_{4b_2} + t_{4b_4} )  \gamma_{\phi{\hskip 0.5pt}1} =  t_{3b_1}  \, \beta_{y{\hskip 0.5pt}2a} +  t_{4b_2} \,  \gamma_{\phi{\hskip 0.5pt}1}\nn \\
  \noalign{\vskip - 2pt}
  &\hskip 0.55cm {}=  t_2  ( \gamma_{\phi{\hskip 0.5pt}3c} - \upsilon_{\phi{\hskip 0.5pt}3c} ) +  t_{3b_2} \, \beta_{y{\hskip 0.5pt}2a} + t_{4a_5} \,  \beta_{\lambda{\hskip 0.5pt}1b}
  =  t_2  ( \gamma_{\phi{\hskip 0.5pt}3c} + \upsilon_{\phi{\hskip 0.5pt}3c} ) +  t_{3b_3} \, \beta_{y{\hskip 0.5pt}2a} + t_{4a_4} \,  \beta_{\lambda{\hskip 0.5pt}1b} \, , \nn \\
  \noalign{\vskip 2pt}
  & a_{5h} = t_2  \, \beta_{y{\hskip 0.5pt}3j} + ( t_{4b_3} + t_{4b_4} )  \gamma_{\psi{\hskip 0.5pt}1} =  t_{3a_1}  \, \beta_{y{\hskip 0.5pt}2a} +  t_{4b_3} \, \gamma_{\psi{\hskip 0.5pt}1}
  =  t_2  \, \gamma_{\psi{\hskip 0.5pt}3b} + ( t_{3a_2} + t_{3a_3} )  \beta_{y{\hskip 0.5pt}2a} \, , \nn \\
  \noalign{\vskip 2pt}
  & a_{5i} = t_2  \, \beta_{y{\hskip 0.5pt}3f} = t_2  \, \beta_{y{\hskip 0.5pt}3l} \, .
  \label{a5yb}
 \end{align}
 
 For the  $\lambda$ dependent contributions to $A^{(5)}$ there is an additional ${\rm O}(\beta^2)$ freedom 
 as well as that corresponding to \eqref{A4beta}
 \begin{align}
& \delta a_{\lambda{\hskip 0.5pt}5a} =  \epsilon_{3\lambda} \, \beta_{\lambda{\hskip 0.5pt}1a}{\!}^2 \, ,\quad 
 \delta a_{\lambda{\hskip 0.5pt}5b} = 8 \, \epsilon_{3\lambda} \, \beta_{\lambda{\hskip 0.5pt}1a}{\!}^2 \, , \quad 
 \delta a_{5a} = 8 \, \epsilon_{3\lambda} \,  \gamma_{\phi{\hskip 0.5pt}1}\, \beta_{\lambda{\hskip 0.5pt}1a} \, , \quad
 \delta a_{5b} = 4 \, \epsilon_{3\lambda} \,  \gamma_{\phi{\hskip 0.5pt}1}{\!}^2 \, , \nn \\
& \delta a_{5c} = 6 \,\epsilon_2 \, \gamma_{\phi{\hskip 0.5pt}1}\gamma_{\phi{\hskip 0.5pt}2a} + 4 \, \epsilon_{3\lambda} \,  \gamma_{\phi{\hskip 0.5pt}1}{\!}^2 \, ,\quad  
 \delta a_{5d} = 6\, \epsilon_2 \, \gamma_{\psi{\hskip 0.5pt}1}\gamma_{\phi{\hskip 0.5pt}2a} \, , \quad
 \delta a_{5e} = 2\,  \epsilon_{3\lambda} \, \beta_{\lambda{\hskip 0.5pt}1a}\, \beta_{\lambda{\hskip 0.5pt}1b} \, , \nn \\
& \delta a_{5f} = 8\,  \epsilon_{3\lambda} \, \beta_{\lambda{\hskip 0.5pt}1a}\,\beta_{\lambda{\hskip 0.5pt}1b} \, , \quad
 \delta a_{5g} =  \epsilon_2 \, \gamma_{\phi{\hskip 0.5pt}1}\,\beta_{y{\hskip 0.5pt}2a} + 4\,  \epsilon_{3\lambda} \, \gamma_{\phi{\hskip 0.5pt}1}\,\beta_{\lambda{\hskip 0.5pt}1b} \, , \quad
 \delta a_{5h} =  2\,\epsilon_2 \, \gamma_{\psi{\hskip 0.5pt}1}\, \beta_{y{\hskip 0.5pt}2a} \, .
 \label{A5beta}
 \end{align}
 Correspondingly in addition to \eqref{vart3} we need to take
 \begin{align}
& \delta t_{4\lambda} = 2\, \epsilon_{3\lambda} \, \beta_{\lambda{\hskip 0.5pt}1a}\, , \qquad \delta t_{4a_1} = 
 \delta t_{4a_4} = \delta t_{4a_5} =  4\,  \epsilon_{3\lambda} \, \gamma_{\phi{\hskip 0.5pt}1} \, ,
  \qquad \delta t_{4b_2} = 4\,  \epsilon_{3\lambda} \, \beta_{\lambda{\hskip 0.5pt}1b}\, ,\nn \\
 & \delta t_{4a_2}  =   
 \delta t_{4a_3} =  \delta t_{4a_6} = 3\,  \epsilon_2 \, \gamma_{\phi{\hskip 0.5pt}2a}\, , \qquad
 \delta t_{4b_1} = \delta t_{4b_3} =  \delta t_{4b_4} =  \epsilon_2 \, \beta_{y{\hskip 0.5pt}2a}\, ,
 \end{align}
 to  ensure \eqref{a5lambda} and \eqref{a5yb} are invariant.  These variations do not preserve
 \eqref{t4sym} as expected.
 
By eliminating the $a_5$'s, $t_4$'s from \eqref{a5lambda}, and then the $t_3$'s using \eqref{Afour} there remain
 10 necessary consistency relations  which become
\begin{align} 
& U_1=  \beta_{y{\hskip 0.5pt}3f}  -  \beta_{y{\hskip 0.5pt}3l}  \, , \qquad
U_2= \gamma_{\phi{\hskip 0.5pt}1} \, \beta_{y{\hskip 0.5pt}3f} -
 3 \, \gamma_{\phi{\hskip 0.5pt}2a} \, \beta_{\lambda{\hskip 0.5pt}2g}   \, , \nn \\
& U_3 = 2 \,  \gamma_{\phi{\hskip 0.5pt}1} (  \beta_{y{\hskip 0.5pt}3b} -  \beta_{y{\hskip 0.5pt}3c} ) - 
3 ( \beta_{\lambda{\hskip 0.5pt}2c} - 2\, \beta_{\lambda{\hskip 0.5pt}2d} ) \,  
 \gamma_{\phi{\hskip 0.5pt}2a}   \, , \nn \\
 & U_4 = 2 \, (  \gamma_{\phi{\hskip 0.5pt}1} \,  \gamma_{\phi{\hskip 0.5pt}3a} - \beta_{\lambda{\hskip 0.5pt}1a}\,  \gamma_{\phi{\hskip 0.5pt}3b}  ) 
 +  3 \, \beta_{\lambda{\hskip 0.5pt}2b} \,  \gamma_{\phi{\hskip 0.5pt}2a}   \, , \nn \\
 &  U_5 = \gamma_{\psi{\hskip 0.5pt}1} \,  \beta_{y{\hskip 0.5pt}3e} +  \gamma_{\phi{\hskip 0.5pt}1} \,  
 \gamma_{\psi{\hskip 0.5pt}3b}  - \beta_{y{\hskip 0.5pt}2a}\,  \gamma_{\psi{\hskip 0.5pt}2a} 
 \, , \nn \\
& U_6 = \gamma_{\psi{\hskip 0.5pt}1} ( \gamma_{\psi{\hskip 0.5pt}1} \, \beta_{y{\hskip 0.5pt}3d} - \gamma_{\phi{\hskip 0.5pt}1} \,\beta_{y{\hskip 0.5pt}3j} )  - 
  \beta_{y{\hskip 0.5pt}2a} (  \gamma_{\psi{\hskip 0.5pt}1} \, \gamma_{\psi{\hskip 0.5pt}2a} -
  \gamma_{\phi{\hskip 0.5pt}1} \, \gamma_{\psi{\hskip 0.5pt}2b} )      \, , \nn \\
  &U_7=  2\, \gamma_{\psi{\hskip 0.5pt}1} \big ( 2\,  \gamma_{\phi{\hskip 0.5pt}1} \, \gamma_{\phi{\hskip 0.5pt}3c } -  \beta_{\lambda{\hskip 0.5pt}1b} \,  \gamma_{\phi{\hskip 0.5pt}3b }  + \gamma_{\phi{\hskip 0.5pt}1}\,   
  \beta_{y{\hskip 0.5pt}3e} \big )  -\gamma_{\phi{\hskip 0.5pt}1} 
  \big (  \beta_{y{\hskip 0.5pt}2a} \, \gamma_{\phi{\hskip 0.5pt}2b}
 - 6\,   \gamma_{\phi{\hskip 0.5pt}2a} \, \beta_{\lambda{\hskip 0.5pt}2e} \big )  \, ,  \nn \\
 & U_8=  \gamma_{\psi{\hskip 0.5pt}1} \big ( \gamma_{\phi{\hskip 0.5pt}1} \, \gamma_{\psi{\hskip 0.5pt}3b }
  - 2\, \gamma_{\psi{\hskip 0.5pt}1}\,   \gamma_{\phi{\hskip 0.5pt}3c} 
  - 3\, \beta_{\lambda{\hskip 0.5pt}1b} \,  \gamma_{\psi{\hskip 0.5pt}3a} 
  +  \beta_{y{\hskip 0.5pt}2a} ( \gamma_{\phi{\hskip 0.5pt}2b} 
  +  \gamma_{\psi{\hskip 0.5pt}2a} ) 
  - 3\,  \beta_{\lambda{\hskip 0.5pt}2e}\, \gamma_{\phi{\hskip 0.5pt}2a}  \big )
   \nn \\    \noalign{\vskip-1pt}  & \hskip 4cm {} 
 -    \gamma_{\phi{\hskip 0.5pt}1} \, \beta_{y{\hskip 0.5pt}2a} \, \gamma_{\psi{\hskip 0.5pt}2b}  \, , 
 \nn \\
 & U_9=  \gamma_{\psi{\hskip 0.5pt}1}{\!}^2 \big  (2\,  \gamma_{\phi{\hskip 0.5pt}1} \,  \beta_{y{\hskip 0.5pt}3b} - \beta_{\lambda{\hskip 0.5pt}1a} \, \beta_{y{\hskip 0.5pt}3e} 
 +  \beta_{y{\hskip 0.5pt}2a}\, \beta_{\lambda{\hskip 0.5pt}2b}  - 3\, \gamma_{\phi{\hskip 0.5pt}2a}   \, 
 \,  \beta_{\lambda{\hskip 0.5pt}2c}  \big ) \nn \\
 \noalign{\vskip-1pt}
 & \hskip 4cm {}
 - \beta_{\lambda{\hskip 0.5pt}1a}\big ( \beta_{y{\hskip 0.5pt}2a}  (  \gamma_{\psi{\hskip 0.5pt}1}
 \gamma_{\psi{\hskip 0.5pt}2a}   - \gamma_{\phi{\hskip 0.5pt}1}\gamma_{\psi{\hskip 0.5pt}2b}   )
 +  3 \,  \gamma_{\psi{\hskip 0.5pt}1}\gamma_{\phi{\hskip 0.5pt}2a} 
 \,  \beta_{\lambda{\hskip 0.5pt}2e} \big )  \, ,
 \label{Uscheme1}
\end{align}
and 
\be
 2 \, \gamma_{\psi{\hskip 0.5pt}1} \,  \gamma_{\phi{\hskip 0.5pt}2a} \, \upsilon_{\phi{\hskip 0.5pt}3c} +
  ( \gamma_{\phi{\hskip 0.5pt}1} \, \gamma_{\psi{\hskip 0.5pt}3a} -  \gamma_{\phi{\hskip 0.5pt}2a} \, 
  \gamma_{\psi{\hskip 0.5pt}2a}) \,   \beta_{y{\hskip 0.5pt}2a} = 0  \, .
\ee
This result for $ \upsilon_{\phi{\hskip 0.5pt}3c} $ is equivalent to the one in \eqref{resup} using $B_1 = U_{11}=0$.
Substituting for one and two loop coefficients the relations reduce to
\begin{align}
&\beta_{y{\hskip 0.5pt}3b} = \beta_{y{\hskip 0.5pt}3l} =2 \, , \quad
\beta_{y{\hskip 0.5pt}3b} - \beta_{y{\hskip 0.5pt}3b} =0 \, , \quad \beta_{y{\hskip 0.5pt}3c} - \beta_{y{\hskip 0.5pt}3b} = \beta_{y{\hskip 0.5pt}3d} - \beta_{y{\hskip 0.5pt}3j} = 1\, , \nn \\
& \beta_{y{\hskip 0.5pt}3b} + \gamma_{\psi{\hskip 0.5pt}3b} = \tfrac32 \, , \quad \gamma_{\phi{\hskip 0.5pt}3a} -2\, \gamma_{\phi{\hskip 0.5pt}3b} = \tfrac14 \, , \quad
 \gamma_{\phi{\hskip 0.5pt}3b} + 3\,  \gamma_{\psi{\hskip 0.5pt}3a}   = - \tfrac12 \, , \nn \\
&2\, \gamma_{\phi{\hskip 0.5pt}3c} -24 \,  \gamma_{\psi{\hskip 0.5pt}3a} - \gamma_{\psi{\hskip 0.5pt}3b} = 3 \, .
\end{align}

There are 14 contributions to $A^{(5)}$ independent of $\lambda$ which are
\begin{align}
A^{(5)} \big |_y = {}& a_{5j} \, \tr( \by_i \, y^j)  \, \tr( \by_j \, y^k) \, \tr( \by_k \, y^l) \,  \tr( \by_l \, y^i) \nn \\
&{} +  a_{5k}\,
\big ( \tr (\by_i \, y^k \by_j \, y^l ) + \tr (\by_i \, y^l \by_j \,  y^k) \big )\, \tr( \by_k \, y^i)\, \tr( \by_l \, y^j) \nn \\
&{} +  a_{5l}\,
\big ( \tr (\by_i \, y^k \by_j \, y^j) + \tr (\by_i \, y^j \by_j\,  y^k) \big )\, \tr( \by_k \, y^l)\, \tr( \by_l \, y^i ) \nn \\
&{} +  a_{5m}\,
\big ( \tr (\by_i \, y^l \by_j \, y^j \by_k \, y^k ) + \tr (\by_i \, y^j \by_j \, y^k \by_k \, y^l ) \big )\, \tr( \by_l \, y^i)  \nn \\
&{} +  a_{5n}\,
\big ( \tr (\by_i \, y^j \by_k \, y^k \by_j \, y^l ) + \tr (\by_i \, y^l \by_j \, y^k \by_k \, y^j) \big )\, \tr( \by_l \, y^i) \nn \\
&{} +  a_{5o}\,  \tr (\by_i \, y^j\by_j \, y^l \by_k \, y^k ) \, \tr( \by_l \, y^i)
+  a_{5p}\,  \tr (\by_i \, y^j \by_k \, y^l \by_j \, y^k ) \, \tr( \by_l \, y^i)    \nn \\
&{} +  a_{5q}\, \big ( \tr (\by_i \, y^j \by_k \, y^l ) + \tr (\by_i \, y^l \by_k \, y^j )\big  )\,   \tr (\by_j\, y^k\by_l \, y^i ) \nn \\
&{} +  a_{5r}\, \big ( \tr (\by_i \, y^i \by_j \, y^l) + \tr (\by_i \, y^k \by_j \, y^i )\big  )\,  
\big ( \tr (\by_k\, y^j\by_l \, y^l ) + \tr (\by_l \, y^j \by_k\, y^l )\big ) \nn \\
&{} +  a_{5s}\,
\big ( \tr (\by_i \, y^i \by_j \, y^j \by_k \, y^k \by_l \, y^l) + \tr (\by_i \, y^j \by_j \, y^k \by_k \, y^l \by_l\, y^i) \big ) \nn \\
&{} +  a_{5t}\,
\big ( \tr (\by_i \, y^i \by_j \, y^j \by_k \, y^l \by_l \, y^k) + \tr (\by_i \, y^j \by_j \, y^k \by_l \, y^l \by_k\, y^i ) \big ) \nn \\
&{} +  a_{5u}\,
\big ( \tr (\by_i \, y^i \by_j \, y^k \by_l \, y^l \by_k \, y^j) + \tr (\by_i \, y^j \by_k \, y^l \by_l \, y^k \by_j\, y^i) \big ) \nn \\
&{} +  a_{5v}\,
\big ( \tr (\by_i \, y^i \by_j \, y^k \by_l \, y^j \by_k \, y^l) + \tr (\by_i \, y^j \by_k \, y^l \by_j \, y^k \by_l\, y^i ) \big ) \nn \\
&{} +  a_{5w}\,
\big ( \tr (\by_i \, y^j \by_k \, y^i \by_l \, y^k \by_j \, y^l) + \tr (\by_j \, y^k \by_l \, y^i \by_k \, y^j \by_i\, y^l ) \big ) \, .
\label{A5loop}
\end{align} 
The ${\rm O}(\beta^2)$ freedom in \eqref{A4beta}, \eqref{A5beta} extends to
\begin{align}
& \delta a_{5j} = ( \epsilon_{3b_1} +  \epsilon_{3b_2} +  2\,\epsilon_{3b_3}) \gamma_{\phi{\hskip 0.5pt}1}{\!}^2 \, , \quad
\delta a_{5k} = 2\, \epsilon_2 \, \gamma_{\phi{\hskip 0.5pt}1} \, \gamma_{\psi{\hskip 0.5pt}2a} 
+ ( \epsilon_{3a_2} + 2\, \epsilon_{3a_3}) \gamma_{\phi{\hskip 0.5pt}1}{\!}^2 \, ,  \nn \\
& \delta a_{5l} = \epsilon_2 \, \gamma_{\phi{\hskip 0.5pt}1} \, \gamma_{\phi{\hskip 0.5pt}2b} 
+  \epsilon_{3a_1} \, \gamma_{\phi{\hskip 0.5pt}1}{\!}^2 + 2
 ( \epsilon_{3b_1} +  \epsilon_{3b_2} +  2\,\epsilon_{3b_3}) \gamma_{\phi{\hskip 0.5pt}1}\,\gamma_{\psi{\hskip 0.5pt}1}
\, , \nn \\
& \delta a_{5m} = 2\, \epsilon_2 \, \gamma_{\psi{\hskip 0.5pt}1} \, \gamma_{\psi{\hskip 0.5pt}2a} 
+ 2  ( \epsilon_{3a_1} +  \epsilon_{3a_2} +  2\,\epsilon_{3a_3})
\gamma_{\phi{\hskip 0.5pt}1}\,\gamma_{\psi{\hskip 0.5pt}1} +  
\epsilon_{3b_1} \, \gamma_{\psi{\hskip 0.5pt}1}{\!}^2 \, , \nn \\ 
& \delta a_{5n} = 2\, \epsilon_2 \,(  \gamma_{\phi{\hskip 0.5pt}1} \, 
\gamma_{\psi{\hskip 0.5pt}2b} + \gamma_{\psi{\hskip 0.5pt}1} \, \gamma_{\psi{\hskip 0.5pt}2b} ) 
+ 2  (  \epsilon_{3a_2} +  2\,\epsilon_{3a_3}) \gamma_{\phi{\hskip 0.5pt}1}\,\gamma_{\psi{\hskip 0.5pt}1} 
 \, , \nn \\
& \delta a_{5o} = 4 \, \epsilon_{3a_1}\,  \gamma_{\phi{\hskip 0.5pt}1}\,\gamma_{\psi{\hskip 0.5pt}1}  +
2 \, \epsilon_{3b_1}\,\gamma_{\psi{\hskip 0.5pt}1}{\!}^2 \, , \quad
\delta a_{5p} = 2\, \epsilon_2 \, \gamma_{\phi{\hskip 0.5pt}1} \, \beta_{y{\hskip 0.5pt}2f} \, , \quad
\delta a_{5q} = \tfrac14\, \epsilon_{3\lambda} \, \beta_{\lambda{\hskip 0.5pt}1b}{\!}^2 \, , \nn \\
& \delta a_{5r} = 2\, \epsilon_2 \, \gamma_{\psi{\hskip 0.5pt}1} \, \gamma_{\phi{\hskip 0.5pt}2b}
 + 2  (  \epsilon_{3b_2} +  2\,\epsilon_{3b_3})\,\gamma_{\psi{\hskip 0.5pt}1} {\!}^2 
 \, , \quad
 \delta a_{5s} = ( \epsilon_{3a_1} +  \epsilon_{3a_2} +  2\,\epsilon_{3a_3}) \gamma_{\psi{\hskip 0.5pt}1}{\!}^2 \, ,
 \nn \\
& \delta a_{5t} = 2\, \epsilon_2 \, \gamma_{\psi{\hskip 0.5pt}1} \, \gamma_{\psi{\hskip 0.5pt}2b}
+  ( 3\, \epsilon_{3a_1} + 2\, \epsilon_{3a_2} +  4\,\epsilon_{3a_3}) \gamma_{\psi{\hskip 0.5pt}1}{\!}^2 \, , \nn \\
& \delta a_{5u} = 2\, \epsilon_2 \, \gamma_{\psi{\hskip 0.5pt}1} \, \gamma_{\psi{\hskip 0.5pt}2b} 
+   (  \epsilon_{3a_2} +  2\,\epsilon_{3a_3}) \gamma_{\psi{\hskip 0.5pt}1}{\!}^2 \, , \quad
\delta a_{5v} = 2\, \epsilon_2 \, \gamma_{\psi{\hskip 0.5pt}1} \, \beta_{y{\hskip 0.5pt}2f} \, .
\end{align}

The contributions to the four loop $T$ involving only the $y{\hskip 0.5pt}\by$ couplings can be obtained by determining inequivalent pairs of vertices in \eqref{A4loop}. There are 36 possible contributions.
There are  36 equations resulting correspond to the number of inequivalent vertices in $A^{(5)} \big |_y$ as given 
in \eqref{A5loop}.
\begin{align}
a_{5j} = {}& \tfrac14 ( t_{4c_1} + t_{4c_2}+ t_{4c_3} + t_{4c_4} + t_{4c_5} ) \, \gamma_{\phi{\hskip 0.5pt}1} \, , \nn \\
\noalign{\vskip 2pt}
a_{5k} = {}& \tfrac14 \, t_2\,  \gamma_{\phi{\hskip 0.5pt}3d} 
+ \tfrac12 ( t_{3b_2}+t_{3b_3} )\, \gamma_{\psi{\hskip 0.5pt}2a} 
+ \tfrac12\, (t_{4e_{11}} + t_{4e_{15 }} ) \, \gamma_{\phi{\hskip 0.5pt}1}  \nn \\
= {}&  \tfrac12\,  t_{3b_1} \,  \gamma_{\psi{\hskip 0.5pt}2a} 
+ \tfrac12 (  t_{4e_1} + t_{4e_6}+ t_{4e_9}) \, \gamma_{\phi{\hskip 0.5pt}1}\, ,  \nn \\
\noalign{\vskip 2pt}
a_{5l} = {}&  t_2\,  \gamma_{\psi{\hskip 0.5pt}3c} + ( t_{4e_2} + t_{4e_3}+ t_{4e_4} + t_{4e_5}) \gamma_{\phi{\hskip 0.5pt}1} 
=   t_{4c_1} \,  \gamma_{\psi{\hskip 0.5pt}1} + ( t_{4e_7} + t_{4e_8}+ t_{4e_{10}})\, \gamma_{\phi{\hskip 0.5pt}1} 
\nn \\
= {}&  \tfrac12 ( t_{3b_1}+t_{3b_3} ) \gamma_{\phi{\hskip 0.5pt}2b} 
+  ( t_{4c_2} + t_{4c_4})  \gamma_{\psi{\hskip 0.5pt}1} + t_{4e_{12}} \, \gamma_{\phi{\hskip 0.5pt}1} \nn \\
=  {} &  \tfrac12 \, t_{3b_2} \,  \gamma_{\phi{\hskip 0.5pt}2b} 
+  ( t_{4c_3} + t_{4c_5})  \gamma_{\psi{\hskip 0.5pt}1} + ( t_{4e_{13}} + t_{4e_{14}})\, \gamma_{\phi{\hskip 0.5pt}1} \, ,  \nn \\
\noalign{\vskip 2pt}
a_{5m} = {}& \tfrac12 \, t_2\,  \gamma_{\phi{\hskip 0.5pt}3f} + ( t_{4e_{11}} + t_{4e_{12}}+ t_{4e_{14}}+ t_{4e_{15}}  ) \,\gamma_{\phi{\hskip 0.5pt}1} 
=   t_{4d_1} \,  \gamma_{\phi{\hskip 0.5pt}1} + ( t_{4e_6} + t_{4e_7}+ t_{4e_9})\, \gamma_{\psi{\hskip 0.5pt}1} \nn \\
= {}&  t_{3a_2} \,  \gamma_{\psi{\hskip 0.5pt}2a}
+  ( t_{4d_3} + t_{4d_5})  \gamma_{\phi{\hskip 0.5pt}1} + ( t_{4e_4} + t_{4e_5} ) \, \gamma_{\psi{\hskip 0.5pt}1} \nn \\
=  {} &  (t_{3a_1}+ t_{3a_3}) \,  \gamma_{\psi{\hskip 0.5pt}2a} 
+  ( t_{4d_2} + t_{4d_4})  \gamma_{\phi{\hskip 0.5pt}1} +  t_{4e_2} \, \gamma_{\psi{\hskip 0.5pt}1} \, ,  \nn \\
\noalign{\vskip 2pt}
a_{5n} = {}& \tfrac12 \, t_2\,  \gamma_{\phi{\hskip 0.5pt}3g} +( t_{3b_2} + t_{3b_3} ) \gamma_{\psi{\hskip 0.5pt}2b}  + ( t_{4e_{11}} +     t_{4e_{15}} ) \,  \gamma_{\psi{\hskip 0.5pt}1} 
=  t_{3a_1}  \,  \gamma_{\psi{\hskip 0.5pt}2a} + t_{4e_1} \, \gamma_{\psi{\hskip 0.5pt}1}  +  ( t_{4f_7} + t_{4f_8})  \,  \gamma_{\phi{\hskip 0.5pt}1} \nn \\
= {}&   t_2\,  \gamma_{\psi{\hskip 0.5pt}3e}+ (  t_{3a_2} + t_{3a_3} )  \gamma_{\psi{\hskip 0.5pt}2a}
+  ( t_{4f_2} + t_{4f_4})  \gamma_{\phi{\hskip 0.5pt}1} \nn \\
= {}&   t_{3b_1}  \,  \gamma_{\psi{\hskip 0.5pt}2b} +(  t_{4e_6} + t_{4e_9}) \, \gamma_{\psi{\hskip 0.5pt}1} +  t_{4f_3} \,  \gamma_{\phi{\hskip 0.5pt}1}  \, ,  \nn \\
\noalign{\vskip 2pt}
a_{5o} = {}&  t_2\,  \gamma_{\psi{\hskip 0.5pt}3d} +( t_{4e_2} + t_{4e_4} ) \gamma_{\psi{\hskip 0.5pt}1}  +(  t_{4f_1} + t_{4f_5} )\,  \gamma_{\phi{\hskip 0.5pt}1} 
=   2\, t_{4e_7} \, \gamma_{\psi{\hskip 0.5pt}1}  +  2\, t_{4f_6}  \,  \gamma_{\phi{\hskip 0.5pt}1} \nn \\
= {}&   t_2\,  \gamma_{\psi{\hskip 0.5pt}3h} + 2 ( t_{4e_{12}} + t_{4e_{14}})  \gamma_{\psi{\hskip 0.5pt}1} \nn  \, ,  \\
\noalign{\vskip 2pt}
a_{5p} = {}& t_2\,  \gamma_{\phi{\hskip 0.5pt}3m} + ( t_{3b_2} + t_{3b_3} ) \beta_{y{\hskip 0.5pt}2f} =  t_2\,  \beta_{y{\hskip 0.5pt}3s} +( t_{4g_1} + t_{4g_3} ) \gamma_{\phi{\hskip 0.5pt}1}  \nn \\
\noalign{\vskip 2pt}
a_{5q} = {}&  \tfrac{1}{8} ( t_2 \, \beta_{y{\hskip 0.5pt}3w}  + t_{4a_2} \, \beta_{\lambda{\hskip 0.5pt}1b} ) \, , \nn \\
\noalign{\vskip 2pt}
a_{5r} = {}&   \tfrac14\, t_2 \, \gamma_{\psi{\hskip 0.5pt}3g}+ \tfrac14\, ( t_{3a_2} + t_{3a_3} )  \,  \gamma_{\phi{\hskip 0.5pt}2b}  
+ \tfrac12 ( t_{4e_3} + t_{4e_5})\, \gamma_{\psi{\hskip 0.5pt}1} \nn \\
={}& \tfrac14\, t_{3a_1} \,  \gamma_{\phi{\hskip 0.5pt}2b} + \tfrac12 ( t_{4e_8} + t_{4e_{10}}+ t_{4e_{13}}) \,\gamma_{\psi{\hskip 0.5pt}1} \, , \nn \\
\noalign{\vskip 2pt}
a_{5s} = {}& \tfrac14 ( t_{4d_1} + t_{4d_2}+ t_{4d_3} + t_{4d_4} + t_{4d_5} ) \, \gamma_{\psi{\hskip 0.5pt}1} \, , \nn \\
\noalign{\vskip 2pt}
a_{5t} = {}& t_2\,  \gamma_{\psi{\hskip 0.5pt}3i} +   ( t_{4f_1} + t_{4f_2}+ t_{4f_4} + t_{4f_5} ) \, \gamma_{\psi{\hskip 0.5pt}1} 
=   ( t_{4d_1} + 2\, t_{4f_6}+ t_{4f_7} + t_{4f_8} ) \, \gamma_{\psi{\hskip 0.5pt}1} \nn \\
= {}& (t_{3a_1} +   t_{3a_3}  )\,  \gamma_{\psi{\hskip 0.5pt}2b} +  ( t_{4d_2} + t_{4d_4}+ t_{4f_1} ) \, \gamma_{\psi{\hskip 0.5pt}1}  \nn \\
= {}&     t_{3a_2} \, \gamma_{\psi{\hskip 0.5pt}2b} + ( t_{4d_3} + t_{4d_5}+ t_{4f_3} + t_{4f_5} ) \, \gamma_{\psi{\hskip 0.5pt}1} \, ,\nn \\
\noalign{\vskip 2pt}
a_{5u} = {}&\tfrac12\,  t_2\,  \gamma_{\psi{\hskip 0.5pt}3j} + \tfrac12 (t_{3a_2} +   t_{3a_3}  )\,  \gamma_{\psi{\hskip 0.5pt}2b} 
+ \tfrac12 ( t_{4f_2} + t_{4f_4}) \, \gamma_{\psi{\hskip 0.5pt}1} \nn \\
={}&    \tfrac12 \, t_{3a_1} \,  \gamma_{\psi{\hskip 0.5pt}2b}  + \tfrac12 ( t_{4f_3} + t_{4f_7} + t_{4f_8} ) \, \gamma_{\psi{\hskip 0.5pt}1}\, , \nn \\
\noalign{\vskip 2pt}
a_{5v} = {}&  t_2\, \beta_{y{\hskip 0.5pt}3{\tilde o}} + ( t_{4g_1} + t_{4g_3}) \, \gamma_{\psi{\hskip 0.5pt}1} 
=  t_2\, \beta_{y{\hskip 0.5pt}3{\tilde p}} + ( t_{4g_2} + t_{4g_3}) \, \gamma_{\psi{\hskip 0.5pt}1}  \nn \\
={}&  t_2\, \gamma_{\psi{\hskip 0.5pt}3p} + ( t_{3a_2} + t_{3a_3}) \, \beta_{y{\hskip 0.5pt}2f}  
=    t_{3a_1} \, \beta_{y{\hskip 0.5pt}2f}   + t_{4g_1} \, \gamma_{\psi{\hskip 0.5pt}1}\, , \nn \\
\noalign{\vskip 2pt}
a_{5w} = {}& \tfrac14\, t_2 \, \beta_{y{\hskip 0.5pt}3{\tilde z}} \, .
\end{align}

This leads to the non planar condition from the relations involving $a_{5p}, \, a_{5v}$
\be
U_{10}= 2 \, \gamma_{\psi{\hskip 0.5pt}1}( \beta_{y{\hskip 0.5pt}3s} - \gamma_{\phi{\hskip 0.5pt}3m}) - 2 \, \gamma_{\phi{\hskip 0.5pt}1} ( \beta_{y{\hskip 0.5pt}{\tilde o}} -
\gamma_{\psi{\hskip 0.5pt}3p}) +  ( \gamma_{\phi{\hskip 0.5pt}2b} - 2\, \gamma_{\psi{\hskip 0.5pt}2a} )\, 
\beta_{y{\hskip 0.5pt}2f}  \, ,
\label{Uscheme2}
\ee
and just two other conditions
\begin{align}
&U_{11}=  \gamma_{\phi{\hskip 0.5pt}1} ( \gamma_{\psi{\hskip 0.5pt}3j} - 2\, \gamma_{\psi{\hskip 0.5pt}3i} )
+  \gamma_{\psi{\hskip 0.5pt}1} (  \gamma_{\psi{\hskip 0.5pt}3d} +  \gamma_{\psi{\hskip 0.5pt}3e}
+  \gamma_{\phi{\hskip 0.5pt}3f} -  \gamma_{\phi{\hskip 0.5pt}3g} -  \gamma_{\phi{\hskip 0.5pt}3h} )+
(  \gamma_{\phi{\hskip 0.5pt}2b} - 2\,  \gamma_{\psi{\hskip 0.5pt}2a}  )  \gamma_{\psi{\hskip 0.5pt}2b}   \, ,
\nn \\
& U_{12}= 2\, \gamma_{\phi{\hskip 0.5pt}1}{\!}^2 \, \gamma_{\psi{\hskip 0.5pt}1} \,\gamma_{\psi{\hskip 0.5pt}3j} 
-   \gamma_{\phi{\hskip 0.5pt}1}\, \gamma_{\psi{\hskip 0.5pt}1}{\!}^2
( 2\, \gamma_{\psi{\hskip 0.5pt}3e }  + \gamma_{\phi{\hskip 0.5pt}3g }  ) 
+  \gamma_{\psi{\hskip 0.5pt}1}{\!}^3  \gamma_{\phi{\hskip 0.5pt}3d } \nn \\
& \hskip 1.5cm {} - ( 2\, \gamma_{\phi{\hskip 0.5pt}1} \, \gamma_{\psi{\hskip 0.5pt}2b}
- \gamma_{\psi{\hskip 0.5pt}1}\, \gamma_{\phi{\hskip 0.5pt}2b }) ( \gamma_{\phi{\hskip 0.5pt}1} \, 
\gamma_{\psi{\hskip 0.5pt}2b } - \gamma_{\psi{\hskip 0.5pt}1} \,  \gamma_{\psi{\hskip 0.5pt}2a } ) \, .
\label{Uscheme3}
\end{align}

The results obtained in \eqref{Uscheme0},  \eqref{Uscheme1},  \eqref{Uscheme2},  \eqref{Uscheme3} 
are a subset of those obtained in section \ref{sec:consis}  in the general case. We list some scheme variations
which are not immediately evident from previous results
\begin{align}
 \delta U_6 =  {} &  - 2 \, \gamma_{\psi1} \, Y_{\phi 1, \psi 1, y2a}\, , \qquad 
\delta U_8 = - 2 \, \gamma_{\psi1} \big (  Y_{\phi 1, \psi 1, y2a} + 3 \,  Y_{\phi 2a, \psi 1, y1} \big ) \, , \nn \\
 \delta U_9 =   {} & 2 \, \gamma_{\psi1} \big ( 
 - \gamma_{\psi 1}\,  Y_{\phi 1,  y2a, \lambda 1a}  -  \beta_{y2a} \,  Y_{\phi 1, \psi 1, \lambda 1 a} 
+  3 \, \gamma_{\phi 2a} \,  Y_{ \psi 1, \lambda 1a , \lambda 1b} +  U_0 \, X_{\psi 1, \lambda 1a}
\big ) \, , \nn \\
 \delta U_{10} = {} & -8\,  Y_{\phi 1, \psi 1, y2f}\, , \qquad \delta U_{11} = -8\,  Y_{\phi 1, \psi 1, \psi 2b}\, ,  \nn \\
 \delta U_{12} = {}&   -2 \, \gamma_{\psi 1}{\!}^2 \big ( 2\,  Y_{\phi 1, \psi 1, \psi 2a}
+  Y_{\phi 1, \phi 2b, \psi 1} \big )\, .
\end{align}

\section{Scheme Variations in General}

\label{varG}

The coordinates in quantum field theories are the couplings. Physical results should be
invariant under reparametrisations of the couplings or in this context changes of renormalisation scheme. 
Of course determining possible invariants is an exercise in differential geometry. Under changes
of scale there is a RG flow in the space of couplings determined by a vector field, the $\beta$-function,
and at any fixed point where the $\beta$-function vanishes
the scale dimensions of operators are determined by the eigenvalues of
the anomalous dimension matrix which, at a fixed point, is a two index tensor under reparametrisations.

In a perturbative context the possible  reparametrisations  of couplings
are naturally restricted to preserve the form
of the $\beta$-function in terms of contributions corresponding to 1PI diagrams which are superficially
divergent. For an expansion in terms of diagrams with increasing loop order there a usually a restricted  set of
possible vertices $\{V_\vv \}$, labelled by $\vv$ and edges $\{E_\ev\}$ labelled by $\ev$.
The various possible  $\ev$ correspond to the different fields in the theory and 
each $\vv$ to the different basic couplings.
For  $n_\vv$  lines meeting at a particular vertex $V_\vv$ there is then an associated coupling 
$(G^\vv)_{i_1 \dots i_{n_\vv}}$ where $i_r$ an index associated with the a diagram line or edge, 
$\ev$, connected to the  vertex $\vv$. 
 In a diagram with a line $\ev$ there is an associated  two index link or propagator
$(P^\ev )_{i j}$ with $i,j= 1, \dots n_\ev$.  
For each coupling $G^\vv$ there there is a symmetry group  ${\sl G}_\vv \subset \S_{n_\vv}$ 
generated by permutations of those lines corresponding to identical particles.
For convenience we may consider a basis in which the couplings are real and $(P^\ev)_{i j }$ is symmetric, otherwise 
for complex couplings they form conjugate pairs. For simplicity we restrict to dimensionless couplings.

It is convenient to adopt a notation where for any set of $\{(R^\ev)_{i j}\}$ and 
$\{(\kappa^\ev)_{i j}  \}$
\begin{align}
( G^\vv \circ R)_{i_1 \dots i_{n_\vv}} ={}& (G^\vv)_{j_1 \dots j_{n_\vv}} \, (R^1)_{j_1 i_1} \dots
(R^{n_\vv})_{j_{n_\vv} i_{n_\vv}} \, , \nn \\
( G^\vv  \kappa)_{i_1 \dots i_{n_\vv}} ={}&   {\textstyle \sum_r} \, 
 (G^\vv)_{i_1 \dots i_{r{-1}} j\hskip 0.5pt  i_{r{+1}} \dots  i_{n_\vv}} \, 
(\kappa^r)_{j i_r} \, .
\end{align}
Clearly $(G^\vv \circ R) \circ R' = G^\vv \circ R\hskip 0.5pt R'$, with $R \hskip 0.5pt R'= \{ R_\ev R'{\!}_\ev\}$,
 $((G^\vv \,\kappa)\,\kappa') - ((G^\vv \,\kappa')\,\kappa) = (G^\vv\, [\kappa,\kappa'])$.
For an overall symmetry $\cG$ then $ G^\vv \circ R = G^\vv$, 
for each $R^\ev $ belonging to the appropriate representation
of $\cG$, and $ (P^\ev)_{k l} (R^\ev)_{k i} (R^\ev)_{l j} = (P^\ev)_{i j}$. For a vacuum diagram there
is a corresponding amplitude formed by joining couplings for each vertex with appropriate propagators
\be
A(G , P) = A (  g  ) \, , \qquad  g^\vv = G^\vv \circ P^{\frac12} \, ,  \qquad  A ( g )=  A (g,\I ) \, , 
\label{ghat}
\ee
with $G=\{G^\vv\}, \ P=\{P_\ev\}$ and $P_\ev{}^\frac12$ is required to be symmetric. In general we require
\be
A ( G \circ \R, \R^T P\, \R ) = A(G, P) \, , \quad A (g\circ \R ) = A(g) \, , \qquad \R = \{ \R^\ev\} \, , \quad
\R^\ev \R^\ev{}^T = \I^\ev \, .
\label{RA}
\ee

Reparametrisations of relevance here are generated by
\begin{align}
 \cD_{v,w}(G,P) = {}& \sum_\vv v^\vv(G,P)\cdot  \frac{\pr}{\pr G^\vv} 
 + 2\,  \sum_\ev \big  (P^\ev\, w^\ev (G,P)  P^\ev \big )\cdot \frac{\pr}{\pr P^\ev} \nn \\
  = {}& \sum_\vv v^\vv(G,P)\cdot  \frac{\pr}{\pr G^\vv} 
 - 2\, \sum_\ev w^\ev (G,P)\cdot \frac{\pr}{\pr P^\ev{}^{-1}}    \, ,
 \label{Dvw}
 \end{align}
 where $v^\vv(G,P), \, (w^\ev (G,P))_{i j}$ are determined in terms of sums of 1PI one and higher loop
 vertex and propagator graphs  with  vertices mapped to the appropriate $G^\vv$ and similarly internal lines to $P^\ev$. 
 The 2 in \eqref{Dvw} is introduced for later convenience.
 For a finite transformation
 $G^\vv \to G^\vv{}' ,  \, P^\ev \to P^\ev{}'$ then
 \be
 G^\vv{}' ( G,P) =  \exp  \big (   \cD_{v,w}(G,P)\big  )\, G^\vv \, , \qquad P^\ev{}' (G,P) 
 = \exp \big (   \cD_{v,w}(G,P)  \big)\, P^\ev\, .
 \ee
 With this definition
 \be
  G^\vv{}' ( G,P) =  G^\vv +  f^\vv (G,P ) \, , \qquad  P^\ev{}'(G,P )^{-1} = P^\ev{}^{-1} - 2\,c^\ev (G,P) \, ,
  \label{reparam}
  \ee
  where $f^\vv , \, c^\ev$ can be expressed in terms of contributions from 1PI vertex and propagator graphs. 
  Following from \eqref{ghat} then
  \be
 f^\vv(G,P) =  f^\vv(g)  \circ P^{-\frac12} \, ,  \qquad 
 c^\ev(G,P) =   P^\ev{}^{-\frac12} c^\ev(g) P^\ev{}^{-\frac12}  \, ,
 \ee
 and 
 \be
 g^\vv{}'(G,P) = G^\vv{}' (G,P) \circ  P'(G,P)^{\frac12}  =  g^\vv{}' ( g )  \circ \R \, ,
 \label{ghg}
\ee
where
\be
g^\vv{}' ( g ) = \big ( g^\vv + f^\vv(g)  \big ) \circ   \big ( \I -  2\,  c(g) \big )^{-\frac12} \, ,
 \ee
with
 \be
 \R^\ev =  \big ( \I^\ev - 2\,  c^\ev( g) \big )^{\frac12} \, P^\ev {\!}^{-\frac12} \,  P^\ev{}'(G,P )^{\frac 12} \, .
 \label{Rspec}
 \ee
 From this definition
 \be
 \R^\ev \, \R^\ev{}^T = \I^\ev \, , 
 \label{Rrot}
 \ee
 The inverse of \eqref{ghg} 
 \be
 g^\vv ( g{\hskip0.5pt}' ) = \big ( g^\vv{}' + f^\vv{}'( g{\hskip0.5pt}')  \big ) \circ 
 \big ( \I -  2\,  c'(g'{\hskip0.5pt}) \big )^{-\frac12} \, ,  \quad 
 \ee
 is obtained by taking
 \begin{align}
 f^\vv{}'(g{\hskip0.8pt}') ={}& -  f^\vv(g) \circ  
 \big ( \I -   c'(g{\hskip0.8pt}') \big )^{\frac12} \, , \nn \\
     c^\ev{}'(g{\hskip0.8pt}') ={}&  \I^\ev - \big ( \I^\ev -  2\, c^\ev(g) \big )^{- 1}   
     =  - \big ( \I^\ev - 2\,  c^\ev(g) \big )^{- \frac12} \, c^\ev(g)   \,
     \big ( \I^\ev -  c^\ev(g) \big )^{- \frac12}  \,,
 \end{align}
from which it follows that  $ f^\vv{}'(g{\hskip0.5pt}') , \ c^\ev{} '(g{\hskip0.5pt}')$ are both expressible
 as expansions in $g{\hskip0.5pt}'$ in terms of contributions corresponding to 1PI diagrams. For $f,\, c$
 infinitesimal the generator of reparametrisations can be reduced, from \eqref{ghg}, \eqref{Rrot}, to the form
 \be
 \cD_{f,c} (G,P) = \cD\raisebox{-1.5 pt}{$\scriptstyle f + (g\hskip 0.5pt c )$}(g)
 + \cD\raisebox{-1.5 pt}{$\scriptstyle (g\hskip 0.5pt \omega )$}(g)\, , \quad
\cD\raisebox{-1.5 pt}{$\scriptstyle h$}(g) =  \sum_\vv h^\vv(g)\cdot  \frac{\pr}{\pr g^\vv } \, , 
\ee
where
\be
\R^\ev = \I^\ev + \omega^\ev \, , \qquad  \omega^\ev{}^T = -  \omega^\ev \, .
 \ee
 
 Under a reparametrisation
 \be
 A'( G',P') = A(G,P)  \qquad \Rightarrow \qquad A'( g{\hskip 0.8pt}' ) = A(g) \, ,
 \ee
 is consistent with \eqref{ghat}.
 
 The essential RG functions ${\tilde \beta}^\vv(G,P), \ \gamma^\ev(G,P)$ are formed from contributions 
 corresponding to 1PI diagrams, with $ \gamma^\ev(G,P)$ symmetric. Corresponding to \eqref{RA}
 \be
{\tilde \beta}^\vv ( G \circ \R, \R^T P\, \R ) =  {\tilde \beta}^\vv (G, P)\circ \R  \, , \quad
\gamma^\ev ( G \circ \R, \R^T P\, \R ) = \R^\ev{}^T  \gamma^\ev (G, P) \, \R^\ev \,  .
\label{RGP}
\ee
 Reduction to the coupling $g$, determined as in \eqref{ghat}, is achieved by taking
 \be
{\tilde \beta}^\vv(G,P) = {\tilde \beta}^\vv(g) \circ  P^{-\frac12} \, , \qquad  \gamma^\ev(G,P)
= P^\ev{}^{-\frac12} \gamma^\ev (g) P^\ev{}^{-\frac12} \, ,
\ee
and then ${\tilde \beta}^\vv(g) , \, \gamma^\ev(g)$  are expressible
just in terms of 1PI contributions. RG flow is then generated by
\be
  \cD\raisebox{-1.5 pt}{$\scriptstyle {\tilde \beta},\gamma $}(G,P)   \, ,
  \ee
  where
  \be
   \cD\raisebox{-1.5 pt}{$\scriptstyle {\tilde \beta},\gamma $}(G,P)\,  A(g) 
   = \cD_\beta (g) \, A(g)  \, , \qquad 
  \beta^\vv(g) = {\tilde \beta}^\vv(g) + (g^\vv  \gamma(g))\, , 
  \label{betad}
\ee
where we make use of 
\be
\cD\raisebox{-1.5 pt}{$\scriptstyle {\tilde \beta},\,\gamma $}(G,P) \, P^\ev{}^\frac12 
= P^\ev{}^\frac12( \gamma^\ev + \omega^\ev) = ( \gamma^\ev - \omega^\ev) P^\ev{}^\frac12 \, ,
\ \ \mbox{for some} \ \ \omega^\ev = - \omega^\ev{}^T \, .
\ee
and 
\be
 \cD\raisebox{-1.5 pt}{$\scriptstyle (g \hskip 0.5pt\omega)$} (g)\, A(g) = 0 \, .
 \ee
In \eqref{betad} the $\beta$-function $\beta^\vv(g)$ then has the standard
form in terms of 1PI contributions and from \eqref{RA}
 
 Under a reparametrisation as in \eqref{reparam}
 \begin{align}
 {\tilde \beta}^\vv{}'(G',P') = {}& 
 \cD\raisebox{-1.5 pt}{$\scriptstyle {\tilde \beta},\gamma $}(G,P)  \, G^\vv{}'(G,P) \, ,\nn \\
 \gamma^\ev{}' (G',P') = {}& 
 - \cD\raisebox{-1.5 pt}{$\scriptstyle {\tilde \beta}, \gamma $}(G,P)  \, P^\ev{}'(G,P)^{-1}
 =  \gamma^\ev (G,P) + \cD\raisebox{-1.5 pt}{$\scriptstyle {\tilde \beta},\gamma $}(G,P)  \, c^\ev(G,P) \, .
 \label{DGP}
 \end{align}
 Defining
 \be
 {\tilde \beta}^\vv{}'(G',P')  \circ P'{}^\frac12 = {\tilde \beta}^\vv{}' (g') \circ \R \, , \qquad
 P^\ev{}'{}^\frac12   \gamma^\ev{}' (G',P')   \circ P^\ev{} '{}^\frac12  =  \R^\ev{}^T \gamma^\ev{}'(g')\, \R^\ev \, ,
 \ee
then  reduces, with $\R$ given by \eqref{Rspec},  for ${\tilde \beta}(g), \ \gamma(g)$ to
 \begin{align}
 {\tilde \beta}^\vv{}' (g')  = {}& \big ( {\tilde \beta}^\vv(g) + \cD_{\beta}(g) \, f^\vv (g) - 
 \big (f_\vv(g) \gamma(g) \big ) \big  ) 
 \circ ( \I - 2\, c(g) )^{-\frac12} \, , \nn \\
\gamma^\ev{}' (g') = {}&  ( \I - 2\, c^\ev(g) )^{-\frac12} \big ( \gamma^\ev(g) +  \cD_{\beta}(g) \, c^\ev (g)
 -  \big \{ \gamma^\ev(g) , c^\ev(g) \big \} \big )  ( \I - 2\,  c^\ev(g) )^{-\frac12} \, ,
 \label{bgt}
 \end{align} 
  To achieve \eqref{bgt} we make use of
 \be
  \cD\raisebox{-1.5 pt}{$\scriptstyle (g \hskip 0.5pt\omega)$} (g) \,f^\vv (g) = ( f^\vv(g)\hskip 0.5pt \omega ) \, , \qquad
  \cD\raisebox{-1.5 pt}{$\scriptstyle (g \hskip 0.5pt\omega)$} (g)\, c^\ev (g)  = \big [ c^\ev(g), \omega \big ] \, .
  \ee
  The expressions obtained in \eqref{bgt}
 ensure that $ {\tilde \beta}^\vv{}' (g'), \,  \gamma^\ev{}' (g') $ expanded in terms
 of $g'$ as in \eqref{ghg}  are expressible in terms of 1PI contributions and furthermore
 \be
 \beta^\vv{}' (g') = {\tilde \beta}^\vv{}' (g') + \big  ( g^\vv{}'  \gamma(g')\big  )
 = \cD_\beta( g) \, g^\vv{}' + ( g^\vv{}' \, \Omega ) \, , \quad \Omega = - \Omega^T \, ,
 \ee
 where
 \begin{align}
&  \Omega^\ev= ( \I - 2\, c^\ev(g) )^{-\frac12} \big ( [ c^\ev(g),  \gamma^\ev(g) ] + {\hat \Omega}^\ev \big ) 
 ( \I - 2\, c^\ev(g) )^{-\frac12} \, , \nn \\
&  \cD_\beta (g) ( \I - 2\, c^\ev(g) )^{-\frac12} \ ( \I - 2\, c^\ev(g) )^{\frac12} = ( \I - 2\, c^\ev(g) )^{-1 } 
\big ( \cD_\beta (g) \, c^\ev(g) - {\hat \Omega}^\ev \big ) \, .
 \end{align}
  
  From \eqref{DGP} 
  \be
   \cD\raisebox{-1.5 pt}{$\scriptstyle {\tilde \beta}',\gamma' $}(G,P)
   = \exp \big ( - \cD_{v,w}(G,P) \big ) \,   \cD\raisebox{-1.5 pt}{$\scriptstyle {\tilde \beta},\gamma $}(G,P)
   \exp \big (  \cD_{v,w} (G,P) \big ) \, ,
   \ee
   or
   \begin{align}
   {\tilde \beta}^\vv{}'(G,P) = {}& \exp \big ({ - \cL_{v,w} (G,P)} \big ) {\tilde \beta}^\vv(G,P) \, , \quad
   \gamma^\ev{}'(G,P) =  \exp \big ({ - \cL_{v,w} (G,P)} \big ) \gamma^\ev(G,P) \, ,
   \end{align}
   with, to lowest order, 
   \begin{align}
   \delta {\tilde \beta}^\vv(G,P) = - \cL_{v,w} (G,P) \,  {\tilde \beta}^\vv(G,P) = {}& 
    \cD\raisebox{-1.5 pt}{$\scriptstyle {\tilde \beta} ,\gamma$}(G,P) \, v^\vv (G,P) 
    - \cD_{v,w}(G,P) \, {\tilde \beta}^\vv(G,P)  \, ,   \nn \\
     \delta \gamma^\ev(G,P) =    - \cL_{v,w} (G,P) \,  \gamma^\ev(G,P) = {}& 
 \cD\raisebox{-1.5 pt}{$\scriptstyle {\tilde \beta} ,\gamma$}(G,P) \, w^\ev (G,P) 
    - \cD_{v,w}(G,P) \, \gamma^\ev(G,P)  \, .
    \label{varbg}
    \end{align}

 For application here we set up a basis of 1PI vertex and propagator graphs so that
 \begin{align}
 v^\vv(G,P) = \sum_{\ell,r} \, \epsilon_{\vv \hskip 0.5pt \ell r}\, 
 \S\raisebox{-1 pt}{$\scriptstyle {p_{\vv\ell r}}$}  \, G^{\vv \hskip 0.5pt \ell r}(G,P)   \, ,
 \qquad w^\ev(G,P) = \sum_{\ell,r} \, \epsilon_{\ev \hskip 0.5pt \ell r}\, 
 \S\raisebox{-1 pt}{$\scriptstyle {p_{\ev\ell r}}$} \, 
 G^{\hskip 0.5pt\ev \hskip 0.5pt \ell r} (G,P)    \, .
 \label{vwexp}
 \end{align}
 In  the expansion of $ v^\vv(G,P) $ the sum is over contributions
 $G^{\vv \hskip 0.5pt \ell r}(G,P)$ corresponding to  particular  $\ell$, $\ell= 1,2,\dots$, 
 loop 1PI vertex graphs  $\cG^{\vv\hskip 0.5pt \ell r}$,  with the same external lines as  $\vv$,    and labelled by $r$.
 In each case $\S\raisebox{-1pt}{$\scriptstyle {p_{\vv\ell r}}$} $ denotes the sum over the $p_{\vv\hskip 0.5pt \ell r}$
 permutations of the external lines of $\cG^{\vv\hskip 0.5pt \ell r}$ necessary to ensure the symmetry under external  line 
 permutations satisfied by $G^\vv$.  Similarly  $G^{\hskip 0.5pt \ev \hskip 0.5pt \ell r}(G,P)$ corresponds 
 to  a 1PI propagator graph $\cG^{\ev\hskip 0.5pt \ell r}$,
  the associated permutations over external lines are 
  $\S\raisebox{-1 pt}{$\scriptstyle {p_{\ev\ell r}}$} $ with $p_{\ev\ell r} =1, \, 2$ 
  according to whether  $\cG^{\hskip 0.5pt\ev \hskip 0.5pt \ell r} $ is symmetric or not. 
 Inserting vertex or propagator graphs generates an algebra which  arises from
 \begin{align}
\S\raisebox{-1 pt}{$\scriptstyle {\hskip 0.5pt p_{\vv' \ell' r'}}$}  \, 
G^{\hskip 0.5pt\vv' \hskip 0.5pt \ell' r'}(G,P)  \cdot \frac{\pr}{\pr G} \
 \S\raisebox{-1 pt}{$\scriptstyle {\hskip 0.5pt p_{\hskip 0.5pt\gv \ell r}}$}  \, 
 G^{\hskip 0.5pt\gv \hskip 0.5pt \ell r}(G,P)  =  {}&
  \sum_s \, N_{\gv \hskip 1pt L s}^{\vv' \skip0.8pt\ell' r'\! ,\, \gv  \hskip0.8pt\ell r}\,
 \S\raisebox{-1 pt}{$\scriptstyle {\hskip 0.5pt p_{\hskip 0.5pt\gv L s}}$}  \, 
   G^{\hskip 0.5pt\gv \hskip 0.5pt L s}(G,P) \, , \nn \\
 -  \S\raisebox{-1 pt}{$\scriptstyle {\hskip 0.5pt p_{\ev' \ell' r'}}$}  \, G^{\hskip 0.5pt\ev' \hskip 0.5pt \ell' r'}(G,P)  
 \cdot \frac{\pr}{\pr P^{-1}} \
 \S\raisebox{-1 pt}{$\scriptstyle {\hskip 0.5pt p_{\hskip 0.5pt\gv \ell r}}$}\, 
 G^{\hskip 0.5pt\gv \hskip 0.5pt \ell r}(G,P)  =  {}&
  \sum_s \, N_{\gv \hskip 1pt L s}^{\ev' \hskip0.8pt\ell' r' \! ,\, \gv \hskip0.8pt\ell r}\,
    \S\raisebox{-1 pt}{$\scriptstyle {\hskip 0.5pt p_{\hskip 0.5pt\gv L s}}$}   \, 
    G^{\hskip 0.5pt\gv \hskip 0.5pt L s}(G,P) \, , \nn \\
   & \gv \in \{ \vv, \, \ev\} \, , \quad L = \ell' + \ell \, .
  \end{align}
  In general $ N_{\gv \hskip 1pt L s}^{\gv' \skip0.8pt\ell' r' \! ,\, \gv  \hskip0.8pt\ell r}$ are integers.
  
  Just as in \eqref{vwexp} there is a similar expansion
  \begin{align}
 {\tilde \beta}^\vv(G,P) = \sum_{\ell,r} \, \beta_{\vv \hskip 0.5pt \ell r}\, 
  \S\raisebox{-1 pt}{$\scriptstyle {p_{\vv\ell r}}$}  \, G^{\vv \hskip 0.5pt \ell r}(G,P) \, ,
 \qquad \gamma^\ev(G,P) = \sum_{\ell,r} \, \gamma_{\ev \hskip 0.5pt \ell r}\, 
  \S\raisebox{-1 pt}{$\scriptstyle {p_{\ev\ell r}}$} \, G^{\ev \hskip 0.5pt \ell r} (G,P)    \, .
 \label{bgexp}
 \end{align}
 The results in \eqref{varbg} then ensure
 \begin{align}
 \delta \alpha_{\gv \hskip 0.5pt L s} =  \sum_ {\genfrac{}{}{0pt}{3}{\gv'r'\ell',r\hskip0.5pt\ell}{\ell'+\ell=L}}
\hskip - 0.1cm n_{\gv'} \hskip 0.1cm {}&
 N_{\gv \skip0.8pt L s}^{\gv' \hskip0.8pt\ell' r'\!,\, \gv \hskip0.8pt\ell r}\; X_{\gv' \hskip0.8pt\ell' r'\!, \, \gv \hskip0.8pt\ell r} \, ,
 \quad \alpha_{\vv \hskip 0.5pt  L s} =  \beta_{\vv \hskip 0.5pt Ls} \, , \ 
 \alpha_{\ev \hskip 0.5pt  L s} =  \gamma_{\ev \hskip 0.5pt L s} \, , \  n_{\vv'}=1, \, n_{\ev'} = 2 \, , \nn \\
 \noalign{\vskip - 6pt}
 &  X_{\gv' \hskip0.8pt\ell' r' \! , \, \gv \hskip0.8pt\ell r} = \alpha_{\gv' \hskip0.8pt\ell' r'}\ \epsilon_{ \gv \hskip0.8pt\ell r}
 -  \epsilon_{\gv' \hskip0.8pt\ell' r'}\ \alpha_{ \gv \hskip0.8pt\ell r} \, .
 \end{align}
 This is then equivalent to \eqref{defX} and \eqref{varA} used in section \ref{scheme}.

\clearpage
\bibliographystyle{jhep.bst}
\bibliography{ref2.bib}

\end{document}